\documentclass[a4paper,fleqn,usenatbib]{mnras}

\usepackage{newtxtext,newtxmath}
\usepackage{mathptmx}
\usepackage{txfonts}

\usepackage[T1]{fontenc}
\usepackage{ae,aecompl}

\usepackage{graphicx}	
\usepackage{amsmath}	
\usepackage{amssymb}	

\usepackage[dvipsnames]{xcolor}
\usepackage{xspace}
\usepackage{lscape}
\usepackage{url}
\PassOptionsToPackage{hyphens}{url}\usepackage{hyperref}
\usepackage{breqn}
\usepackage{natbib}
\usepackage[english]{babel}

\newcommand{\kms}{km\,s$^{-1}$\xspace}
\newcommand{\com}{\mbox{$^{12}$CO(1-0)}\xspace}
\newcommand{\coi}{\mbox{$^{13}$CO(1-0)}\xspace}
\newcommand{\cou}{\mbox{$^{12}$CO(2-1)}\xspace}
\newcommand{\htwo}{H$_{\rm 2}$\xspace}

\newcommand{\msun}{M$_{\odot}$\xspace}

\newcommand{\sightwo}{$\Sigma_{\rm H_2}$\xspace}
\newcommand{\sigmol}{$\Sigma_{\rm mol}$\xspace}
\newcommand{\sigsfr}{$\Sigma_{\rm SFR}$\xspace}

\newcommand{\xco}{$X_{\rm CO}$\xspace}
\newcommand{\aco}{$\alpha_{\rm CO}$\xspace}
\newcommand{\hers}{{\it Herschel}\xspace}
\newcommand{\spit}{{\it Spitzer}\xspace}
\newcommand{\sigsfrunit}{M$_{\odot}$\,yr$^{-1}$\,kpc$^{-2}$\xspace}

\title[\coi in nearby spirals]{Full-disc \coi mapping across nearby galaxies of the EMPIRE survey and the CO-to-\htwo conversion factor}

\author[D. Cormier et al.]{
D. Cormier$^{1,2}$\thanks{E-mail: diane.cormier@cea.fr},
F. Bigiel$^{2}$,
M.~J. Jim{\'e}nez-Donaire$^{2}$,
A.~K. Leroy$^{3}$,
M. Gallagher$^{3}$, \newauthor
~A. Usero$^{4}$,
K. Sandstrom$^{5}$,
A. Bolatto$^{6}$,
A. Hughes$^{7,8}$,
C. Kramer$^{9}$,
M.~R. Krumholz$^{10}$, \newauthor
~D.~S. Meier$^{11}$,
E.~J. Murphy$^{12}$,
J. Pety$^{13,14}$,
E. Rosolowsky$^{15}$,
E. Schinnerer$^{16}$,  \newauthor
~A. Schruba$^{17}$,
K. Sliwa$^{16}$,
F.~Walter$^{16}$
\vspace{5pt}
\\
$^{1}$Laboratoire AIM, CEA/DSM - CNRS - Universit\'e Paris
  Diderot, Irfu/Service d'Astrophysique, CEA Saclay, F-91191
  Gif-sur-Yvette, France \\
$^{2}$Zentrum f\"ur Astronomie der Universit\"at Heidelberg,
  Institut f\"ur theoretische Astrophysik,
  Albert-Ueberle-Str. 2, D-69120 Heidelberg, Germany\\
$^{3}$Department of Astronomy, The Ohio State University,
  140 W 18th Street, Columbus, OH 43210, USA\\
$^{4}$Observatorio Astron{\'o}mico Nacional,
  Alfonso XII 3, E-28014 Madrid, Spain\\
$^{5}$Center for Astrophysics and Space Sciences,
  Department of Physics, University of California,
  San Diego, 9500 Gilman Drive, La Jolla, CA 92093, USA\\
$^{6}$Department of Astronomy and Laboratory for
  Millimeter-Wave Astronomy, University of Maryland,
  College Park, MD 20742, USA\\
$^{7}$CNRS, IRAP, 9 Avenue Colonel Roche,
  BP 44346, F-31028 Toulouse cedex 4, France \\
$^{8}$Universit{\'e} de Toulouse, UPS-OMP, IRAP,
  F-31028 Toulouse cedex 4, France\\
$^{9}$Instituto de Radioastronom{\'i}a Milim{\'e}trica,
  Av. Divina Pastora 7, N{\'u}cleo Central, E-18012 Granada, Spain\\
$^{10}$Research School of Astronomy and Astrophysics,
  Australian National University, Canberra, ACT 2611, Australia\\
$^{11}$Department of Physics, New Mexico Institute of Mining
  and Technology, 801 Leroy Place, Soccoro, NM 87801, USA\\
$^{12}$National Radio Astronomy Observatory,
  520 Edgemont Road, Charlottesville, VA 22903, USA\\
$^{13}$Institut de Radioastronomie Millim{\'e}trique,
  300 Rue de la Piscine, F-38406 Saint Martin d'H{\`e}res, France\\
$^{14}$Observatoire de Paris, 61 Avenue de l'Observatoire,
  F-75014 Paris, France\\
$^{15}$Department of Physics, University of Alberta,
  Edmonton, AB T6G 2E1, Canada\\
$^{16}$Max-Planck-Institut f{\"u}r Astronomie,
  K{\"o}nigstuhl 17, D-69117 Heidelberg, Germany\\
$^{17}$Max-Planck-Institut f{\"u}r extraterrestrische Physik,
  Giessenbachstrasse 1, D-85748 Garching, Germany
}

\date{Accepted 2018 January 4. Received 2018 January 4; in original form 2017 November 8}

\pubyear{2018}

\begin{document}
\label{firstpage}
\pagerange{\pageref{firstpage}--\pageref{lastpage}}
\maketitle

\begin{abstract}
Carbon monoxide (CO) provides crucial information about
the molecular gas properties of galaxies. While $^{12}$CO
has been targeted extensively, isotopologues such as
$^{13}$CO have the advantage of being less optically thick
and observations have recently become accessible across
full galaxy discs.
We present a comprehensive new dataset of \coi observations
with the IRAM 30-m telescope of the full discs of 9 nearby
spiral galaxies from the EMPIRE survey at a spatial resolution
of $\sim$1.5\,kpc. \coi is mapped out to $0.7-1\,r_{25}$
and detected at high signal-to-noise throughout our maps.
We analyse the \com-to-\coi ratio ($\Re$) as a function
of galactocentric radius and other parameters such as
the \cou-to-\com intensity ratio, the 70-to-160\,$\mu$m flux density
ratio, the star-formation rate surface density, the star-formation
efficiency, and the CO-to-\htwo conversion factor. We find
that $\Re$ varies by a factor of $2$ at most within and
amongst galaxies, with a median value of $11$ and
larger variations in the galaxy centres than in the discs.
We argue that optical depth effects, most likely
due to changes in the mixture of diffuse/dense gas, are
favored explanations for the observed $\Re$ variations,
while abundance changes may also be at play. 
We calculate a spatially-resolved \coi-to-\htwo conversion
factor and find an average value of
$1.0\times10^{21}$\,cm$^{-2}$\,(K\,\kms)$^{-1}$ over
our sample with a standard deviation of a factor of 2.
We find that \coi does not appear to be a good predictor
of the bulk molecular gas mass in normal galaxy discs
due to the presence of a large diffuse phase,
but it may be a better tracer of the mass than \com in the
galaxy centres where the fraction of dense gas is larger.
\end{abstract}




\section{Introduction}
\label{sect:intro}
Since stars form out of the cold, dense regions of the interstellar
medium (ISM) where conditions favor the presence of molecules,
the low-level rotational transitions of $^{12}$CO are commonly
used to study star-formation properties in galaxies
\citep[e.g.,][]{solomon-1988,fukui-2010,kennicutt-2012}.
CO has been extensively targeted because it is the most
abundant molecule after molecular hydrogen (\htwo) and cold
\htwo cannot be observed directly in emission.
CO starts to form at visual extinctions of 1-3\,mag corresponding
to column densities of $1-3 \times 10^{21}$\,cm$^{-2}$ at solar
metallicity, while hydrogen becomes mostly molecular for
column densities greater than a few $10^{20}$\,cm$^{-2}$.
Under normal metallicity and moderate radiation field
conditions, most of the cold, dense ISM is not expected
to be dark in CO, and CO and its rarer isotopologues are
expected to trace well the \htwo column density of clouds
\citep{tielens-1985,vandishoeck-1988,sternberg-2014}.

Over the past decades, the $^{12}$CO(J=1-0) emission has been
calibrated to provide a measure of the total mass of molecular
hydrogen via the CO-to-\htwo conversion factor \xco or \aco
\citep[e.g.,][]{bolatto-2013}. 
In external galaxies, the main calibration techniques employed
are based on the virial method, dust emission, optically
thin molecular tracers, or radiative transfer of multiple
molecules/transitions. These techniques often rely on making
strong assumptions regarding, e.g., abundances, grain properties,
filling factors, or the virialisation of molecular clouds.
In addition, dependencies of the \xco factor on physical conditions
within clouds, such as density, temperature, turbulence or metallicity
are expected \citep[e.g.,][]{wolfire-2010,shetty-2011,narayanan-2011}.
As a consequence, the \xco factor varies, as seen from galaxy
to galaxy and across individual galaxies in observations
\citep[e.g.,][]{arimoto-1996,downes-1998,leroy-2011,papadopoulos-2012,
sandstrom-2013,cormier-2014,kamenetzky-2014}.
In normal, star-forming disc galaxies, the amplitude of those
variations is up to an order of magnitude
\citep{bolatto-2013,sandstrom-2013}.

%
In addition to environmental dependencies of a luminosity-mass
conversion factor, the high abundance and densities of
$^{12}$CO makes the $J=(1-0)$ transition optically thick
in most molecular clouds, which complicates interpretation
and can hamper accurate determination of cloud properties.
Rarer isotopologues of the most abundant molecules containing
carbon, such as $^{13}$CO, are on the one hand less
abundant and hence more difficult to observe in galaxies.
On the other hand, they have the advantage of being more
optically thin, allowing us to access the full column density
of the material they arise from. 
Beyond our Galaxy \citep[see][for a review]{heyer-2015},
$^{13}$CO has mainly been observed in the centres or in
small, targeted regions of nearby galaxies or integrated
over entire, bright galaxies \citep[e.g.,][]{encrenaz-1979,
young-1986,casoli-1992,aalto-1995,wilson-1997,paglione-2001,
krips-2010,tan-2011,danielson-2013,alatalo-2015,vila-vilaro-2015,
sliwa-2017a,sliwa-2017b}.
With the ``EMIR Multiline Probe of the ISM Regulating Galaxy Evolution''
survey (EMPIRE; \citealt{bigiel-2016}), we have obtained
complete and high signal-to-noise maps of $^{13}$CO as well as
the main dense molecular gas tracers (HCN, HCO$^+$, HNC)
in the $J=1\rightarrow0$ transition across the discs of nine nearby
spiral galaxies with the IRAM 30-m telescope
\citep{bigiel-2016,jimenez-2017a,jimenez-2017b,gallagher-2017}.
We also obtained full maps of \com emission for those
nine galaxies in follow-up programs.
These galaxies are drawn from the HERACLES \cou survey
\citep{leroy-2009a}. They are selected to have diverse
structural properties (barred/unbarred, flocculent/grand-design
spiral arms) and to reside in different environments
(field/Virgo Cluster galaxy) in order to test whether these
parameters influence their observable ISM properties.

In this paper, we present an analysis of the \coi and \com
observations from EMPIRE. Since $^{13}$CO has not been
observed or mapped as extensively as $^{12}$CO in galaxies,
our goal is to investigate variations in the \com-to-\coi ratio
among and within those galaxies, and to understand if/under
which conditions $^{13}$CO may be a better tracer of the
molecular gas mass than $^{12}$CO. Throughout this paper,
the \com-to-\coi integrated intensity ratio is denoted $\Re$.
Section~\ref{sect:data} describes the observations. 
Section~\ref{sect:analysis} presents an analysis of $\Re$ with
radial profiles and correlation diagrams, as well as a qualitative
comparison to models and a derivation of column densities.
The physical origin of $\Re$ variations and the ability of the
CO lines to trace the molecular gas mass are discussed
in section~\ref{sect:discussion}.
Finally, we summarize our conclusions in section~\ref{sect:concl}.

\section{Observations}
\label{sect:data}

\subsection{EMPIRE observations of \coi}
\subsubsection{Data reduction}
Observations of the IRAM 30-m large program EMPIRE (PI Bigiel)
were carried out in 2012 (pilot program for NGC\,5194) and
between December 2014 and December 2016 (program 206-14,
for the other galaxies).
We mapped the full discs of 9 nearby spiral galaxies with
the EMIR E0 receiver in the on-the-fly mapping mode.
The half-power beam width (HPBW) at 110\,GHz is 22\,arcsec
and the adopted spectral resolution is 4\,\kms.

The data were reduced with our in-house pipeline.
The main steps include:
baseline subtraction with a polynomial function of order~2,
rejection of spectra above $3$ times the theoretical noise,
conversion to main beam temperature assuming main beam
and forward efficiencies of $0.78$ and $0.94$,
projection of the spectra onto grids of pixel size 4\,arcsec.
After gridding, the full width half-maximum (FWHM)
of the \coi data is 27\,arcsec. This corresponds to a linear
resolution of $\sim$1.5\,kpc for our sample of galaxies.
We refer to \cite{jimenez-2017b,jimenez-2017c} for a
detailed description of the data acquisition and reduction.
Line calibrators were observed during each run of the
campaign and their intensities vary by about 5\,per cent only.
Table~\ref{table:list} provides the list of our targets and noise
levels achieved at the frequency of the \coi line (110.20\,GHz).

\subsubsection{Moment maps}
\label{sect:mapmaking}
We used the \cou data from HERACLES \citep{leroy-2009a}
as a guide to create integrated intensity maps for the \coi line.
The data were retrieved from the HERACLES repository\footnote{
\protect\url{http://www.iram-institute.org/EN/content-page-242-7-158-240-242-0.html}},
convolved to a common resolution of 27\,arcsec using Gaussian
kernels, and put on the same spatial grid
as the EMPIRE data using the IDL procedure \texttt{hastrom}.
At each position in the map, we fitted the \cou line with
a single Gaussian. For pixels below a signal-to-noise ratio
of 5 for the velocity-integrated intensity, we interpolated
central velocities and line widths from well-detected
neighboring pixels by fitting a plane to the maps.
In regions where most pixels are not detected in \cou,
the interpolated central velocities are not allowed to take values
lower (higher) than the minimum (maximum) velocity measured
in the well-detected pixels, and the interpolated line widths
are set to the average line width measured in the well-detected
pixels.

The central velocities and line widths of the \cou line were
used as initial guesses for those of the \coi line. We created
intensity maps for \coi both by fitting and integrating a single
Gaussian and by integrating directly the signal in specific
velocity windows. We defined the windows as
$\sim3\times$\,FWHM of the \cou line, thus the windows
vary for each line of sight. The two methods yield
differences in integrated intensities that are typically less
than 7\,per cent. Since the CO line profiles are not always Gaussian,
especially in the galaxy centres, we prefer to use the direct
integration maps and not the line-fitted maps.
We also produced error maps. For each pixel, the error
on the integrated intensity is calculated as the standard
deviation in the line-free parts of each spectrum, multiplied
by the square root of the number of (4\,\kms wide) channels
inside the FWHM given by the Gaussian fit. 
Figure~\ref{fig:maps} in the Appendix shows final integrated
intensity maps for each galaxy.

\begin{table*}
  \caption{General properties of the EMPIRE galaxies.} 
\begin{center}
\begin{tabular}{llccccccccc}
    \hline\hline
     \vspace{-8pt}\\
    \multicolumn{1}{l}{Name} & 
    \multicolumn{1}{l}{Type} & 
    \multicolumn{1}{c}{RA} & 
    \multicolumn{1}{c}{Dec} & 
    \multicolumn{1}{c}{Dist} &
    \multicolumn{1}{c}{$D_{25}$} &
    \multicolumn{1}{c}{$i$} & 
    \multicolumn{1}{c}{PA} & 
    \multicolumn{1}{c}{$\sigma_{\rm rms}$} &
    \multicolumn{1}{c}{$I_{13}$} &
    \multicolumn{1}{c}{Map size}  \\
    \multicolumn{1}{l}{} & 
    \multicolumn{1}{l}{} & 
    \multicolumn{1}{c}{(J2000)} & 
    \multicolumn{1}{c}{(J2000)} & 
    \multicolumn{1}{c}{(Mpc)} &
    \multicolumn{1}{c}{(arcmin)} & 
    \multicolumn{1}{c}{(deg)} & 
    \multicolumn{1}{c}{(deg)} & 
    \multicolumn{1}{c}{(mK)} & 
    \multicolumn{1}{c}{(K\,\kms)} & 
    \multicolumn{1}{c}{(arcmin$^2$)}  \\
    \hline
{NGC\,0628}	& SAc			& 01:36:41.8	& 15:47:01	& 9.6		& 9.8		& 7		& 20	 	& 4.8 		& $0.25\pm0.01$	& $4.5\times4.5$ \\
{NGC\,2903}	& SABbc HII		& 09:32:10.1	& 21:30:04	& 8.9		& 11.8	& 65		& 204	& 3.5 		& $1.27\pm0.01$	& $2.4\times4.1$ \\
{NGC\,3184}	& SABcd group		& 10:18:17.0	& 41:25:28	& 11.8	& 7.4		& 16		& 179	& 3.8 		& $0.36\pm0.01$	& $3.5\times3.5$ \\
{NGC\,3627}	& SABb liner AGN	& 11:20:15.0	& 12:59:30	& 9.4		& 10.3	& 62		& 173	& 4.4 		& $1.24\pm0.01$	& $2.9\times5.2$ \\
{NGC\,4254}	& SAc HII		& 12:18:49.6	& 14:24:59	& 14.4	& 5.0		& 32		& 55	 	& 3.0 		& $0.87\pm0.01$	& $3.4\times3.2$ \\
{NGC\,4321}	& SABbc AGN	& 12:22:54.9	& 15:49:21	& 14.3	& 6.0		& 30		& 153	& 3.6 		& $0.91\pm0.01$	& $4.3\times2.9$ \\
{NGC\,5055}	& SAbc liner AGN	& 13:15:49.2	& 42:01:45	& 7.9		& 11.9	& 59		& 102	& 4.6 		& $1.02\pm0.01$	& $6.4\times3.5$ \\
{NGC\,5194}	& SAbc Seyfert 2	& 13:29:52.7	& 47:11:43	& 7.6		& 7.7		& 20		& 172 	& 4.5		 	& $1.07\pm0.01$	& $6.8\times6.8$ \\
{NGC\,6946}	& SABcd HII		& 20:34:52.2	& 60:09:14	& 6.8		& 11.4	& 33		& 243	& 4.6 		& $0.88\pm0.01$	& $6.3\times5.3$ \\
    \hline \hline
\end{tabular}
\end{center}
    \vspace{-8pt}
\begin{minipage}{16cm}
Notes. See \cite{jimenez-2017a} for references on the galaxy parameters.
The distance to NGC\,0628 has been updated to the value from \cite{kreckel-2017}.
Morphological types are from the NASA Extragalactic Database.
Column~9: rms noise in the \coi data, calculated as the median
noise level at the rest frequency of 110.2\,GHz,
for a spectral resolution of 4\,\kms and spatial resolution of 27\,arcsec.
Column~10: average \coi integrated intensity and its statistical uncertainty
measured by stacking all spectra across the entire map size given in Column~11.\\
\end{minipage}
  \label{table:list}
\end{table*}

\subsection{Ancillary data}
\subsubsection{Reference $^{12}$CO data}
In order to make homogeneous and matched-quality
measurements of $\Re$, we performed new observations
of 8 galaxies of the EMPIRE survey (all except NGC\,5194)
in the \com line, with the IRAM 30-m telescope.
Maps of the entire discs were obtained as part of the
programs {061-15, 059-16} (PI Jim{\'e}nez-Donaire)
and {D15-12} (PI Cormier).
The data were reduced with the same pipeline as for
EMPIRE and we produced final cubes at a spatial resolution
of $\simeq$25\,arcsec and a spectral resolution of 4\,\kms.
We reached sensitivities of 17-30\,mK ($T_{\rm mb}$)
per 4\,\kms channel.
For NGC\,5194, the \com and \coi data are taken from
the PAWS survey \citep[30-m cubes\footnote{\protect\url{
http://www.mpia.de/PAWS/PAWS/Data.html}};][]{schinnerer-2013,pety-2013}.

All datasets are convolved to a common resolution of 27\,arcsec
using Gaussian kernels and put on the same spatial grid as
the EMPIRE data. Integrated intensity maps are created
following the same steps as for the \coi data, described in
section~\ref{sect:mapmaking}.

\subsubsection{Ultraviolet and infrared photometry}
All of the EMPIRE targets were observed with GALEX
as well as with the MIPS instrument onboard \spit as part
of the programs LVL \citep{dale-2009} and SINGS
\citep{kennicutt-2003}. 
\hers photometry exists for all of our targets except NGC\,2903.
We use user-provided products from the key programs
VNGS and KINGFISH (data release 3) \citep{bendo-2012,kennicutt-2011}.
No PACS 100\,$\mu$m observations are available for NGC\,5194.
The reduction and map-making were done in HIPE
versions 9 and 8 and Scanamorphos versions 21 and 16.9,
respectively. Convolutions are done using the kernels
from \cite{aniano-2011}.

\subsection{Deriving physical quantities}

\subsubsection{\sigsfr, \sigmol, $N(\rm{H_2})$}
From the photometry and spectroscopy, we derive
physical quantities such as star-formation rate surface
densities (\sigsfr), molecular gas surface densities (\sigmol),
and \htwo column densities ($N(\rm{H_2})$).
All surface densities are corrected for inclination.

Our SFR estimates are based on IR data.
For all galaxies overlapping with KINGFISH, we use
the TIR maps calculated with dust models from
\cite{galametz-2013}. 
For NGC\,5194 and NGC\,2903, we compute TIR
surface brightness using the generic calibration
from \cite{galametz-2013}. We combine the bands
MIPS 24\,$\mu$m, PACS 70, 160\,$\mu$m and
SPIRE 250\,$\mu$m for NGC\,5194 and the bands
MIPS 24 and 70\,$\mu$m for NGC\,2903.
For the galaxies in KINGFISH, the comparison of
TIR maps calculated from SED models and from
the generic calibration yield differences of about 10\,per cent
for the first combination (MIPS, PACS and SPIRE)
and 20\,per cent for the second combination (MIPS only).
The generic calibration tends to systematically 
overpredict slightly the TIR flux in the brightest regions.
Uncertainty maps are generated by adding in quadrature
errors on the fluxes and errors on the calibration
coefficients and from the choice of method as quoted
above in the case of the TIR calibration. The TIR maps
are then converted to star-formation rate surface
density maps using the calibration from \cite{murphy-2011}.

Molecular gas mass surface densities and \htwo column
densities are commonly derived using the \com line as a proxy
for \htwo and a CO-to-\htwo conversion factor. For \sigmol,
the standard conversion factor is \aco of 4.4~\msun\,pc$^{-2}$\,(K\,\kms)$^{-1}$
which includes helium \citep{bolatto-2013}.
For $N(\rm{H_2})$, the standard conversion factor is
\xco of $2\times10^{20}$~cm$^{-2}$\,(K\,\kms)$^{-1}$ which
does not account for helium.
These are considered as reference, Milky-Way values.
Variations of the conversion factors are discussed in
section~\ref{sect:xcodisc}.

\subsubsection{Isotope abundance ratio and $^{12}$CO abundance}
\label{sect:abund}
The $^{12}$C/$^{13}$C isotope abundance ratio results from
stellar processing and evolution \citep[e.g.,][]{wilson-1994}.
The $^{12}$CO/$^{13}$CO abundance ratio further depends
on chemical processes within molecular clouds.
While we discuss possible sources of variations in those
abundances in section~\ref{sect:discussion}, in our analysis we
make no difference between the isotope ($^{12}$C/$^{13}$C)
abundance ratio and the isotopologue ($^{12}$CO/$^{13}$CO)
abundance ratio.
The abundance ratio is largely unconstrained for external galaxies
but has been observed to vary within galaxies and from galaxy
to galaxy. In the Milky Way, it increases from $25$ in the centre
to $100$ in the outer disc, with a value of $\sim$70 in the solar
neighborhood \citep{langer-1990,wilson-1994,milam-2005}.
In the Large Magellanic Cloud, it is $\sim$50 \citep{wang-2009}.
In starburst galaxies, it is found to be $>$40, and in some ULIRGs
it is even $>100$ \citep[e.g.,][]{martin-2010,henkel-2014,sliwa-2017b}.
Here, we adopt a fiducial value for the isotope abundance
ratio of $60$. In the Milky Way, this value corresponds to
a distance of $\simeq 6.7$\,kpc or $0.6\,r_{25}$.

The $^{12}$CO/\htwo abundance ratio is also sensitive to
radiative processes and gas chemistry. It can vary by orders
of magnitudes from translucent to dense lines-of-sight, with
a typical scatter of 0.5\,dex at any given \htwo column density
\citep{van-dishoeck-1992,sheffer-2008}.
In our large telescope beam, such different lines
of sights are mixed and beam-to-beam variations of
the abundance are expected to be smaller. In the following,
we take the canonical value of $10^{-4}$ for the
$^{12}$CO/\htwo abundance ratio, which corresponds
to a $^{13}$CO/\htwo abundance ratio of
$1.7\times10^{-6}$ \citep[e.g.,][]{dickman-1978} and could
be uncertain by a factor of a few.

\begin{figure*}
\centering
\includegraphics[clip,width=3.4cm]{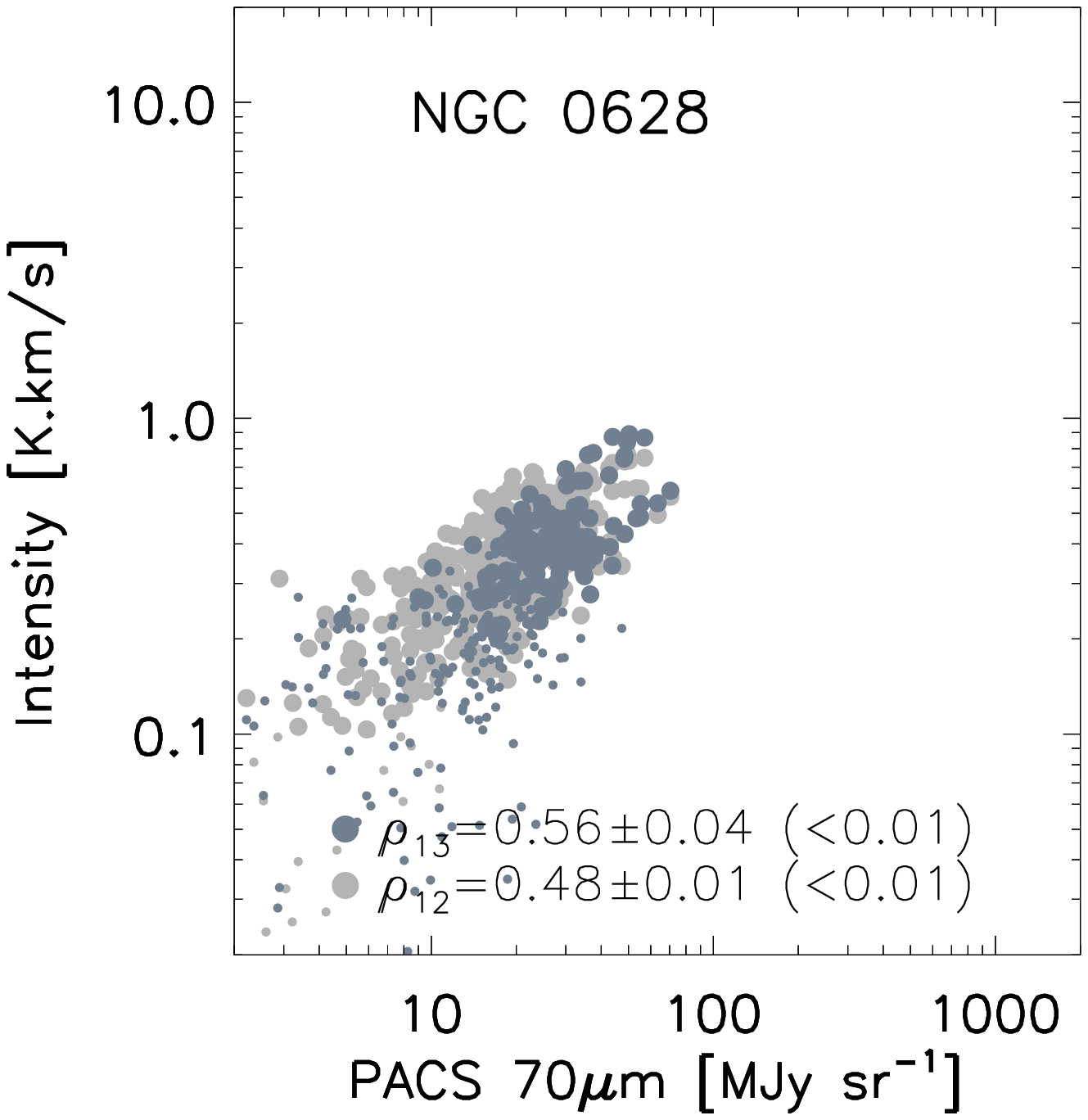}
\includegraphics[clip,width=3.4cm]{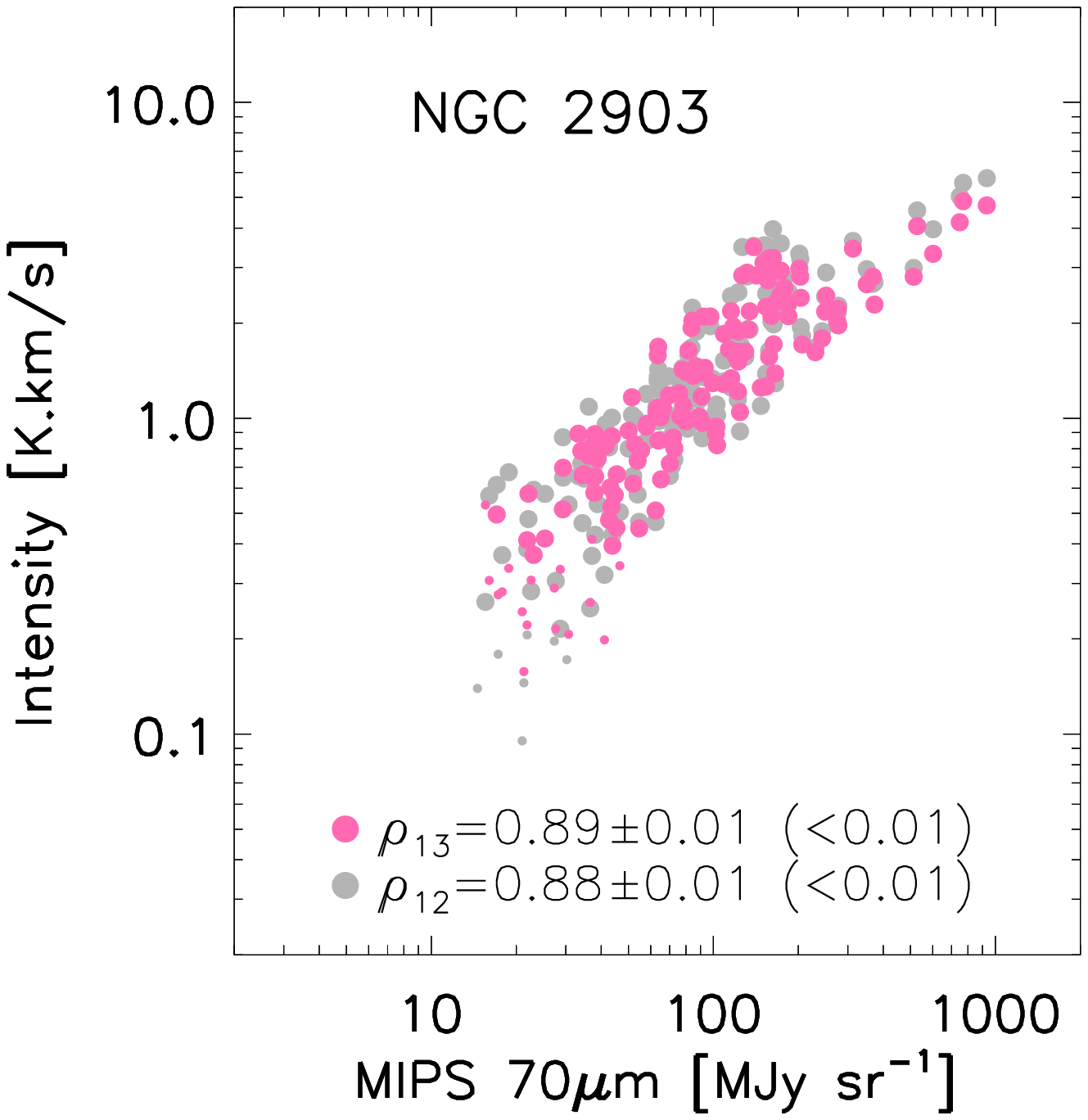}
\includegraphics[clip,width=3.4cm]{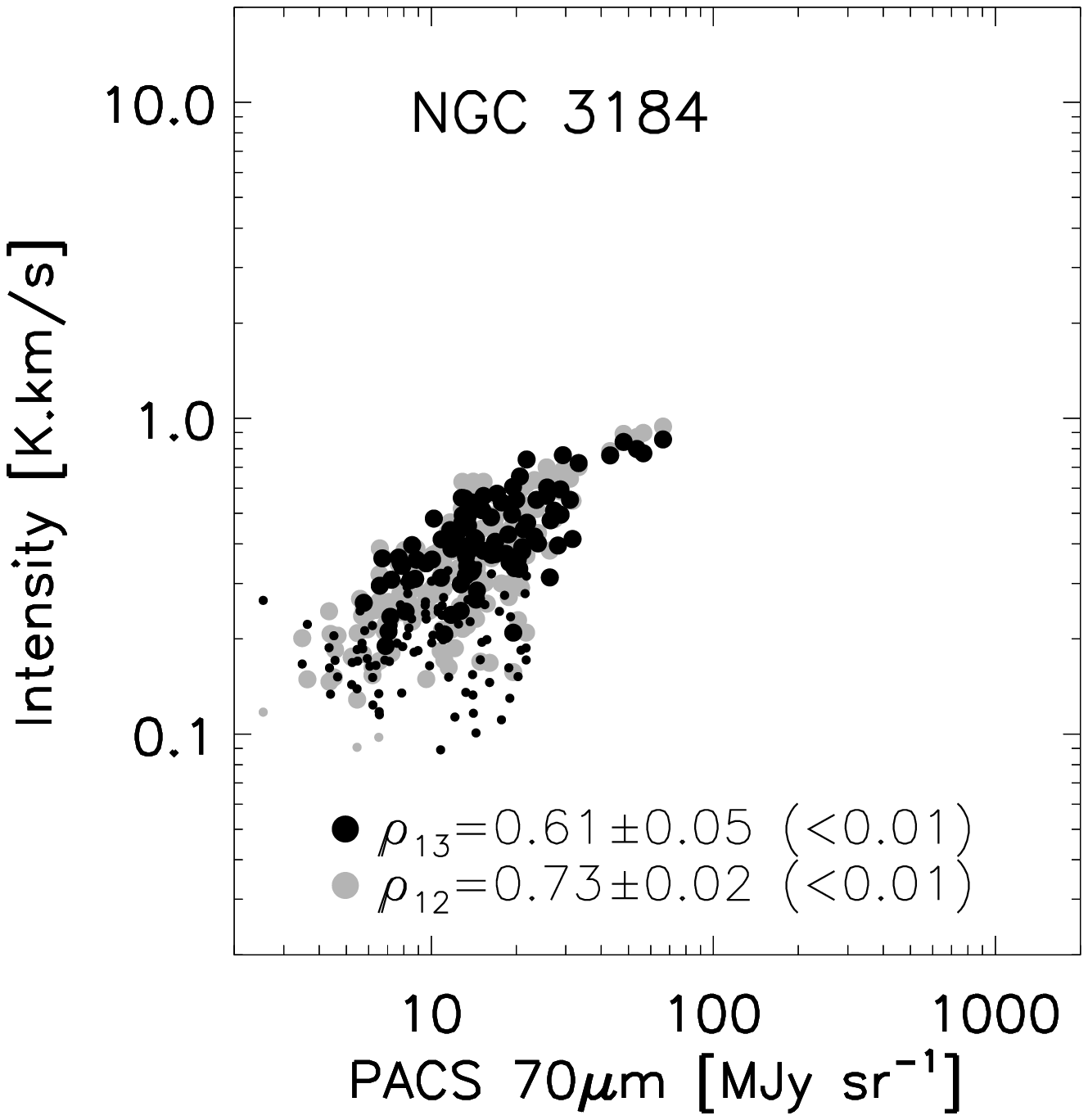}
\includegraphics[clip,width=3.4cm]{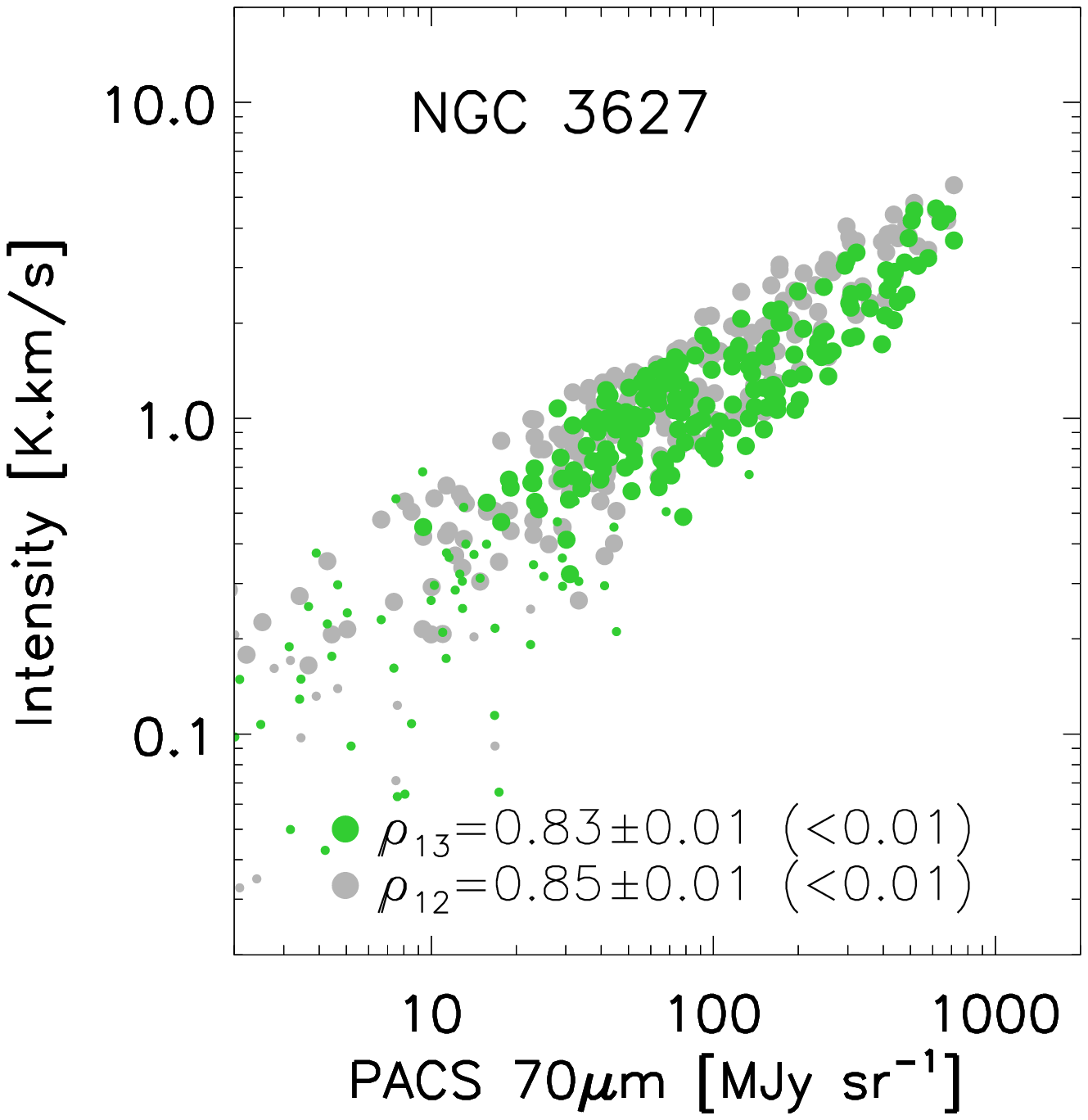}
\includegraphics[clip,width=3.4cm]{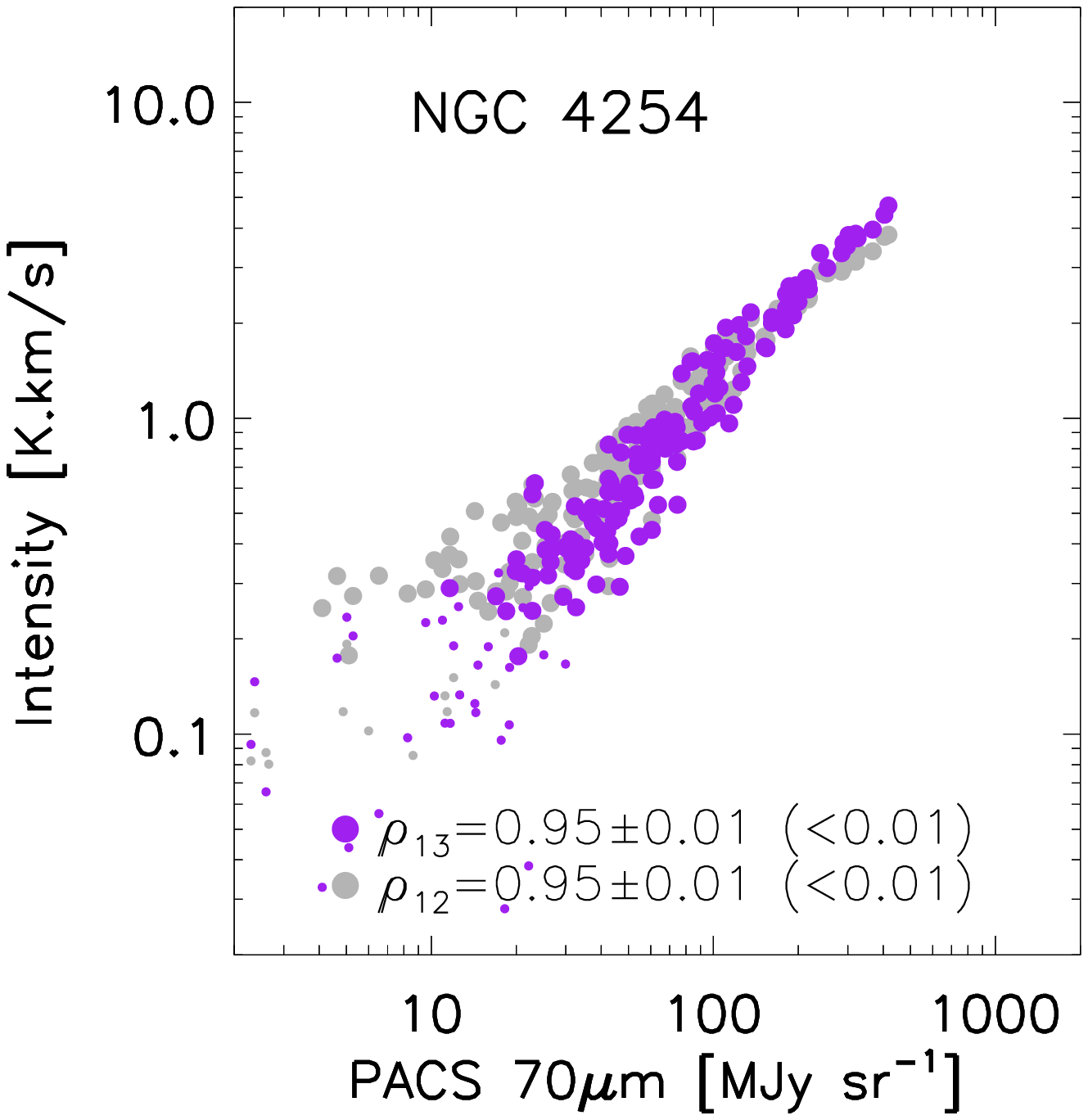} \\
\includegraphics[clip,width=3.4cm]{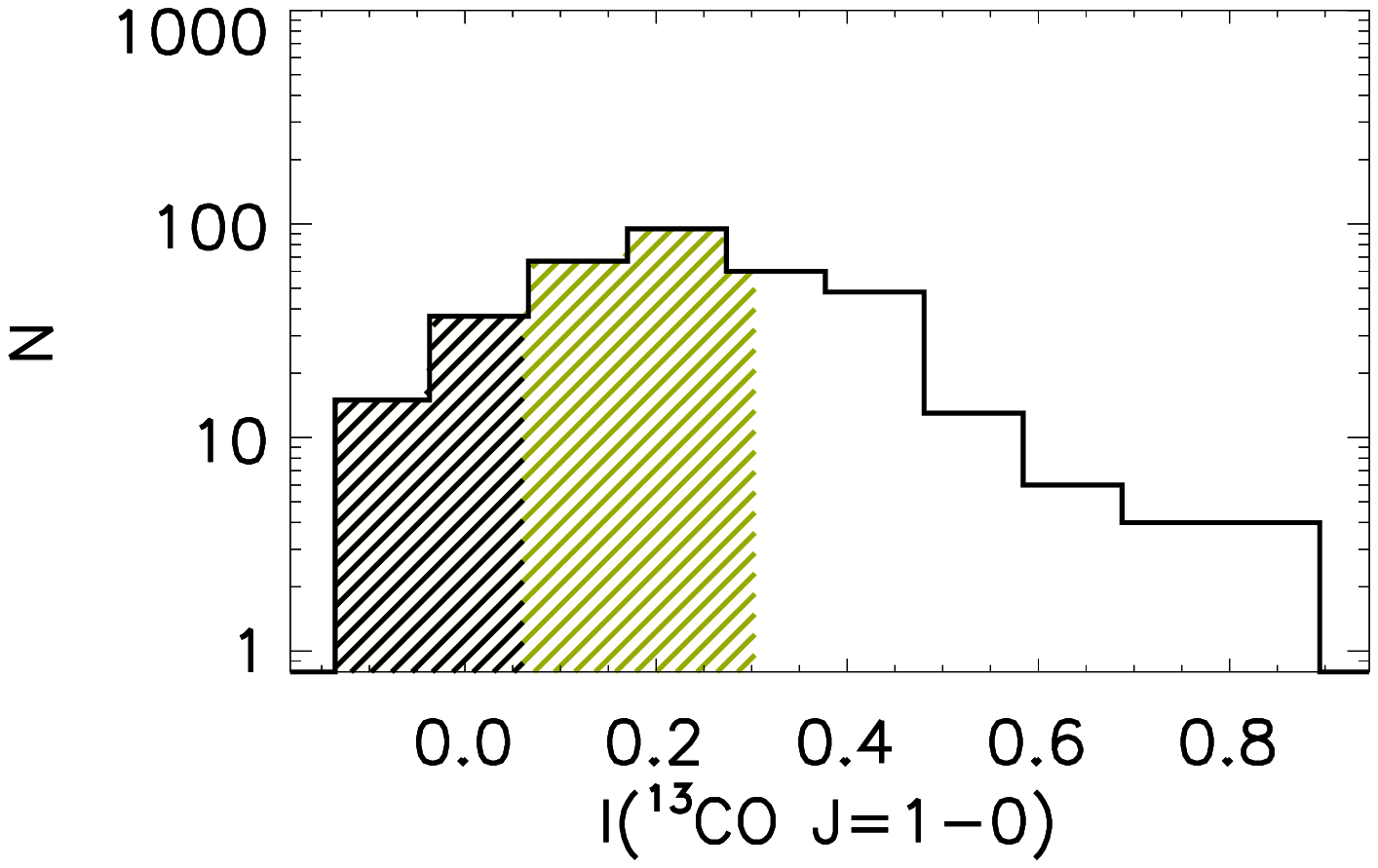}
\includegraphics[clip,width=3.4cm]{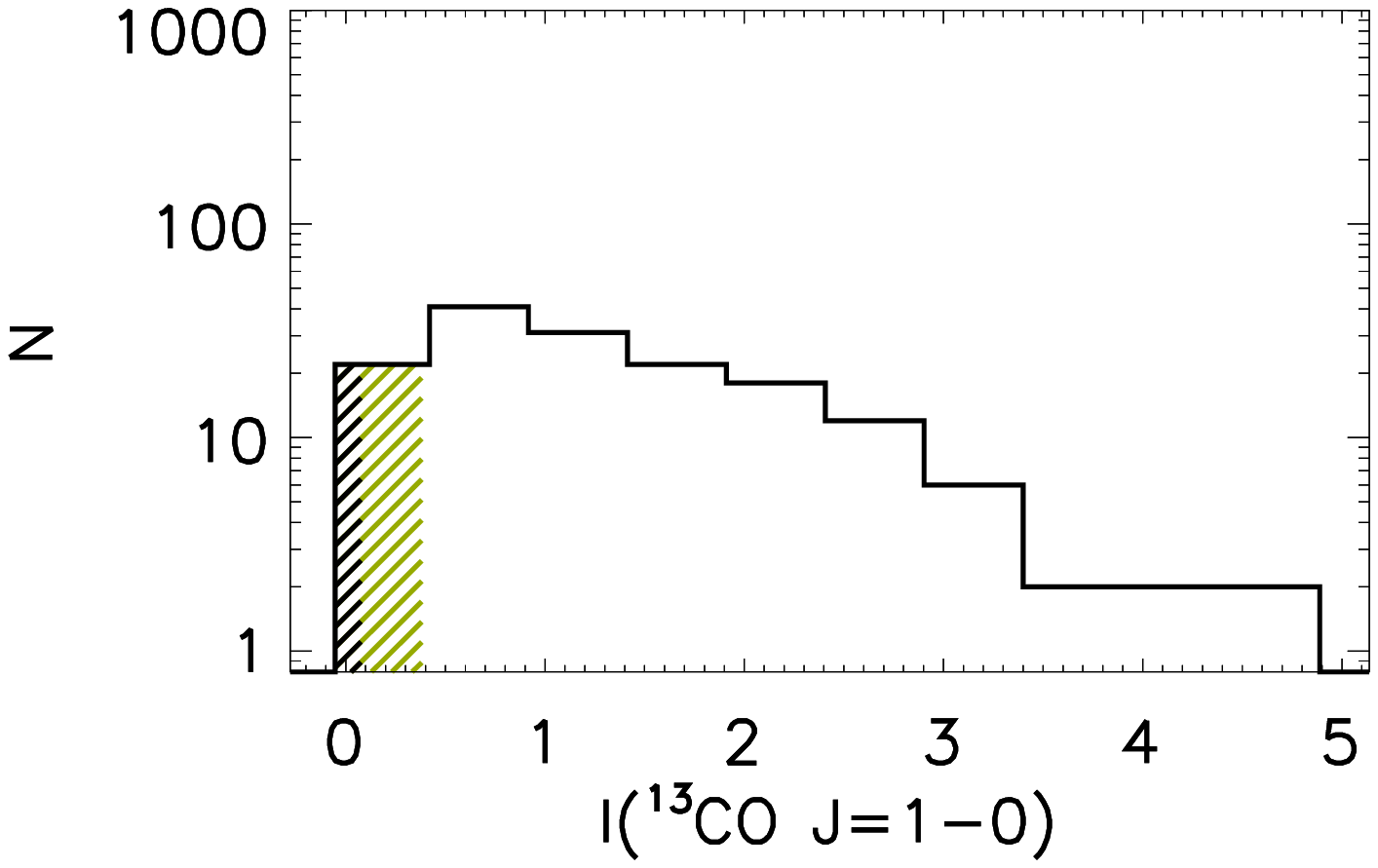}
\includegraphics[clip,width=3.4cm]{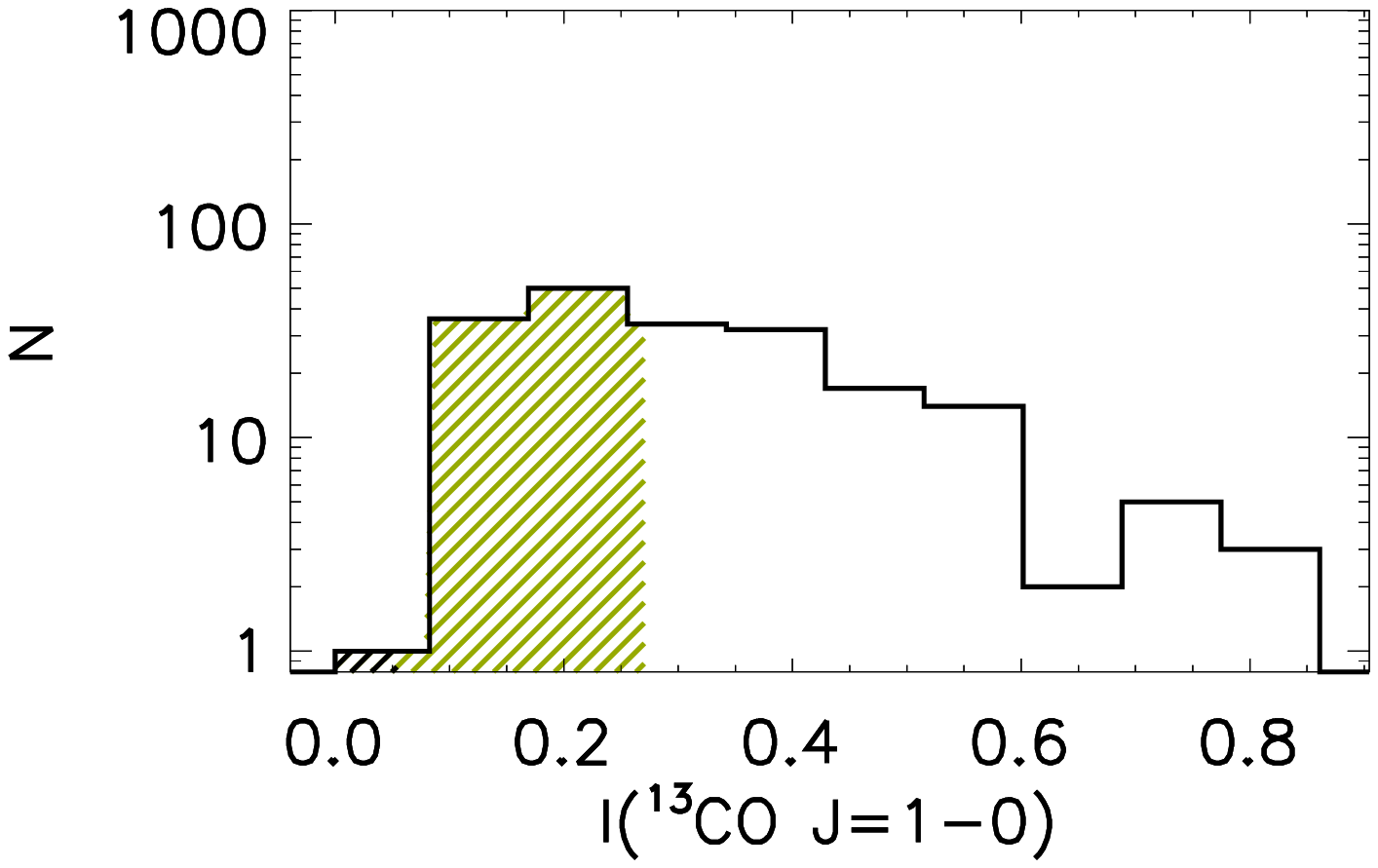}
\includegraphics[clip,width=3.4cm]{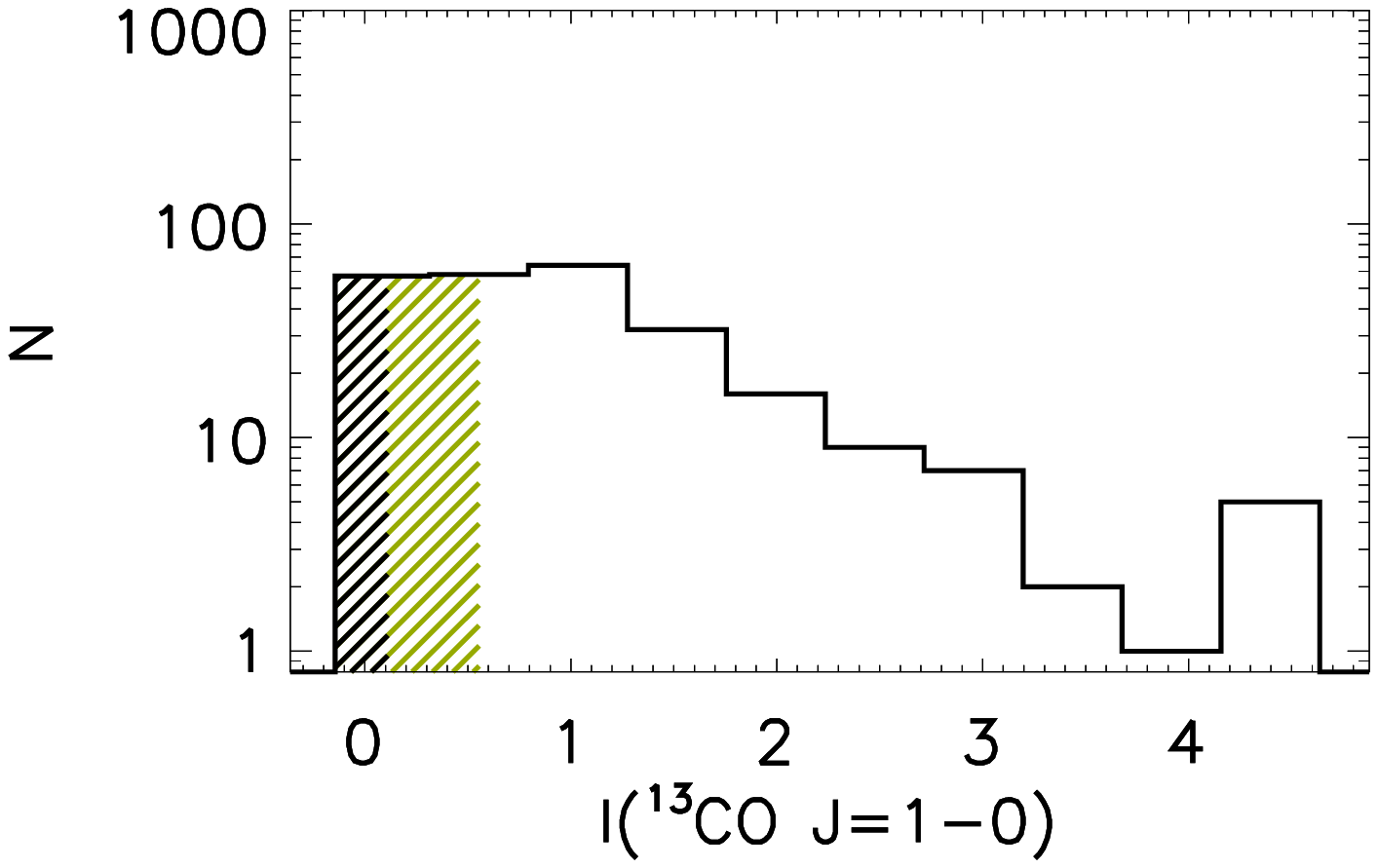}
\includegraphics[clip,width=3.4cm]{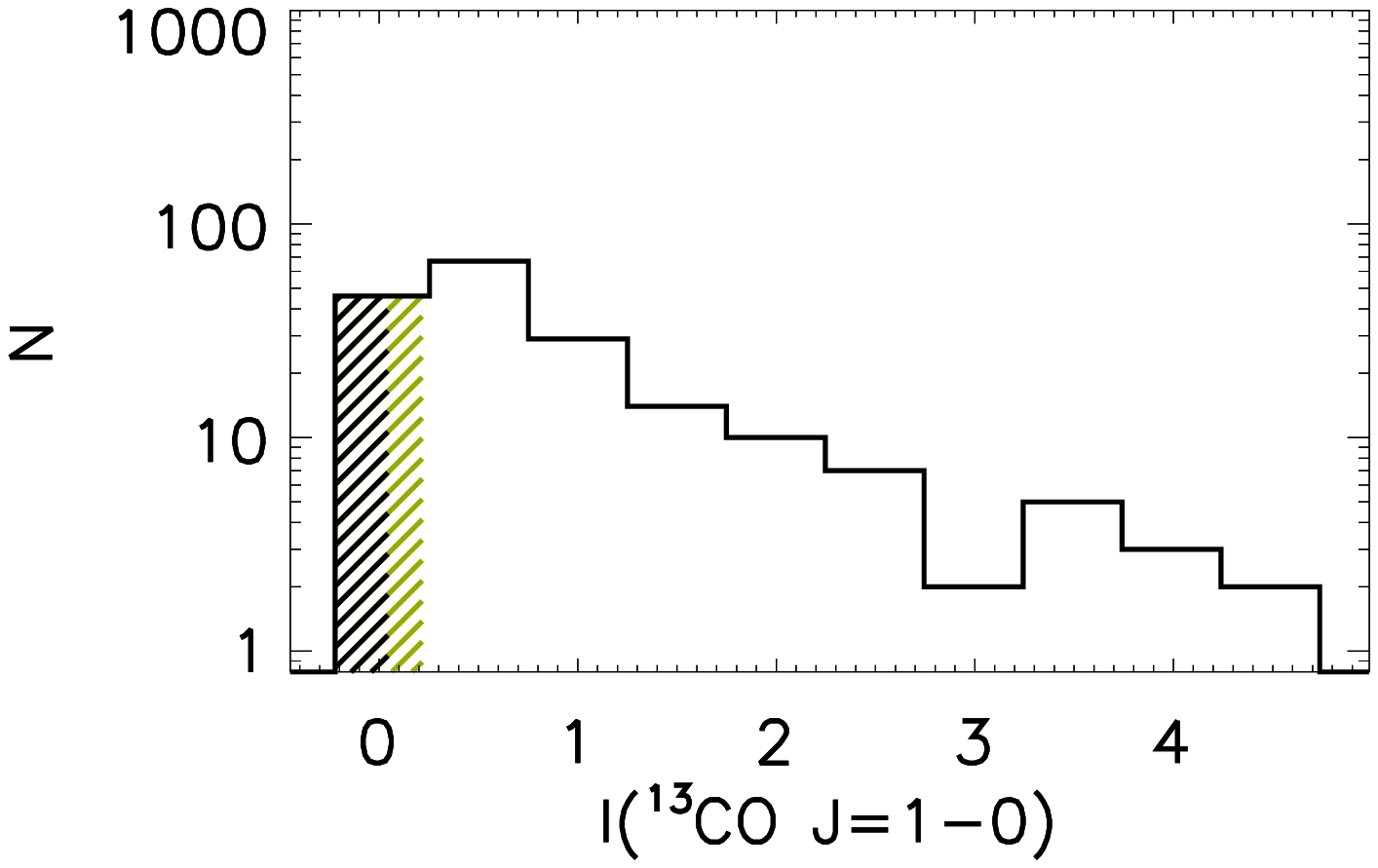} \\
\includegraphics[clip,width=3.4cm]{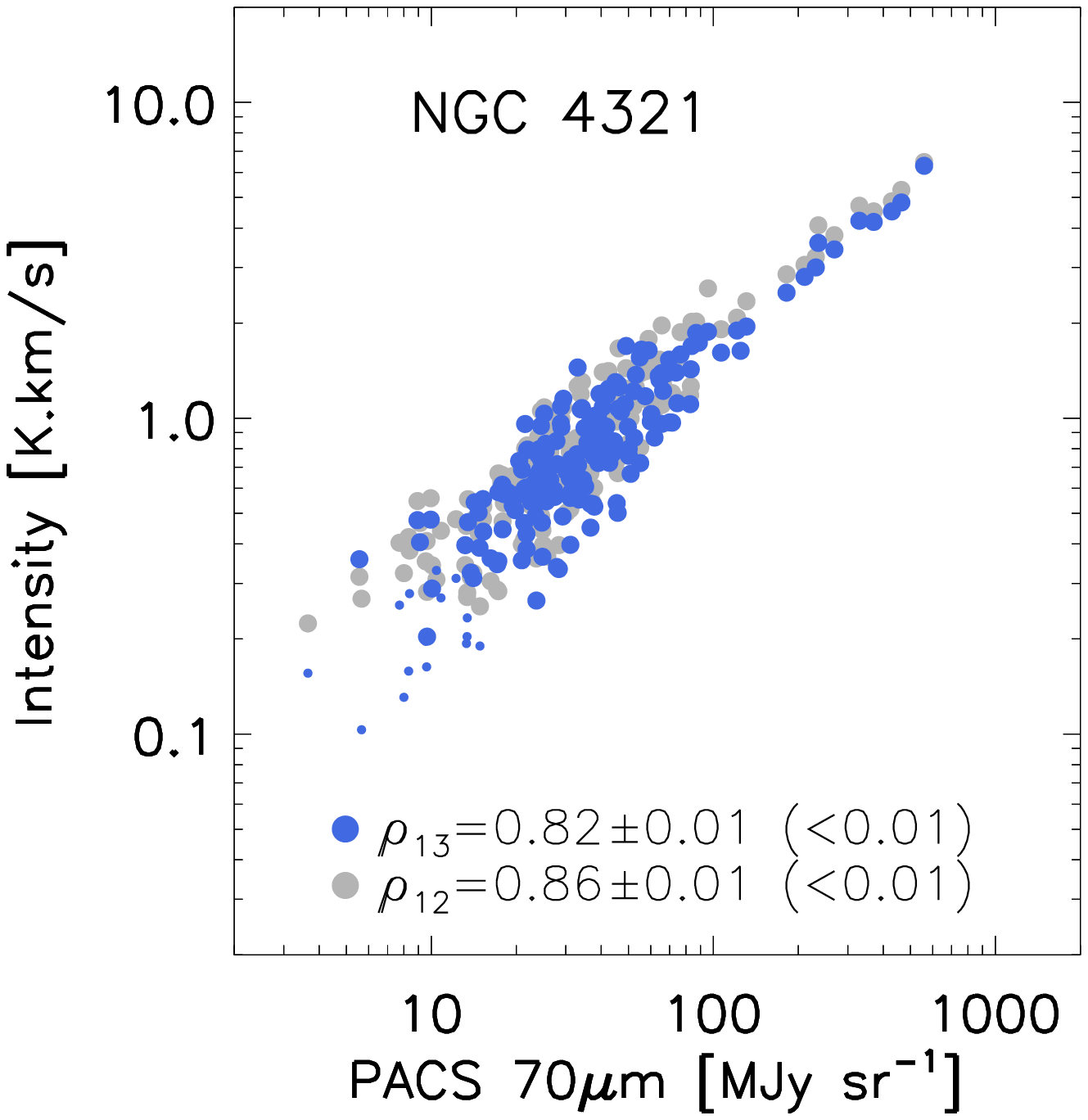}
\includegraphics[clip,width=3.4cm]{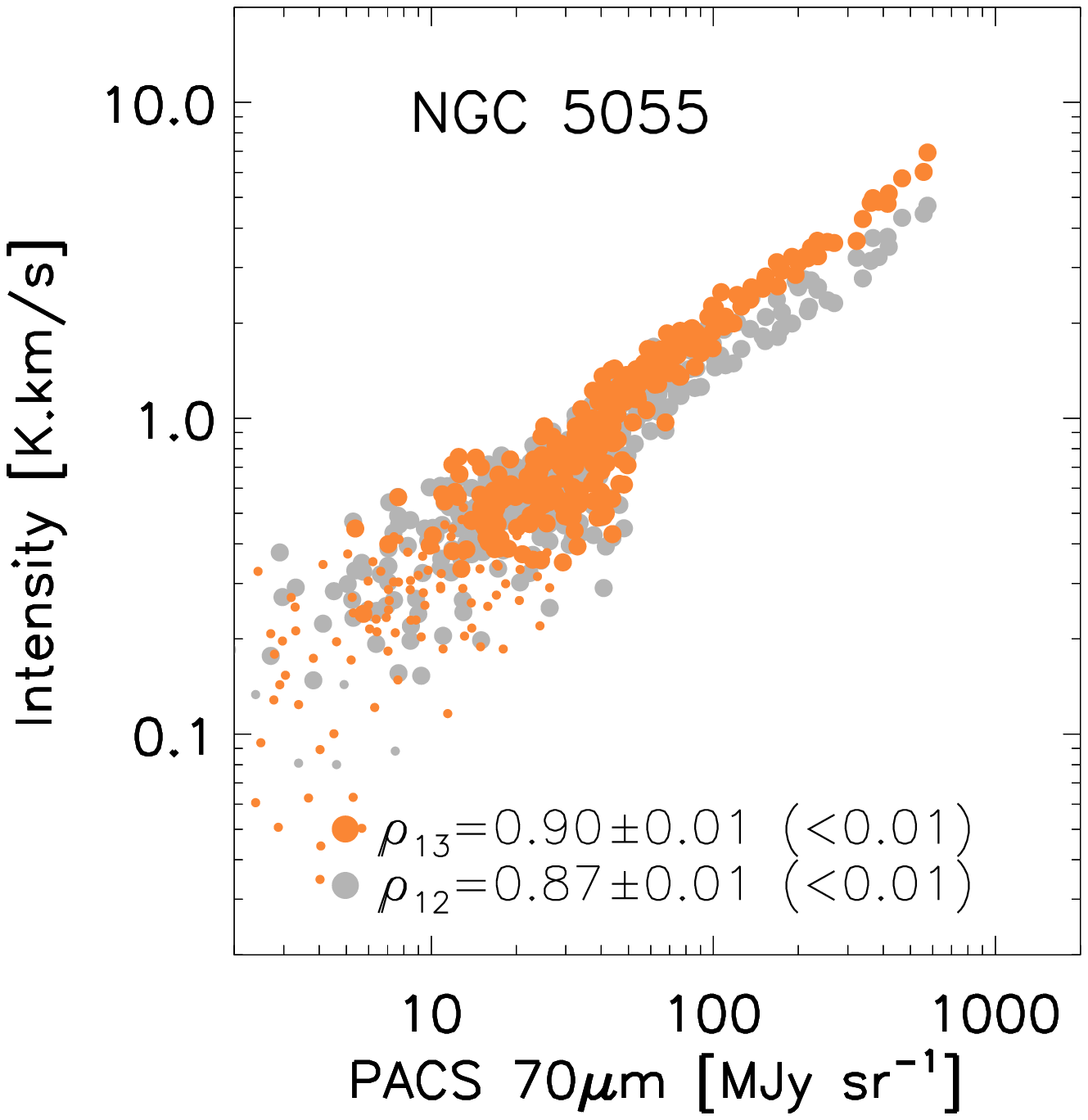}
\includegraphics[clip,width=3.4cm]{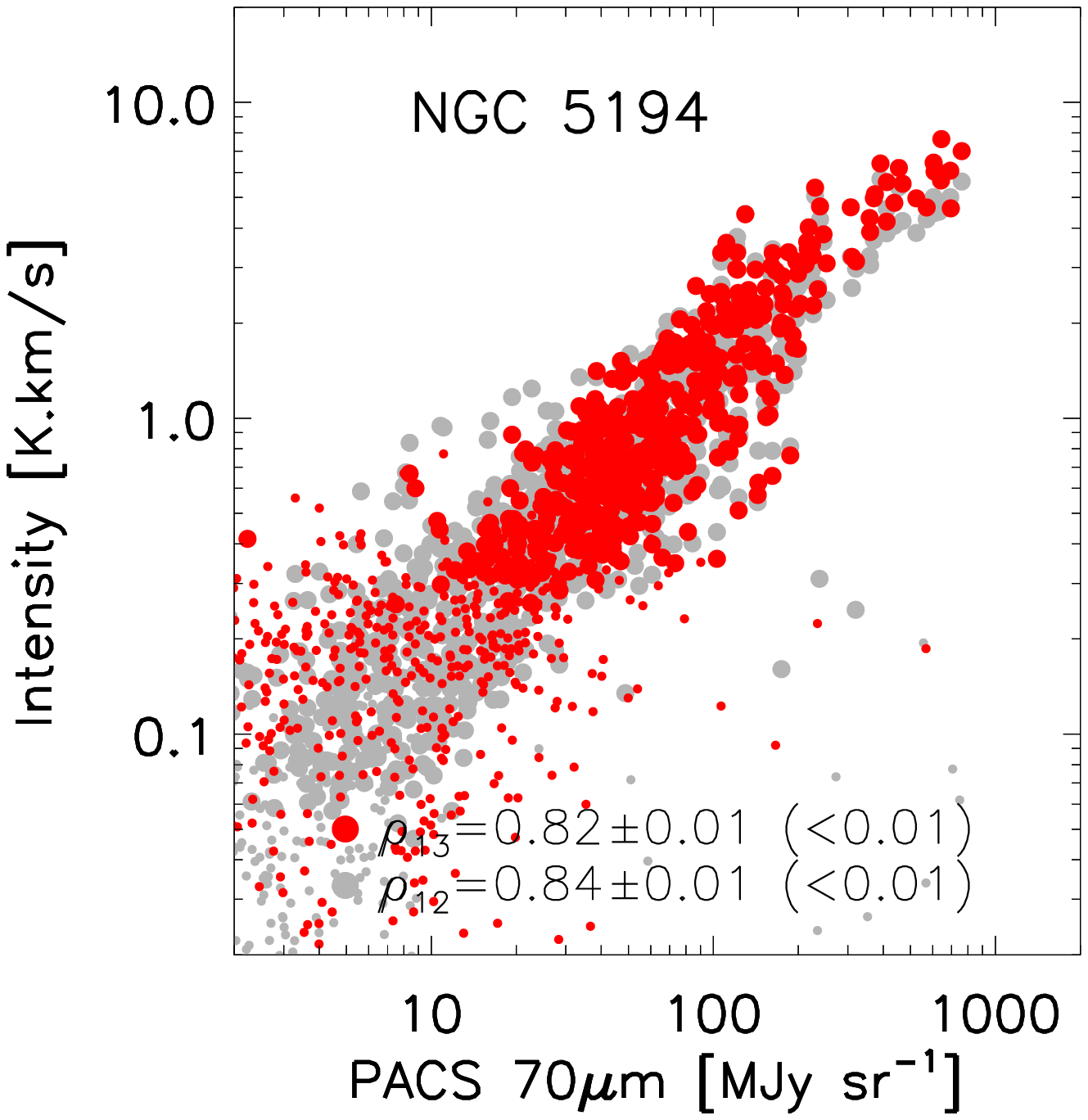}
\includegraphics[clip,width=3.4cm]{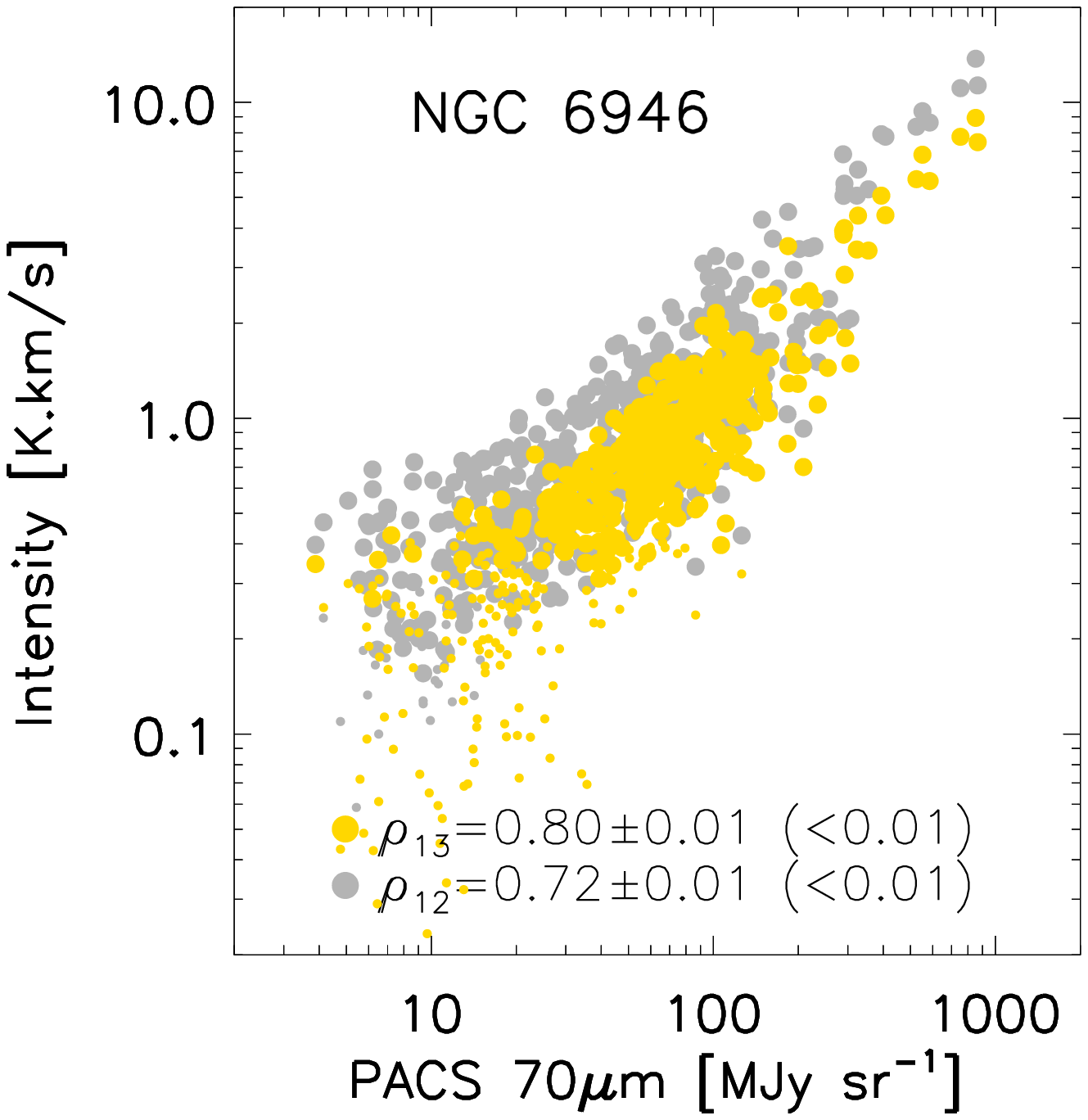}\\
\includegraphics[clip,width=3.4cm]{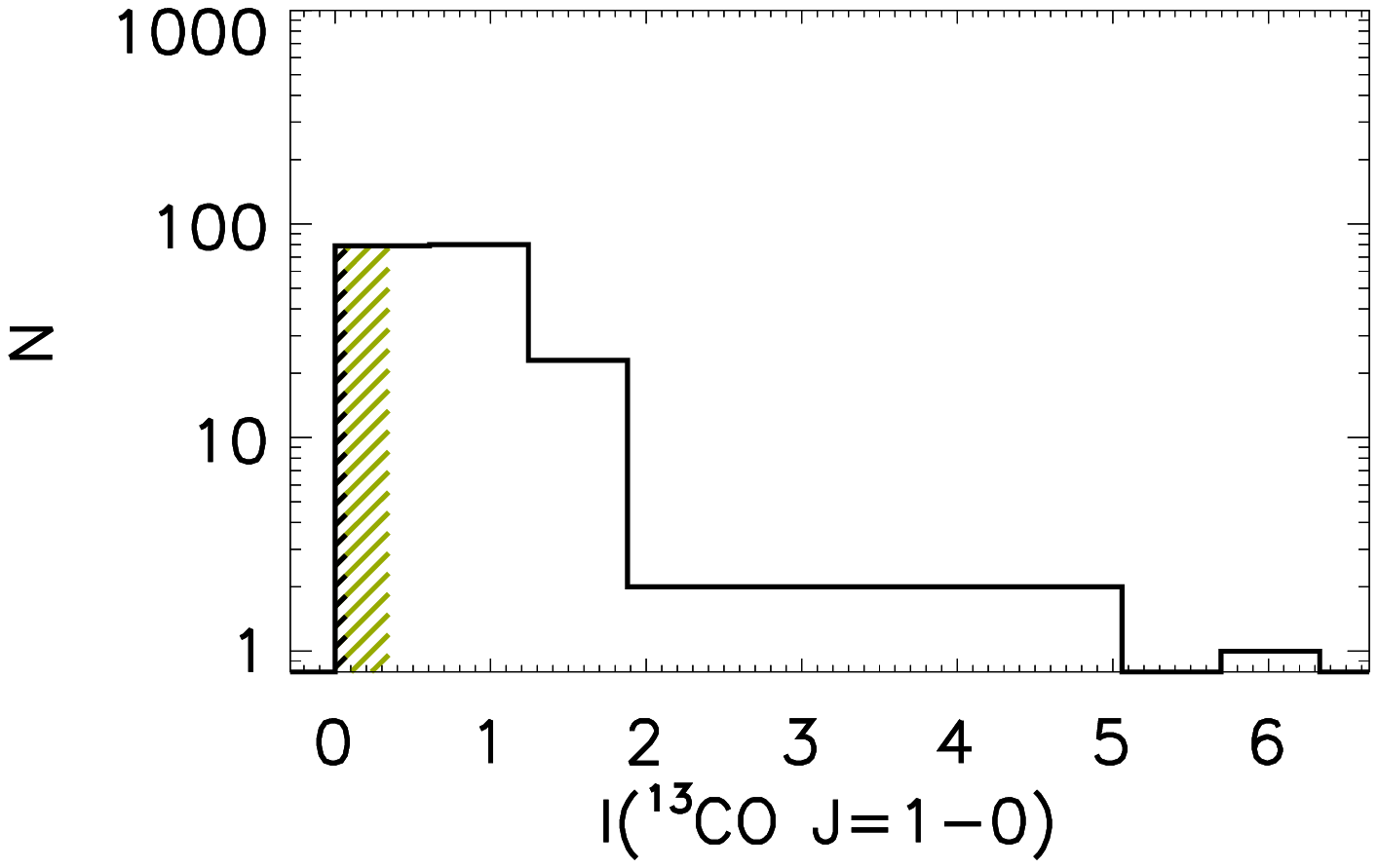}
\includegraphics[clip,width=3.4cm]{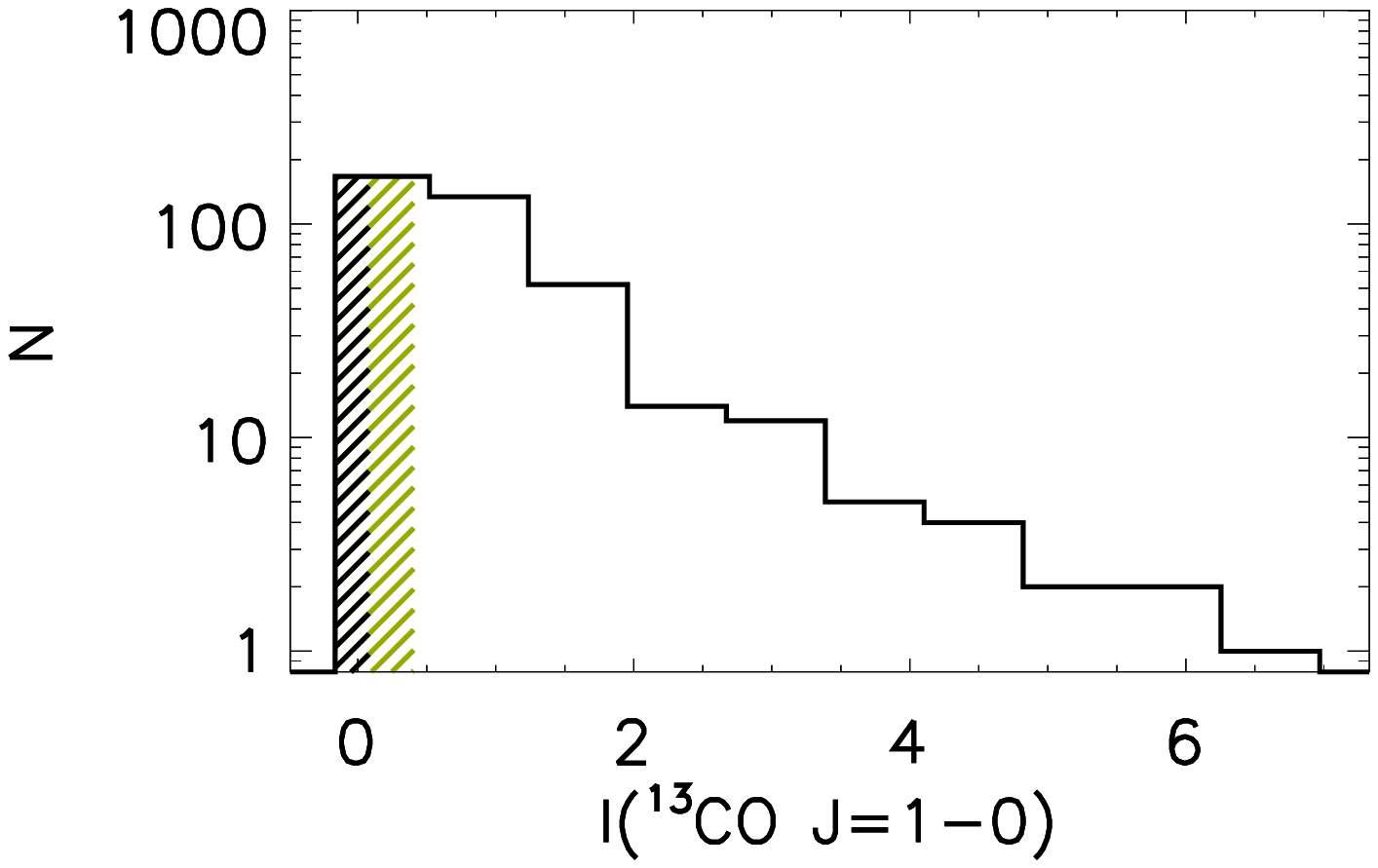}
\includegraphics[clip,width=3.4cm]{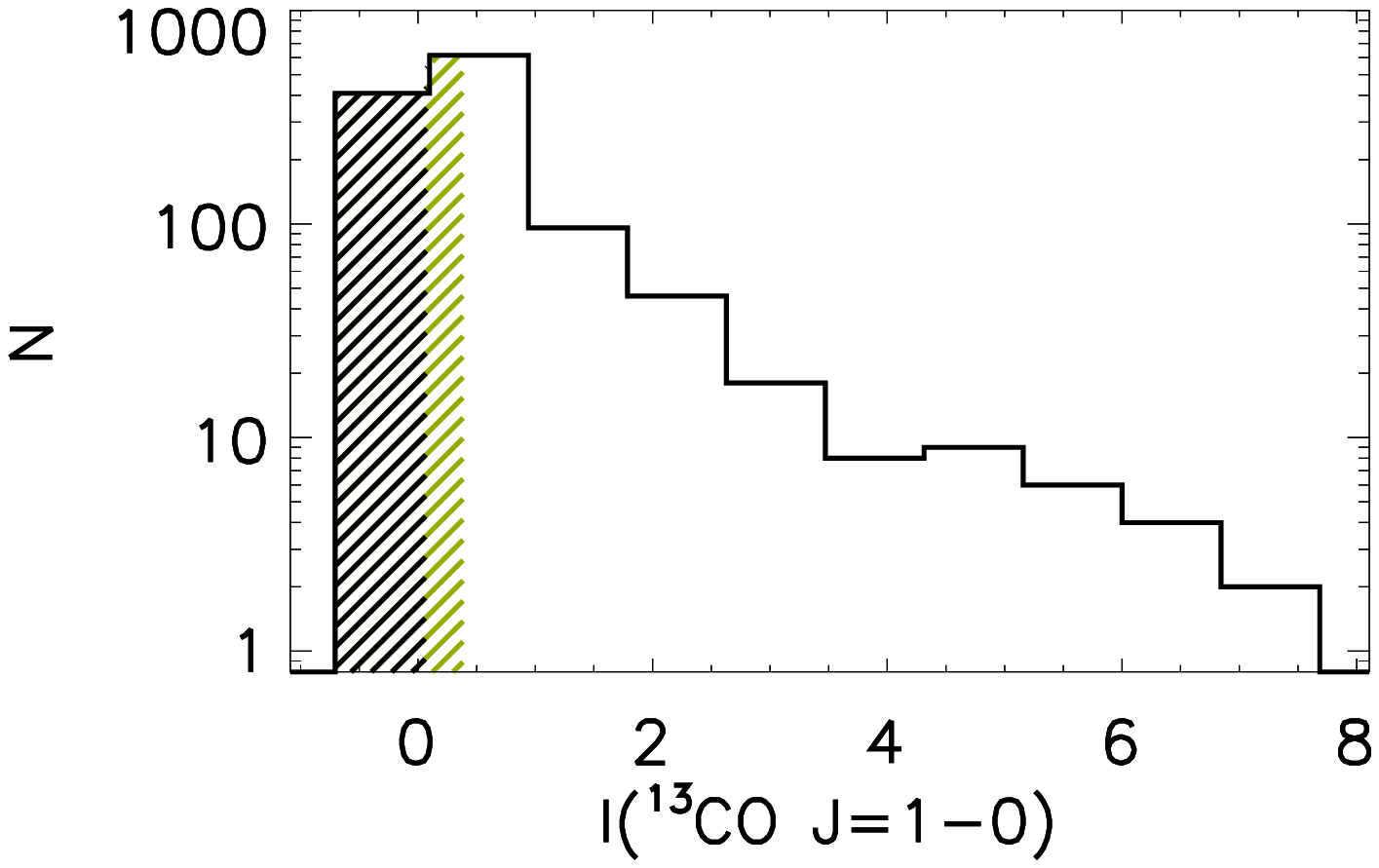}
\includegraphics[clip,width=3.4cm]{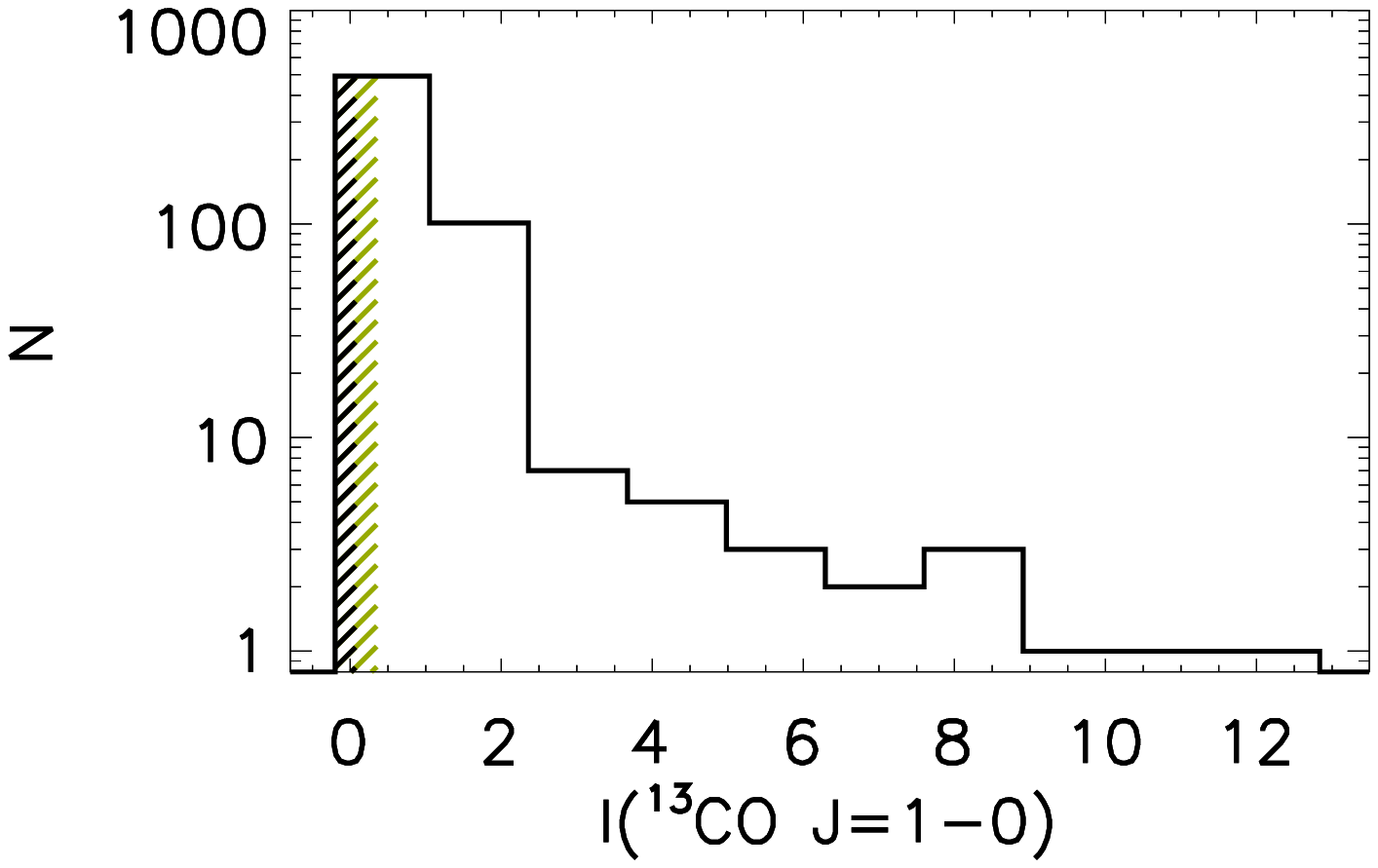}
\vspace{-8pt}
\caption{
\underline{Top panels:} 
EMPIRE observations of the CO line intensity (in K\,\kms)
on the y-axis following well the broadband 70\,$\mu$m intensity
(in MJy\,sr$^{-1}$) which is a proxy for the star-formation rate
surface density on the x-axis.
\coi data are in colour and \com data, scaled by a factor of 10, are in grey.
For both CO lines, smaller circles correspond to pixels below
a signal-to-noise ratio of 5 in the respective maps. 
The Spearman's rank correlation coefficients between
each CO line intensity and the 70\,$\mu$m intensity,
their uncertainty, and their significance (in parenthesis),
are indicated. Those are measured on data with
signal-to-noise ratio above 5.
\underline{Bottom panels:} 
Histogram of the \coi intensities.
Only few pixels are below the 1$\sigma$ (5$\sigma$)
noise level as indicated by the hashed black (beige)
parts of the histograms.
}
\label{fig:histo}
\end{figure*}

\section{Analysis}
\label{sect:analysis}

\subsection{\coi intensities across galaxies}
Figure~\ref{fig:histo} shows the \coi and scaled \com intensities
as a function of the PACS 70\,$\mu$m intensity, which can be
seen as a proxy of the star-formation rate surface density,
as well as histograms of the \coi intensities for each galaxy.
\coi is detected throughout our maps with signal-to-noise ratio
$>10$ in the brightest regions, and the majority of
the pixels in our maps are detected at a $>5\sigma$ level.
\coi peaks on the galaxy centres, except in NGC\,3627
where it peaks on the star-forming knots at the end of
the bar (though this may be somewhat an artifact of resolution).
In the inter-arm/outer-disc regions, the \coi intensity has
values $<$1\,K\,\kms ($<$0.3\,K\,\kms for the two faintest
galaxies NGC\,0628 and NGC\,3184).

We quantify correlations with the Spearman's rank
correlation coefficient and its significance using the IDL
procedure \texttt{r\_correlate.pro}.
The correlation coefficient is computed for data points
with signal-to-noise ratio of the \coi intensity greater than 5.
With this condition, the signal-to-noise of the ancillary
data sets described above is generally not a concern.
The significance corresponds to the p-value or probability
of null hypothesis. It is reported in parenthesis in the figures.
The uncertainty ($\pm$) on the coefficient is estimated
with a Monte-Carlo simulation.
We find that \coi and \com are generally well correlated
with the dust emission from PACS 70\,$\mu$m. Both CO
lines show very similar distributions and scatter, with
departure from each other in some galaxy centres
(e.g., NGC\,5055).
At the resolution of our data, their line profiles and
line widths are also similar (see Fig.~\ref{fig:specs}).

Given the high quality of the data, we perform a line-of-sight
based analysis. Only in the following section~\ref{sect:envir},
we perform a stacking analysis to extract representative
spectra in the different environments of our
sample of galaxies.

\subsection{$\Re$ in the different environments of spirals}
\label{sect:envir}
$\Re$ is the ratio of the \com intensity and the \coi intensity,
which have units of K~\kms.
We measure $\Re$ in the galaxy centres, arm, inter-arm
regions and total emission in our maps by means of stacking.
In the stacking step, spectra corresponding to a given region
are aligned in velocity using the \cou data as reference
and averaged. \cite{jimenez-2017a} describes the method
in detail and Figure~\ref{fig:specs} shows stacked
spectra of \coi and \com for the galaxy centres and entire
galaxies. CO intensities are then measured by direct integration
of the stacked spectra over a velocity window large enough
to encompass all the signal.
Table~\ref{table:reval} reports the $\Re$ values obtained this way
for each galaxy as well as averages and dispersions
over the sample.
Figure~\ref{fig:maps} in the Appendix shows maps of $\Re$
with contours delineating the centre and arm regions.
The regions were defined using cuts in intensity of the
\com line (at our working resolution of 27\,arcsec or $\simeq$1.5\,kpc),
and the cut levels were chosen by eye.
'Centre' refers to the galaxy centres (distance to the centre
of the galaxy $<$16\,arcsec or inner 0.8\,kpc, typically) where
the \com emission is brightest (first cut at CO
intensity levels of: 5.5, 35, 7, 30, 32, 22, 30, 32, 40~K\,\kms,
for NGC\,0628, 2903, 3184, 3627, 4254, 4321, 5055, 5194,
6946, respectively).
'Arm' refers to the galaxy discs where the \com emission is
bright (second cut at CO intensity levels of: 3, 18, 3.5,
11, 11, 8, 10, 10, 10~K\,\kms, for NGC\,0628, 2903, 3184,
3627, 4254, 4321, 5055, 5194, 6946, respectively).
'Inter-arm' refers to the inter-arm and outer parts of the maps
where the \com emission is fainter but detected at a signal-to-noise
ratio $>3$. Finally, 'total' refers to the entire map where \com
is detected.
We note that we also defined contours by hand, identifying
centres and arm regions based on 24\,$\mu$m and 70\,$\mu$m
continuum images. Values of $\Re$ obtained by stacking
with these hand contours vary by at most 10\,per cent compared
to values reported in Table~\ref{table:reval}. This 10\,per cent
discrepancy can be seen as a methodology uncertainty.

Thanks to our sensitive, full maps, we can achieve much
better galaxy-integrated and environment-specific measurements
than previous investigations of $\Re$ in nearby galaxies.
In all galaxies and all environments, the noise
of the data indicates that we could have measured ratios
up to $150$ in the stacks (but the observed ratios are much lower).
The mean global (i.e. full-galaxy) $\Re$ value that we measure
over our sample is $11$ with a standard deviation of $1.5$.
When the galaxies are divided into different environments
(centre, arm, inter-arm regions), we obtain similar average
$\Re$ values, with standard deviation $\sim$2. 
We notice that the dispersion in $\Re$ is highest in the
galaxy centres, with values varying between 7 and 15
(see Table~\ref{table:reval}).
The dispersion is also high in the inter-arm regions, but
with larger error bars on individual measurements.
Studying, also at a kpc-scale, centres of $\sim$10 nearby
galaxies with AGN activity, signs of a recent merger, or an
intense central starburst, \cite{israel-2009a,israel-2009b}
found values of $\Re$ in the range $8-16$.
In the EMPIRE survey, we find that the galaxies with
bright, starburst-dominated nuclei (NGC\,2903, NGC\,3627,
NGC\,4321, and NGC\,6946) have the largest central
$\Re$ values. Two galaxies in our sample have
strong bars (NGC\,2903 and NGC\,3627) and three
other galaxies have weaker bars (NGC\,3184, NGC\,4321,
NGC\,6946, and possibly NGC\,5194). Those barred
galaxies also show generally higher central $\Re$ values
than the non-barred galaxies. Both properties of
having a bar and a starburst-dominated nucleus might
be related as bars may help to funnel gas towards the
galaxy center -- leading to high dense gas fractions
\citep{gallagher-2017} -- and to fuel star formation \citep[e.g.,][]{ho-1997}.
Finally, two galaxies with AGN activity have low central
$\Re$ values (NGC\,5055, NGC\,5194) but the trend is
not systematic (e.g., NGC\,3627 also has AGN activity).

\begin{table}
  \caption{$\Re$ values in different regions of spirals.} 
\begin{center}
\begin{tabular}{lcccc}
    \hline\hline
     \vspace{-8pt}\\
    \multicolumn{1}{l}{Name} & 
    \multicolumn{1}{c}{$\Re_{\rm total}$} & 
    \multicolumn{1}{c}{$\Re_{\rm centre}$} & 
    \multicolumn{1}{c}{$\Re_{\rm arm}$} & 
    \multicolumn{1}{c}{$\Re_{\rm inter-arm}$} \\
    \hline
	{NGC\,0628}			& $13.2\pm0.6$	& $9.9\pm0.4$		& $13.5\pm0.7$	& $14.4\pm1.6$ \\
	{NGC\,2903}			& $10.3\pm0.1$	& $12.3\pm0.3$	& $11.0\pm0.1$		& $9.4\pm0.2$ \\
	{NGC\,3184}			& $10.2\pm0.3$	& $11.2\pm0.7$		& $10.7\pm0.4$	& $9.3\pm0.5$ \\
	{NGC\,3627}			& $12.0\pm0.1$	& $15.2\pm0.5$	& $12.2\pm0.1$	& $10.8\pm0.4$ \\
	{NGC\,4254}			& $11.0\pm0.2$		& $8.4\pm0.1$		& $10.5\pm0.1$	& $13.3\pm0.5$ \\
	{NGC\,4321}			& $10.3\pm0.2$	& $10.9\pm0.1$	& $10.3\pm0.2$	& $9.8\pm0.3$ \\
	{NGC\,5055}			& $9.3\pm0.2$		& $7.2\pm0.1$		& $8.5\pm0.2$		& $10.9\pm0.4$ \\
	{NGC\,5194}			& $10.3\pm0.3$	& $8.2\pm0.1$		& $9.9\pm0.1$		& $12.8\pm0.9$ \\
	{NGC\,6946}			& $13.8\pm0.2$	& $15.2\pm0.1$	& $13.5\pm0.2$	& $13.2\pm0.4$ \\ \hline
	{average all}			& $11.2~(1.5)$		& $10.9~(2.9)$		& $11.1~(1.7)$		& $11.5~(1.9)$ \\
	{average nuc.$^{(a)}$}	& $11.6~(1.7)$		& $13.4~(2.2)$		& $11.8~(1.4)$		& $10.8~(1.7)$ \\
    \hline \hline
\end{tabular}
\end{center}
    \vspace{-8pt}
\begin{minipage}{8.5cm}
Notes.
The $inter-arm$ region refers to the entire map disregarding
the centre and the arms, i.e. ${[total - (centre + arms)]}$.
Errors on the stacked values of $\Re$ correspond to statistical errors.
For averages, we indicate the standard deviation of $\Re$ in parenthesis.
$(a)$~Only galaxies with nuclear starbursts (NGC\,2903,
NGC\,3627, NGC\,4321, and NGC\,6946).
\end{minipage}
    \vspace{8pt}
  \label{table:reval}
\end{table}

\begin{figure*}
\centering
\includegraphics[clip, trim=2mm 0 2mm 2mm,width=8.8cm]{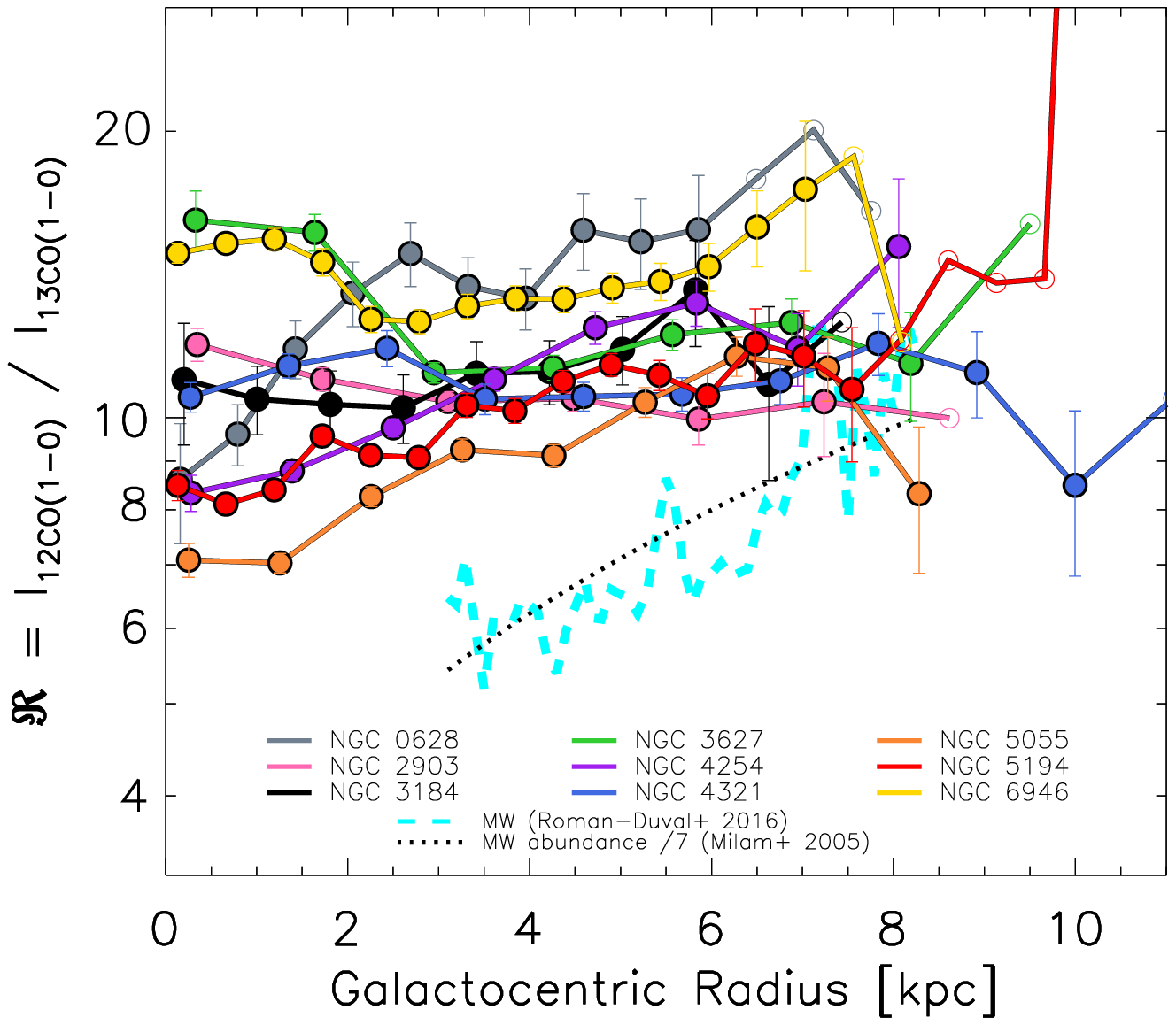}
\includegraphics[clip, trim=4mm 0 0 2mm,width=8.8cm]{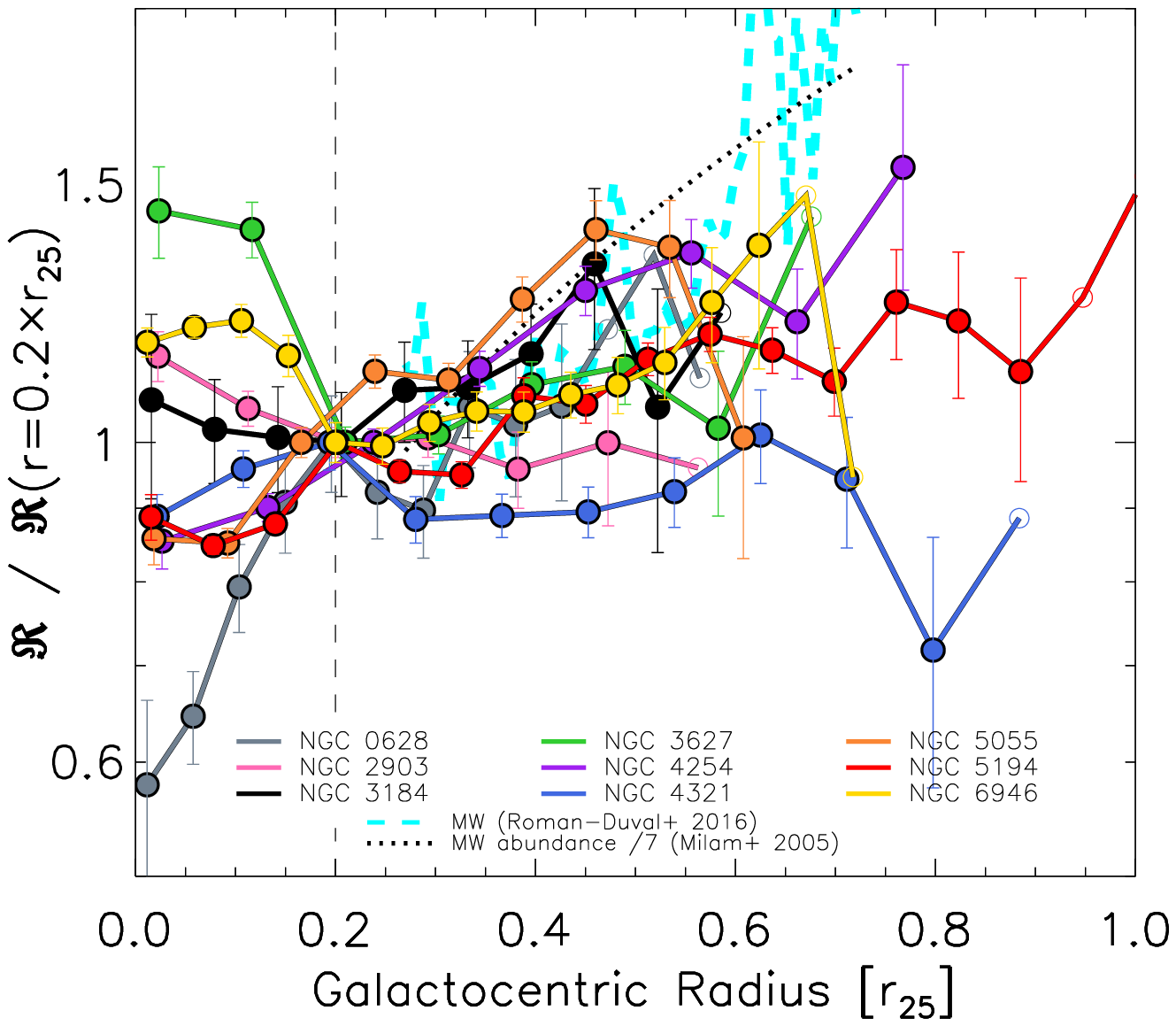}
\vspace{-5pt}
\caption{
\underline{Left panel:}
Profiles of the \com-to-\coi intensity ratio, $\Re$, as a function of
galactocentric radius for each EMPIRE galaxy.
Profile points for which the average \coi or \com intensity within
the ring is below 5$\sigma$ are shown with open circles.
We overlay for reference the Milky Way measurements
of the $\Re$ profile (dashed cyan curve) from
\protect\cite{roman-duval-2016} and of the $^{12}$C/$^{13}$C
abundance gradient (black dotted line, divided by
a factor of 7 for display) from \protect\cite{milam-2005}.
\underline{Right panel:} same as the left panel with radii
normalized to $r_{25}$ (x-axis) and the profiles
normalized to their value at a radius of $r=0.2\,r_{25}$.
We assume $r_{25} = 11.5$\,kpc for the Milky Way
\protect\citep{devaucouleurs-1978}.
}
\label{fig:ratioprofiles}
\end{figure*}

\subsection{Radial Profiles}
\label{sect:profiles}

\subsubsection{Method}
We generate radial profiles for the main tracers discussed
in this paper: \coi, \com, and \cou. 
We choose a step size in radius of 13.5\,arcsec, corresponding
to half of the spatial resolution. At a given radius, we measure
the average of the intensities within a tilted ring. The width
of the rings is taken as the chosen step size (defined along
the minor axis), so that the rings are not overlapping, but
they are correlated because the beam is oversampled.
Table~\ref{table:list} gives the assumed inclinations and
position angles. Error bars on those measurements
are calculated as the root-mean-squared error within each
ring of our error maps and multiplied by the square root
of the oversampling factor $N_{\rm s}$, defined as
$N_{\rm s}$= 1.13$\times$(map resolution/pixel size)$^2$.
All intensity profiles are multiplied by $cos(i)$ to correct for
inclination. Profiles of intensity ratios, such as $\Re$, are
built by dividing the radial profile of the quantity at the numerator
with the radial profile of the quantity in the denominator.
Figure~\ref{fig:allprofiles} shows radial profiles of the individual
CO line intensities for each galaxy and Figure~\ref{fig:ratioprofiles}
shows radial profiles of $\Re$ for all galaxies.

\subsubsection{Description of the profiles}
Radial profiles of \coi, \com, and \cou globally follow each
other very well for all galaxies (Fig.~\ref{fig:allprofiles}).
Intensities peak in the centre and decrease by an order
of magnitude at $r\simeq0.6$\,$r_{25}$, and even more
for NGC\,6946 which has a very prominent centre.

Differences in the radial behavior of \coi and \com are made
more obvious by inspecting $\Re$ (Fig.~\ref{fig:ratioprofiles}).
We have reliable measurements of $\Re$ out to
$r\simeq8$\,kpc or $r\simeq0.7\,r_{25}$.
\begin{itemize}
\item
The profiles of $\Re$ in NGC\,2903, NGC\,3627, and
NGC6946 decrease by a factor $1.2-1.4$ from centre
to disc ($r=3$\,kpc) and stay mostly flat at larger radii.
For NGC\,3627, we notice a clear suppression
of the \coi peak intensities in the galaxy centre
(see Figures~\ref{fig:maps} and \ref{fig:allprofiles}
in the Appendix).
\item
The profiles of NGC\,3184 and NGC\,4321 are flat
at all radii.
\item
The profiles of NGC\,4254, NGC\,5194, and NGC\,5055
increase steadily by a factor of $1.4$ from centre to
outer disc ($r\simeq7$\,kpc).
\item
The profile of NGC\,0628 is peculiar. It increases by
a factor of $1.8$ from centre to $r=2$\,kpc and stays
mostly flat at larger radii.
\end{itemize}

Centres aside, we find that the profiles increase slightly
as a function of radius on average. The increase is mild
compared to that observed in the Milky Way
\citep{roman-duval-2016}. At large radii ($r=7-8$\,kpc),
our sample of galaxies and the Milky Way
have similar $\Re$ values, but at lower radii, the Milky Way
shows systematically lower values, between 5 and 8.
$\Re$ increases by a factor of $\sim2$ from the inner disc
to the outer disc of the Milky Way. Such measurements,
along with observations of rarer isotopologues of CO, have
been used to infer abundance gradients in the Milky Way
\citep[][shown as the black dotted line in Fig.~\ref{fig:ratioprofiles}]{milam-2005}.
$\Re$ profiles are different in our sample of galaxies,
implying that the Milky Way as a massive, rather quiescent
galaxy \citep[e.g.,][]{chomiuk-2011} that may be compact
for its stellar mass, might represent a different physical regime
than the galaxies in our survey
(e.g., different abundance patterns), or that
there could be issues (e.g., geometric) in comparing
Milky Way and extragalactic work.
For example, the study of \cite{roman-duval-2016} was
limited to regions less than 50\,pc away from the Galactic
plane and it could be missing substantial $^{12}$CO
emission at high latitude \citep{dame-2001} that we
do capture in our sample.

\subsubsection{Comparison to the literature on nearby spirals}
\cite{paglione-2001} observed \coi and \com along the major
axes of NGC\,3184, NGC\,3627, NGC\,5055, NGC\,5194 and
NGC\,6946 with the FCRAO 14-m telescope (beam size $\sim$47\,arcsec).
Their central and outer values globally agree with our observations
within errors, except for NGC\,3184 for which \cite{paglione-2001}
find $\Re$ values two times lower but with marginal detections.
For NGC\,3627, we find a lower average in the disc (12 instead
of 17), but they have much lower statistics.
For NGC\,6946, we find similar central values but larger values
in the disc (13-14 instead of 10). At high resolution (5\,arcsec or
$\sim$150\,pc), \cite{meier-2004} find a range of values between
7-20 in the nucleus of NGC\,6946, which is compatible with
our central average of $15$.
\cite{muraoka-2016} mapped NGC\,2903 with the NRO 45-m
telescope (beam size $\sim$14\,arcsec). They find values of
$\Re$ that are around 10, which is globally consistent with
our results. They divide the galaxy in ten distinct regions
and find that $\Re$ varies by a factor of about two
in the different environments. Although our resolution is coarser,
we do not find as large spatial variations within NGC\,2903
as they do. However, we achieve much better signal-to-noise
ratios for individual \coi measurements.
\cite{garcia-burillo-1993} mapped NGC\,5194 with the IRAM 30-m
and find similar values as \cite{pety-2013} (and therefore as us)
in the centre and disc of this galaxy. \cite{tosaki-2002} also mapped
part of the disc and the centre of NGC\,5194 with the NRO telescope
and found larger values ($\ge20$) in the centre and inter-arm
regions and attributed discrepancies in $\Re$ to differences
in beam size ($\simeq$17\,arcsec versus $\simeq$25\,arcsec).
In addition, \cite{tan-2011} observed the centre of NGC\,2903 and
\cite{li-fc-2015} observed the centres of NGC\,3184, NGC\,3627,
NGC\,4254, and NGC\,4321 with the PMO 14-m telescope
(beam size $\sim$55-60\,arcsec). Both studies found $\Re$ values
consistent with ours.
\cite{vila-vilaro-2015} also observed the centres of NGC\,0628,
NGC\,2903, NGC\,4254, and NGC\,5055 with the ARO KP
12-m telescope (beam size $\sim$56\,arcsec). We find similar
values as theirs, except for the centre of NGC\,4254 
(8 instead of 11.5). The discrepancy could be attributed to
beam size differences, as we find a global average for NGC\,4254
closer to 11, or to calibration uncertainties. Our value
in the centre of NGC\,4254 is closer to that reported by
\cite{li-fc-2015}.

In high-resolution mapping observations of $23$ nearby
disc galaxies selected to lie on the blue sequence and to be
actively star-forming, IR-bright galaxies (CARMA STING
survey\footnote{\protect\url{http://www.astro.umd.edu/~bolatto/STING/}}),
probing scales of $300-500$\,pc, \cite{cao-2017} find
similar $\Re$ ratios and flat $\Re$ profiles. They report that
$\Re$ varies mostly from galaxy to galaxy, with values between
5 and 15. This behavior resembles what we find in the
centres of our sample of galaxies,
but our profiles in the discs show less scatter.
While they are limited by sensitivity and focus on the inner
bright, molecular gas-rich regions, we detect fainter and more
diffuse regions with EMPIRE, enabling us to probe a wider
range of environments (centre, arm, inter-arm regions).
gal
\cite{sakamoto-1997} performed strip-scan observations of
the edge-on spiral NGC\,891 at a resolution of 14\,arcsec.
They find a high value of $\Re\simeq15$ in the centre and
a general increase of $\Re$, from $4$ at $r=3$\,kpc to $20$
at $r=10$\,kpc. The increase of $\Re$ with radius is less
pronounced in our sample of galaxies.

Overall, there is generally good agreement between our
$\Re$ measurements and those reported in the literature
for the EMPIRE galaxies or for similar types of galaxies
and at similar scales. When there are discrepancies, these
tend to be due to noise in the data. The uniqueness of
the EMPIRE survey lies in the large, homogeneous spatial
coverage (out to $r\simeq0.5\,r_{25}$) and high
signal-to-noise ratios achieved.

\subsection{Correlation with tracers of star formation and ISM properties}
\label{sect:correlations}
%
\begin{figure*}
\centering
\includegraphics[clip, trim=08mm 0 6mm -1mm,width=4.4cm]{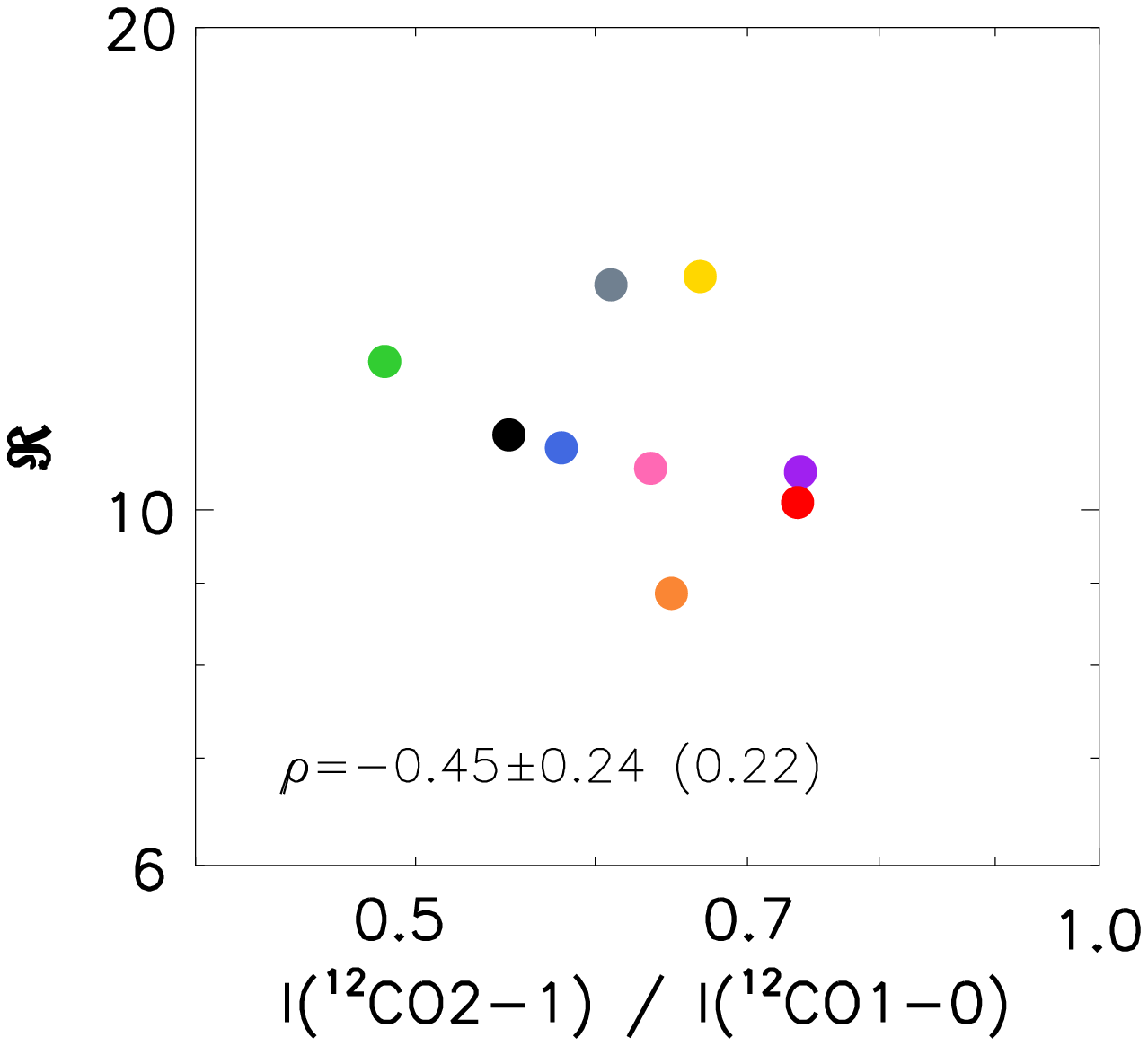}
\includegraphics[clip, trim=30mm 0 6mm 1mm,width=3.65cm]{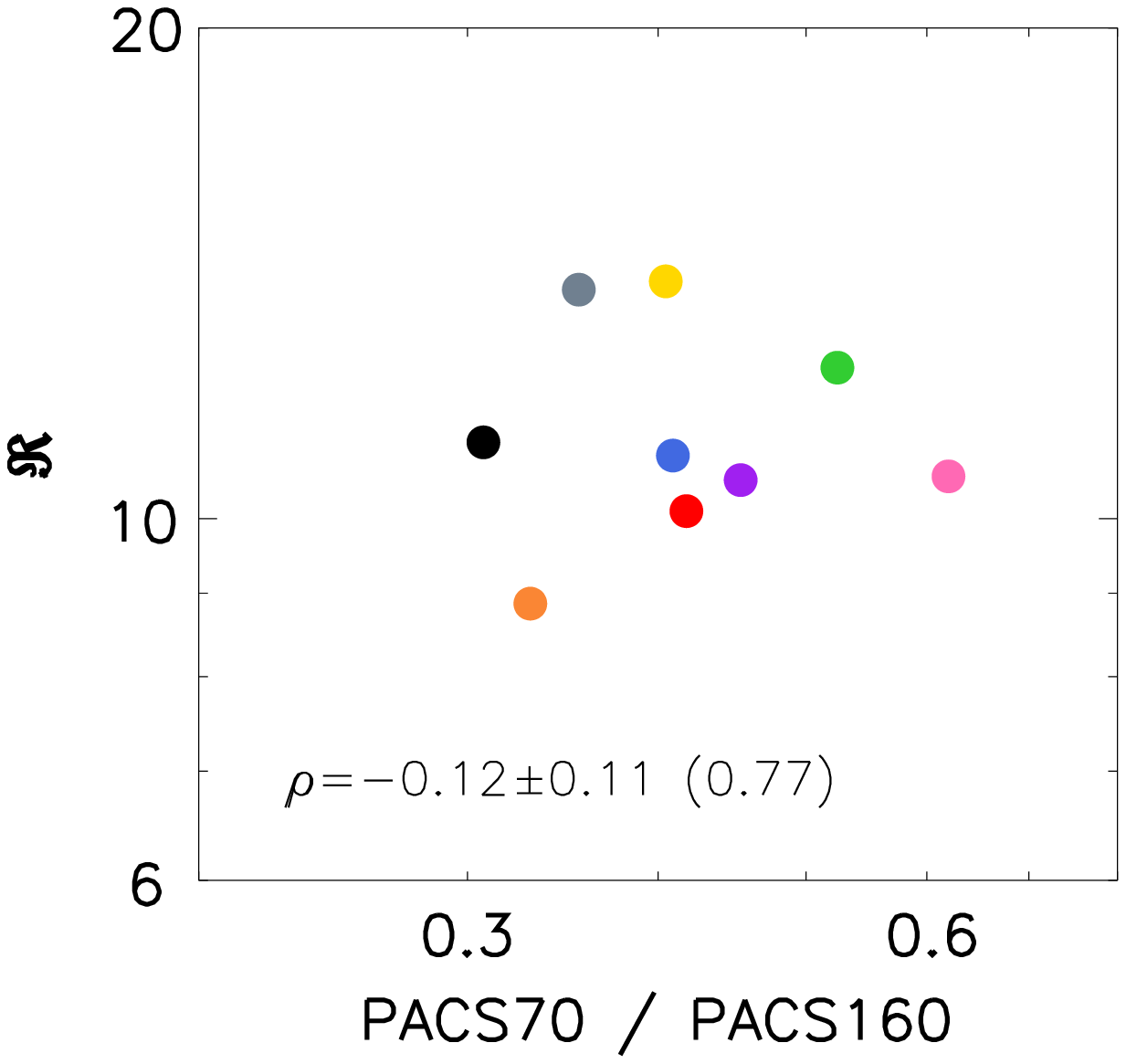}
\includegraphics[clip, trim=30mm 0 6mm 1mm,width=3.65cm]{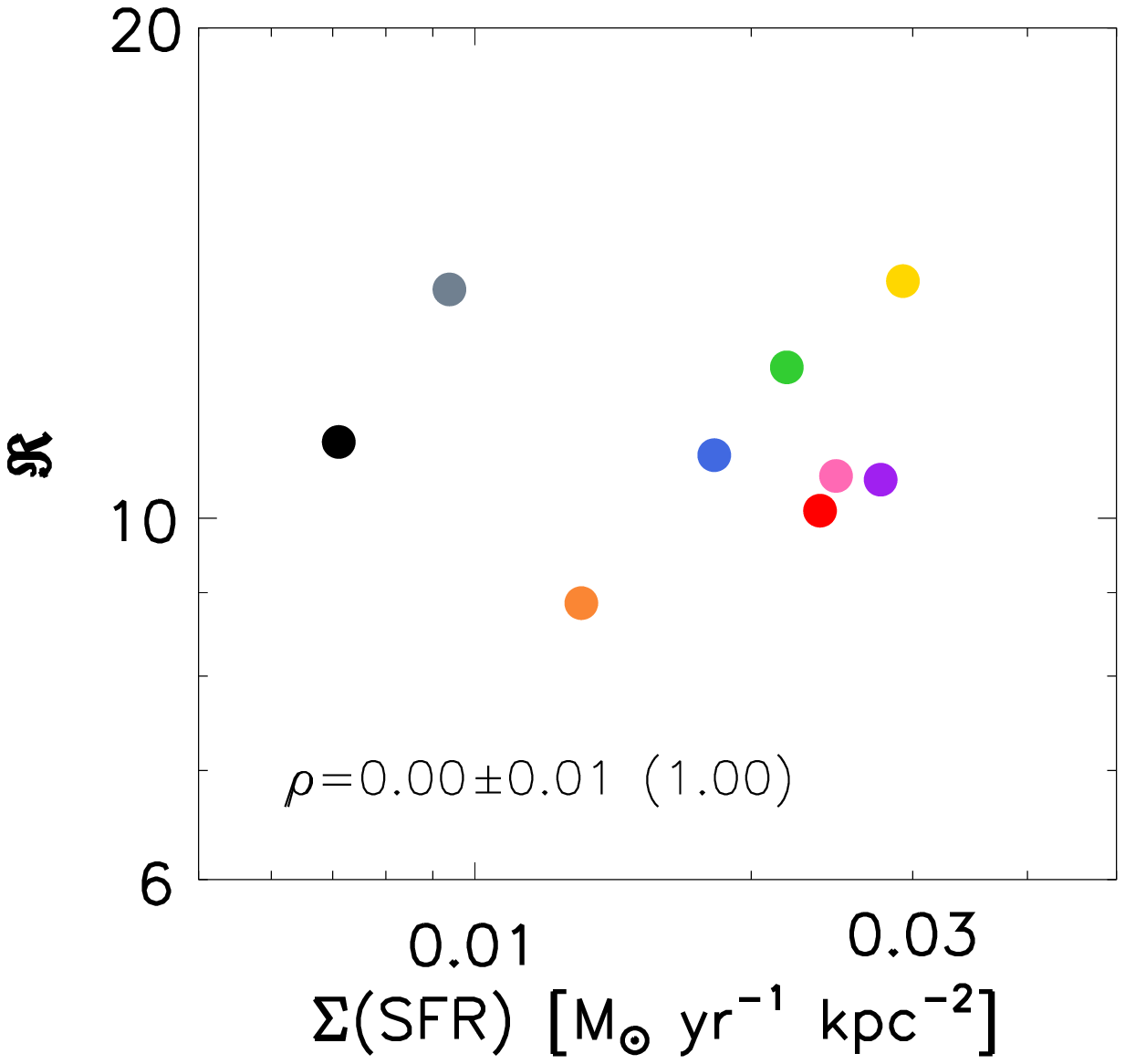}
\includegraphics[clip, trim=30mm 0 -4cm 1mm,width=5.25cm]{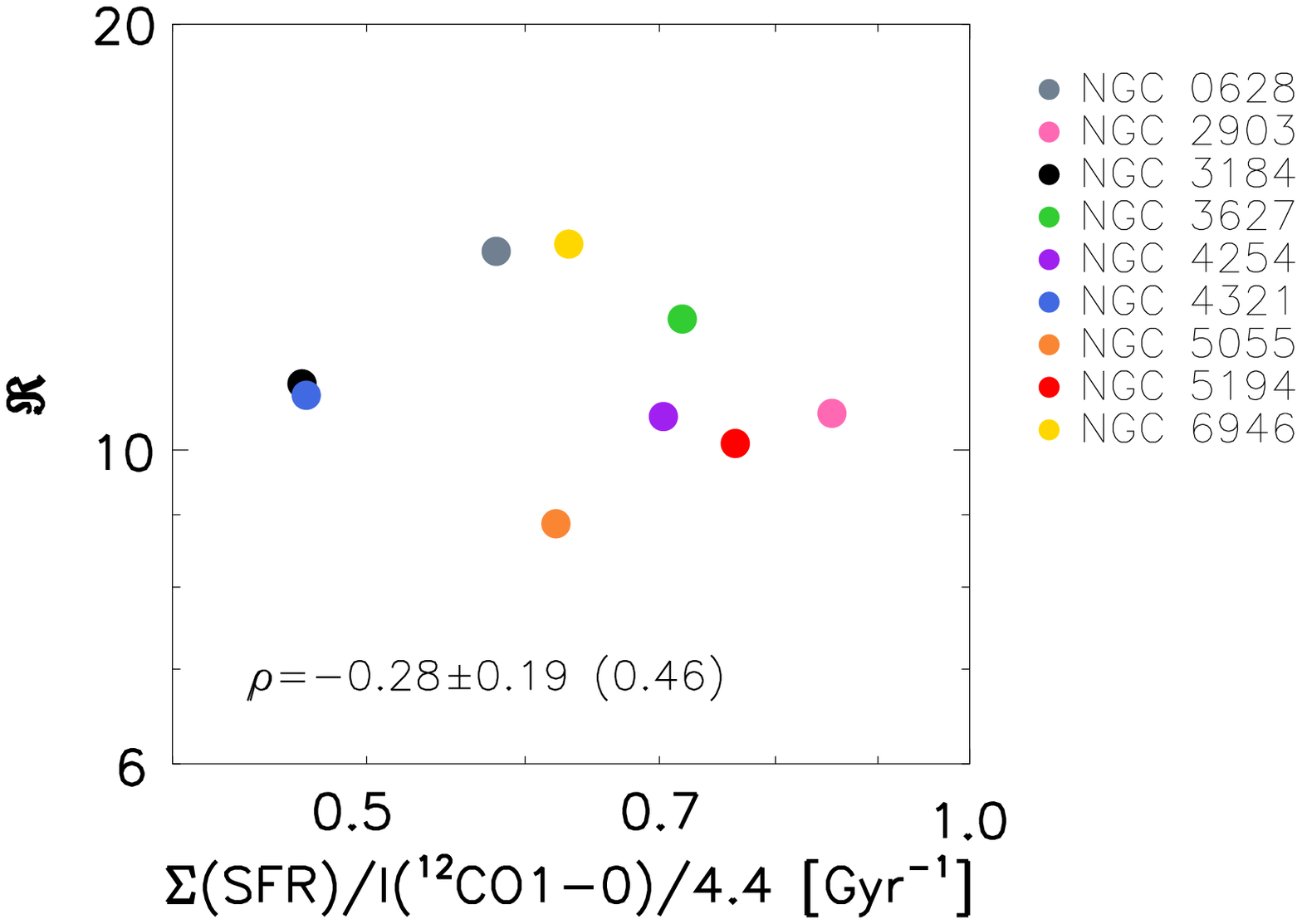}
\caption{
Correlation between the \com-to-\coi intensity ratio
$\Re$ and several quantities:
the \cou/\com ratio probing the gas physical conditions,
the PACS70/PACS160 flux density ratio probing the dust temperature,
the star-formation rate surface density \sigsfr, and
the \sigsfr/CO(1-0) ratio which is a proxy for the star-formation efficiency.
Quantities are averaged over entire galaxies
and do not show significant correlations.
Uncertainties on the galaxy averages are plotted
but smaller than the symbol sizes.
We also indicate the Spearman's rank correlation
coefficients, their uncertainty, and their significance
(in parenthesis).
}
\vspace{3pt}
\label{fig:allcorrav}
\end{figure*}
\begin{figure*}
\centering
\includegraphics[clip, trim=08mm 24mm 6mm 1mm,width=4.4cm]{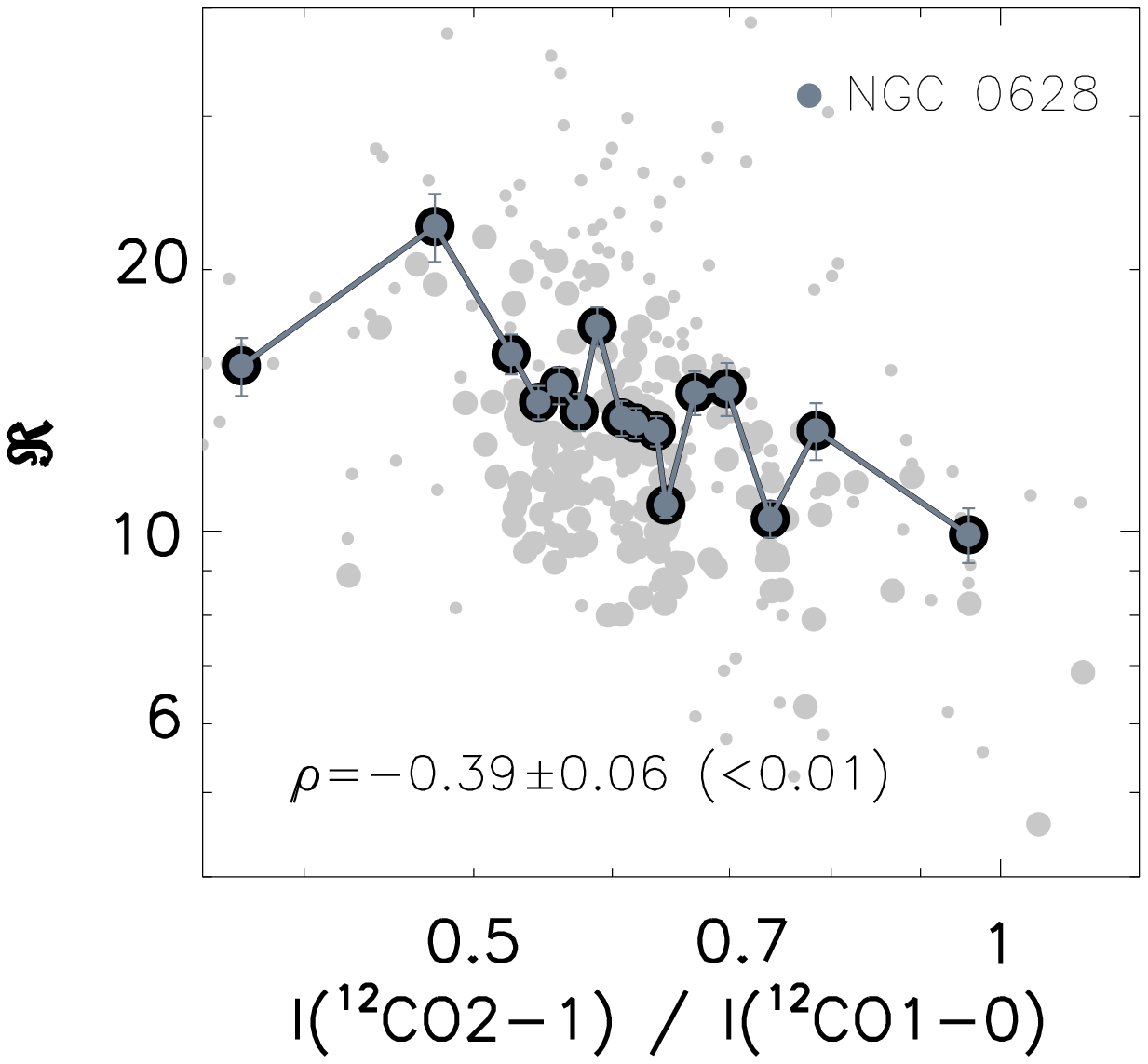}
\includegraphics[clip, trim=30mm 24mm 6mm 1mm,width=3.6cm]{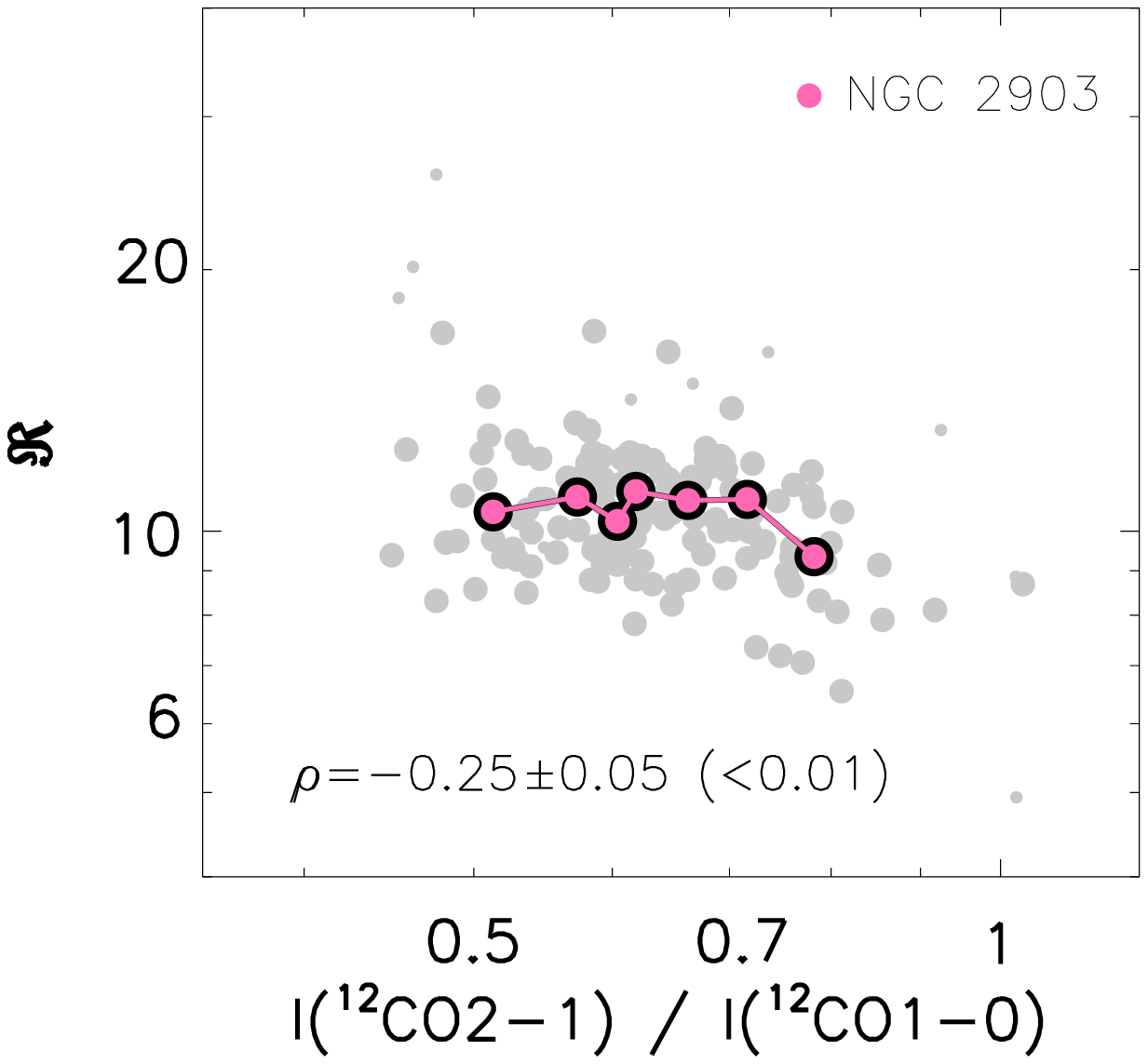}
\includegraphics[clip, trim=30mm 24mm 6mm 1mm,width=3.6cm]{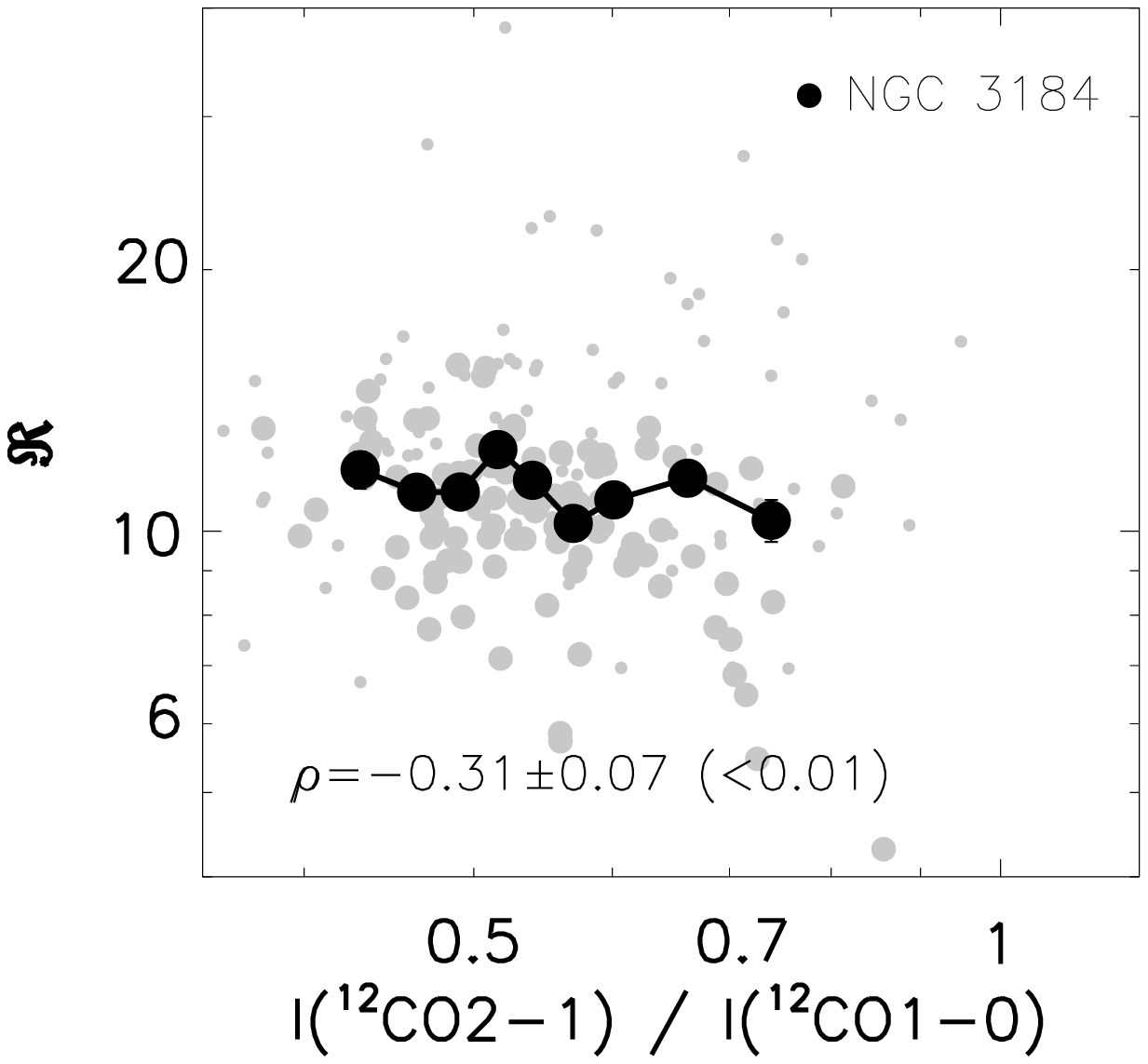}\\
\includegraphics[clip, trim=08mm 24mm 6mm 1mm,width=4.4cm]{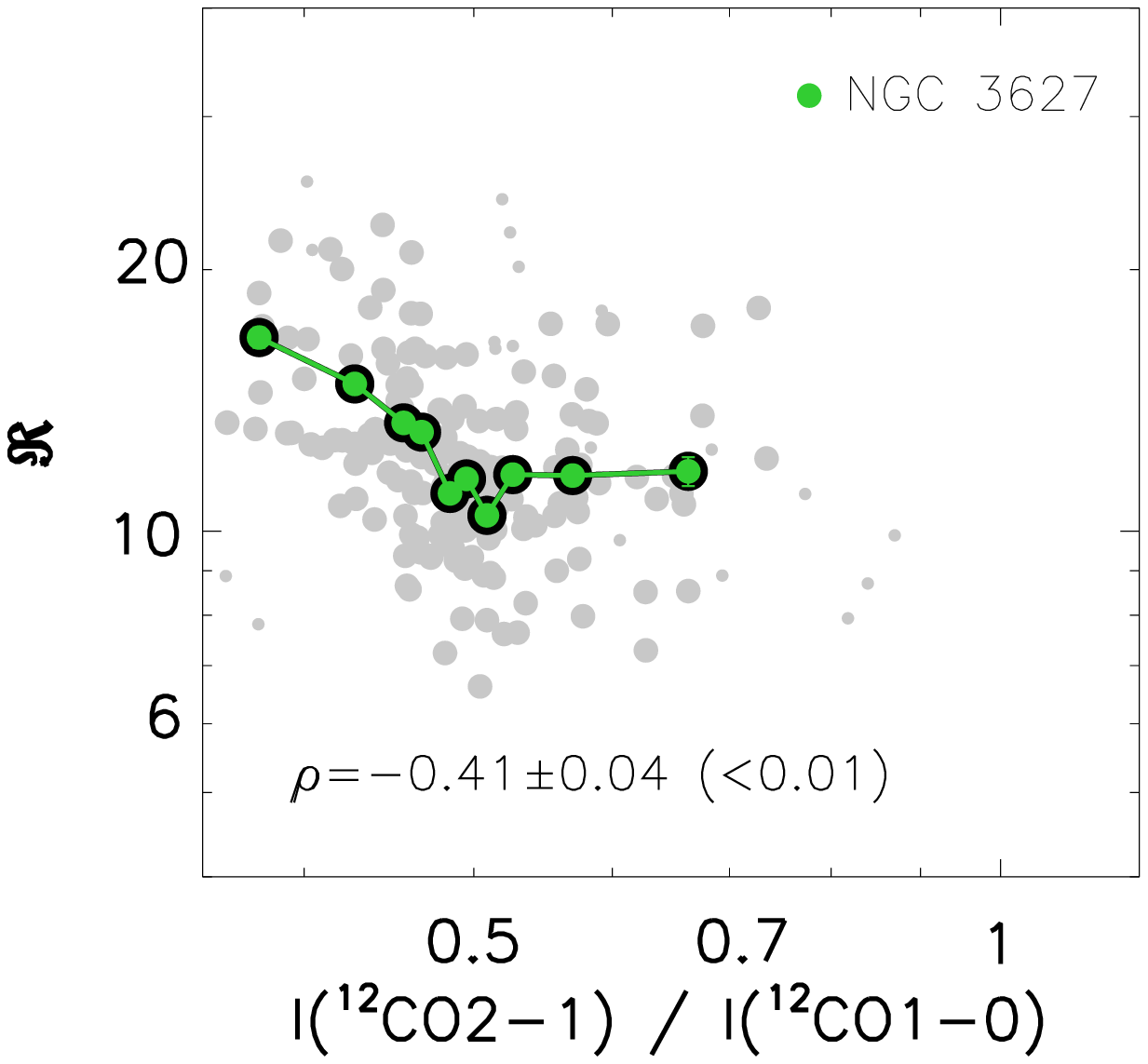}
\includegraphics[clip, trim=30mm 24mm 6mm 1mm,width=3.6cm]{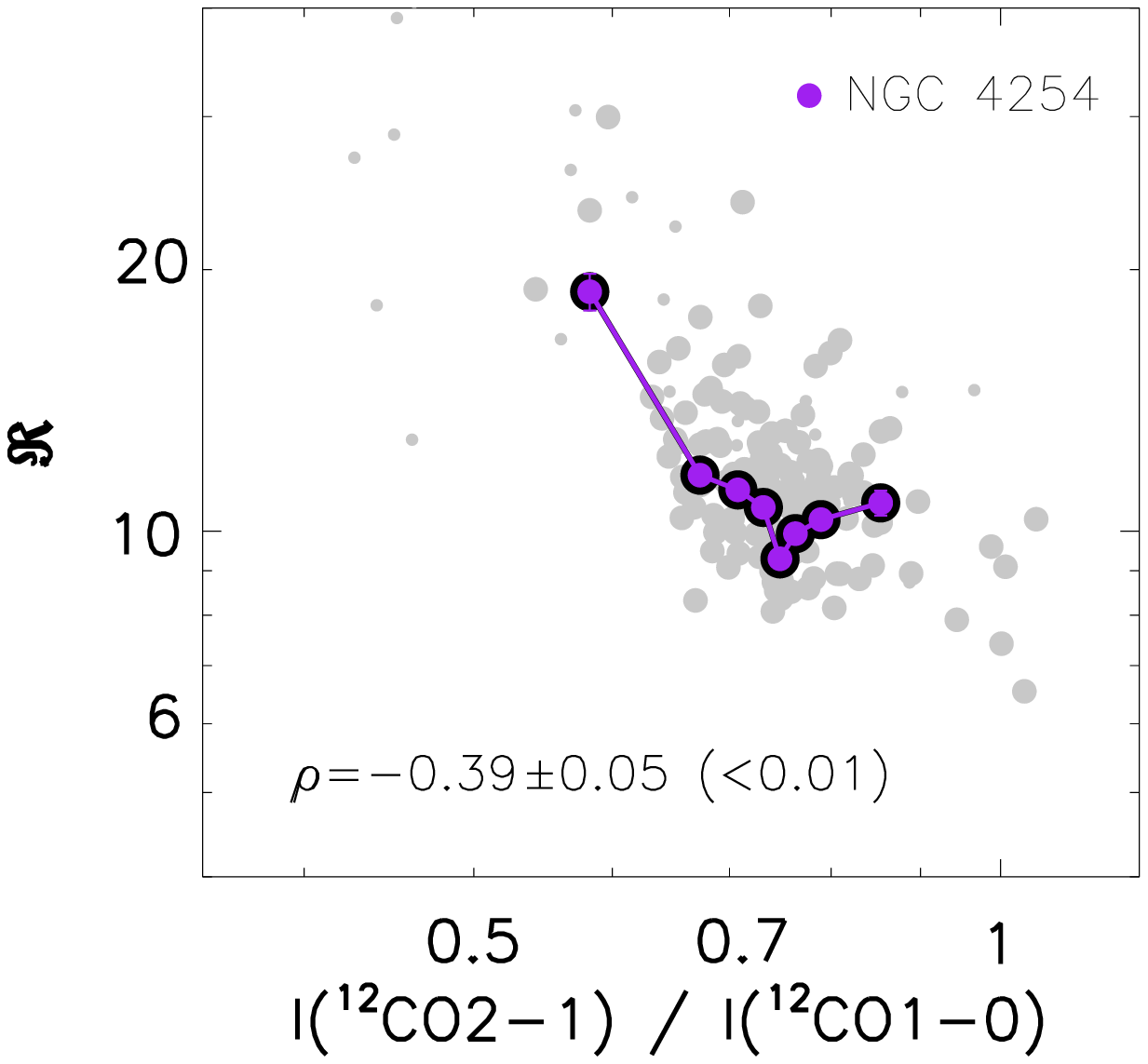}
\includegraphics[clip, trim=30mm 24mm 6mm 1mm,width=3.6cm]{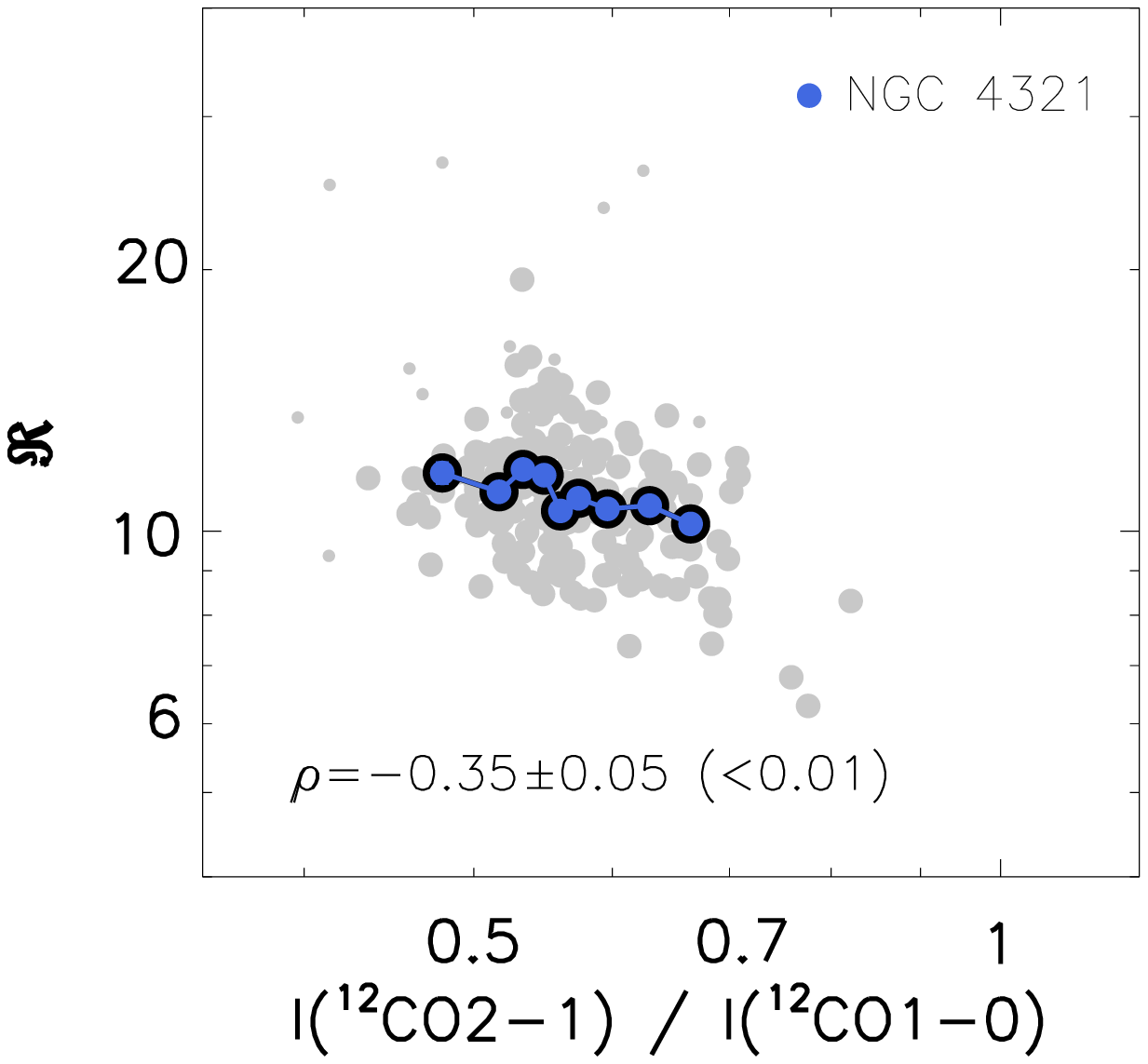}\\
\includegraphics[clip, trim=08mm 0 6mm 1mm,width=4.4cm]{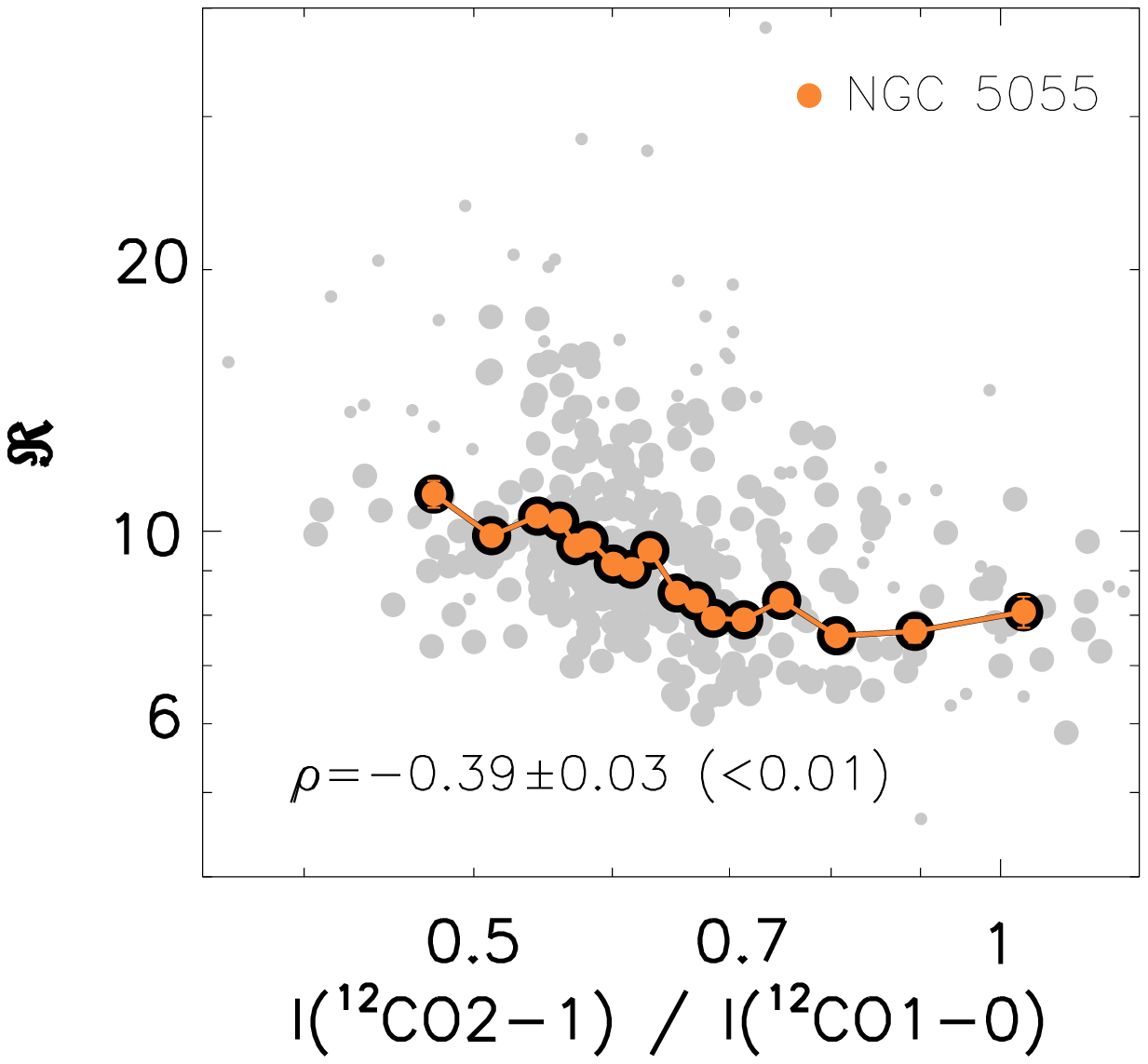}
\includegraphics[clip, trim=30mm 0 6mm 1mm,width=3.6cm]{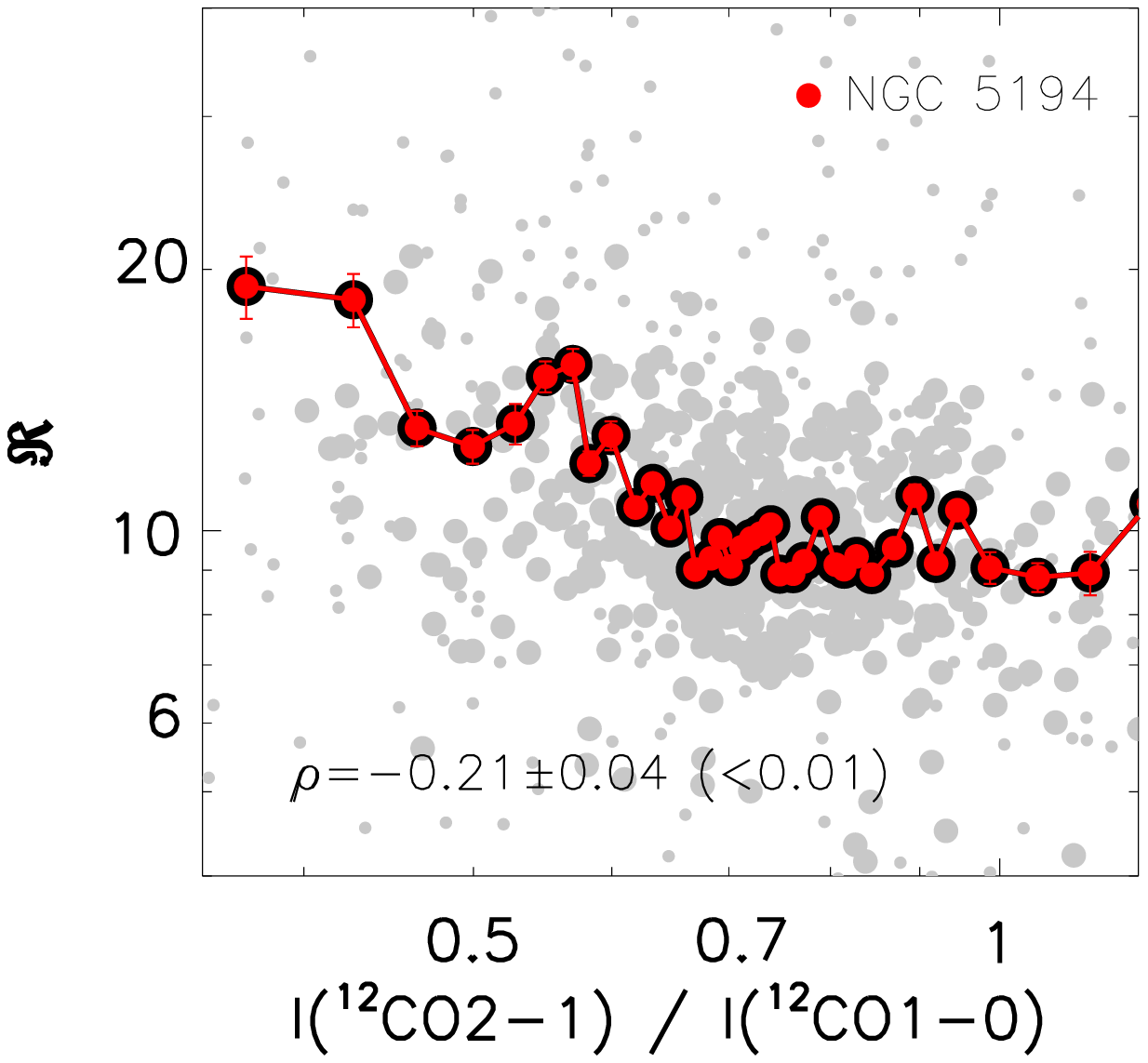}
\includegraphics[clip, trim=30mm 0 6mm 1mm,width=3.6cm]{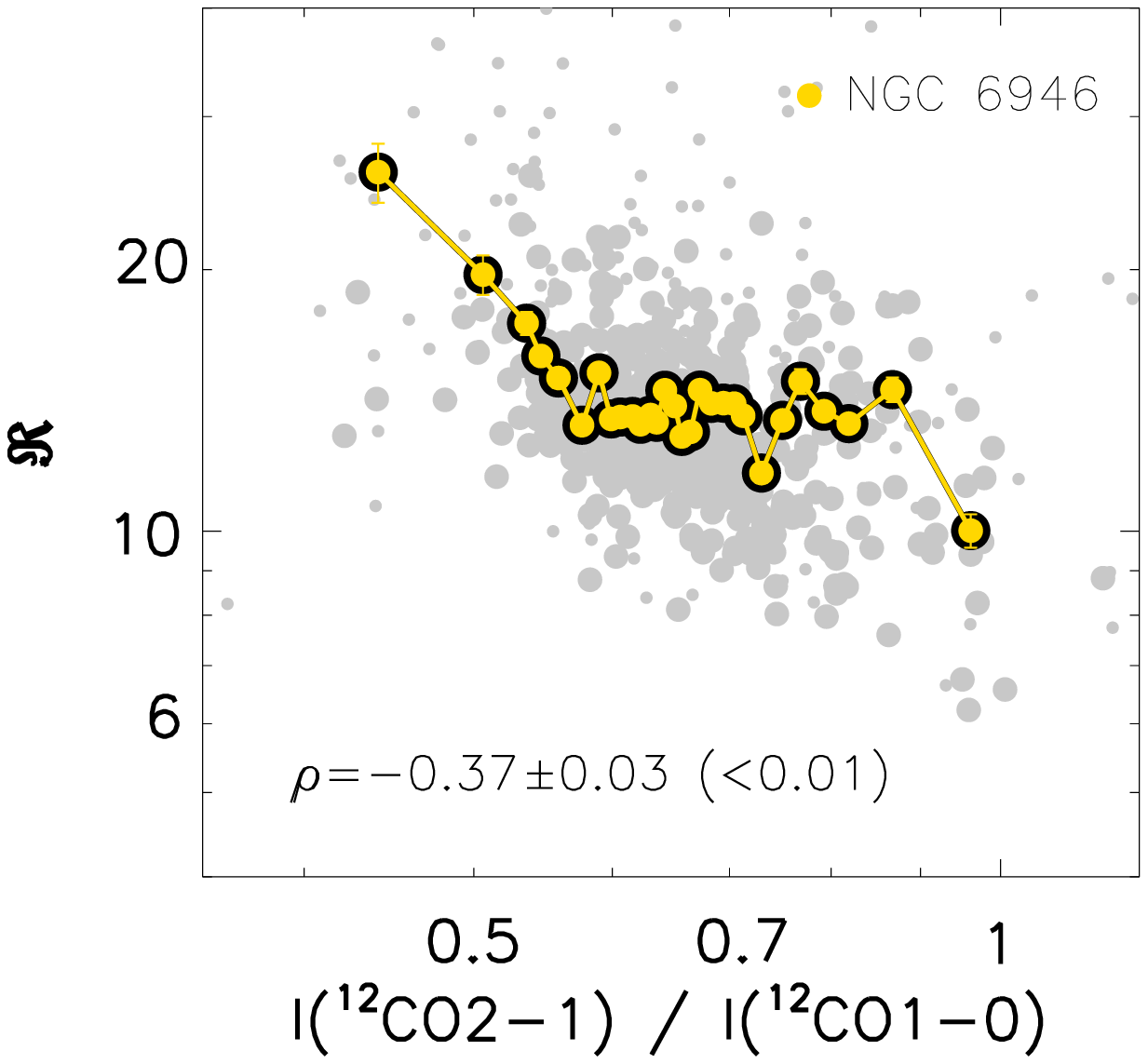}\\
\vspace{-8pt}
\caption{
$(a)$ -- Individual measurements of $\Re$ as a function
of the CO(2-1)/CO(1-0) intensity ratio. We show one panel
per galaxy. The grey data correspond to all kpc-size data
points with smaller symbol size for \coi data with
signal-to-noise ratio below 5. The binned values, in colour,
consider all data points in the bin. We also indicate
Spearman's rank correlation coefficients, their uncertainty,
and their significance (in parenthesis), which
are measured on data with signal-to-noise ratio above 5. 
}
\vspace{3pt}
\label{fig:allcorr}
\end{figure*}
\begin{figure*}
\centering
\includegraphics[clip, trim=08mm 24mm 6mm 1mm,width=4.4cm]{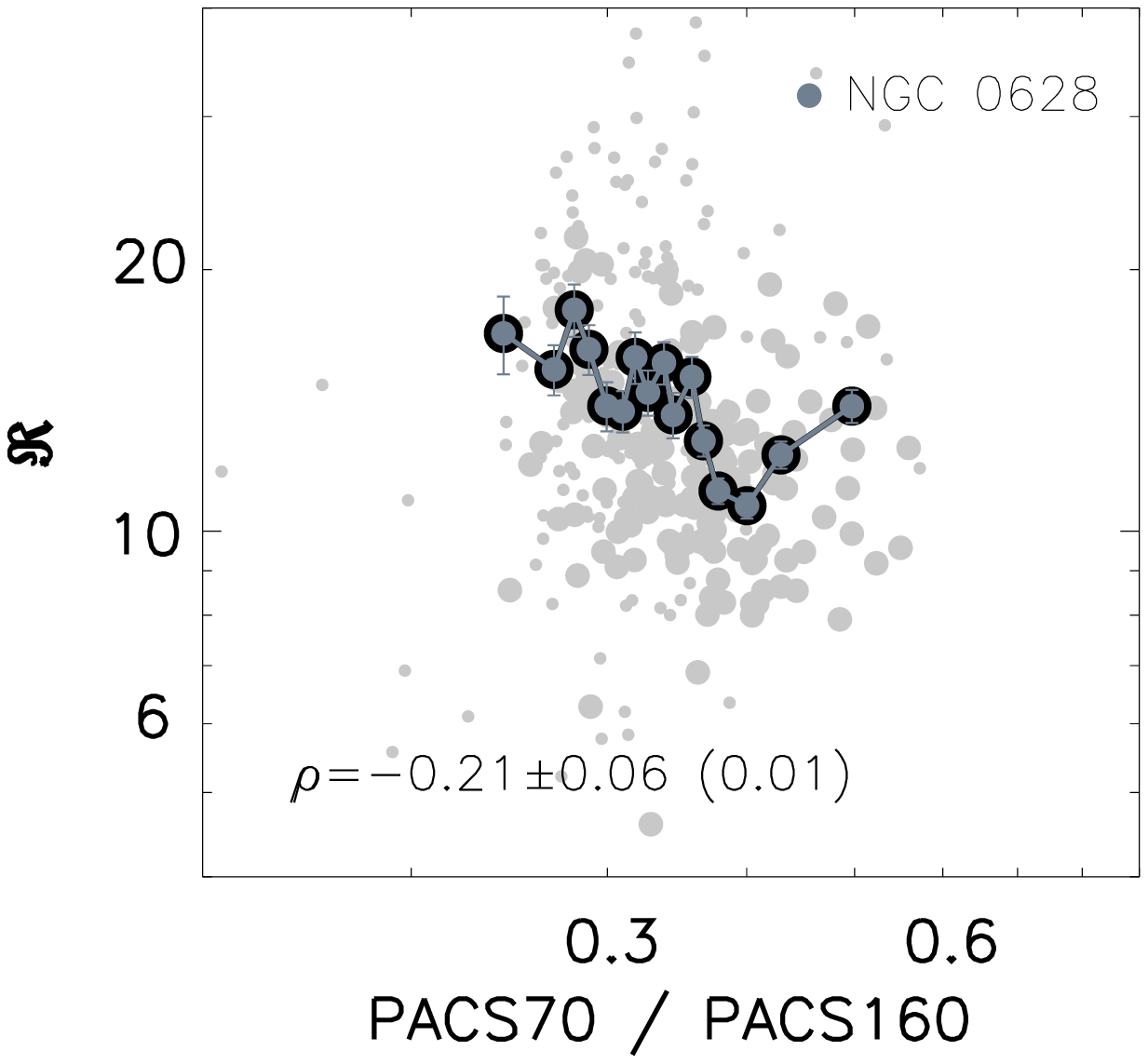}
\includegraphics[clip, trim=30mm 24mm 6mm 1mm,width=3.6cm]{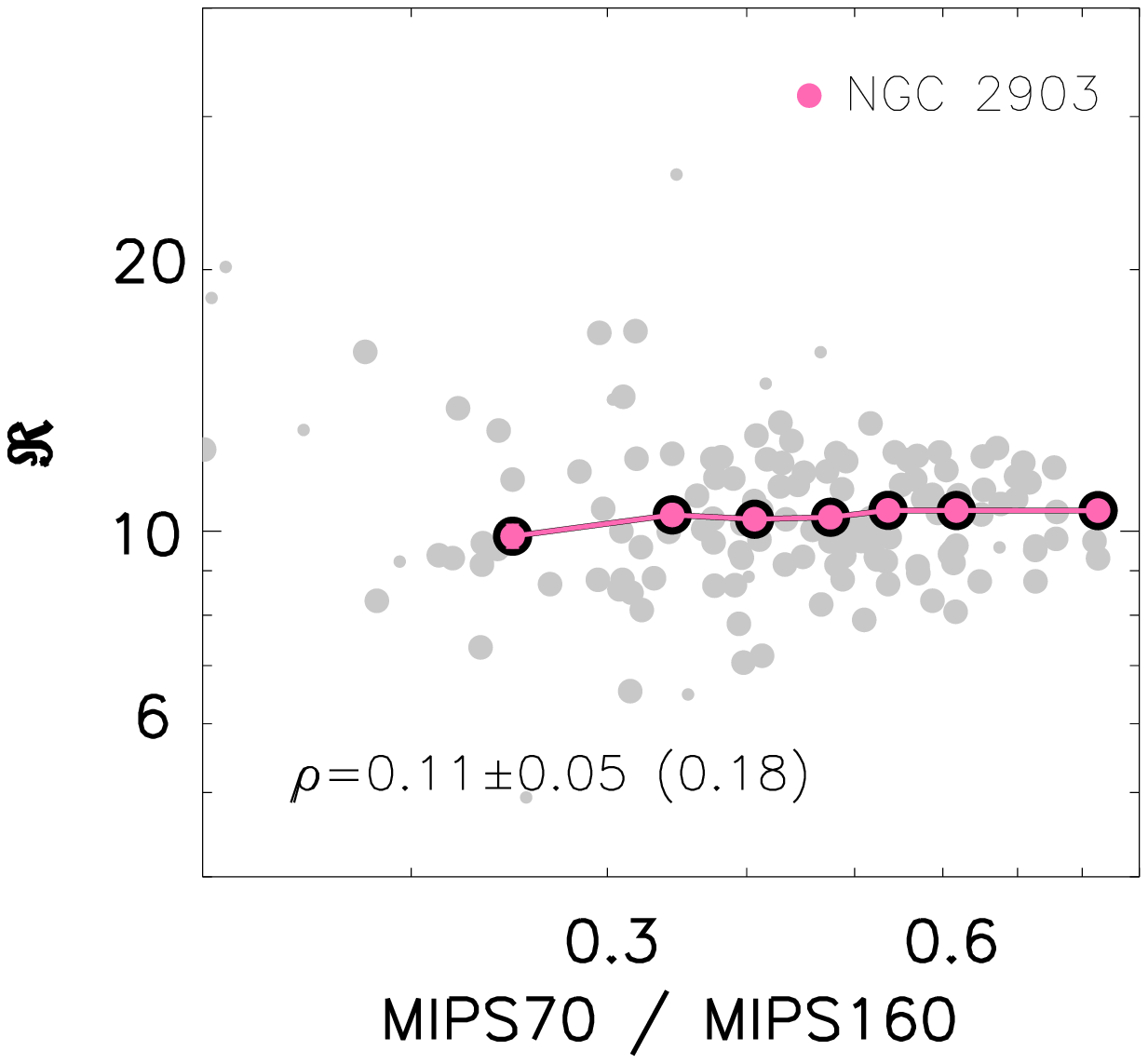}
\includegraphics[clip, trim=30mm 24mm 6mm 1mm,width=3.6cm]{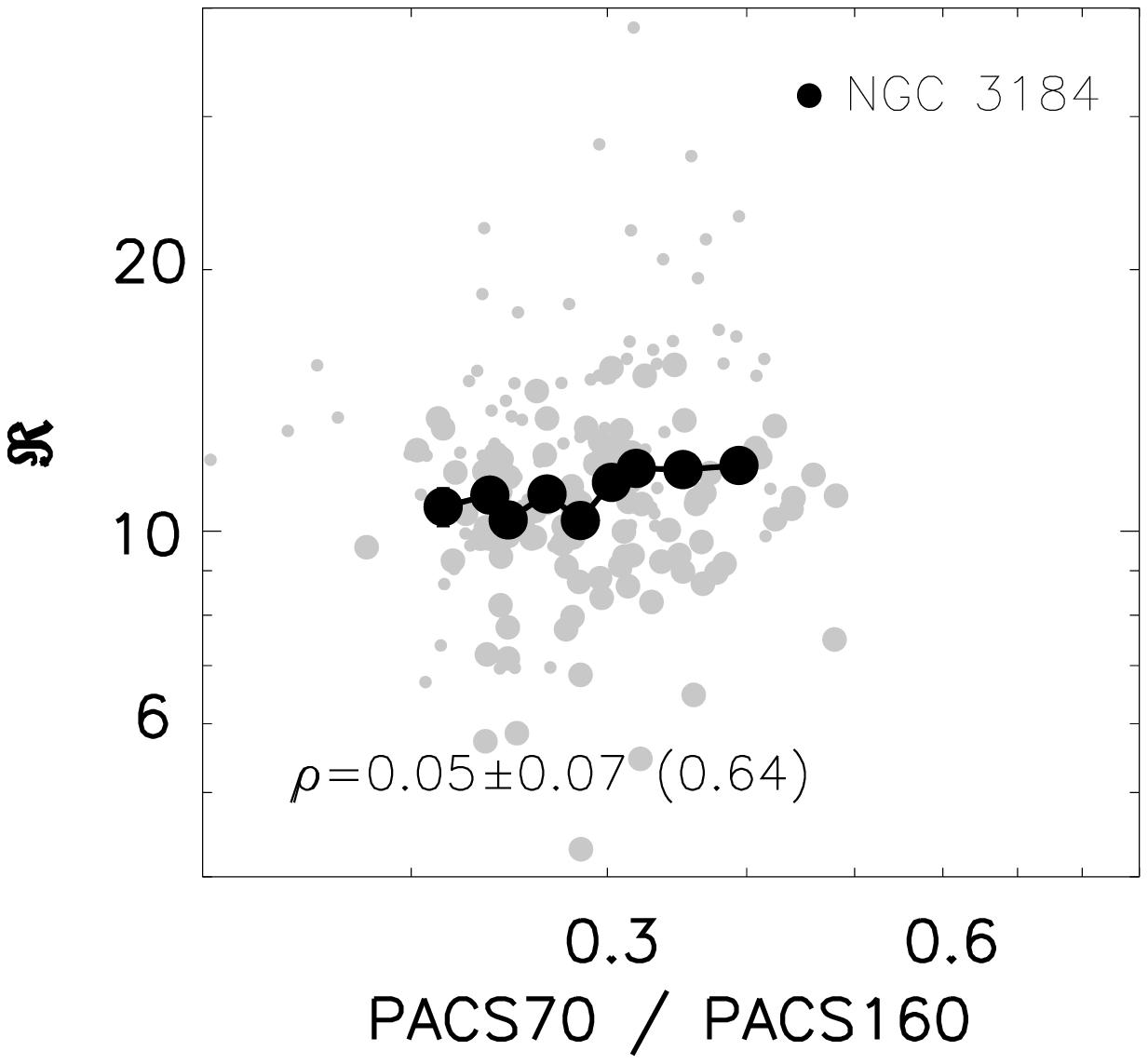}\\
\includegraphics[clip, trim=08mm 24mm 6mm 1mm,width=4.4cm]{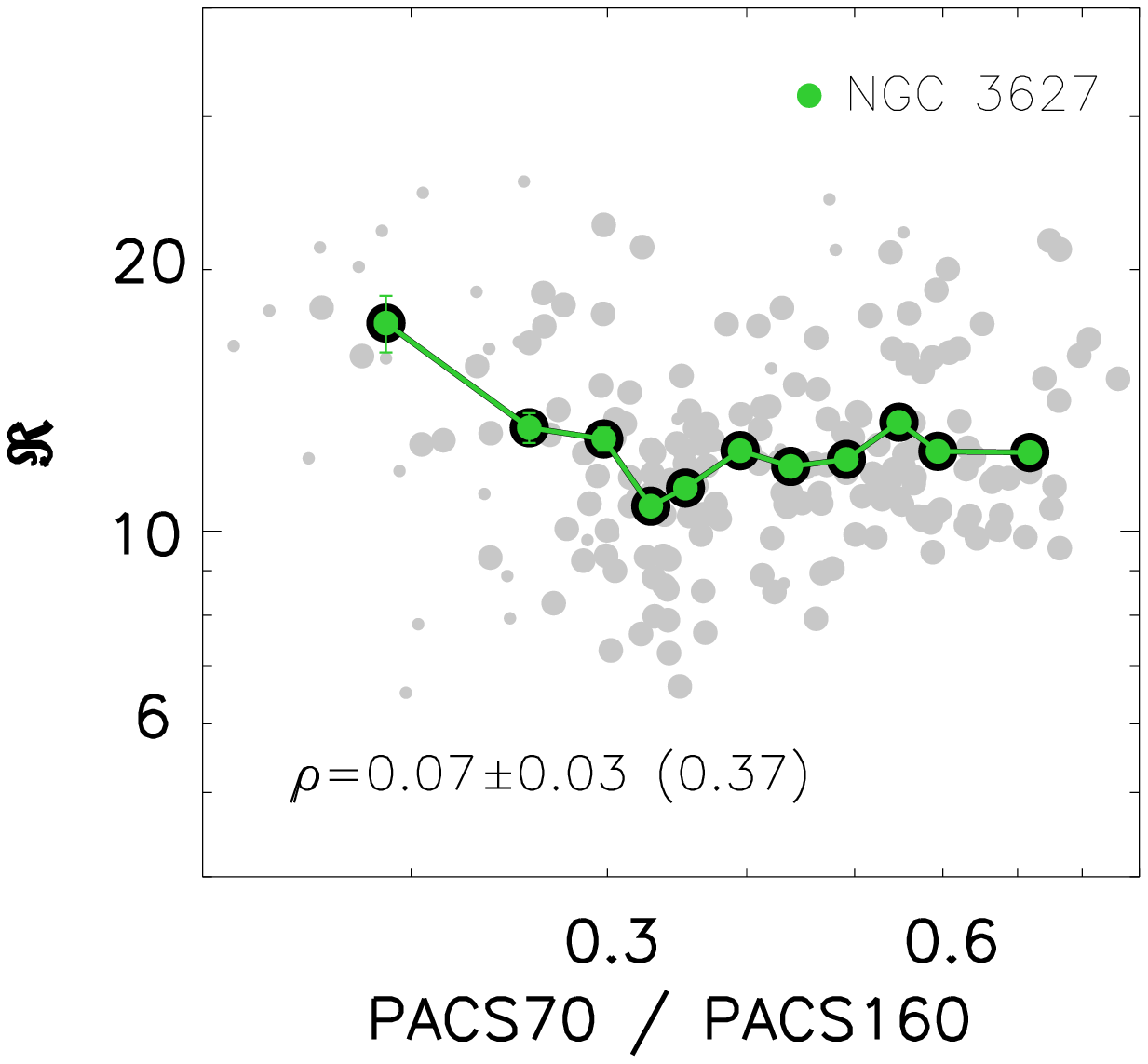}
\includegraphics[clip, trim=30mm 24mm 6mm 1mm,width=3.6cm]{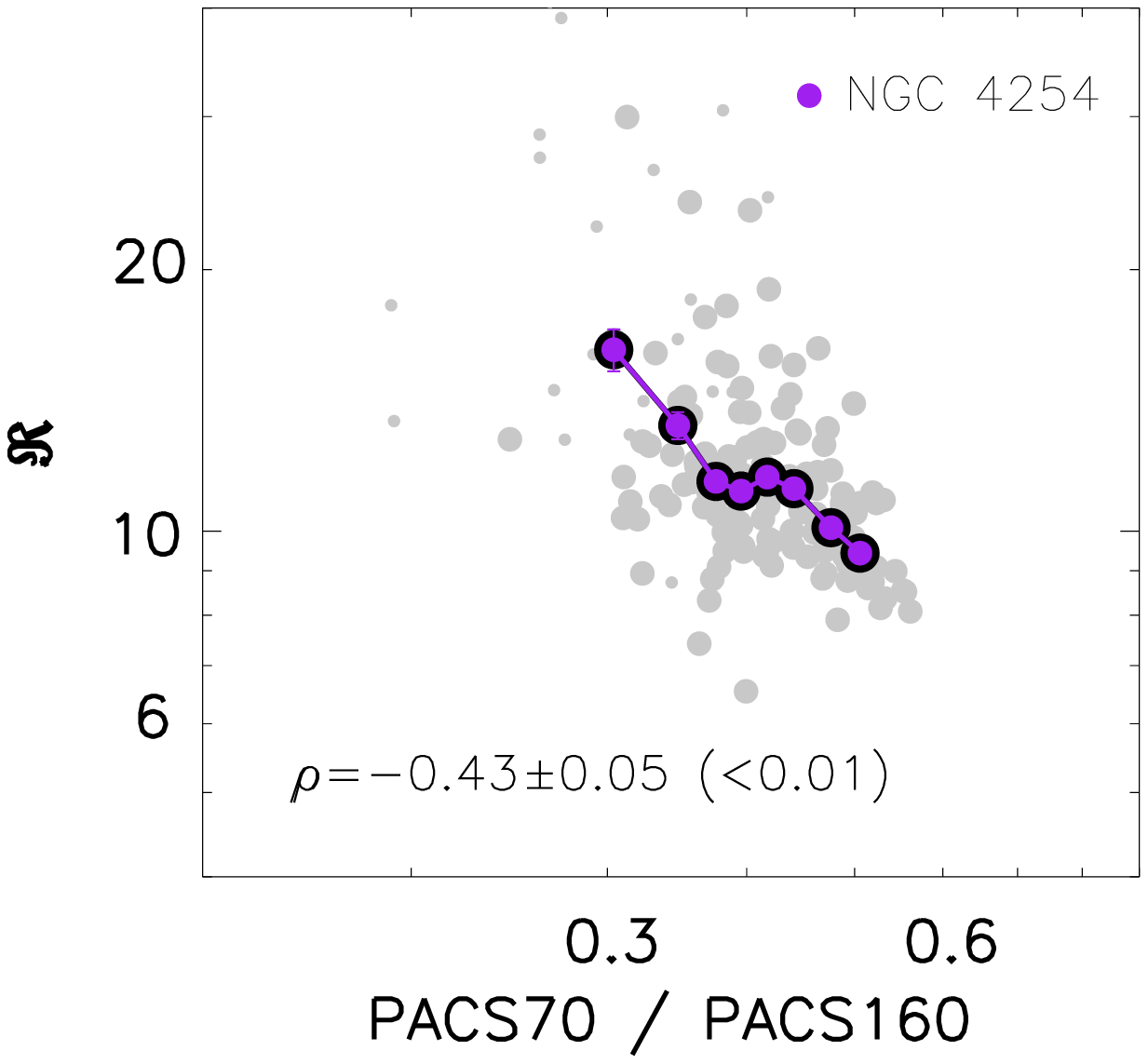}
\includegraphics[clip, trim=30mm 24mm 6mm 1mm,width=3.6cm]{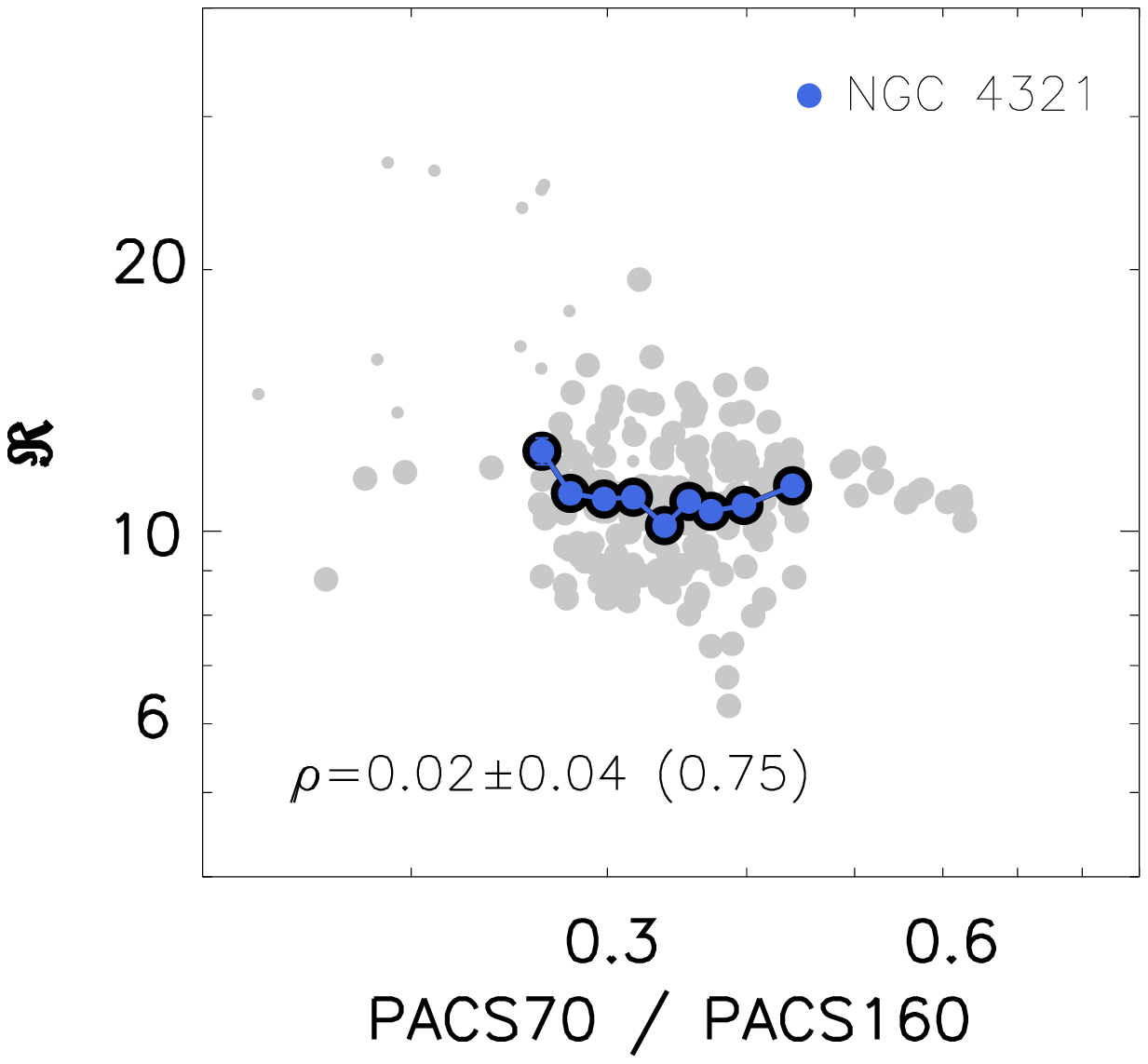}\\
\includegraphics[clip, trim=08mm 0mm 6mm 1mm,width=4.4cm]{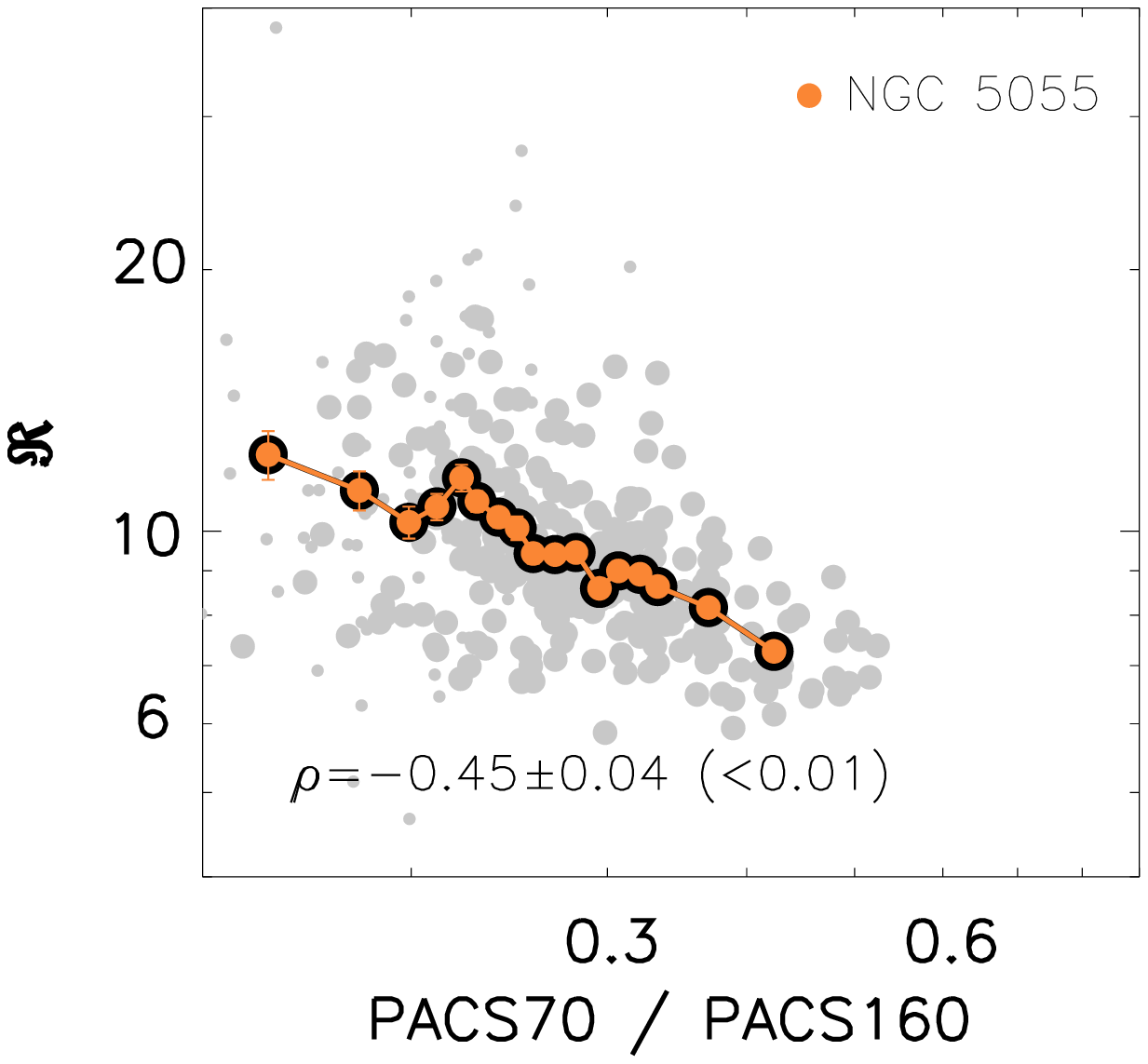}
\includegraphics[clip, trim=30mm 0mm 6mm 1mm,width=3.6cm]{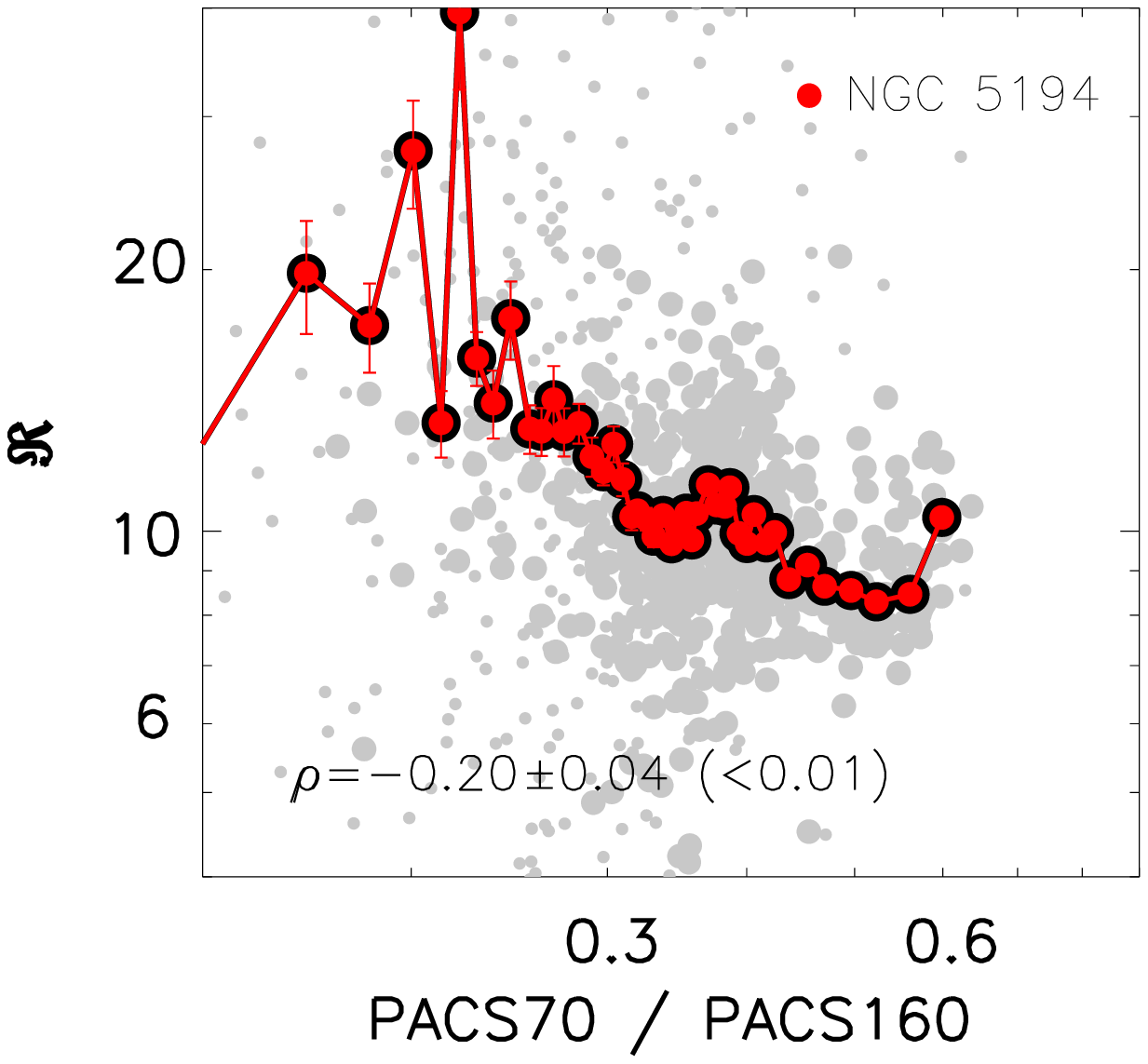}
\includegraphics[clip, trim=30mm 0mm 6mm 1mm,width=3.6cm]{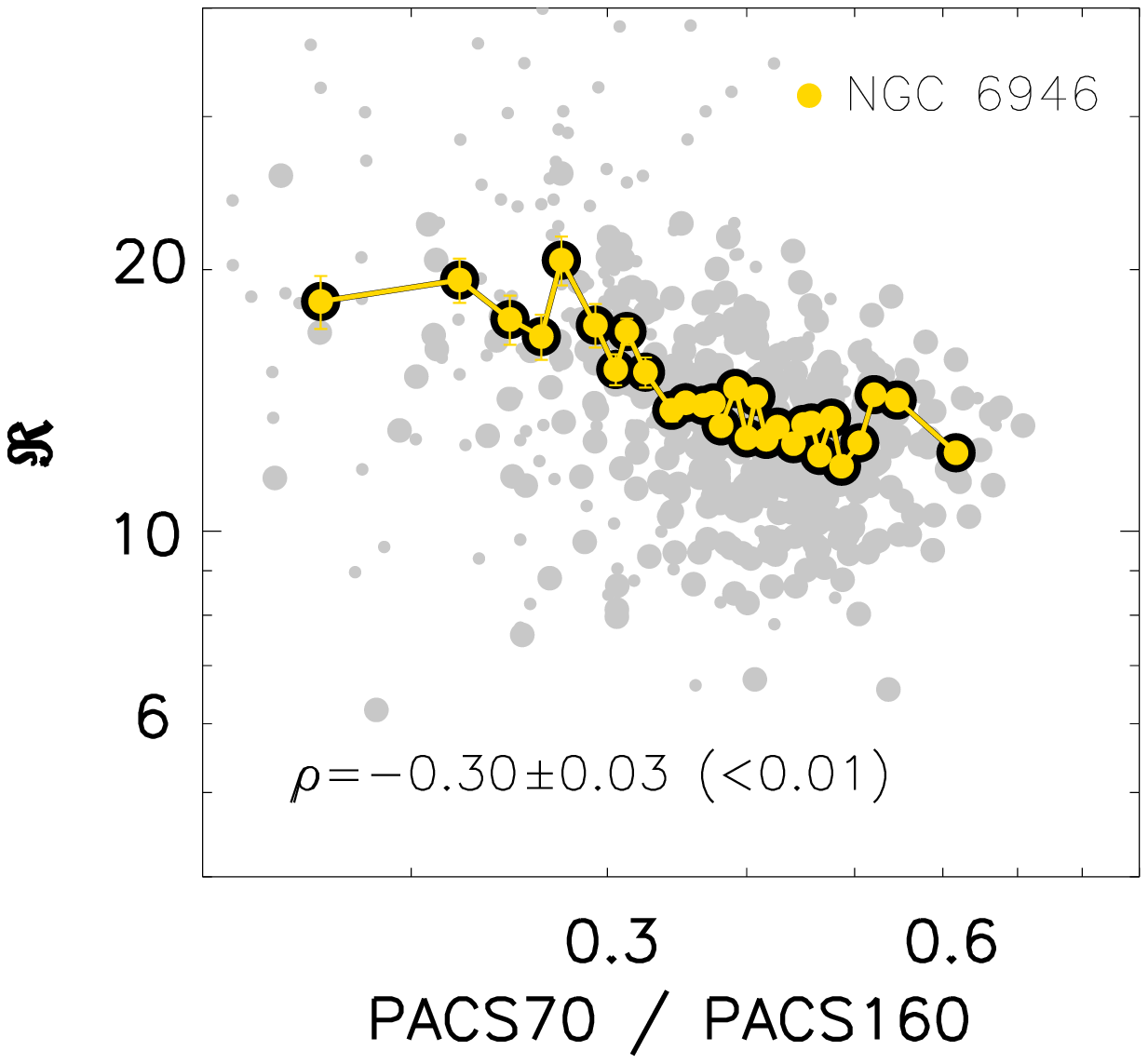}\\
\vspace{-8pt}
\contcaption{
$(b)$ --  Individual measurements of $\Re$ as a function
of the PACS70/PACS160 flux density ratio. For NGC\,2903,
we use the MIPS70/MIPS160 flux density ratio.
}
\vspace{3pt}
\end{figure*}
\begin{figure*}
\centering
\includegraphics[clip, trim=08mm 24mm 6mm 1mm,width=4.4cm]{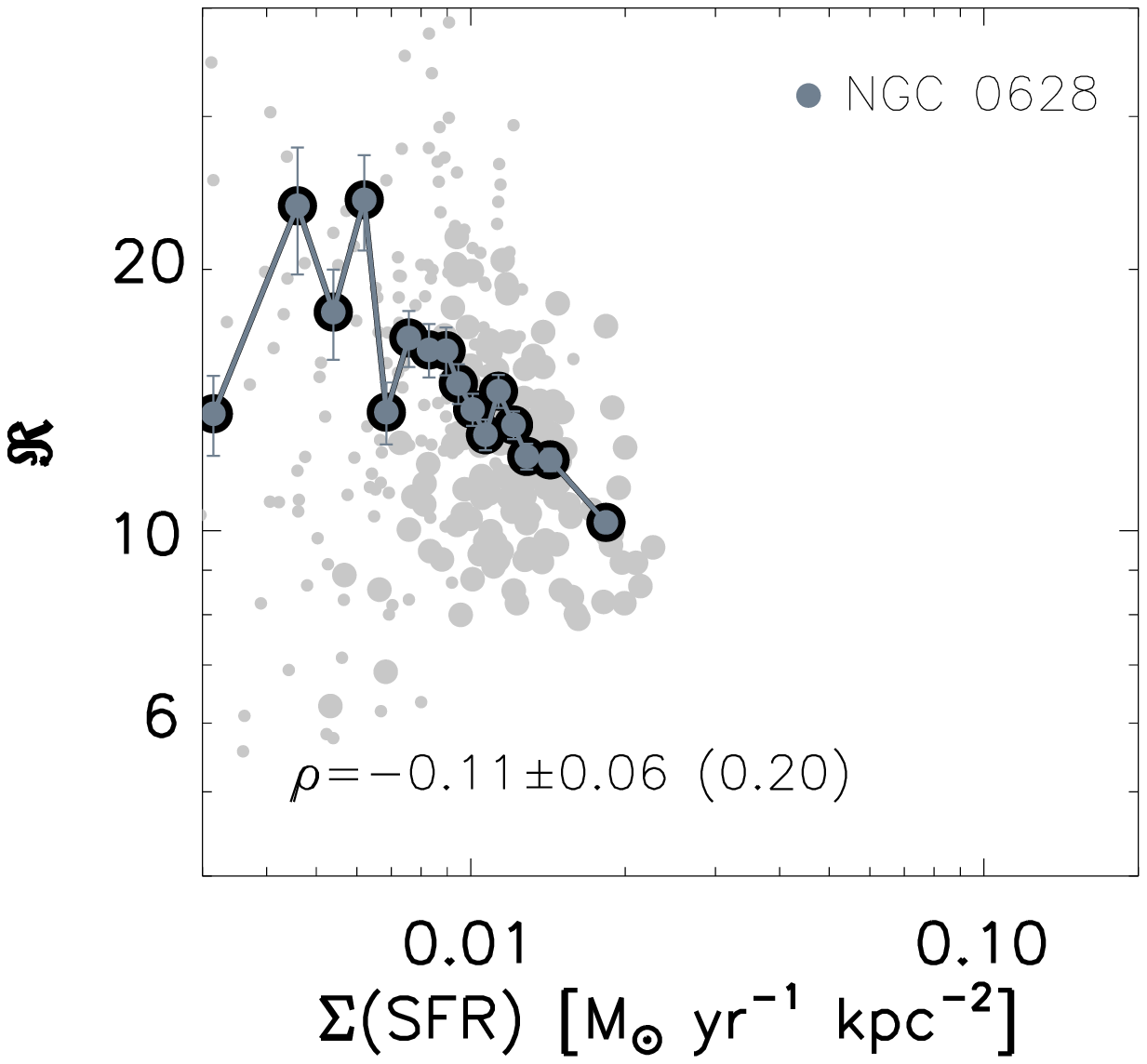}
\includegraphics[clip, trim=30mm 24mm 6mm 1mm,width=3.6cm]{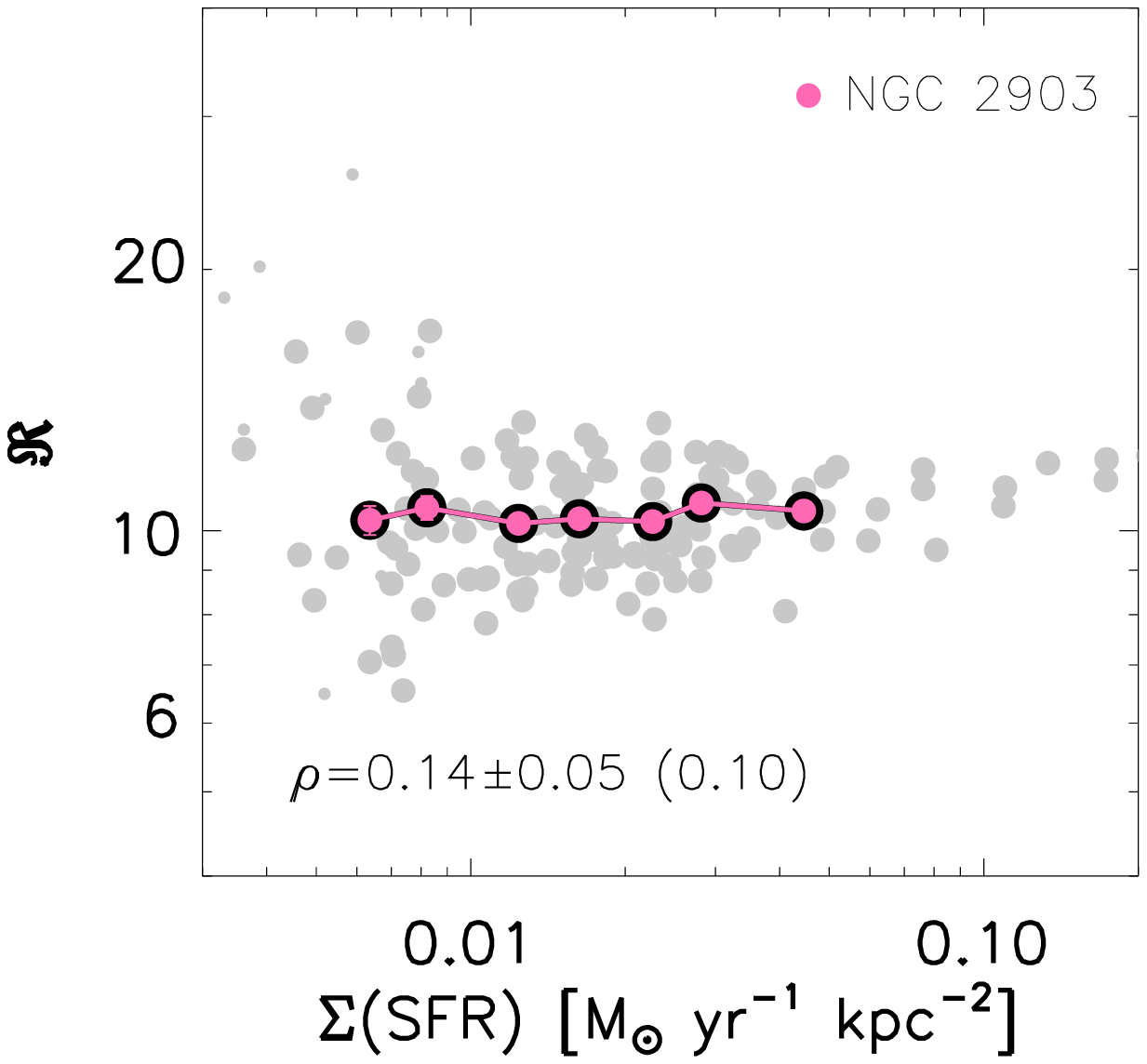}
\includegraphics[clip, trim=30mm 24mm 6mm 1mm,width=3.6cm]{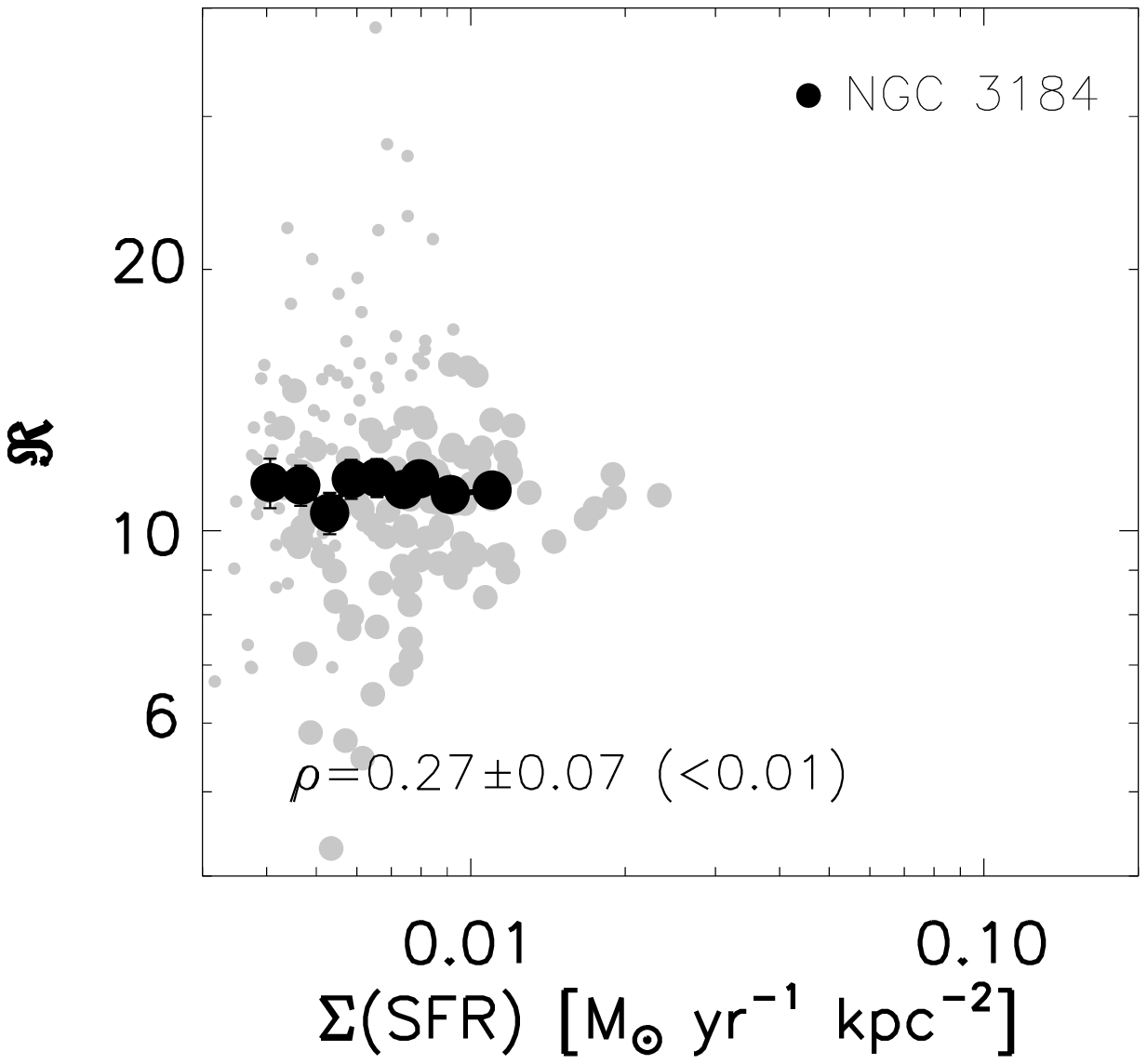}\\
\includegraphics[clip, trim=08mm 24mm 6mm 1mm,width=4.4cm]{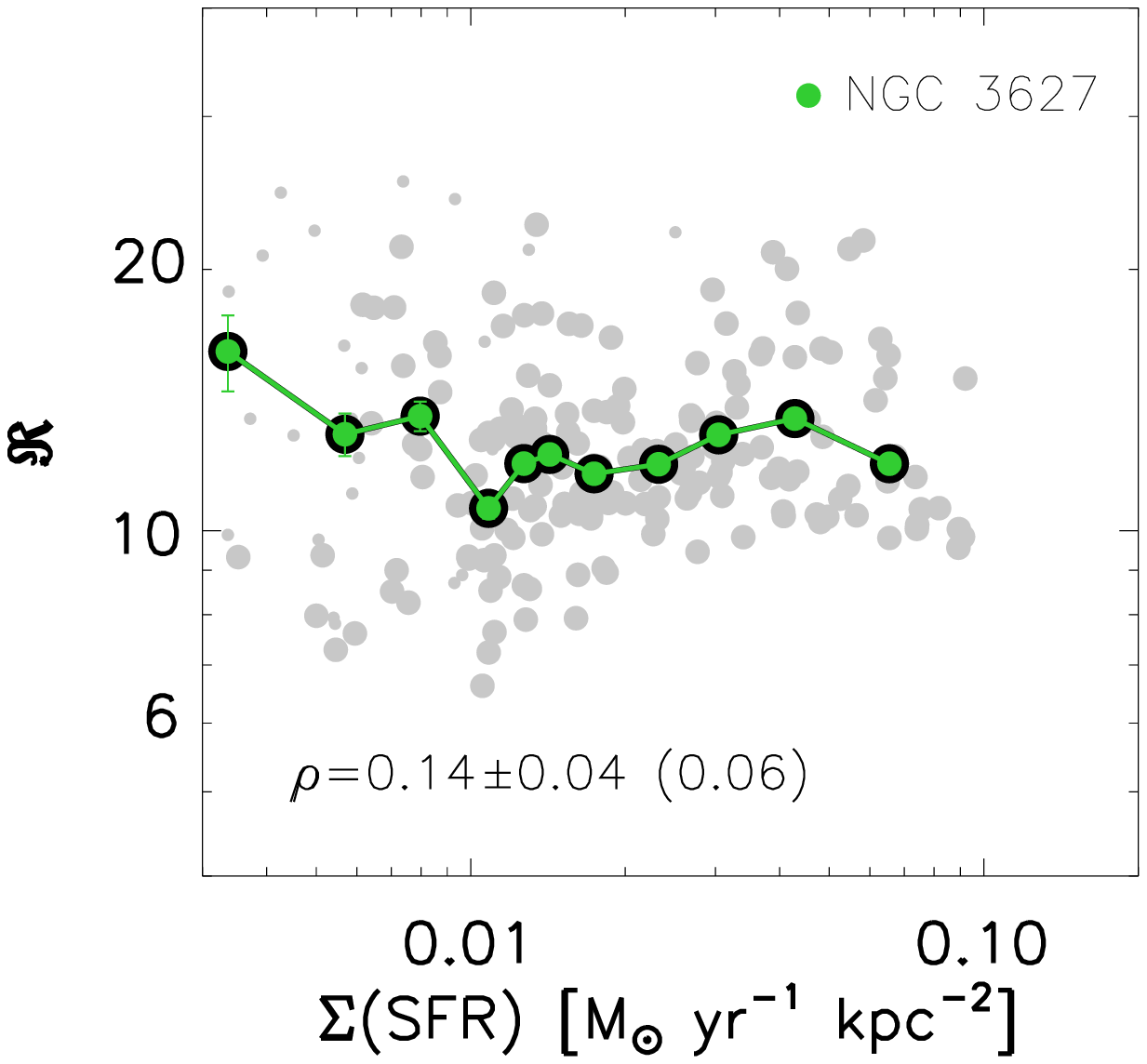}
\includegraphics[clip, trim=30mm 24mm 6mm 1mm,width=3.6cm]{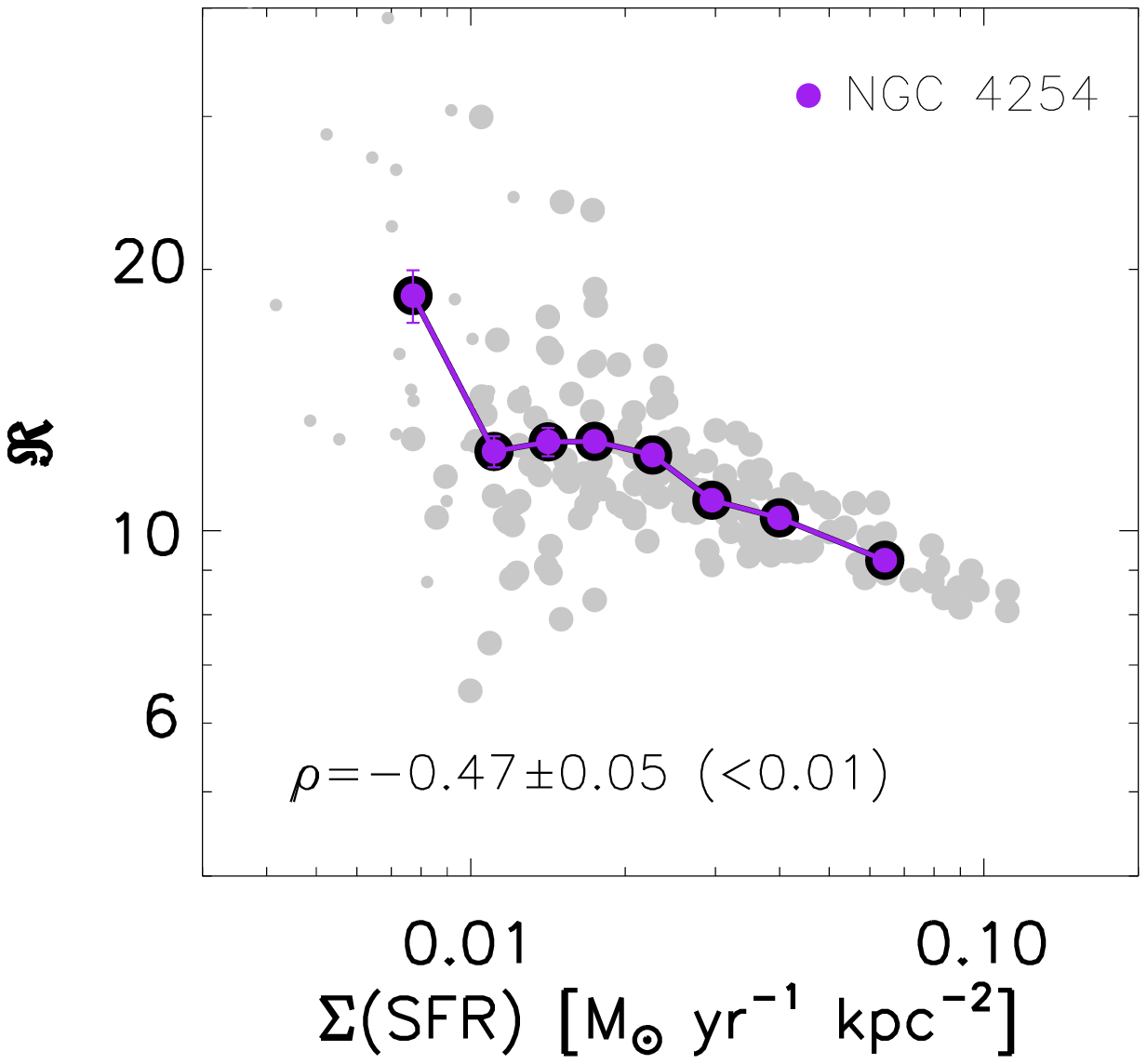}
\includegraphics[clip, trim=30mm 24mm 6mm 1mm,width=3.6cm]{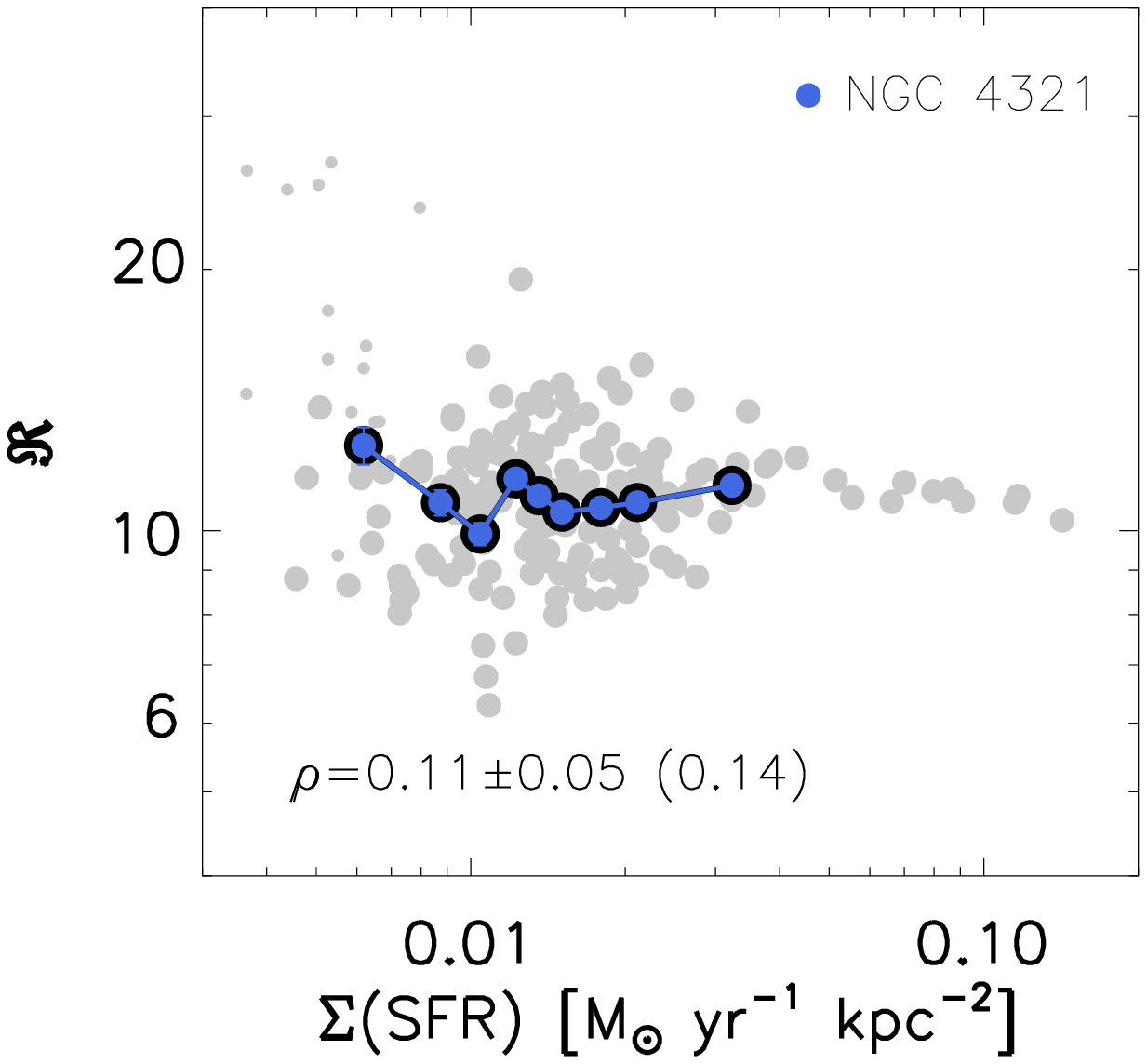}\\
\includegraphics[clip, trim=08mm 0mm 6mm 1mm,width=4.4cm]{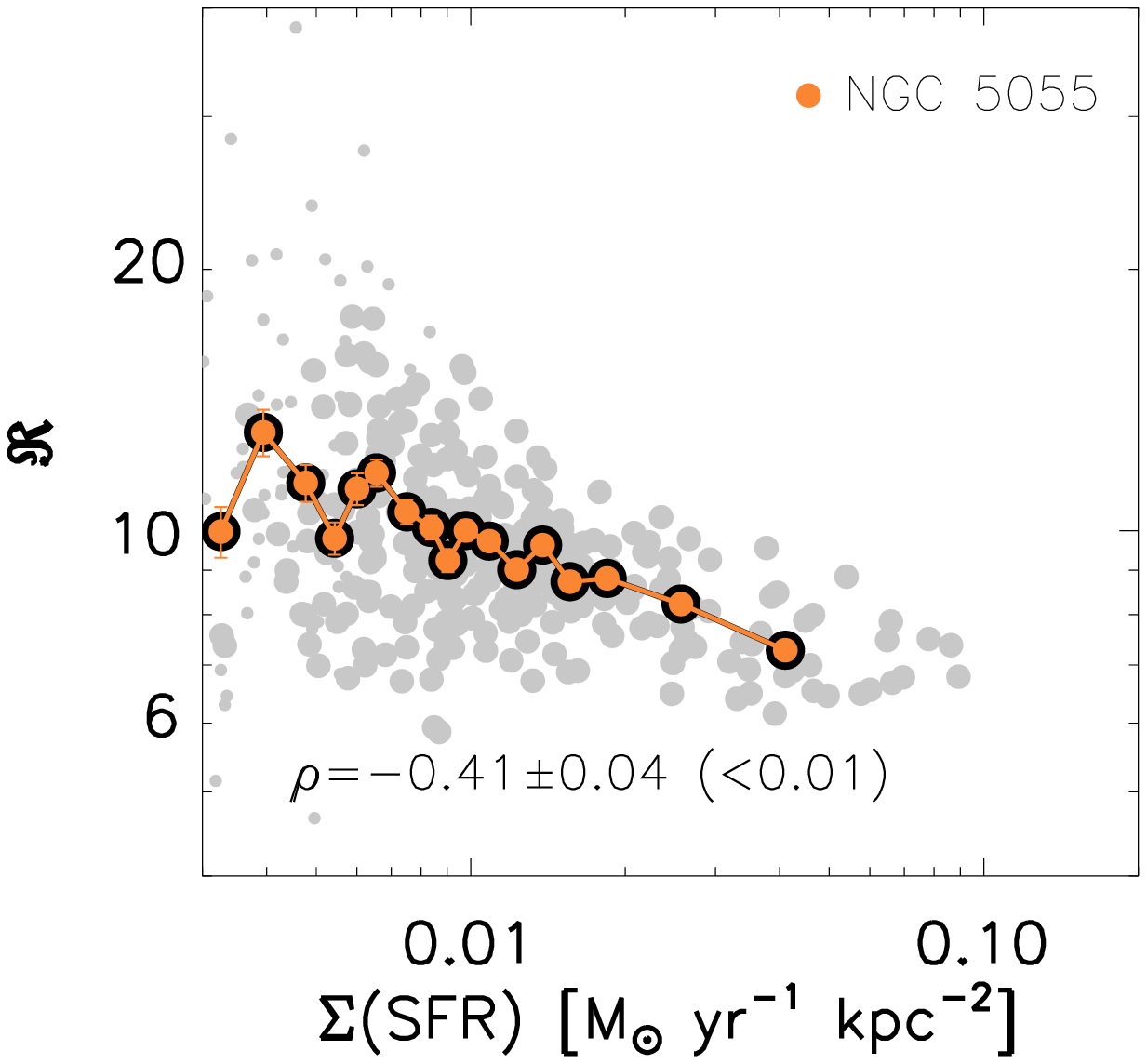}
\includegraphics[clip, trim=30mm 0mm 6mm 1mm,width=3.6cm]{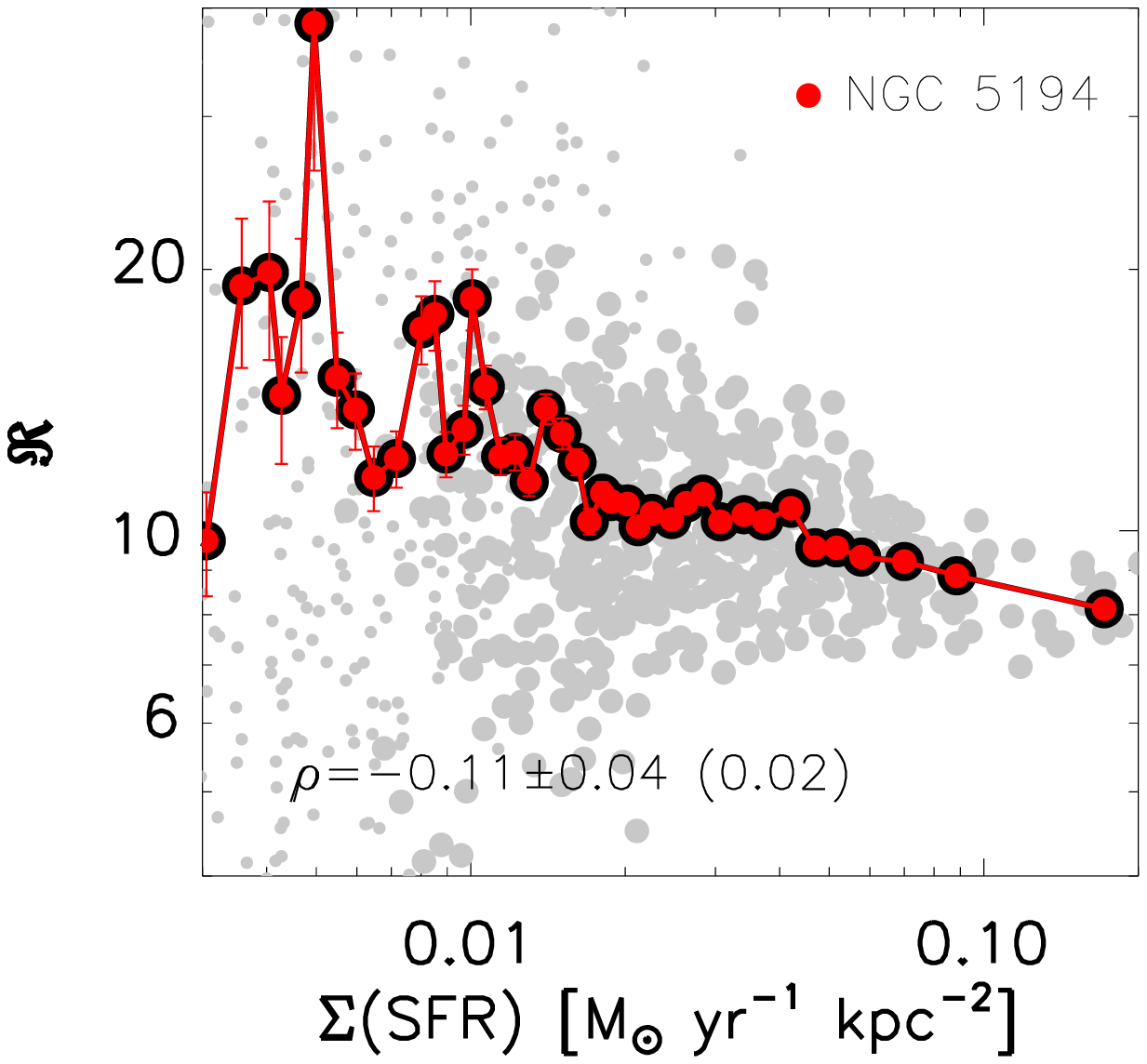}
\includegraphics[clip, trim=30mm 0mm 6mm 1mm,width=3.6cm]{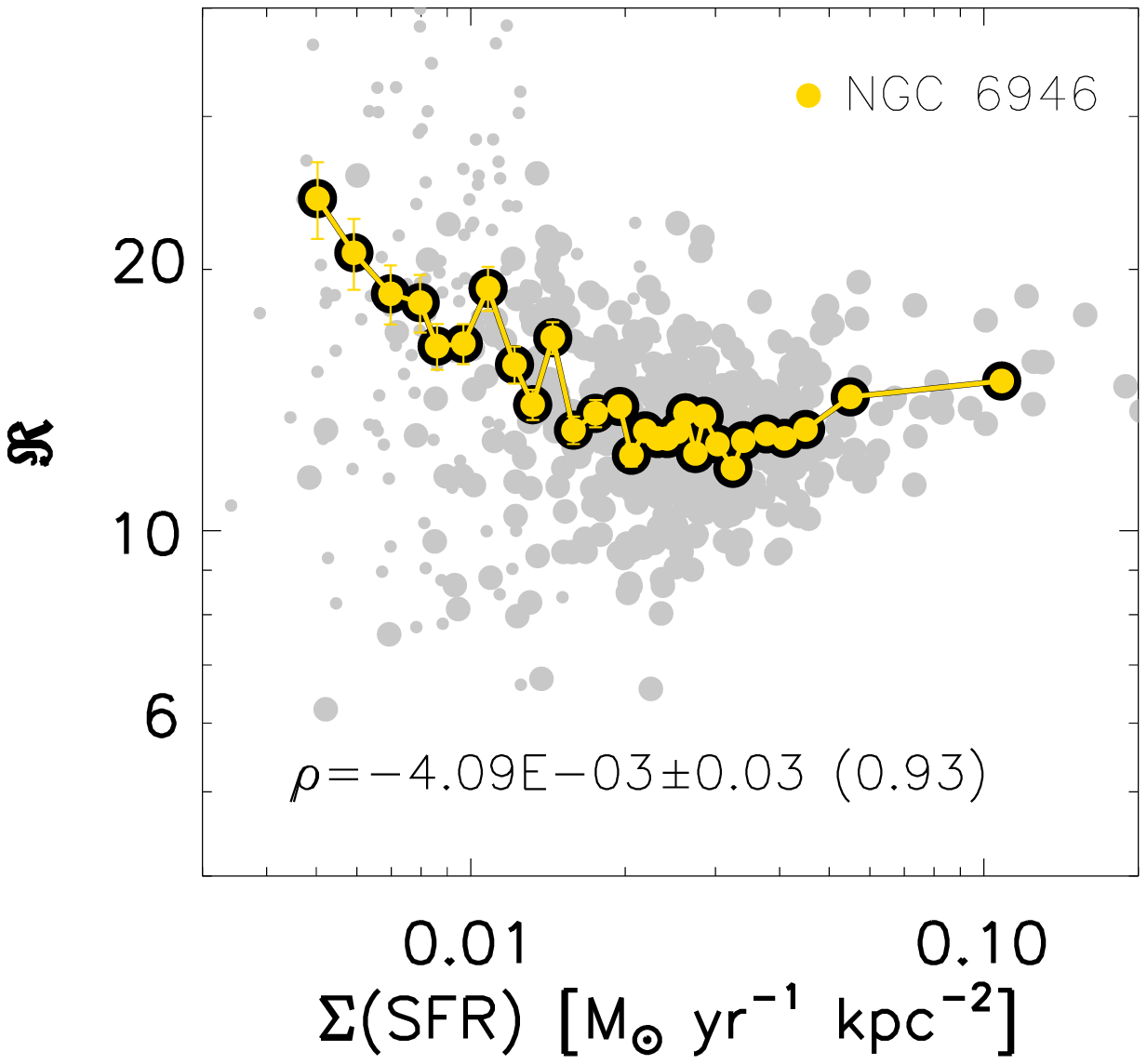}\\
\vspace{-8pt}
\contcaption{
$(c)$ --  Individual measurements of $\Re$ as a function of the star-formation rate surface density, \sigsfr.
}
\vspace{3pt}
\end{figure*}
\begin{figure*}
\centering
\includegraphics[clip, trim=08mm 24mm 6mm 1mm,width=4.4cm]{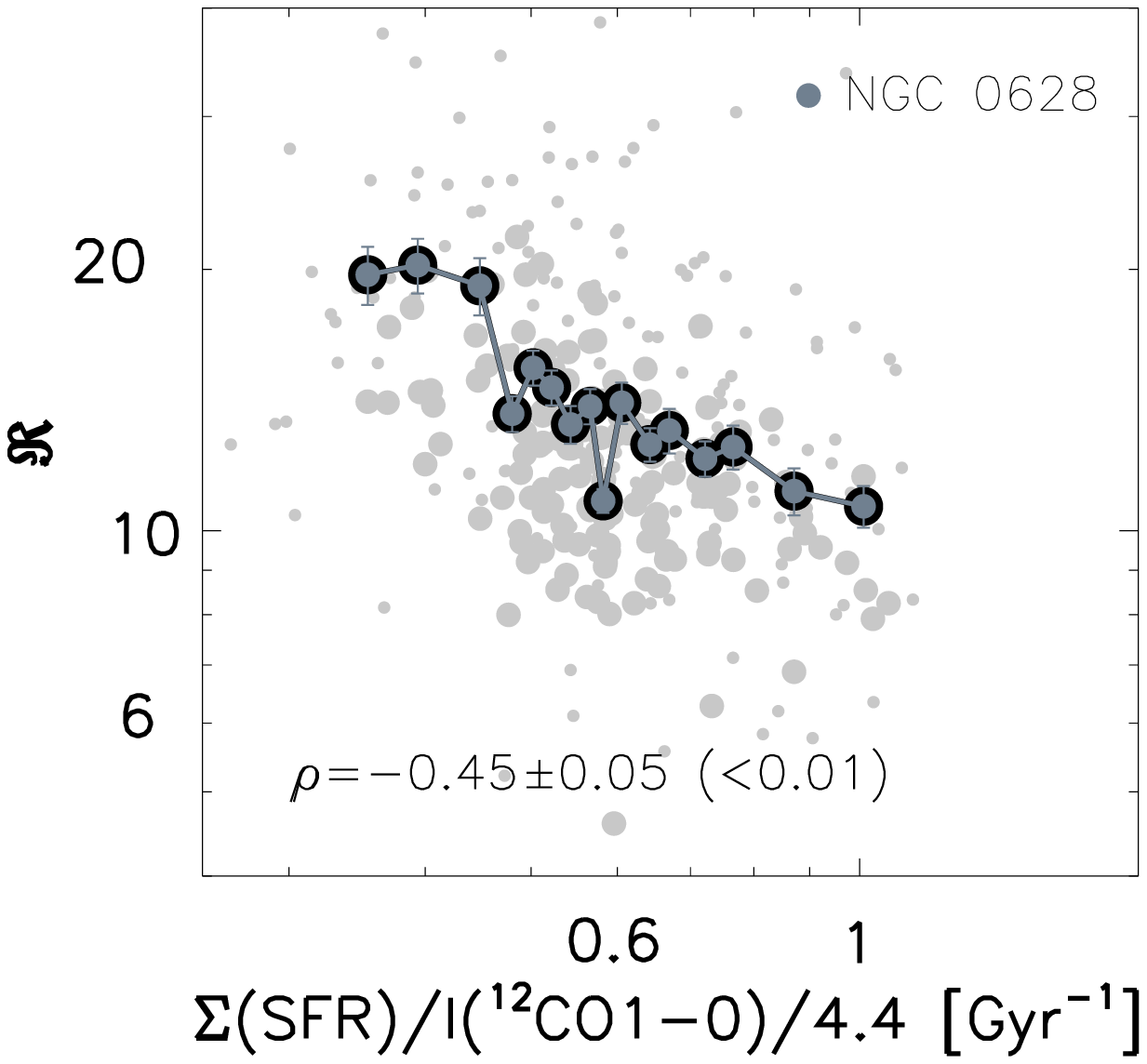}
\includegraphics[clip, trim=30mm 24mm 6mm 1mm,width=3.6cm]{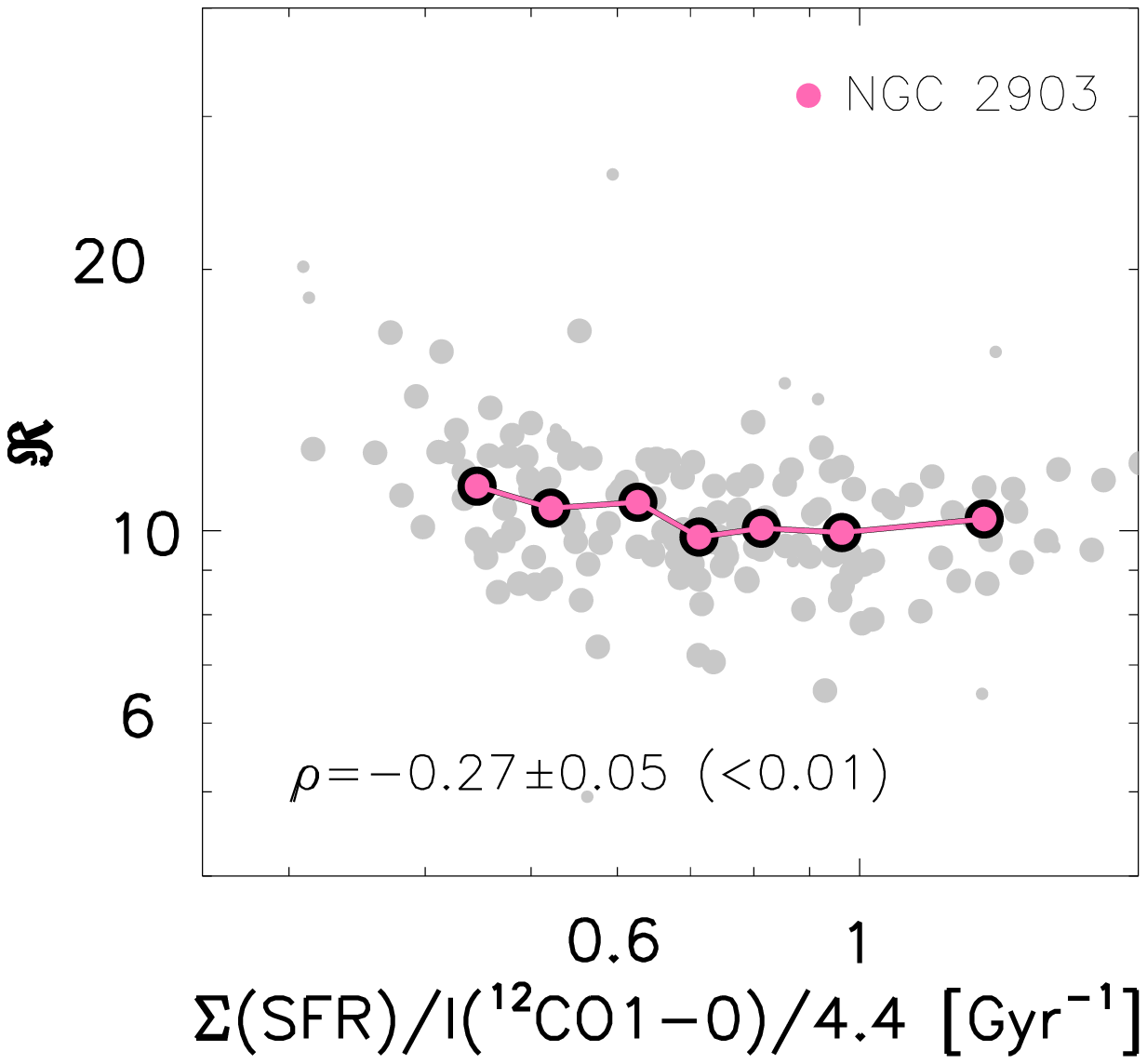}
\includegraphics[clip, trim=30mm 24mm 6mm 1mm,width=3.6cm]{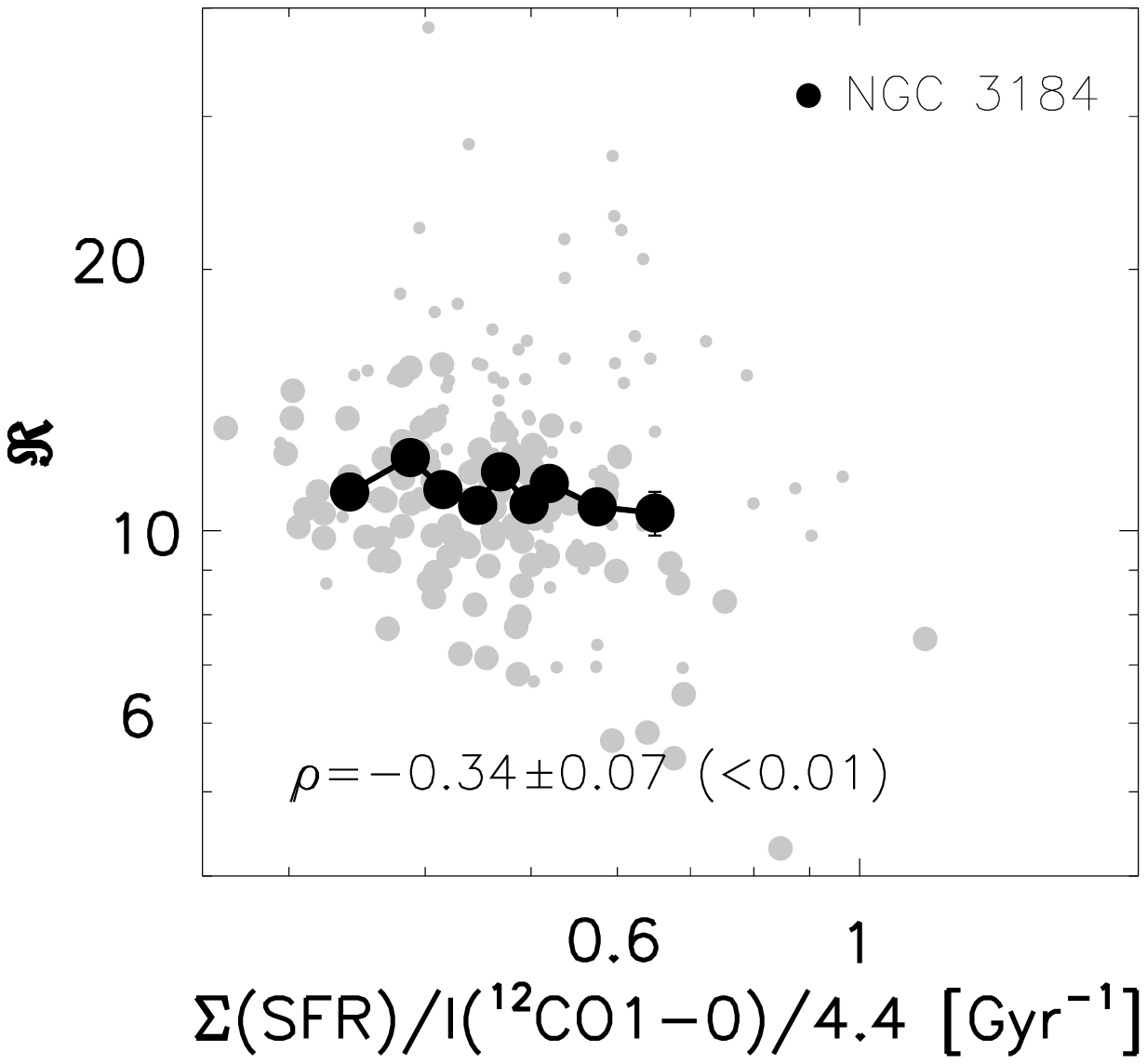}\\
\includegraphics[clip, trim=08mm 24mm 6mm 1mm,width=4.4cm]{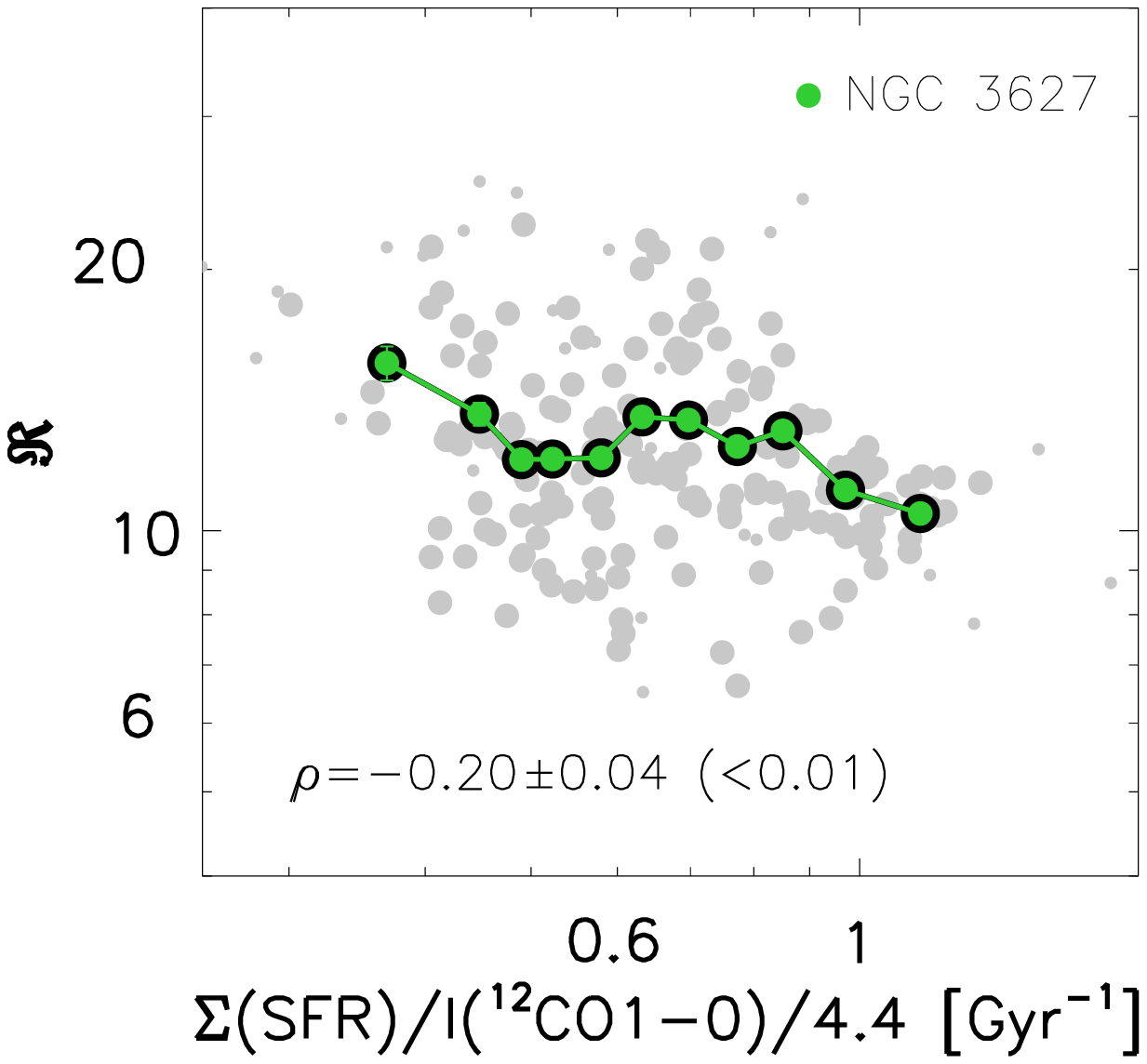}
\includegraphics[clip, trim=30mm 24mm 6mm 1mm,width=3.6cm]{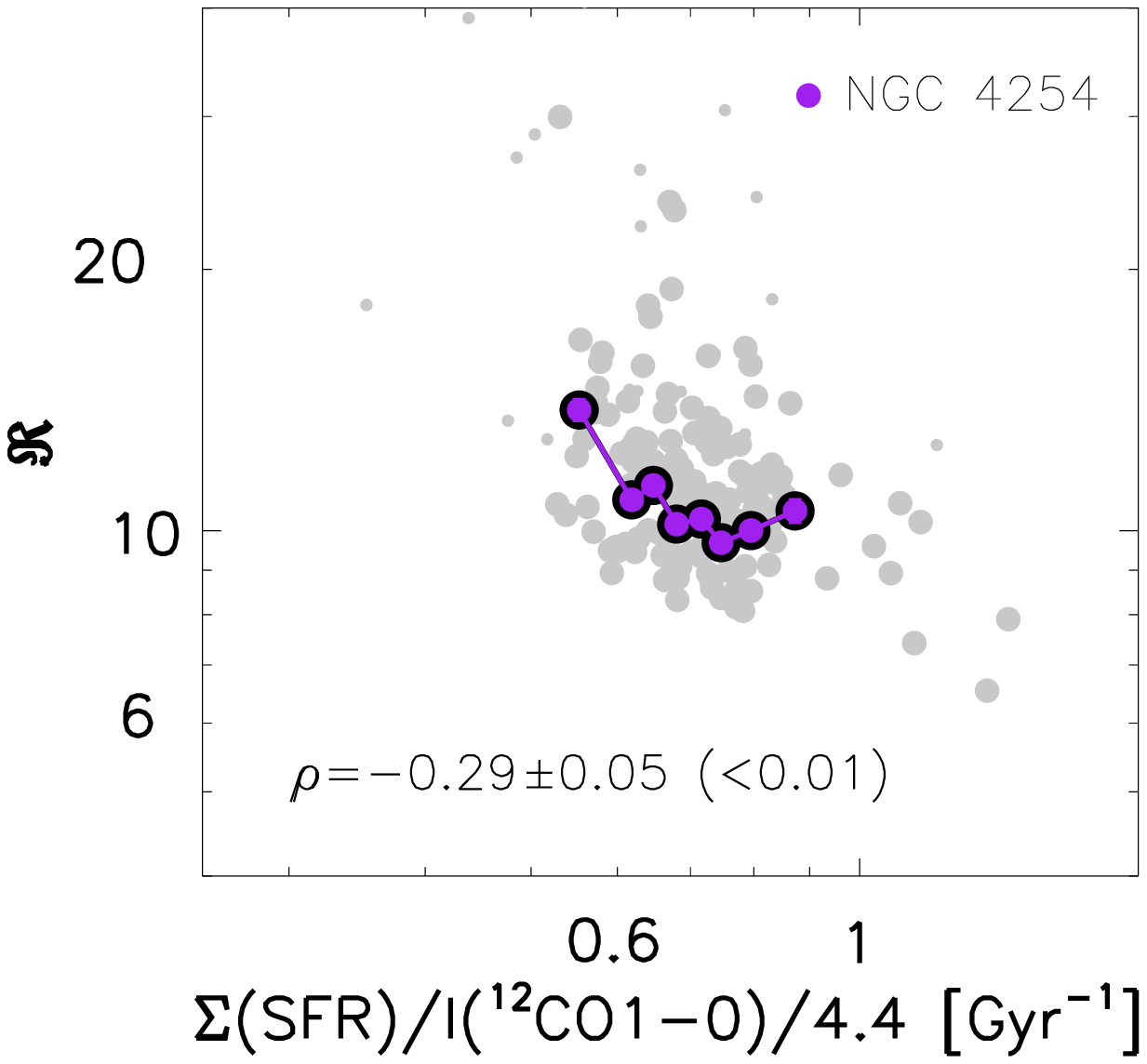}
\includegraphics[clip, trim=30mm 24mm 6mm 1mm,width=3.6cm]{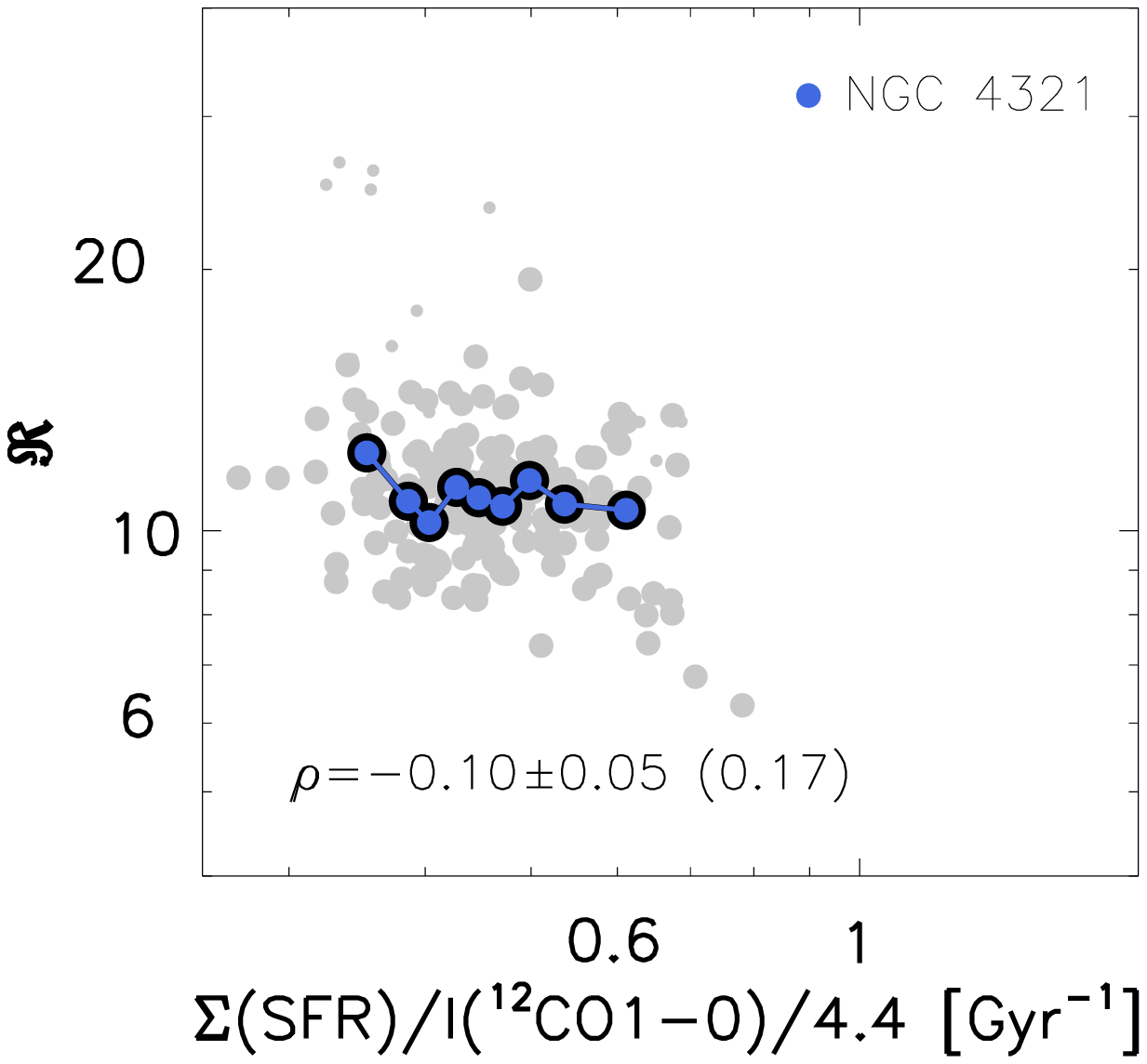}\\
\includegraphics[clip, trim=08mm 0 6mm 1mm,width=4.4cm]{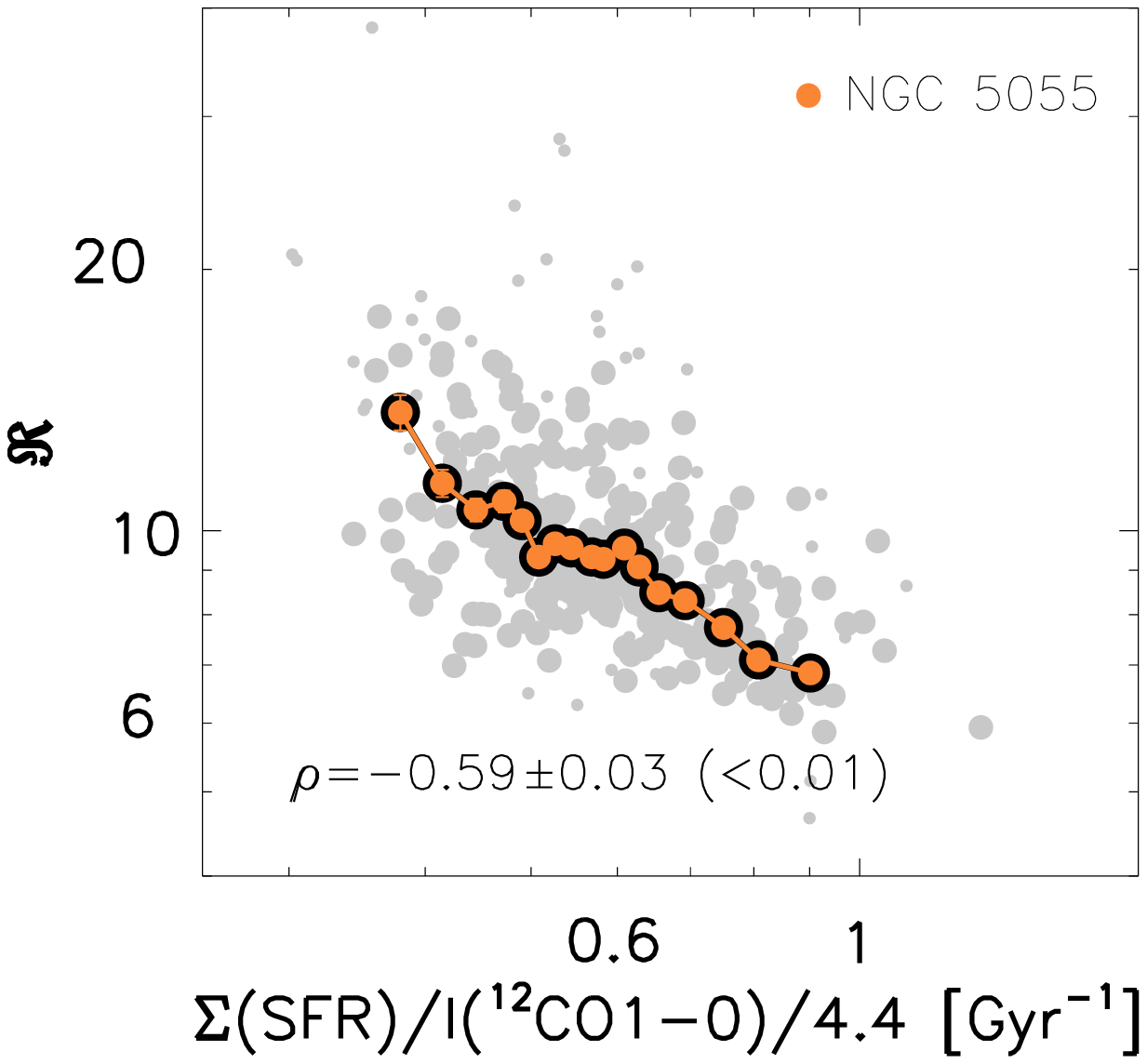}
\includegraphics[clip, trim=30mm 0 6mm 1mm,width=3.6cm]{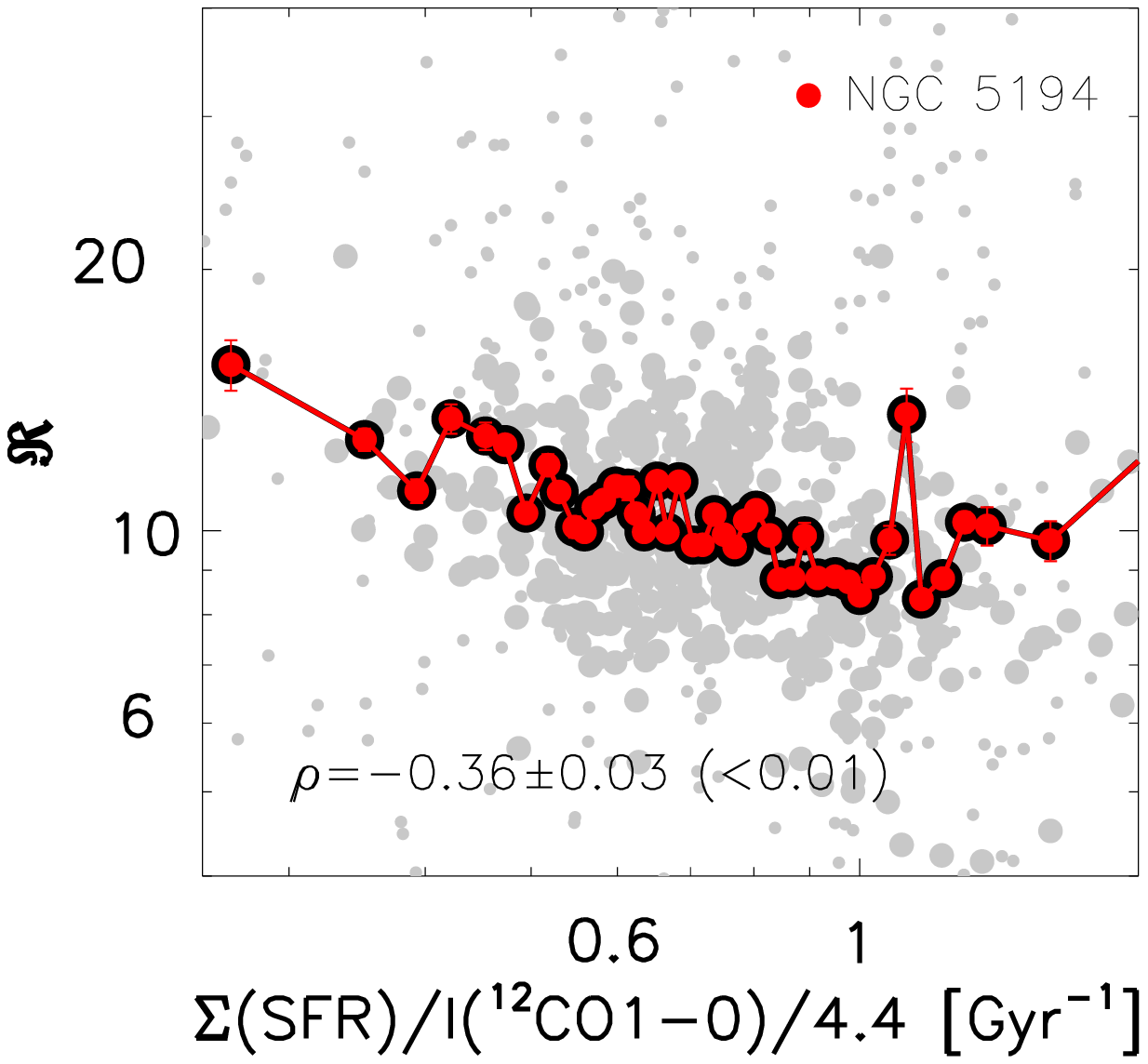}
\includegraphics[clip, trim=30mm 0 6mm 1mm,width=3.6cm]{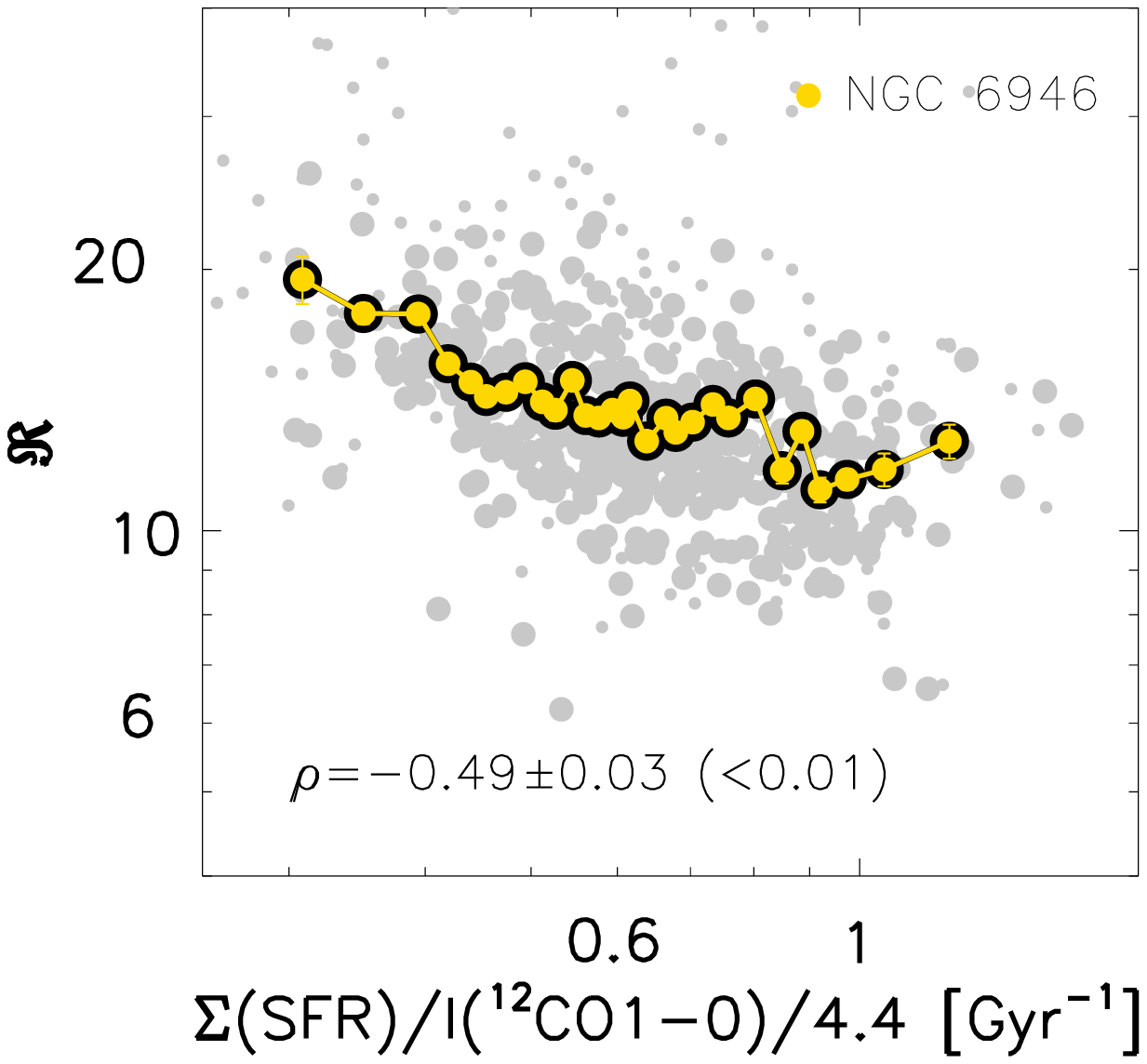}\\
\vspace{-8pt}
\contcaption{
$(d)$ --  Individual measurements of $\Re$ as a function of the \sigsfr/CO(1-0) ratio ($\propto$ star-formation efficiency).
}
\vspace{3pt}
\end{figure*}

In this section, we investigate how $\Re$ correlates with
physical parameters such as the line ratio CO(2-1)/CO(1-0),
the IR colour PACS70/PACS160, the star-formation rate
surface density (\sigsfr), and the \sigsfr/CO(1-0) ratio
($\propto$ star-formation efficiency).
We use the CO(2-1)/CO(1-0) ratio as a probe of gas
conditions (temperature, density, opacity), and
the IR colour PACS70/160 as an indicator of dust temperature
that could be coupled to the gas temperature in the
case where densities are high enough that collisions
between dust grains and the gas particles lead to equal
gas and dust temperatures.

Figure~\ref{fig:allcorrav} shows values averaged over entire
galaxies while Figure~\ref{fig:allcorr} shows the individual
$\sim$kpc-sized regions as well as binned values for each galaxy.
To produce these measurements, we first sample our maps with
a hexagonal grid of spacing 13.5\,arcsec (half of the spatial resolution).
For total averages (Fig.~\ref{fig:allcorrav}), we measure the
average of each observed quantity (CO intensity, PACS flux
density, \sigsfr, etc.) by considering all sampling points, and
for ratios, we divide these averages with each other.
For the bins, we construct irregular bins of the
physical quantities such that each bin contains $20$ sampling
points. Similarly (Fig.~\ref{fig:allcorr}), we measure the intensity
of each CO line and its error, either by considering the individual
sampling points (for the kpc-size regions) or by averaging
intensities of the sampling points falling in the bin (binned averages).
We then calculate $\Re$ by taking the ratio of these intensities.
We do not mask sampling points with low signal-to-noise when
doing the binned averages.
Error bars are calculated as the dispersion in $\Re$
measurements obtained with a Monte-Carlo simulation.

Inspecting Fig.~\ref{fig:allcorrav}, we find that the global
averaged values of $\Re$ are not correlated with any quantity.
Looking at integrated galaxies, \cite{aalto-1995} find that
the warmest galaxies (F60/F100$\ge$0.7), which are absent
in our EMPIRE sample, display systematically high $\Re$
values ($>10$), while $\Re$ and F60/F100 do not correlate
for cold-intermediate IR colours (F60/F100$\le$0.6),
which is in line with our results.
Moreover, \cite{davis-2014} report a correlation between $\Re$
and \sigsfr for integrated early-type galaxies. The \sigsfr range
that they investigate is much larger than ours (5 orders of magnitude
as opposed to a dynamic range in the average \sigsfr of less
than 1\,dex in our case). This may be why we do not find a
significant trend with \sigsfr.
We discuss Figure~\ref{fig:allcorr} in detail next.

\begin{figure*}
\centering
\includegraphics[clip, trim=8mm 0 0 0,width=8cm]{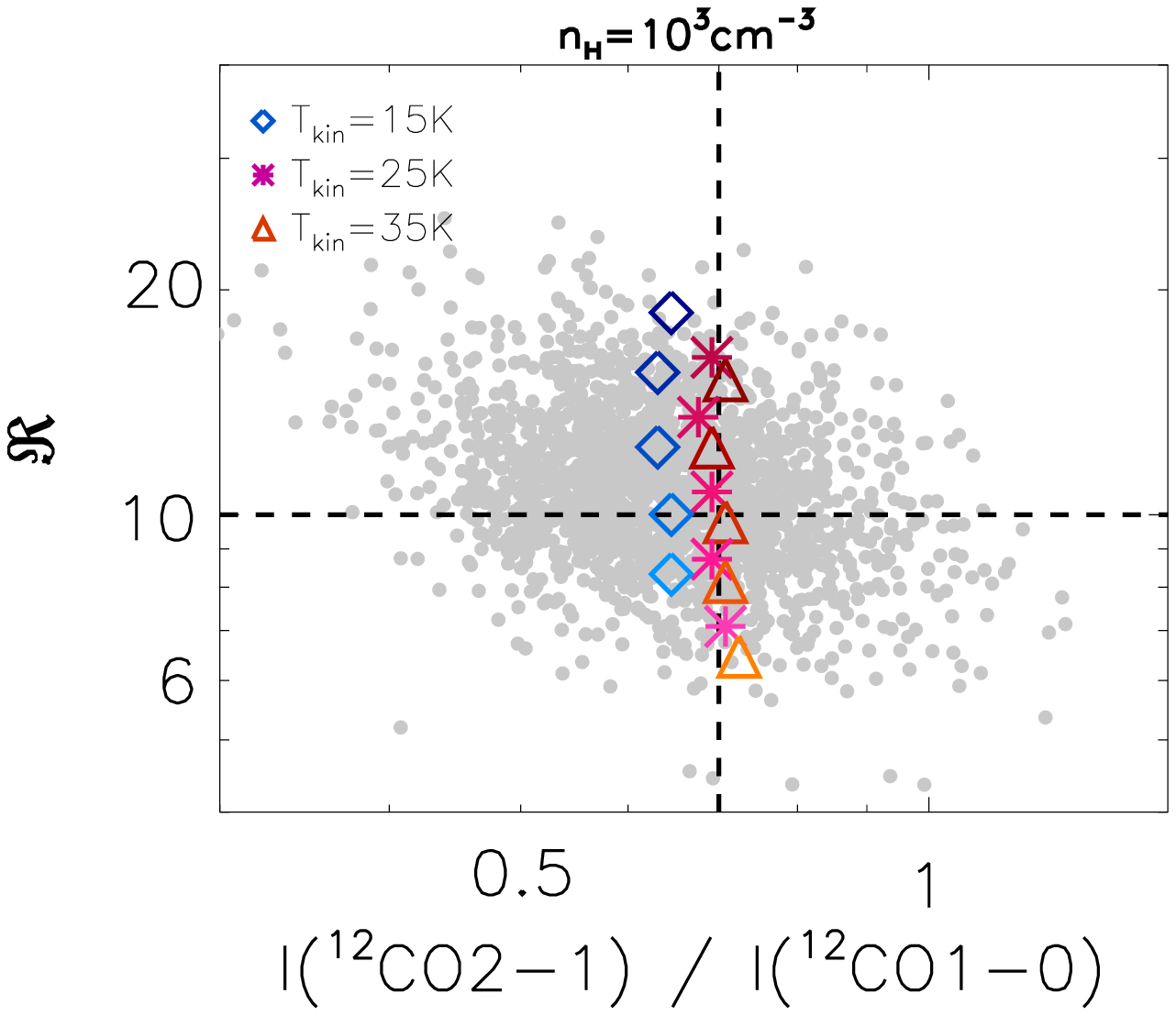}
\includegraphics[clip, trim=8mm 0 0 0,width=8cm]{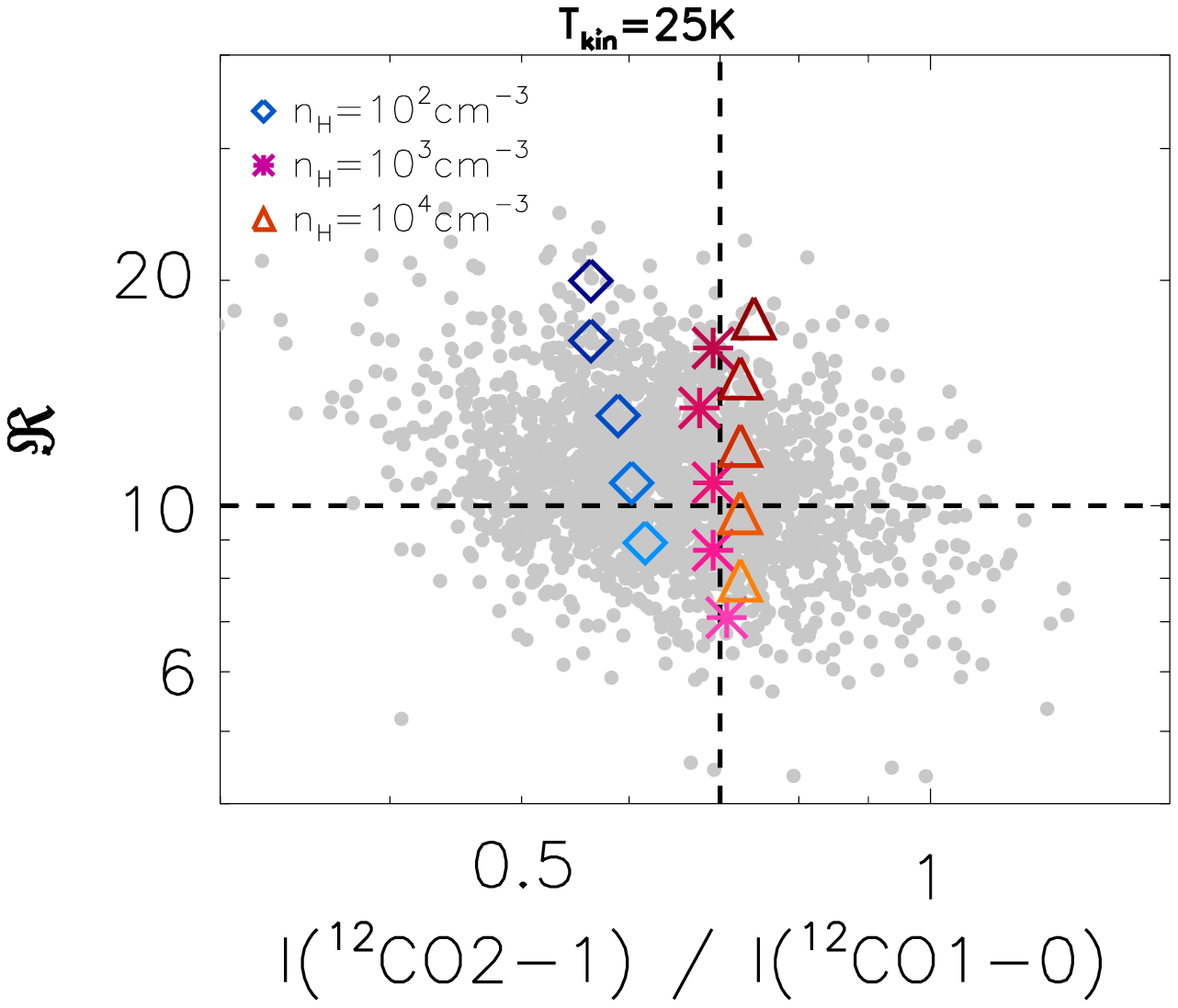}
\includegraphics[clip, trim=8mm 0 0 0,width=8cm]{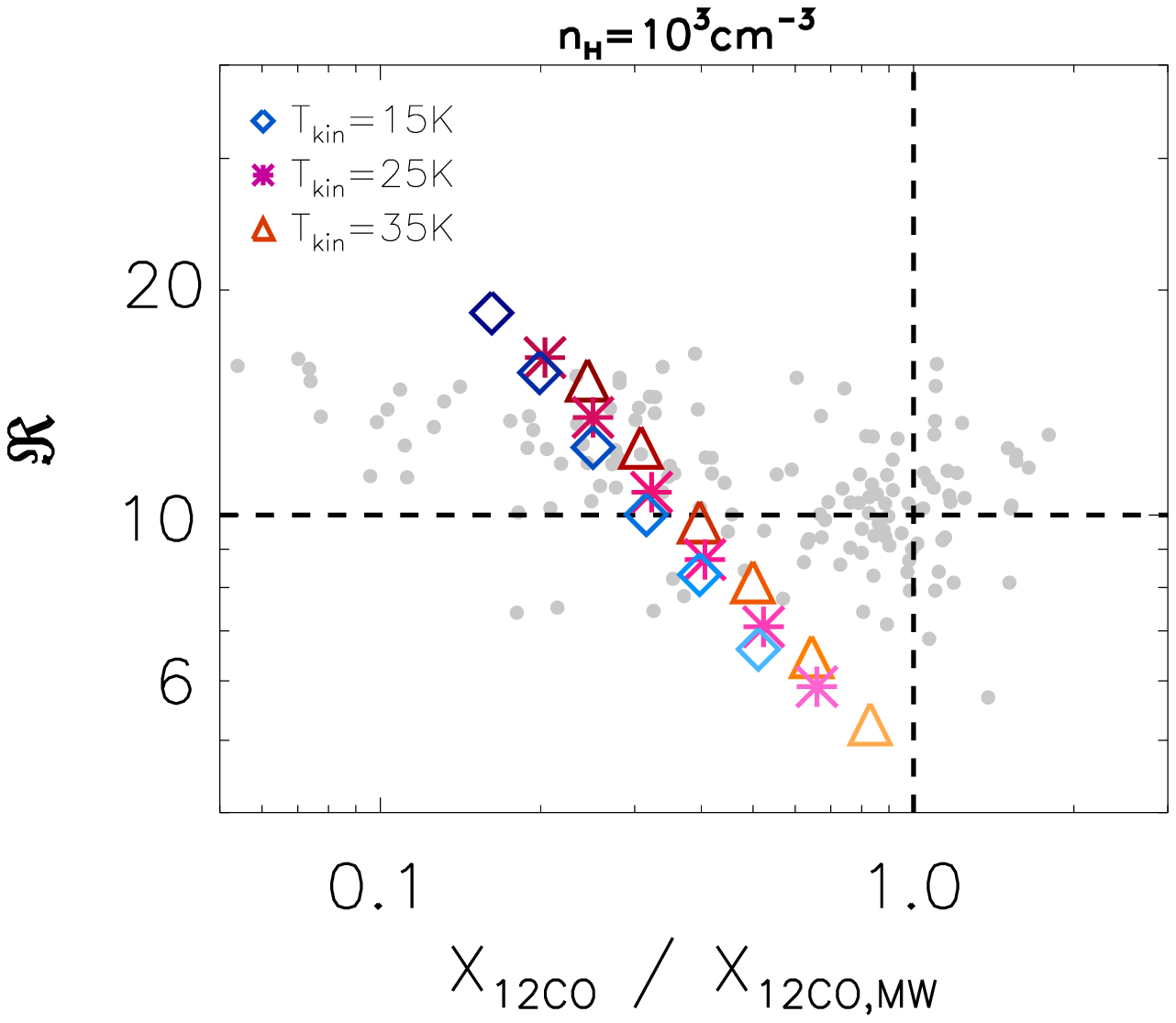}
\includegraphics[clip, trim=8mm 0 0 0,width=8cm]{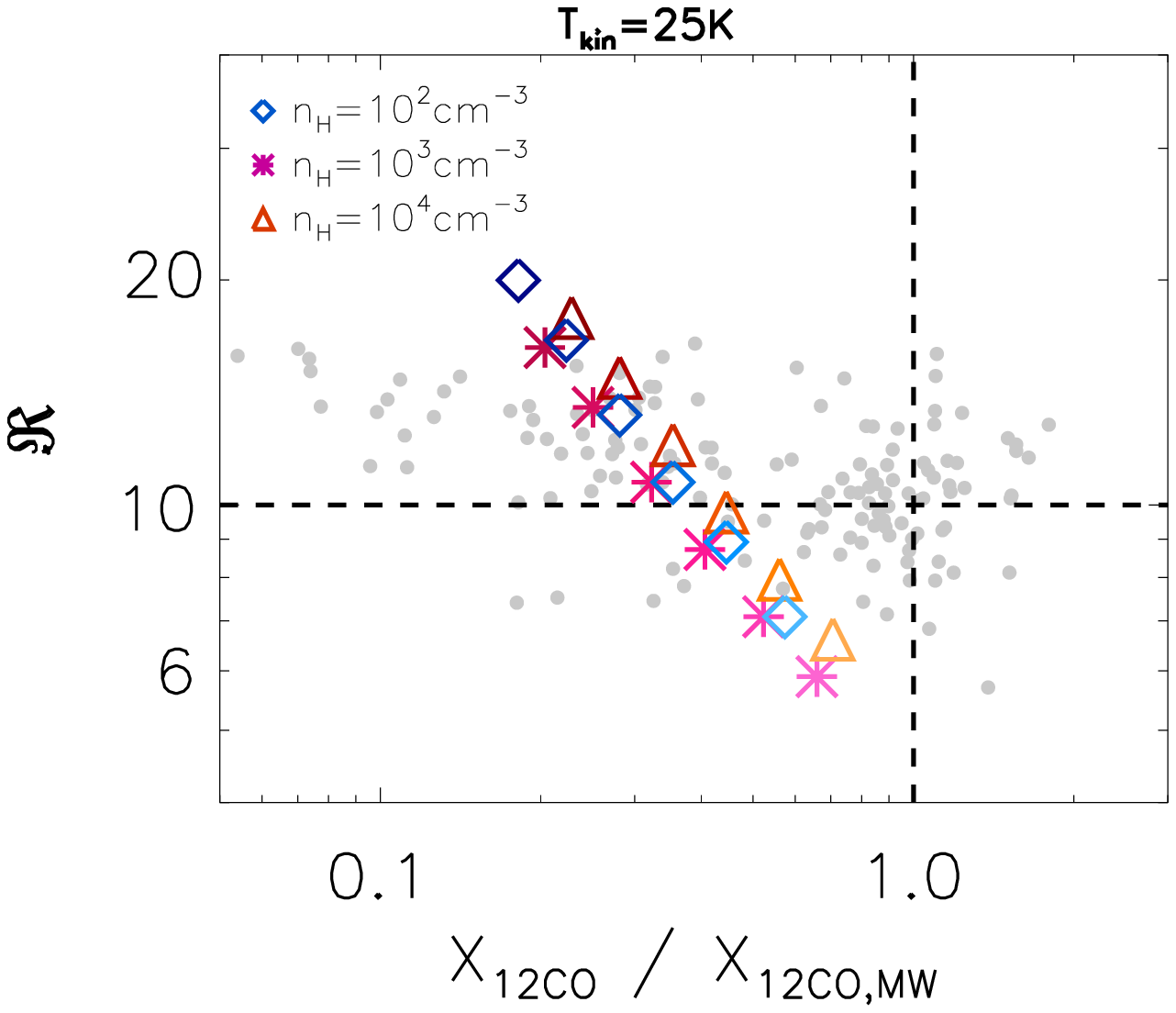}
\vspace{-8pt}
\caption{
Correlation between $\Re$ and the \cou/\com ratio (top)
and between $\Re$ and the $X_{\rm 12CO}$
conversion factor (bottom) predicted by the RADEX
grid of models from \protect\cite{leroy-2017}.
Observations are plotted in grey.
The models show the expected trend, in LTE with
filling factors of unity and fixed abundances, of $X_{\rm 12CO}$
being inversely proportional to $\Re$ (see section~\ref{sect:opacity}).
Optical depths $\tau_{12}$ are varied from 3 (dark-color symbols)
to 10 (light-color symbols) in steps of 0.1\,dex, while the ratio
$\tau_{13}/\tau_{12}$ is fixed to 0.03.
\underline{Left panels:} model predictions for kinetic temperatures of
$T_{\rm kin}=15$\,K (blue diamonds),
$T_{\rm kin}=25$\,K (purple asterisks), and
$T_{\rm kin}=35$\,K (red triangles).
The mean volume density is set to $10^3$\,cm$^{-3}$.
\underline{Right panels:} model predictions for mean volume densities of
$n_{\rm H}=10^2$\,cm$^{-3}$ (blue diamonds),
$n_{\rm H}=10^3$\,cm$^{-3}$ (purple asterisks), and
$n_{\rm H}=10^4$\,cm$^{-3}$ (red triangles).
The kinetic temperature is set to $25$\,K.
}
\label{fig:xcomodels}
\end{figure*}

{\em \underline{CO(2-1)/CO(1-0) intensity ratio (Fig.~\ref{fig:allcorr}a):}}
We find that $\Re$ systematically decreases with increasing
CO(2-1)/CO(1-0) ratio within galaxies. Moreover, for a given
value of CO(2-1)/CO(1-0), there are significant offsets in $\Re$
from galaxy to galaxy.
The \coi, \com and \cou lines have upper energy levels $E/k$
of 5.3\,K, 5.5\,K and 16.6\,K, and critical densities of
$2\times10^2$\,cm$^{-3}$, $1\times10^3$\,cm$^{-3}$ and
$6.7\times10^3$\,cm$^{-3}$ (for $T_{\rm kin}=20$\,K and
$\tau_{line}\simeq1$), respectively.
We note that the critical density is a function of optical depth
when line trapping effects are important \citep[e.g.,][]{scoville-1974,shirley-2015}.
In the case of $^{12}$CO, the \cou-to-\com ratio would not
only be sensitive to the temperature and density of the gas
but also to the optical depth of $^{12}$CO which may play
a role in dictating where sub-thermal excitation happens
\citep{penaloza-2017}.
On kpc-scales, our sample of galaxies span a range
of CO(2-1)/CO(1-0) ratios between 0.3 and 2 and show a weak-to-moderate
anti-correlation between $\Re$ and CO(2-1)/CO(1-0), indicating
that $\Re$ decreases for increasing temperature/density/opacity
(see Fig.~\ref{fig:xcomodels} for model predictions).
Both $\Re$ and CO(2-1)/CO(1-0), i.e. the y-axis and x-axis
in Fig.~\ref{fig:allcorr}a, are correlated via \com.
We investigate if the correlations are real or not by performing
two Monte-Carlo tests that are detailed in Appendix~\ref{sect:mctest}.
Those tests indicate that the observed scatter in the line ratios
is physical, it cannot be explained purely by noise.
They also indicate that the correlation coefficients that we
measure are robust and not driven by the correlated axes.
Hence both the variations in the line ratios and the
observed correlations are real.

{\em \underline{PACS70/160 IR colour (Fig.~\ref{fig:allcorr}b):}}
All galaxies in our sample span a similar range of
PACS70/PACS160 values ($0.2-0.8$). This IR colour can be
used as a proxy for the dust temperature. We observe a weak-to-moderate
anti-correlation of $\Re$ with PACS70/PACS160 for 5 galaxies: 
NGC\,0628, NGC\,4254, NGC\,5055, NGC\,5194, and NGC\,6946.
The high PACS70/PACS160 and low $\Re$ values are
found in the centre of those galaxies, except for NGC\,6946,
where they are found in more diffuse regions of the disc.
Comparing ratio behaviors between Fig.~\ref{fig:allcorr}a and
Fig.~\ref{fig:allcorr}b, temperature effects may be at play in
those five galaxies, but it is probably not the dominant/unique
condition affecting $\Re$.

{\em \underline{SFR surface density (Fig.~\ref{fig:allcorr}c):}}
All galaxies in our sample span a similar range of \sigsfr
values, with NGC\,0628 and NGC\,3184 being a bit less active.
We observe a weak correlation between $\Re$ and \sigsfr
for NGC\,3184 and a moderate anti-correlation for NGC\,4254
and NGC\,5055.
Those two galaxies have $\Re$ profiles steadily increasing
with radius and also show an anti-correlation with
the dust temperature.
For NGC\,3627, the behavior of $\Re$ with \sigsfr is not
immediately apparent from the radial profiles (Fig.~\ref{fig:ratioprofiles}),
probably because the star-forming knots at the end of the bar
and the centre have high \sigsfr but different $\Re$ values.
For the CARMA STING survey, \cite{cao-2017} also find
no trend on global scales and moderately decreasing $\Re$
with increasing \sigsfr for some galaxies, their galaxies
probing the range \sigsfr$\simeq 0.01-1.0$\sigsfrunit.

{\em \underline{\sigsfr/CO(1-0) (Fig.~\ref{fig:allcorr}d):}}
All galaxies in our sample span a similar range of \sigsfr/CO(1-0)
values ($0.3-2$\,Gyr$^{-1}$), which we have normalized
such that they have unit of star-formation efficiency
(SFE$_{\rm mol}$) under a constant, Galactic \aco value.
We find that $\Re$ is moderately anti-correlated with
\sigsfr/CO(1-0) within all galaxies, i.e. that $\Re$ is
lower at high efficiencies, though the two quantities
are correlated by construction.
As for the CO(2-1)/CO(1-0) intensity ratio, we test those
correlations with Monte-Carlo simulations that are detailed
in Appendix~\ref{sect:mctest}. The test indicates that
the scatter in the \sigsfr/CO(1-0) values is physical and cannot be
purely explained by noise. The correlation coefficients
are also robust and highest for NGC\,5055 and NGC\,6946.
Previous works have suggested that variations in the
star-formation efficiencies of massive galaxies could be
linked to, e.g., a change in the relative fractions of diffuse
and dense molecular gas \citep{saintonge-2012,shetty-2014},
or to enhanced CO excitation in galaxy centres \citep{leroy-2013}.
Both effects have implications on the gas opacity
and, though moderate, our trends indicate that the optical
depth of CO, through $\Re$, could indeed account for some
changes in star-formation efficiencies.

We also explored how $\Re$ correlates with UV/TIR (not shown).
This ratio can be viewed as an indicator of the
visible/obscured star-formation activity and, for a given
geometry, of the ISM photodissociation/shielding ability.
The EMPIRE galaxies span a wide range of values for
the UV/TIR luminosity ratio. However, we find no clear
correlation between $\Re$ and UV/TIR within galaxies.

Overall, most of the trends or lack of trends of $\Re$ with
the physical parameters (Fig.~\ref{fig:allcorrav} and
Fig.~\ref{fig:allcorr}) resemble the trends with galactocentric
radius (Fig.~\ref{fig:ratioprofiles}). None of the parameters
is revealing a strong positive or negative correlation,
indicating that the kpc-scale resolution may be too coarse
to isolate and identify the local effects affecting $\Re$.

\begin{table}
  \caption{New average \aco values using \com intensities
  and comparison to original values from \protect\cite{sandstrom-2013}.
  Standard deviations of values taken in the maps are indicated in dex
  in parenthesis. Helium is included.} 
\begin{center}
\begin{tabular}{lcc}
    \hline\hline
     \vspace{-8pt}\\
    \multicolumn{1}{l}{} & 
    \multicolumn{1}{c}{$\alpha_{\rm CO, new}$} &
    \multicolumn{1}{c}{$\alpha_{\rm CO, new}$/$\alpha_{\rm CO, S13}$} \\
    \hline
	{NGC\,0628}	& 3.6~~(0.2)		& 0.84~~(0.04)  \\
	{NGC\,3184}	& 4.3~~(0.3)		& 0.73~~(0.03)  \\
	{NGC\,3627}	& 0.7~~(0.2)		& 0.59~~(0.03)  \\
	{NGC\,4254}	& 4.3~~(0.1)		& 0.98~~(0.02)  \\
	{NGC\,4321}	& 1.3~~(0.3)		& 0.79~~(0.04)  \\
	{NGC\,5055}	& 3.1~~(0.2)		& 0.86~~(0.04)  \\
	{NGC\,6946}	& 1.2~~(0.2)		& 0.97~~(0.05)  \\
	{all galaxies}	& 2.3~~(0.4)		& 0.84~~(0.08)  \\
    \hline \hline
\end{tabular}
\label{table:xconew}
\end{center}
\end{table}

\begin{figure}
\centering
\includegraphics[clip,trim=8mm 5mm 0 0,width=8.4cm]{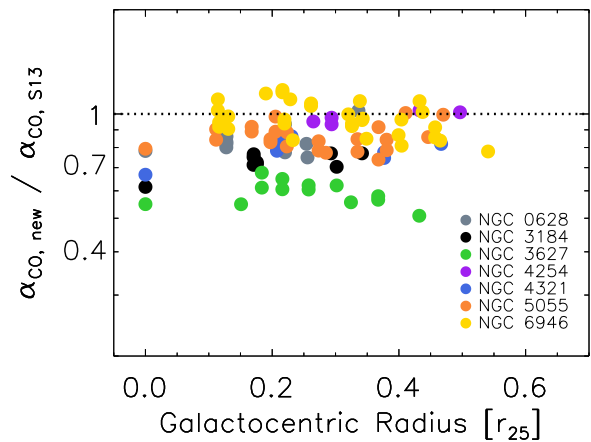}
\caption{
Pixel-by-pixel comparison of the \aco values
from \protect\cite{sandstrom-2013}
and those derived in this paper.
There are no large trends within galaxies
but the new \aco values are slightly lower
on average.}
\label{fig:xconew}
\end{figure}

\begin{figure*}
\centering
\includegraphics[clip, trim=08mm 24mm 6mm 2mm,width=6.6cm]{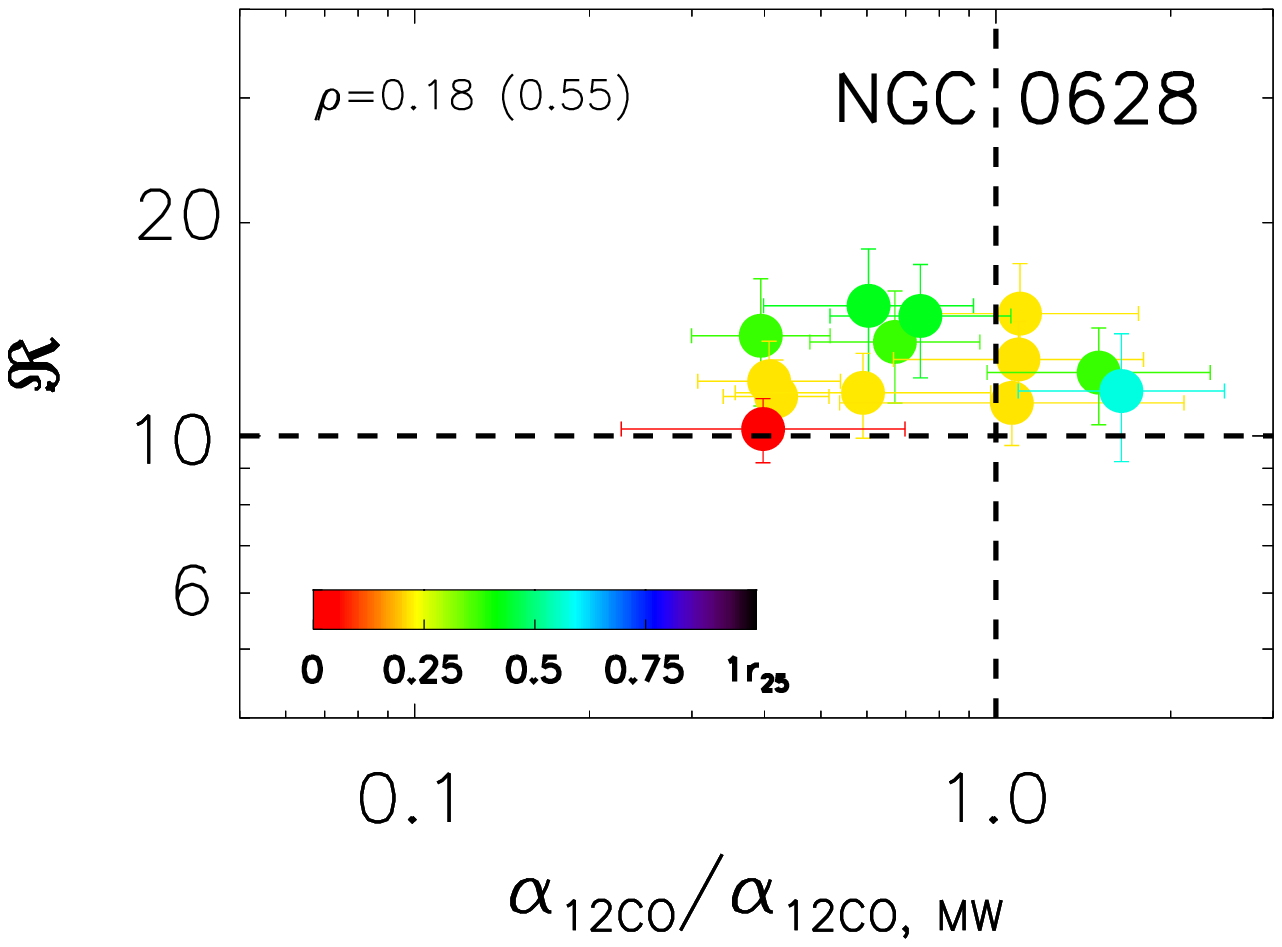}
\hspace{5.2cm}
\includegraphics[clip, trim=37mm 24mm 6mm 2mm,width=5.2cm]{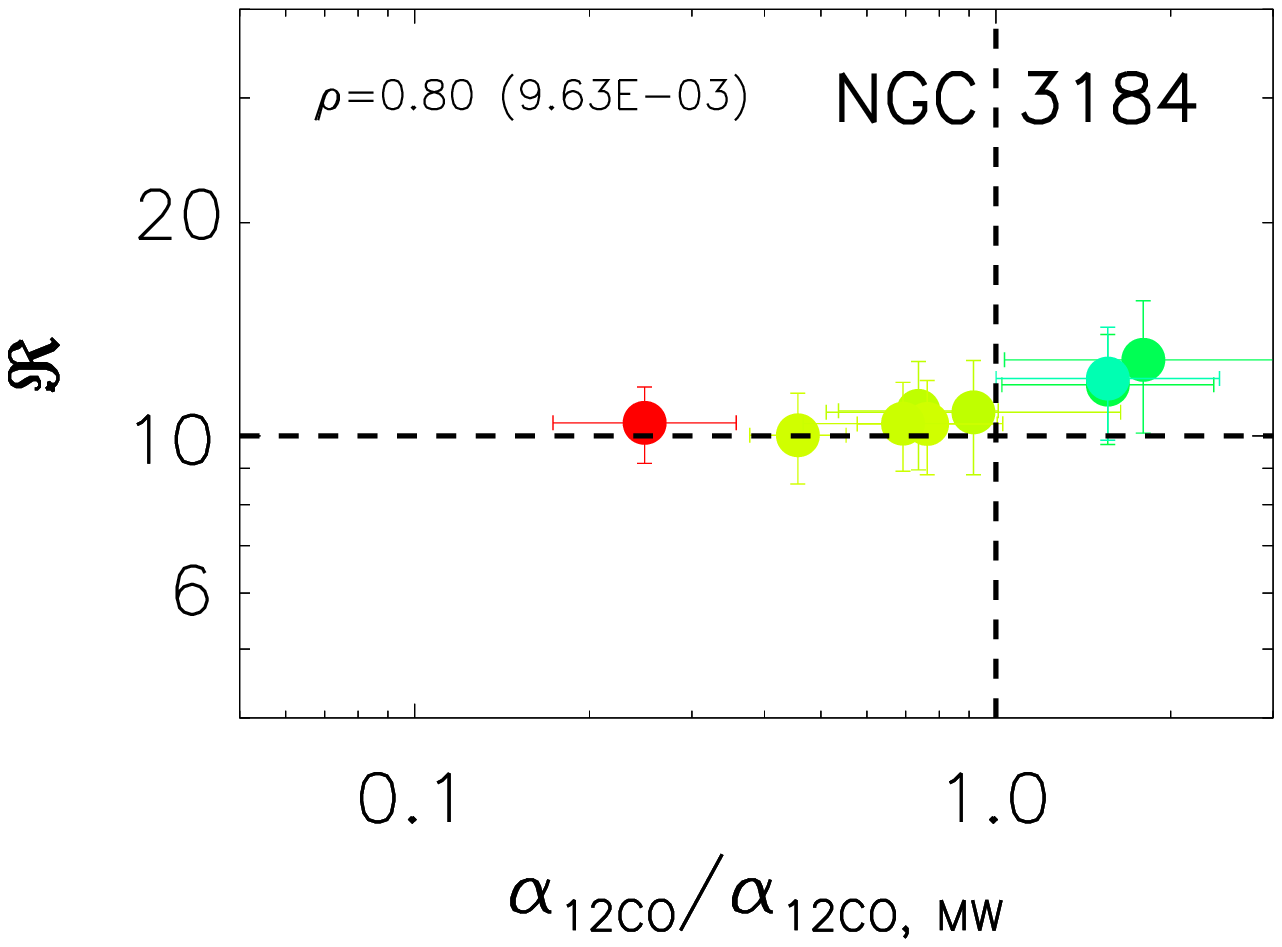}\\
\includegraphics[clip, trim=08mm 24mm 6mm 2mm,width=6.6cm]{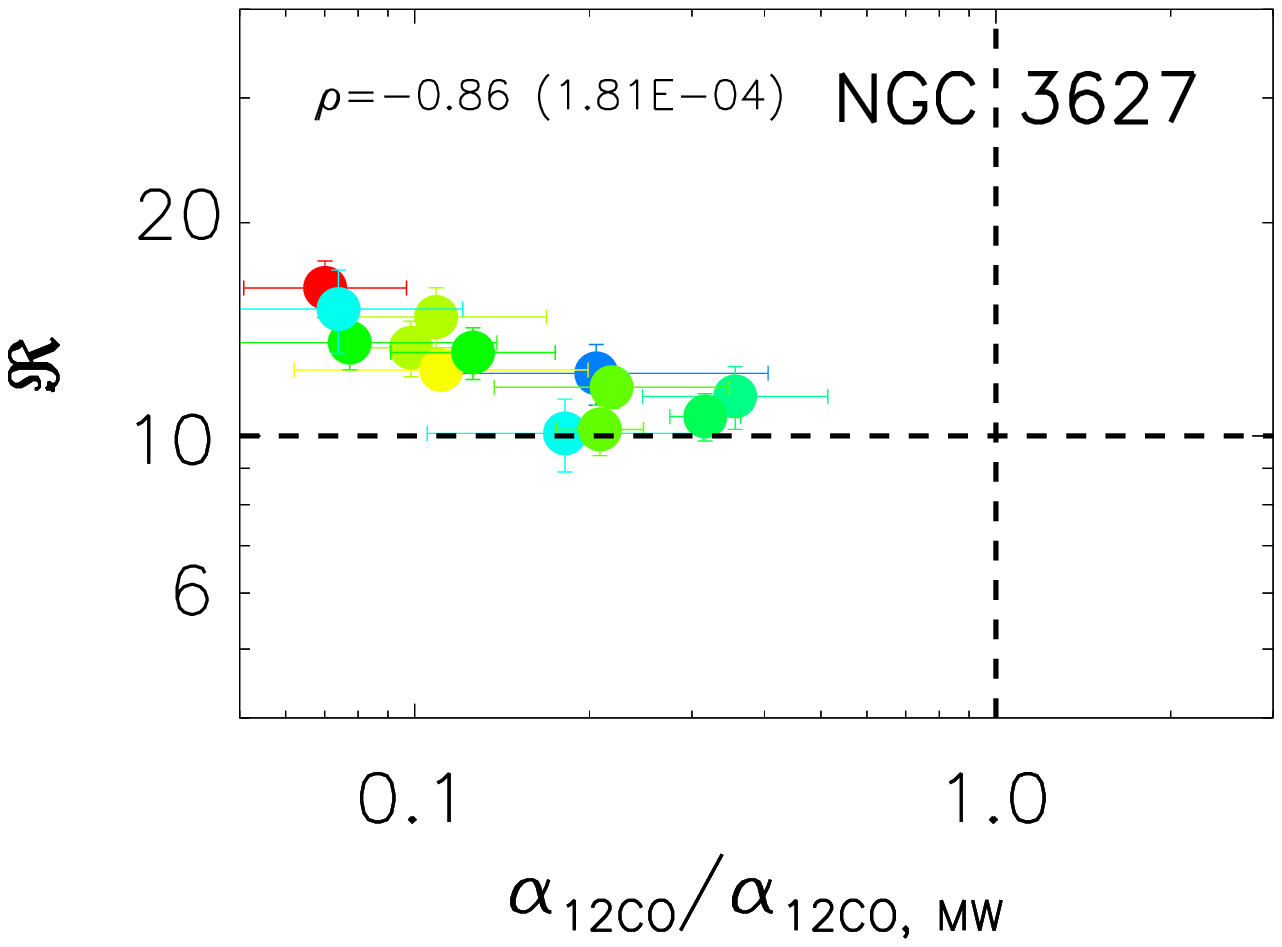}
\includegraphics[clip, trim=37mm 24mm 6mm 2mm,width=5.2cm]{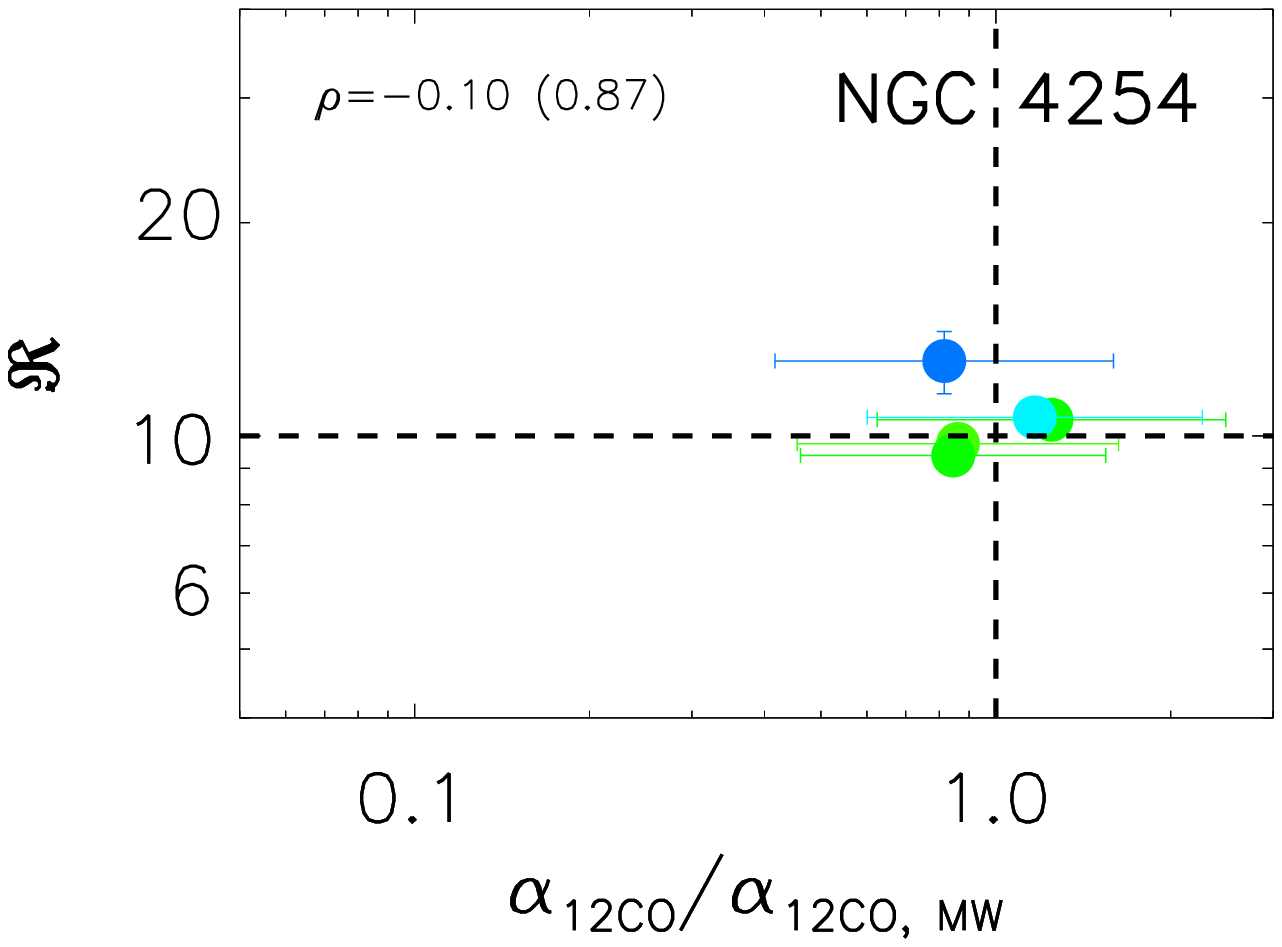}
\includegraphics[clip, trim=37mm 24mm 6mm 2mm,width=5.2cm]{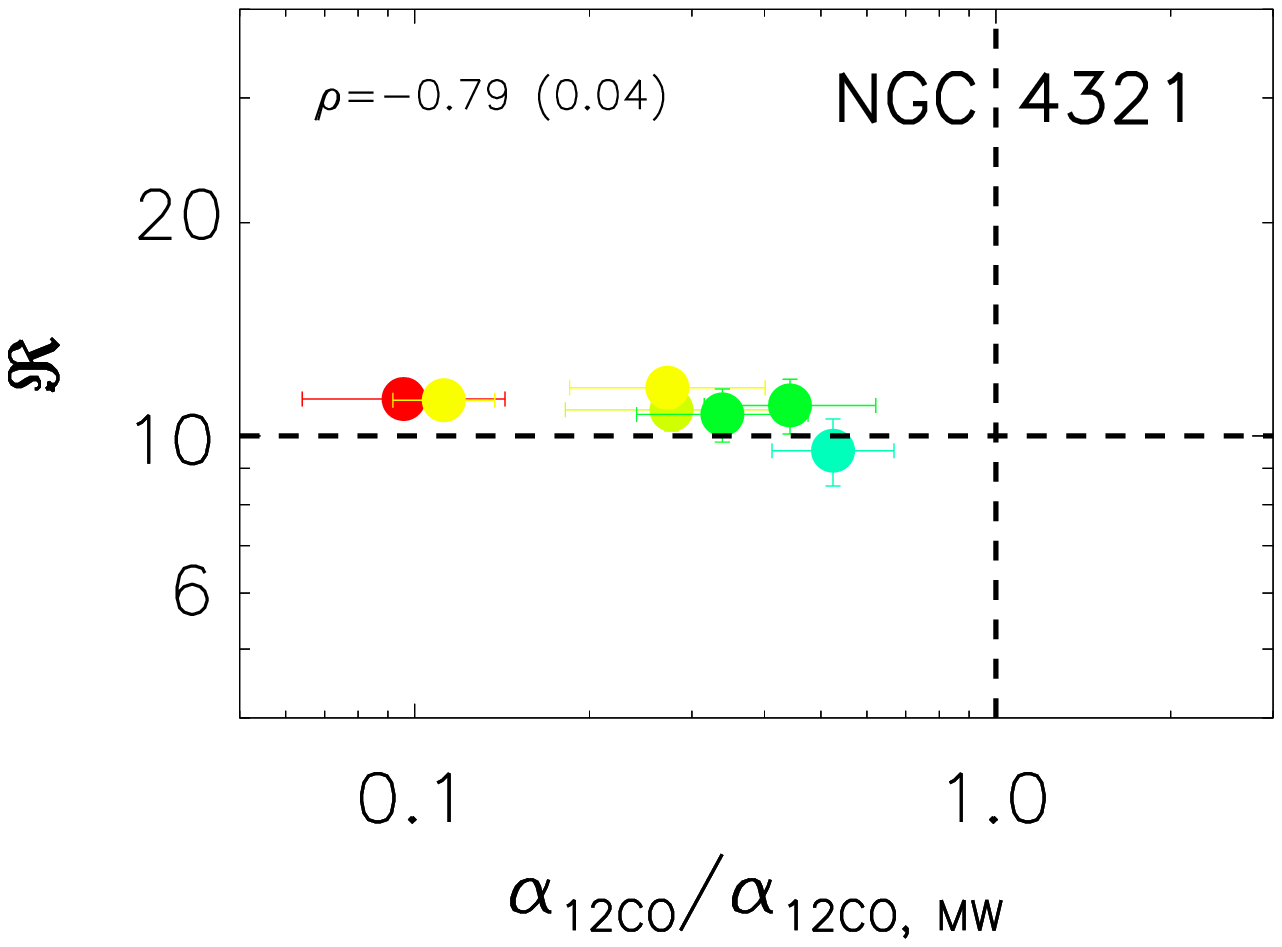}\\
\includegraphics[clip, trim=08mm 0 6mm 2mm,width=6.6cm]{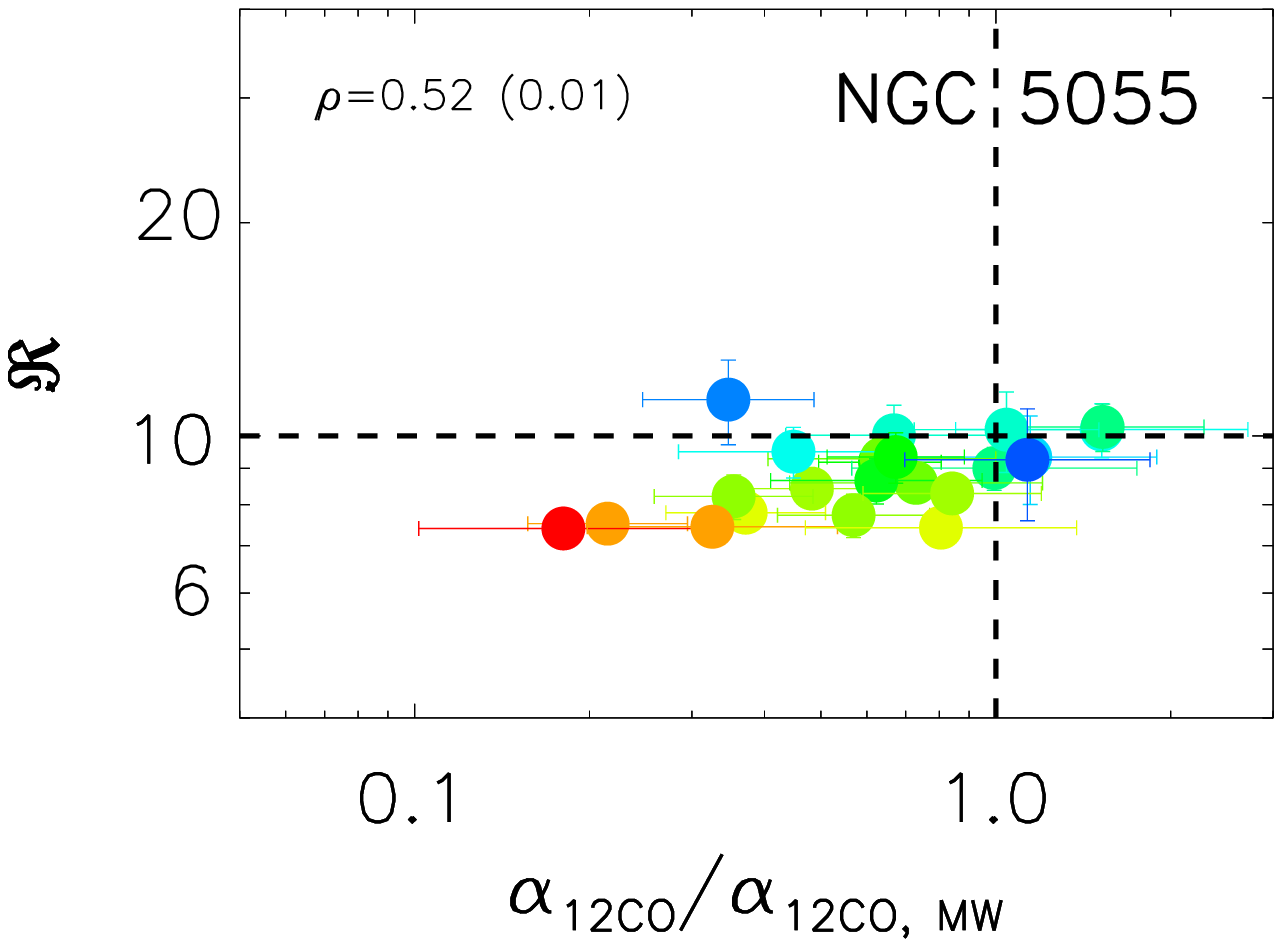}
\includegraphics[clip, trim=37mm 0 6mm 2mm,width=5.2cm]{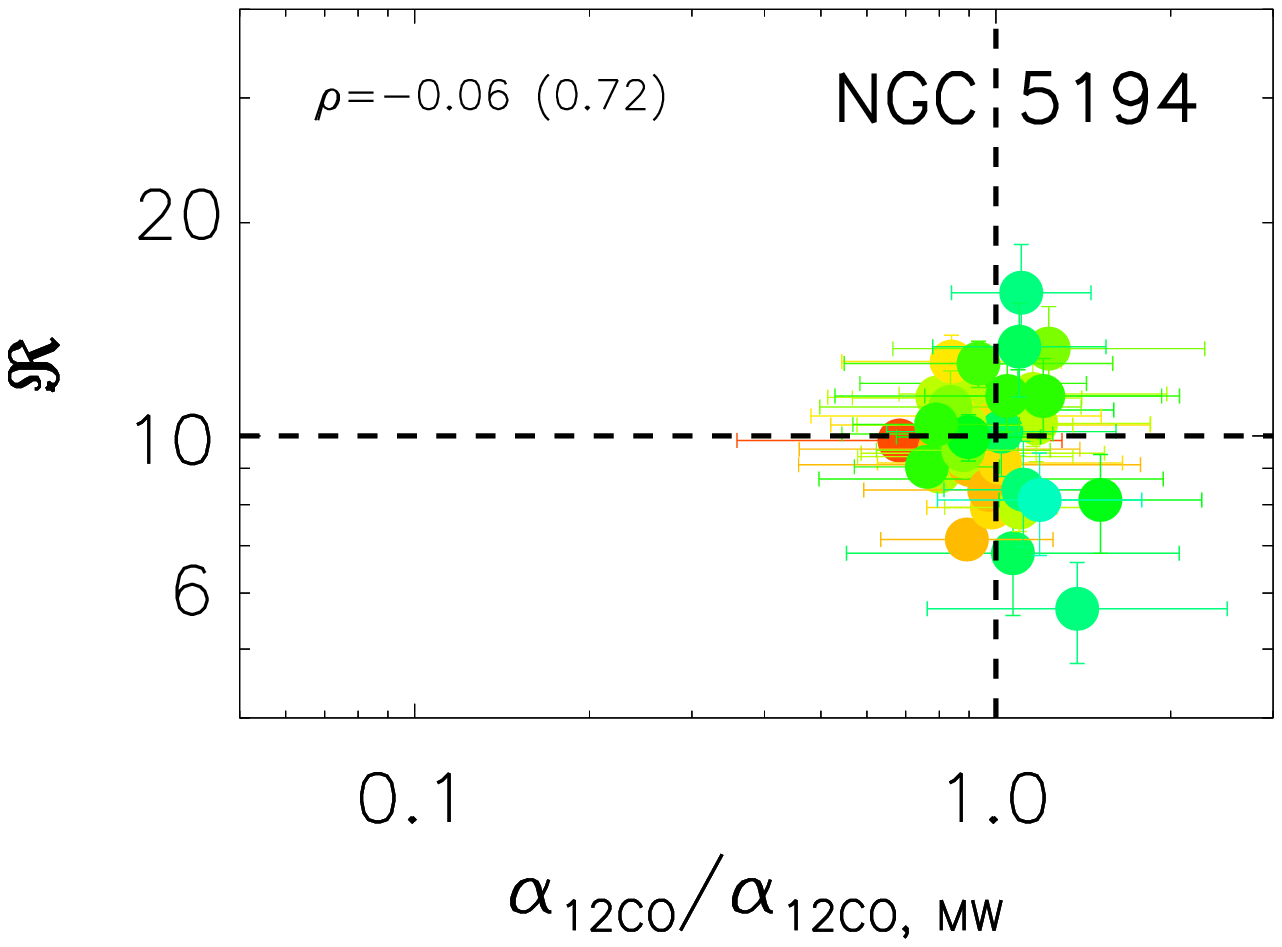}
\includegraphics[clip, trim=37mm 0 6mm 2mm,width=5.2cm]{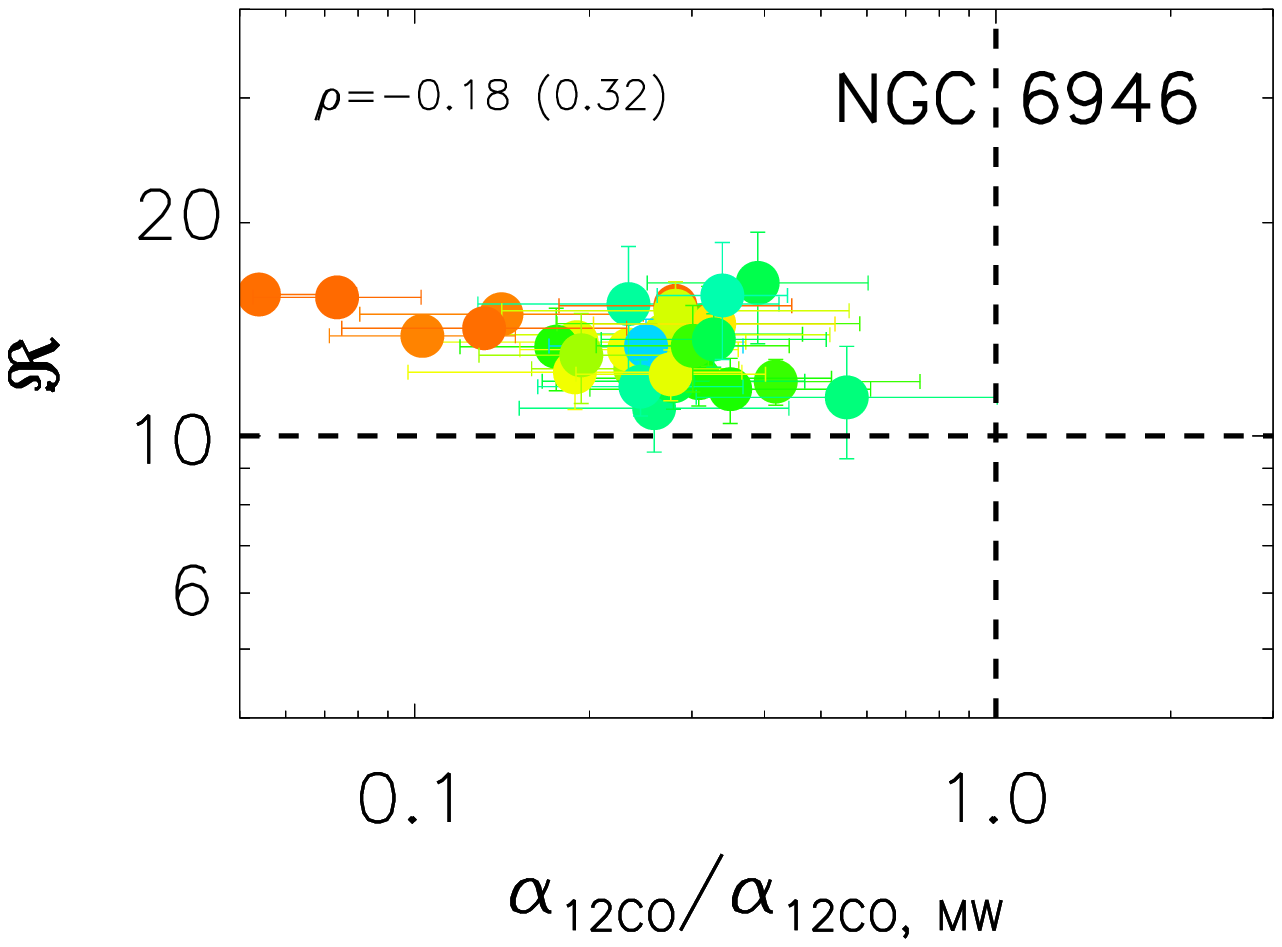}\\
\vspace{-3pt}
\caption{
Correlation within galaxies of $\Re$ and the conversion
factor \aco, which is calibrated on dust emission.
The trends are quite different from galaxy to galaxy.
The colour-code corresponds to distance to
the galaxy centre and it is the same for all panels.
The dashed lines are shown to guide the eye.
We also indicate the Spearman's rank correlation
coefficients and their significance in parenthesis.
}
\label{fig:xcovar}
\end{figure*}

\subsection{Correlation between the \xco factor and $\Re$}
\label{sect:rexco}
Since the physical quantities described above may
also influence the conversion of \com intensity to molecular
gas mass, we investigate empirically and with models
how the variations in $\Re$ and in the \xco factor, calibrated
on the dust reference, are linked.

The \xco factor can be determined indirectly from dust
emission. Modeling of the dust emission provides a
dust mass that is converted to a total gas mass with a
dust-to-gas ratio and to a molecular mass by subtracting
the mass of atomic gas. This molecular gas mass
or surface density (denoted below $\Sigma({\rm mol, dust})$)
is then divided by the CO intensity to determine \xco or \aco.
\cite{leroy-2011} first applied this by solving for the dust-to-gas
ratio and \aco simultaneously on spatially-resolved scales
of a few kpc for the Local Group.
That method was further employed in \cite{sandstrom-2013}
for the HERACLES galaxies (including 8 out of the 9 galaxies
of our sample) and is similarly applied to NGC\,5194
in \cite{leroy-2017b} and Groves et al. (in prep.).
In \cite{sandstrom-2013}, \aco is measured for hexagonal
pixels of size 37.5\,arcsec, using molecular surface densities
from dust emission and HERACLES \cou observations.
They formally measure $\alpha_{\rm CO(2-1)}$ that they
express as $\alpha_{\rm CO(1-0)}$ because \com is more
commonly used.
However, they adopt a fixed CO line ratio of $0.7$.
Here, with measurements of the \cou/\com ratio for each
location, we update those \aco values. 
As we find \cou/\com ratios spanning a range of values
between $0.3$ and $2$ on a kpc-scale in our sample
of galaxies, we expect the \aco values to change a bit.
We calculate the average CO intensities and $\Re$ value
in a circular aperture (roughly matched radius of 34.1\,arcsec)
centered on each of the pixels.
The new conversion factor \aco is equal to:
\begin{equation}
\alpha_{\rm CO, new} = \Sigma({\rm mol, dust})~/~{\rm I(^{12}CO(1-0)}).
\end{equation}
We show a pixel-to-pixel comparison of the old and
new \aco values in Figure~\ref{fig:xconew}, and we
report averages of the new values, of ratios with the old values
and dispersions in Table~\ref{table:xconew}.
The new \aco values are about 15\,per cent lower than
the old values. There is a significant offset for NGC\,3627
because its \cou/\com ratio is the lowest, with an average
value of 0.5 as suggested by our new IRAM 30-m
observations as opposed to the canonical value of
0.7 assumed by \cite{sandstrom-2013}.

Figure~\ref{fig:xcovar} shows $\Re$ as a function of \aco,
normalized to the Milky Way value. $\Re$ and \aco are
positively correlated in NGC\,5055 and NGC\,3184, and
more weakly in NGC\,0628. $\Re$ and \aco are anti-correlated
in NGC\,3627, NGC4321, and rather weakly in NGC\,6946,
and the lowest \aco values ($\le 0.1\,\alpha_{\rm CO, MW}$)
are reached in those galaxies.
In galaxy centres, one possible explanation for the low
\aco values observed is a change in the $^{12}$CO optical depth
\citep[e.g.,][]{sandstrom-2013}. Those optical depth effects
seem to be at work in NGC\,3627, NGC\,6946, and to
a lesser extent, NGC\,4321.

\subsection{Comparison to non-LTE models}
\label{sect:radexmodels}
To visualize how changes in physical conditions,
specifically optical depth, density, and temperature,
can affect $\Re$ quantitatively, we compare
our observations to models from the non-LTE code
RADEX \citep{vandertak-2007}.
We note that those physical conditions, especially density
and optical depth, could be correlated in real molecular clouds.

We use the grid of models built by \cite{leroy-2017} to
predict $\Re$, the \cou/\com intensity ratio, and \xco.
These models consider a distribution of volume densities
(as an approximation for sub-beam density variations)
and take into account sub-thermal excitation, but they
assume fixed abundances and the spatial coexistence
of $^{12}$CO and $^{13}$CO that fill the beam and are
at a common kinetic temperature. In that sense, the models
are representative of a single-phase medium, not of
a multi-phase medium.
Model predictions are scaled to our adopted $^{13}$CO
and $^{12}$CO abundances (section~\ref{sect:abund}).
\xco is computed as the inverse of the \com emissivity,
$\epsilon_{12}$, where $\epsilon_{12} = I_{12} / N_{12}$,
$I_{12}$ is the integrated intensity of \com and
$N_{12}$ is the column density of $^{12}$CO.
We vary the optical depth of the \com line, $\tau_{12}$,
from $3$ to $10$ in steps of 0.1\,dex. The ratio of
the \coi and \com line optical depths, $\tau_{13}/\tau_{12}$,
is set to 0.03 and the effects of varying this value are
discussed below. We assume a lognormal distribution
of \htwo volume densities with a width $\sigma=0.8$\,dex.
Several values of mean volume density ($n_{\rm H}=10^2,
~10^3, 10^4$\,cm$^{-3}$) and temperature ($T_{\rm kin}=
15,~25,~35$\,K) are tested.
The predicted quantities that we investigate and the trends
with density are somewhat sensitive to the selected
width of the distribution, but not enough to change
the results reported below.

Figure~\ref{fig:xcomodels} shows predictions of $\Re$,
\cou/\com, and \xco. We discuss the effects of optical depth
and density separately, but these could be correlated.
As expected, the \cou/\com line ratio (top panels) increases
for increasing temperature and density, and it is rather
constant for increasing $\tau_{12}$ (same symbols, different shade).
Concerning \xco (bottom panels), $\Re$ decreases and
\xco increases with increasing $\tau_{12}$ because
the $^{12}$CO emissivity goes down.
$\Re$ is inversely proportional to \xco (see also
equation~\ref{eq:xco} in section~\ref{sect:opacity}),
as long as $\tau_{13}$ remains small. The models predict
an increase of \xco or decrease of $^{12}$CO emissivity
-- for a fixed $\tau$ -- with increasing temperature.
Increasing the temperature reduces the opacity per unit
mass of $^{12}$CO. But to keep $\tau$ constant, the increase
in $^{12}$CO intensity is not as large as the increase in
gas column density because $^{12}$CO is optically thick,
hence its emissivity goes down.
The models predict only mild variations of $\Re$ and \xco
with density: $\Re$ decreases most for a density of
$10^3$\,cm$^{-3}$ and \xco increases linearly for the densities
considered. This can be understood as the emissivity of $^{12}$CO
peaks at densities of $10^2$\,cm$^{-3}$ (high $\Re$ and low \xco)
while that of $^{13}$CO peaks for densities $10^3$\,cm$^{-3}$
\citep{leroy-2017}.

Overall, we find that the models can only partially reproduce
the observations. Density and temperature have a weaker
impact on $\Re$ predictions than optical depth.
The range of $\Re$ values found in the observations
is covered by the models, but the range of \cou/\com
ratios and \xco values is not.
While one can imagine that a broader range of temperatures
or densities (or a different opacity ratio) than tested
could explain the range of observed \cou/\com ratios,
the trends of $\Re$ with \xco require additional
modifications.
Focusing on the isotopic abundance ratio (which is
proportional to the optical depth ratio in LTE), here
we discuss qualitatively which parameter space in
the models is compatible with our observations:
\begin{enumerate}
\vspace{-5pt}
\item
Galaxies with values in the disc of $\Re \simeq 10$ and
\aco~$\simeq \alpha_{\rm CO, MW}$
(NGC\,0628, NGC\,3184, NGC\,4254, NGC\,5194),
require the models in Fig.~\ref{fig:xcomodels} to shift
by a factor of $2$ either to the right, which can be
accomplished by decreasing the $^{12}$CO/\htwo
abundance (keeping the isotopic ratio constant), or up,
which can be accomplished by increasing the isotopic
abundance ratio (keeping the $^{12}$CO/\htwo abundance
constant).
\item
The trend of higher $\Re$ values and lower \aco, noted
above for NGC\,3627, NGC\,6946, and to
a lesser extent, NGC\,4321, can naturally be
explained with the models having low $\tau_{12}$ values
($\simeq4$). In their centres, the lowest \aco values
still require the models to shift by a factor of $2-3$ either
to the left, which can be accomplished by increasing the
$^{12}$CO/\htwo abundance (keeping isotopic ratio constant),
or down for very low $\tau_{12}$ values ($^{12}$CO
becoming optically thin), which can be accomplished by
lowering the isotopic abundance ratio (to $\simeq20$).
\item
The low $\Re$ and low \aco values in the centre of NGC\,5055
can be accounted for with the same modifications as
for (ii), but for higher $\tau_{12}$ values.
\vspace{3pt}
\end{enumerate}
In summary, for the discs of most of galaxies in our
sample, the range of $\Re$ and \xco values observed in
can be explained by variations in optical depths (for a fixed
optical depth ratio or isotopic abundance ratio). 
However some values, especially in galaxy centres, require
a change by a factor of $2-3$ in the optical depth ratio (or
isotopic abundance ratio), which seems reasonable.
We should keep in mind that physical conditions may
also be changing within the model grid (i.e. a multi-phase
model would be more representative).

\subsection{Optical depths and column densities}
\label{sect:opacity}
Since \coi emission remains optically thin over larger parts
of molecular clouds, it is often advocated as a more accurate
tracer of the molecular gas mass than \com in the
intermediate-density regime ($n_H \simeq 10^3$\,cm$^{-3}$).
In this section, we present the optical depth of the \coi line
($\tau_{13}$), $^{13}$CO column densities ($N_{13}$), and
\htwo column densities ($N({\rm H_2})_{\rm 13CO}$)
assuming local thermodynamic equilibrium (LTE).
We compare the \htwo column densities to those obtained
from \com emission.
We opt for a simple framework as an alternative to the
non-LTE models such as those presented in the previous
section to have more flexibility on the choice of conditions/parameters
and, in particular, explore the possibility of $^{12}$CO and
$^{13}$CO having different filling factors.
Ultimately, more lines and transitions would be needed for
a complete, multi-phase modeling.

\subsubsection{Framework}
\label{sect:assum}
Given LTE, the equations of radiative transfer give the general
expression for the observed brightness temperature
of a line ($T_{\rm obs}$):
\begin{equation}
T_{\rm obs} = \eta_{\rm bf} [J_{\nu}(T_{\rm ex}) - J_{\nu}(T_{\rm bg})] ({1-\exp(-\tau)})~~[{\rm K}]
\end{equation}
where $\eta_{\rm bf}$ is the beam filling factor,
$J_{\nu}$ is the line intensity given by the Planck function,
$T_{\rm ex}$ is the excitation temperature,
$T_{\rm bg}$ is the background temperature equal to 2.7\,K,
and $\tau$ is the optical depth of the line.
To reduce this equation, we make the following assumptions
for the \coi and \com lines:\\
- $\tau_{12} > 1$ while $\tau_{13} \le 1$, \\
- $\eta_{\rm bf,12} = 1$ ($^{12}$CO fills the beam) while $\eta_{\rm bf,13}$ is let free. \\

\noindent
The optical depth $\tau_{13}$ and the column density $N_{13}$
are given by:
\begin{equation}
\tau_{13}  = -\ln\left(1-\frac{1}{\eta_{\rm bf,13}} 
	\left[\frac{J_{\nu,12}(T_{\rm ex,12}) - J_{\nu,12}(T_{\rm bg})}{J_{\nu,13}(T_{\rm ex,13}) - J_{\nu,13}(T_{\rm bg})}\right]
	\times \frac{I_{13}}{I_{12}}\right)
\end{equation}
\vspace{-3mm}
\begin{equation}
N_{13} = \frac{3.0\times10^{14}}{1-\exp(-5.29/T_{\rm ex,13})} \times \frac{\tau_{13}}{1-\exp(-\tau_{13})} \times I_{13}~~[{\rm cm^{-2}}]
\end{equation}
where $I$ is the integrated CO line intensity (in K\,\kms),
obtained by integrating the line profiles which have
similar shape and width for both CO lines.
Equation~2 corresponds to equation (15.36) from
\cite{radio-book}. At the adopted $T_{\rm ex}$ values
($\ge20$\,K), we are in the Rayleigh-Jeans regime.
The temperature term is an approximation of the partition
function of CO with all energy levels populated in LTE.
This simplification is valid for $T_{\rm ex} \ge 8$\,K.
If $\tau_{13}$ is always small, $N_{13}$ is
an increasing function of $T_{\rm ex}$ because upper
states get preferentially populated. We also
assume that the continuum is weak (no IR pumping).
We refer to \cite{jimenez-2017a} for details of the calculations.
In case the beam filling factor of $^{13}$CO is lower
than unity (hence lower than that of $^{12}$CO), the
quantities $\tau_{13}$ and $N_{13}$ represent local values
as opposed to beam-averaged values.
To be able to compare \htwo column densities derived
from \com and from \coi, we shall then consider beam-averaged
quantities. The beam-averaged column density
of \htwo can be deduced from \coi with:
\begin{equation}
N({\rm H_2})_{\rm 13CO} = N_{13} \times \left[\frac{{\rm H_2}}{{\rm ^{13}CO}}\right] \times \frac{1}{\eta_{\rm bf,13}} ~~[{\rm cm^{-2}}].
\end{equation}

From these formulae, for $\tau_{13} < 1$, the $^{12}$CO-to-\htwo
conversion factor will depend on $\Re$, which captures
the optical depth of the $^{12}$CO line. In this case:
\begin{equation}
\label{eq:xco}
\begin{aligned}
X_{\rm {^{12}CO}} = \left[\frac{{\rm H_2}}{{\rm ^{13}CO}}\right] \times \frac{1}{\eta_{\rm bf,13}} \times
	\frac{3.0\times10^{14}}{1-\exp(-5.29/T_{\rm ex,13})} \times \Re^{-1} \\~~[{\rm cm^{-2} (K.km/s)^{-1}}].
\end{aligned}
\end{equation}

\subsubsection{Results for $\tau_{13}$, $N_{13}$, $N({\rm H_2})$:
	simplest LTE case (Case~1)}
To start with \textit{(Case~1)}, we assume thermalization of
the lines and make the most simplifying assumptions,
i.e. that both lines have the same excitation temperature,
a beam filling factor equal to unity, abundances equal to
the fiducial values, and \xco equal to the Milky Way value.
The assumed parameters 
as well as the results for the $^{13}$CO optical depths
($\tau_{13}$) and column densities ($N_{13}$ and
$N({\rm H_2})$) are reported in Table~\ref{table:tau}.
We give average values of $\tau_{13}$ and $N_{13}$
and their dispersion within the maps, for entire galaxies,
their centres, and their discs.

Since our sample of galaxies display quite a narrow
range of $\Re$ values, the average values for $\tau_{13}$ are
very similar in all galaxies and around $0.07-0.11$. Those are
beam-averaged values and correspond to
$\tau_{12} \simeq 6$ (for an isotope abundance ratio
of $60$) and to a $^{12}$CO effective critical density
of $n_{\rm crit} = n_{\rm crit,thin} / \tau \simeq 2\times10^2$\,cm$^{-3}$.
For $N_{13}$, we find values in the range 
$0.4-2\times10^{15}$\,cm$^{-2}$ with a factor of two
dispersion in the maps. Those correspond to $^{12}$CO
column densities of $0.2-1.2\times10^{17}$\,cm$^{-2}$
and are averages over large beams, not local quantities.
The implied optically thin $^{13}$CO-to-\htwo conversion
factor is $8\times10^{20}$\,cm$^{-2}$\,(K\,\kms)$^{-1}$.

We find that values of $N({\rm H_2})_{\rm 13CO}$
are systematically lower by a factor of $\sim$3 compared
to $N({\rm H_2})_{\rm 12CO}$ values. If our assumptions
hold, either $^{13}$CO traces poorly the total \htwo column
densities, or $^{12}$CO overpredicts the total \htwo column
densities.
Similar offsets in column densities were found by
\cite{meier-2001,meier-2004,meier-2008}.
Taking the $^{12}$CO-based estimates as reference,
we can speculate about mechanisms to increase the
$^{13}$CO-based estimate.
To increase $N({\rm H_2})_{\rm 13CO}$ by a factor of 3,
we would need either an average excitation temperature
of 60\,K, an isotopic abundance ratio of 180, or a difference
in beam filling factors of about $3$.
The maximum values allowed for each of those parameters
and for each galaxy are reported in Table~\ref{table:tau}.
We note that the beam filling factor can reflect variations
in both abundances and emissivities of the lines but here
we aim to separate the two effects and we consider that
beam filling factors are dominated by emissivity variations.
Although uncertainties associated to the adopted abundances
are large, and temperatures in the galaxy centres may approach
the required 60\,K value \citep[e.g., for NGC\,6946,][]{walsh-2002,
meier-2004}, these requirements seem unlikely to be valid
throughout the discs and in all discs of our sample.
Moreover, if the temperature is indeed higher than
the typical temperature of Galactic molecular clouds, then
a Galactic conversion factor would also no longer apply
for $^{12}$CO.
Hence, the discrepancies between \htwo column densities
derived from \coi and those derived from \com are most likely
due to differences in beam filling factor of the two lines.
Presumably, \com emission fills the beam and traces
a more diffuse phase ($n < 10^3$\,cm$^{-3}$) while \coi
emission is confined to a denser phase. In our
sample of galaxies,
the \htwo column densities can be explained by a
filling factor fraction of diffuse versus dense gas that varies
between $2.2$ and $3.5$, with a possible trend of higher
fraction in galaxies with more clumpy star formation (e.g.,
NGC\,0628, NGC\,6946) and lower fraction in galaxies
with strong spiral modes (e.g., NGC5055, NGC\,5194).

\begin{footnotesize}
\begin{table*}
  \caption{Results on optical depths and column densities.} 
\begin{center}
\begin{tabular}{lccccccccccccc}
    \hline\hline
     \vspace{-8pt}\\
    \multicolumn{1}{l}{Name} & 
    \multicolumn{5}{c}{Assumed conditions} & 
    \multicolumn{1}{c}{} & 
    \multicolumn{3}{c}{Derived quantities} & 
    \multicolumn{1}{c}{} & 
    \multicolumn{3}{c}{Limits on conditions} \\ \cline{2-6}\cline{8-10}\cline{12-14} 
    \multicolumn{1}{l}{} & 
    \multicolumn{1}{c}{$T_{ex,12}$} & 
    \multicolumn{1}{c}{$T_{ex,13}$} & 
    \multicolumn{1}{c}{$\frac{\eta_{bf,12}}{\eta_{bf,13}}$} & 
    \multicolumn{1}{c}{$\frac{[^{12} {\rm CO}]}{[^{13} {\rm CO}]}$} & 
    \multicolumn{1}{c}{$\frac{X_{\rm CO}}{X_{\rm CO,MW}}$} & 
    \multicolumn{1}{c}{} & 
    \multicolumn{1}{c}{$\overline{\tau}_{13}$} & 
    \multicolumn{1}{c}{$\overline{N}_{13}$} & 
    \multicolumn{1}{c}{$\frac{N_{\rm H_2, 13CO}}{N_{\rm H_2, 12CO}}$} &
    \multicolumn{1}{c}{} & 
    \multicolumn{1}{c}{$T_{ex,13}^{l}$} & 
    \multicolumn{1}{c}{$\frac{[^{12} {\rm CO}]}{[^{13} {\rm CO}]}^{l}$} &
    \multicolumn{1}{c}{$\frac{\eta_{bf,12}}{\eta_{bf,13}}^{l}$} \\
    \multicolumn{1}{l}{} & 
    \multicolumn{1}{c}{(K)} & 
    \multicolumn{1}{c}{(K)} & 
    \multicolumn{1}{c}{} & 
    \multicolumn{1}{c}{} & 
    \multicolumn{1}{c}{} & 
    \multicolumn{1}{c}{} &
    \multicolumn{1}{c}{} & 
    \multicolumn{1}{c}{($10^{15}$\,cm$^{-2}$)} & 
    \multicolumn{1}{c}{} & 
    \multicolumn{1}{c}{} & 
    \multicolumn{1}{c}{(K)} & 
    \multicolumn{1}{c}{} & 
    \multicolumn{1}{c}{} \\
    \hline \vspace{-10pt} \\
    \multicolumn{9}{l}{\textit{Case~1: simple LTE conditions}}\\
    \hline \vspace{-10pt} \\
	{NGC\,0628}	& 20	& 20	& 1	& 60	& 1.0		& & $0.07$ 	(0.33\,dex)	& $0.36$	(0.38\,dex)	& 0.29	& & $80$	& $210$	& $3.4$ \\
	{NGC\,2903}	& 20	& 20	& 1	& 60	& 1.0		& & $0.10$ 	(0.10\,dex)	& $1.96$ 	(0.31\,dex)	& 0.38	& & $60$	& $160$	& $2.6$ \\
	{NGC\,3184}	& 20	& 20	& 1	& 60	& 1.0		& & $0.09$ 	(0.14\,dex)	& $0.43$ 	(0.23\,dex)	& 0.36	& & $65$	& $170$	& $2.8$ \\
	{NGC\,3627}	& 20	& 20	& 1	& 60	& 1.0		& & $0.09$ 	(0.18\,dex)	& $1.65$ 	(0.36\,dex)	& 0.33	& & $70$	& $190$	& $3.1$ \\
	{NGC\,4254}	& 20	& 20	& 1	& 60	& 1.0		& & $0.09$ 	(0.18\,dex)	& $1.36$ 	(0.43\,dex)	& 0.38	& & $60$	& $160$	& $2.6$ \\
	{NGC\,4321}	& 20	& 20	& 1	& 60	& 1.0		& & $0.10$ 	(0.09\,dex)	& $1.27$ 	(0.30\,dex)	& 0.37	& & $60$	& $170$	& $2.7$ \\
	{NGC\,5055}	& 20	& 20	& 1	& 60	& 1.0		& & $0.11$ 	(0.15\,dex)	& $1.48$ 	(0.35\,dex)	& 0.46	& & $50$	& $130$	& $2.2$ \\
	{NGC\,5194}	& 20	& 20	& 1	& 60	& 1.0		& & $0.09$ 	(0.30\,dex)	& $1.01$ 	(0.52\,dex)	& 0.40	& & $55$	& $150$	& $2.5$ \\
	{NGC\,6946}	& 20	& 20	& 1	& 60	& 1.0		& & $0.07$ 	(0.20\,dex)	& $1.20$ 	(0.41\,dex)	& 0.29	& & $80$	& $210$	& $3.5$ \\
    \hline \vspace{-10pt} \\
    \multicolumn{9}{l}{\textit{Case~2: motivated choice of conditions}}\\
    \hline \vspace{-10pt} \\
	{NGC\,0628 - centre}	& 60	& 30	& 1	& 30	& 0.4		& & $0.24$ 	(0.09\,dex)	& $1.40$	(0.11\,dex)	& 0.75	& & $45$	& $40$	& $1.3$ \\
	{NGC\,0628 - disc}	& 30	& 20	& 2	& 60	& 0.7		& & $0.26$ 	(0.35\,dex)	& $0.37$	(0.39\,dex)	& 0.86	& & $25$	& $70$	& $2.3$ \\
	{NGC\,2903 - centre}	& 60	& 30	& 1	& 30	& 0.4$^*$	& & $0.20$ 	(0.03\,dex)	& $7.93$ 	(0.08\,dex)	& 0.65	& & $50$	& $50$	& $1.5$ \\
	{NGC\,2903 - disc}	& 30	& 20	& 2	& 60	& 1.0$^*$	& & $0.37$ 	(0.11\,dex)	& $2.12$ 	(0.28\,dex)	& 0.88	& & $25$	& $70$	& $2.3$ \\
	{NGC\,3184 - centre}	& 60	& 30	& 1	& 30	& 0.3		& & $0.22$ 	(0.03\,dex)	& $1.61$ 	(0.05\,dex)	& 0.98	& & $35$	& $40$	& $1.0$ \\
	{NGC\,3184 - disc}	& 30	& 20	& 2	& 60	& 0.9		& & $0.35$ 	(0.17\,dex)	& $0.46$ 	(0.22\,dex)	& 0.93	& & $25$	& $70$	& $2.1$ \\
	{NGC\,3627 - centre}	& 60	& 30	& 1	& 30	& 0.1		& & $0.14$ 	(0.08\,dex)	& $4.96$ 	(0.11\,dex)	& 2.30	& & $15$	& $20$	& $0.4$ \\
	{NGC\,3627 - disc}	& 30	& 20	& 2	& 60	& 0.2		& & $0.31$ 	(0.19\,dex)	& $1.81$ 	(0.37\,dex)	& 2.07	& & $15$	& $20$	& $0.4$ \\
	{NGC\,4254 - centre}	& 60	& 30	& 1	& 30	& 0.4$^*$	& & $0.29$ 	(0.02\,dex)	& $8.87$ 	(0.04\,dex)	& 0.93	& & $35$	& $40$	& $1.1$ \\
	{NGC\,4254 - disc}	& 30	& 20	& 2	& 60	& 0.8		& & $0.32$ 	(0.20\,dex)	& $1.40$ 	(0.42\,dex)	& 1.12	& & $20$	& $60$	& $1.8$ \\
	{NGC\,4321 - centre}	& 60	& 30	& 1	& 30	& 0.1		& & $0.21$ 	(0.04\,dex)	& $7.15$ 	(0.16\,dex)	& 2.47	& & $15$	& $20$	& $0.4$ \\
	{NGC\,4321 - disc}	& 30	& 20	& 2	& 60	& 0.4		& & $0.35$ 	(0.11\,dex)	& $1.16$ 	(0.26\,dex)	& 2.17	& & $15$	& $30$	& $0.9$ \\
	{NGC\,5055 - centre}	& 60	& 30	& 1	& 30	& 0.2		& & $0.34$ 	(0.05\,dex)	& $1.16$ 	(0.09\,dex)	& 2.31	& & $15$	& $20$	& $0.4$ \\
	{NGC\,5055 - disc}	& 30	& 20	& 2	& 60	& 0.7		& & $0.41$ 	(0.17\,dex)	& $1.60$ 	(0.34\,dex)	& 1.50	& & $15$	& $50$	& $1.3$ \\
	{NGC\,5194 - centre}	& 60	& 30	& 1	& 30	& 0.4$^*$	& & $0.30$ 	(0.06\,dex)	& $11.4$ 	(0.09\,dex)	& 0.95	& & $35$	& $40$	& $1.0$ \\
	{NGC\,5194 - disc}	& 30	& 20	& 2	& 60	& 1.0		& & $0.33$ 	(0.27\,dex)	& $1.19$ 	(0.44\,dex)	& 0.93	& & $25$	& $70$	& $2.2$ \\
	{NGC\,6946 - centre}	& 60	& 30	& 1	& 30	& 0.2$^*$	& & $0.15$ 	(0.05\,dex)	& $11.2$ 	(0.22\,dex)	& 0.87	& & $40$	& $40$	& $1.2$ \\
	{NGC\,6946 - disc}	& 30	& 20	& 2	& 60	& 0.4		& & $0.27$ 	(0.20\,dex)	& $1.08$ 	(0.35\,dex)	& 1.31	& & $15$	& $40$	& $1.2$ \\
    \hline \hline
\end{tabular}
\end{center}
    \vspace{-8pt}
\begin{minipage}{17.3cm}
Notes. Column 1: galaxy name.
Columns 2-3: assumed excitation temperatures of the \com and \coi lines.
Level population of the two levels of interest corresponding
to the level population as predicted by LTE at the specific temperature.
Column 4: assumed ratio of beam filling factors.
Column 5: assumed isotope abundance ratio.
Column 6: assumed value of the \com-to-\htwo
conversion factor, normalized to the Milky Way value.
The values are based on the results from \protect\cite{sandstrom-2013}
when available and otherwise (as indicated by the symbol $^*$)
motivated by \protect\cite{bolatto-2013}.
Column 7: mean optical depth of the \coi line and
 dispersion in the map in parenthesis.
Column 8: column density of the $^{13}$CO molecule and
 dispersion in the map in parenthesis.
Column 9: ratio of \htwo column densities obtained
from \coi emission with the conditions given in columns~3-5
and from \com emission using the conversion factor
given in column~6.
Columns 10-11-12: maximum values allowed for the
physical conditions such that column~9 equals to unity.
\end{minipage}
    \vspace{8pt}
  \label{table:tau}
\end{table*}
\end{footnotesize}

\subsubsection{Considerations on our assumptions}
\label{sect:xco}
%
The discrepancy between the \htwo column densities derived
from \coi and those derived from \com may be due
to oversimplifying assumptions, probably because, in our large
beam size, we are probing clouds with a range of properties
\citep[densities, temperatures, optical depths;][]{szucs-2016,leroy-2017}.
For example, abundance variations dependent on column
density can change $N({\rm H_2})_{\rm 13CO}$ by
a factor of $2-3$ in Galactic molecular clouds
\citep{goldsmith-2008}, though not sufficiently enough
to reconcile LTE masses with the higher virial masses
\citep{heyer-2009}.
In the limit of co-existing $^{12}$CO and $^{13}$CO emission,
and for a given set of physical conditions as chosen for
\textit{Case~1}, considering a sub-beam density distribution
(section~\ref{sect:radexmodels}) would not be
sufficient to reconcile column densities (models predict \xco
below the Milky-Way value).
In numerical simulations of realistic molecular clouds,
\cite{szucs-2016} investigate how standard methods
($^{13}$CO and LTE, $^{12}$CO and \xco, the virial method)
perform in recovering the true molecular mass/column density.
They find that the $^{13}$CO method is the worst predictor
and systematically underpredicts the true mass by a factor
of $2-3$ because of chemical and optical depth issues.
Throughout our maps and with a beam size larger than the
typical size of a molecular cloud, $\tau_{13}$ remains small,
below 0.2, and the variations from beam to beam are not
significant enough to change $N({\rm H_2})_{\rm 13CO}$.
It is possible that our $N({\rm H_2})_{\rm 13CO}$
values are slightly underestimated because $^{13}$CO is not
completely optically thin and hides dense gas within the beam,
although this explanation is unlikely. Indeed, in Galactic
molecular clouds, $\tau_{13}$ can be locally enhanced
on local, sub-parsec scales \citep[e.g.,][]{kramer-1999,kramer-2004,
jakob-2007}, but the impact on the global \htwo column
densities derived is marginal \citep[e.g.,][]{wong-2008}.
Our low values for $\tau_{13}$ are indicative of dense gas
mixed with large amounts of diffuse gas within our (kpc-scale)
beam.

\com being optically thick leads to line trapping with the effect
of lowering the critical density of the $^{12}$CO line.
$^{12}$CO can emit strongly in low-density gas that makes
up a large part of molecular clouds, and where $^{13}$CO
emission is weak because it is sub-thermally excited
\citep[e.g.][]{goldsmith-2008,leroy-2017}.
In that case, the fact that the bulk of the $^{12}$CO and 
$^{13}$CO emission does not trace the same gas
can naturally explain the lower $N({\rm H_2})_{\rm 13CO}$
values compared to the $^{12}$CO-based estimates.
To reasonable approximation, for optically thin emission
the emissivity per molecule scales with the density for
$n < n_{\rm crit}$, and is constant for $n > n_{\rm crit}$.
The CO critical density in the optically thin case is
$\sim2,000$\,cm$^{-3}$, so a factor of 3 correction in
column densities would be expected if the volume density
of the gas dominating the $^{13}$CO emission were
$\sim700$\,cm$^{-3}$.
Conversely, one would only expect that the $^{13}$CO
would produce LTE-like levels of emission if the density
were above $\sim2,000$\,cm$^{-3}$, which is clearly a
large density for the bulk of molecular clouds.
The \com emission is presumably more extended than 
the \coi emission, as seen in resolved (pc-scales) studies
\citep[e.g.,][]{pety-2013}.
In the Milky Way, \cite{roman-duval-2016} quantify the
fraction of diffuse gas (gas detected in \com but not in \coi)
and dense gas (gas detected in both \com and \coi).
In terms of luminosity, they find that half of the gas is diffuse
and half is dense in the outer disc, while in the inner disc,
most of the gas ($\simeq80$\,per cent) is dense.

Non-LTE considerations, such as different excitation
temperatures, abundances, and filling factors for the
two lines (mimicking a two-phase model) are explored
in the following \textit{(Case~2)}.

\subsubsection{Results for $\tau_{13}$, $N_{13}$, $N({\rm H_2})$:
	motivated choice of conditions (Case~2)}
In the second case, we allow for the lines to be
non-thermalized and we adopt reasonable physical
parameters for the centre and discs of our sample
based on Galactic studies and on results from
section~\ref{sect:rexco}. In the centres, the temperatures
and filling factor of dense gas are assumed higher than
in the discs, and the isotopic abundance is assumed lower.
The assumed parameters and resulting quantities
are listed in Table~\ref{table:tau}.

By allowing the excitation temperatures and the beam
filling factors of the two CO lines to differ, the resulting
$\tau_{13}$ values generally increase with respect
to \textit{Case~1}. $\tau_{13}$ is found around $0.14-0.34$
in the centres and around $0.26-0.41$ in the discs.
$N_{13}$ has similar values than in \textit{Case~1}
for the discs, but larger values for the centres, around
$1-11\times10^{15}$\,cm$^{-2}$, because of the higher
temperatures used.

When we adopt more realistic conditions for the centres
and discs of our sample, the values of $N({\rm H_2})_{\rm 13CO}$
get closer to the values of $N({\rm H_2})_{\rm 12CO}$
than in \textit{Case~1}.
The results are also very sensitive to the adopted \xco
factors. The \xco values found by \cite{sandstrom-2013},
based on dust emission, are generally lower than the
standard Milky Way value.
For the discs, the need for a higher filling factor of diffuse
versus dense gas (by a factor of about $2$) still persists
in most galaxies.
In the centres, the effect of a low \xco value is somewhat
compensated by adopting a lower isotope abundance ratio,
without necessarily requiring a change in the beam filling factors.
However, in NGC\,3627, NGC\,4321, and NGC\,5055, the \xco
values are so low that this leads to $^{13}$CO predicting
twice more \htwo than $^{12}$CO. In those galaxies,
reconciling $N({\rm H_2})_{\rm 13CO}$ with $N({\rm H_2})_{\rm 12CO}$
requires either lower temperatures than assumed, or lower
abundance ratios, or larger filling factors for $^{13}$CO
than for $^{12}$CO. Both conditions on the temperatures
and filling factors seem unrealistic and the need for changing
abundances is in line with results from section~\ref{sect:rexco}.
It is also possible that uncertainties on \xco in those
galaxies dominate here.

\section{Discussion}
\label{sect:discussion}

\subsection{The $\Re$ ratio: origin of variations}
%
The main trends of $\Re$ as a function of galactocentric radius
that we find are:
(1)~a slight increase from centre to outer disc (by a factor
of 2 at most);
(2)~offsets between galaxies (by a factor of 2 at most);
(3)~a high value in the centres of NGC\,3627 and NGC\,6946.
Causes for $\Re$ variations are essentially linked to
isotopic (or isotopolgue) abundances or gas physical conditions.
They are discussed extensively for external galaxies in, e.g.,
\cite{paglione-2001,tan-2011,danielson-2013,davis-2014}
and briefly examined here:
\begin{itemize}
 \renewcommand{\labelitemi}{\scriptsize$\blacksquare$} 
\item
\textit{Changes in isotope abundance due to stellar nucleosynthesis}
\citep[e.g.,][]{henkel-1993,casoli-1992}.
After a recent burst, one expects the $^{12}$C and $^{18}$O
abundances to be enhanced relative to the $^{13}$C
abundance \citep[e.g.,][]{meier-2014,sliwa-2017b} and
$\Re$ to increase, although the actual abundances are
sensitive to the star-formation history and chemical effects
within a galaxy.
In the Milky Way, observations show an increase of the
$^{12}$C/$^{13}$C and $^{16}$O/$^{18}$O abundance
ratios with galactocentric radius \citep[e.g.,][]{milam-2005}
that can be reproduced by time-dependent models
\citep[e.g.,][]{romano-2017}.
A mild radial increase of the isotope abundance ratio
(similar to or shallower than the gradient observed
in the Milky Way) would naturally explain the steadily
increasing $\Re$ profiles of some of the galaxies
in our sample.
\item
\textit{Changes in abundance due to selective photo-dissociation}
\citep[e.g.,][]{bally-1982,visser-2009}.
Since $^{13}$CO is less abundant than $^{12}$CO and
their abundances are coupled, one expects $^{13}$CO
to be preferentially photo-dissociated under hard radiation
fields. The effect is less clear for C$^{18}$O as it can be
formed in a separate way \citep{bron-2017}.
Simulating a suite of molecular clouds, \cite{szucs-2014}
showed that the effect of selective photo-dissociation on
the $^{12}$CO/$^{13}$CO abundance ratio is minimal.
\cite{jimenez-2017b} also argue that shielding from dust
and \htwo dominates over self-shielding of CO molecules.
Therefore, we do not consider selective photo-dissociation
as a dominant effect.
\item
\textit{Changes in abundance due to chemical fractionation}
\citep[e.g.,][]{liszt-2007}.
This process can enhance the abundance of $^{13}$CO by
a factor of $\sim$2-3 at low temperature and low optical depth
in molecular clouds \citep{szucs-2014}.
It is believed to have a non-dominant effect on large (kpc)
scales in the discs of galaxies \citep{paglione-2001}, though
our C$^{18}$O(1-0) observations do not rule out fractionation
as an important effect \citep{jimenez-2017b}.
The lowest $\Re$ ratios are found in the centres of a few galaxies
and $\Re$ anti-correlates with dust temperature and \sigsfr
for those galaxies, the inverse of what is expected from chemical
fractionation. Hence we consider chemical fractionation
as an unlikely explanation for the lowest $\Re$ values.
\item
\textit{Changes in physical conditions such as gas density,
temperature, and opacity} \citep[e.g.,][]{japineda-2008,wong-2008}.
Theoretically, we expect \coi emission to trace denser and
cooler gas than \com. As explored in section~\ref{sect:correlations},
temperature/excitation may drive some of the lower $\Re$
values observed but it is not the only condition affecting $\Re$.
The presence of diffuse emission or increased turbulence
\textbf{(for example, due to a stellar bar)}
would lower the optical depth of \com and boost its emission
relative to \coi (high $\Re$).
Using several J transitions of $^{12}$CO and $^{13}$CO,
\cite{israel-2009a,israel-2009b} model the centres of galaxies
that have active nuclei with two ISM components, one hot and
tenuous component with low optical depth, and one cooler
and denser component. A different mixing of such
two components (with different physical conditions or in different
proportions) may explain the range of $\Re$ values observed
in our sample of galaxies.
The centre of NGC\,3627 displays the largest \com line widths
(noting that this includes resolution and inclination effects, and its strong
stellar bar could add significant non-circular motions as well).
In the centre of NGC\,6946, \cite{wu-yl-2017} also find larger
velocity dispersions by analysing cloud-scale \cou observations,
although \cite{meier-2004} do not find a correlation between
line width and $\Re$.
\end{itemize}

We conclude that changes in the isotopic abundance due
to nucleosynthesis or/and changes in the gas physical conditions,
such as a different mixture of dense, cold gas and diffuse, warm
gas that impacts on the mean opacity, could account for
the trends with radius and offsets between galaxies.
In the galaxy centres, \textbf{turbulence/bars} could account for
the highest $\Re$ values that we observe, but their low
\aco values still argue for changes in isotopic abundances
(section~\ref{sect:rexco}).
To disentangle the effects of abundance and physical
conditions on $\Re$, at least another $^{13}$CO transition
would be needed for a future, non-LTE modeling analysis.

\subsection{$^{12}$CO, $^{13}$CO, and the \xco conversion factor}
\label{sect:xcodisc}
In this section, we aim to understand under which circumstances
\coi can be used to improve estimates of the molecular gas mass.

\begin{figure}
\centering
\includegraphics[clip,width=8cm]{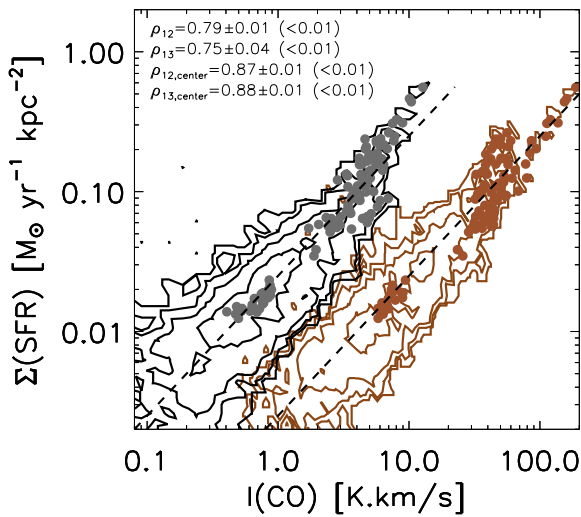}
\caption{
Correlation of \sigsfr with both \coi (black) and
\com (brown) intensities of all EMPIRE galaxies.
Density of data points are shown with contours
and data points corresponding to galaxy centres
are overplotted with filled circles.
The dashed lines indicate a slope of $1$.
We also report the Spearman's rank correlation
coefficients, their uncertainty, and their significance
(in parenthesis), which are measured for a given CO line
on all data points above a signal-to-noise ratio of 5
and measured on galaxy centres only.
}
\label{fig:sfr-co}
\end{figure}

\subsubsection{\com and \coi as tracers of star-formation and molecular gas mass}
%
Galaxy centres are particularly interesting because they
show some of the strongest variations in observations
(including in $\Re$) and in physical properties in spiral galaxies.
Several studies have found that the \xco factor that converts
\com intensity to molecular gas surface density is lower
in the centres of galaxies \citep[e.g.,][]{oka-2001,zhu-2009,
sandstrom-2013}.
This could be due to factors such as an excess of
\com emission from increased temperatures and large
velocity dispersions lowering the optical depth of $^{12}$CO
\citep[e.g.,][]{bolatto-2013}. In our sample,
high \cou/\com ratios, tracing higher temperatures/densities,
are indeed observed in some galaxy centres
\citep{leroy-2009a,leroy-2013}.
The lowest \xco factors are also found for galaxies with the
highest $\Re$ (see section~\ref{sect:rexco}). In those cases,
knowledge on both \com and \coi emission can help
reduce uncertainties on \xco if the later is unknown.
However, we do not find a systematic increase of \com
emission relative to \coi in the galaxy centres.
This implies that, at kpc-scales in normal star-forming
disc galaxies, the issues encountered when using \com
as a tracer of the molecular gas mass are not systematically
solved by using \coi.

Figure~\ref{fig:sfr-co} shows the correlation between \sigsfr
and the \coi and \com intensities of all EMPIRE
galaxies. Density of data points are plotted as contours and
data points corresponding to galaxy centres are overplotted
(filled circles, centres are defined by a cut in CO
intensity and radial distance, see section~\ref{sect:envir}).
The strengths of the correlations as well as
the contour shapes are similar for both lines. The correlation
between \com and \sigsfr is marginally tighter than with \coi
when considering all data points and it is the same when
considering centre points only. In the brightest
regions (\sigsfr $\ge0.1$\,\sigsfrunit; corresponding to
the centres of all galaxies except NGC\,0628 and NGC\,3184),
the \coi data points show slightly less scatter with \sigsfr
than \com, and the star-formation efficiency from \coi
also has less scatter. 
In those bright regions, the distribution of \coi data points
tend to follow a super-linear relation with \sigsfr while
the distribution of \com data points appears to be bimodal
with points (corresponding to NGC\,6946) following a
linear relation and points (corresponding to NGC\,5194)
following a super-linear relation. Hence $^{13}$CO does
not appear to be a more stable tracer of the SFR than
$^{12}$CO in our sample of disc galaxies, except at
the high \sigsfr end.

The analysis of \htwo column densities under the LTE assumption
in section~\ref{sect:opacity} has shown that \coi systematically
underpredicts masses of \htwo, both in all pixels of our maps and
if we consider integrated measurements. The behavior of the
central pixels is obscured by the behavior of the more numerous
disc pixels when looking at galaxies in their entirety.
We have attributed the low \htwo column densities derived
from \coi in the discs to, mainly, the presence of diffuse gas
that is traced by \com or dust emission but not by $^{13}$CO.
General methods to convert the dust and \com emission
of galaxies to mass, however, are not calibrated on those
diffuse phases and may overestimate the molecular gas
mass. In turn, the presence of CO-dark gas could
compensate somewhat for this overestimation
\citep[see][]{liszt-2010,liszt-2012}.
While $^{13}$CO fails in retrieving the bulk molecular gas
mass, its emission may vary less than $^{12}$CO in the galaxy
centres where the fraction of dense gas is larger
and $^{13}$CO traces better the intermediate-density
regime ($\sim1,000$\,cm$^{-3}$).

\subsubsection{The \xco factor from \coi emission}

\begin{figure}
\centering
\includegraphics[clip,width=8.4cm]{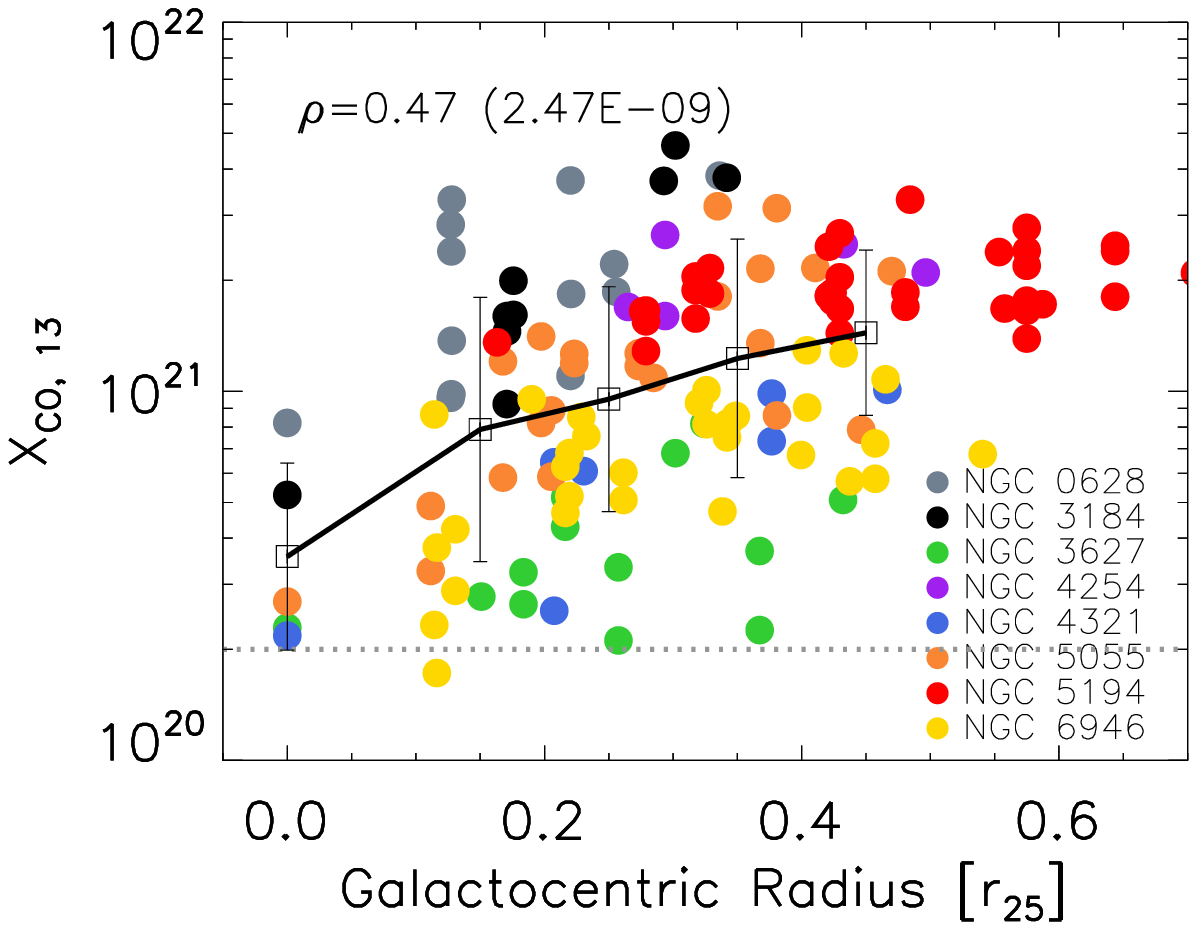}
\includegraphics[clip,width=8.4cm]{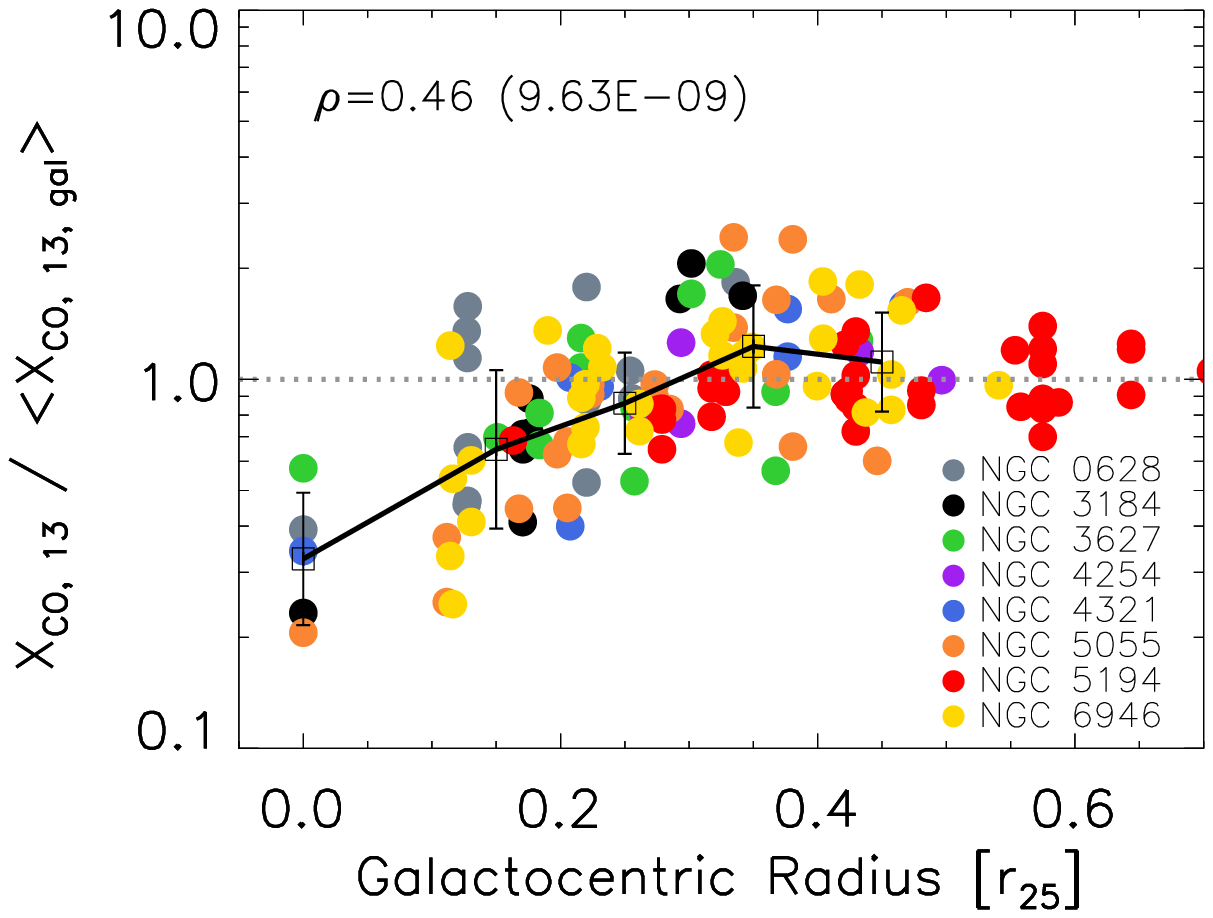}
\includegraphics[clip,width=8.4cm]{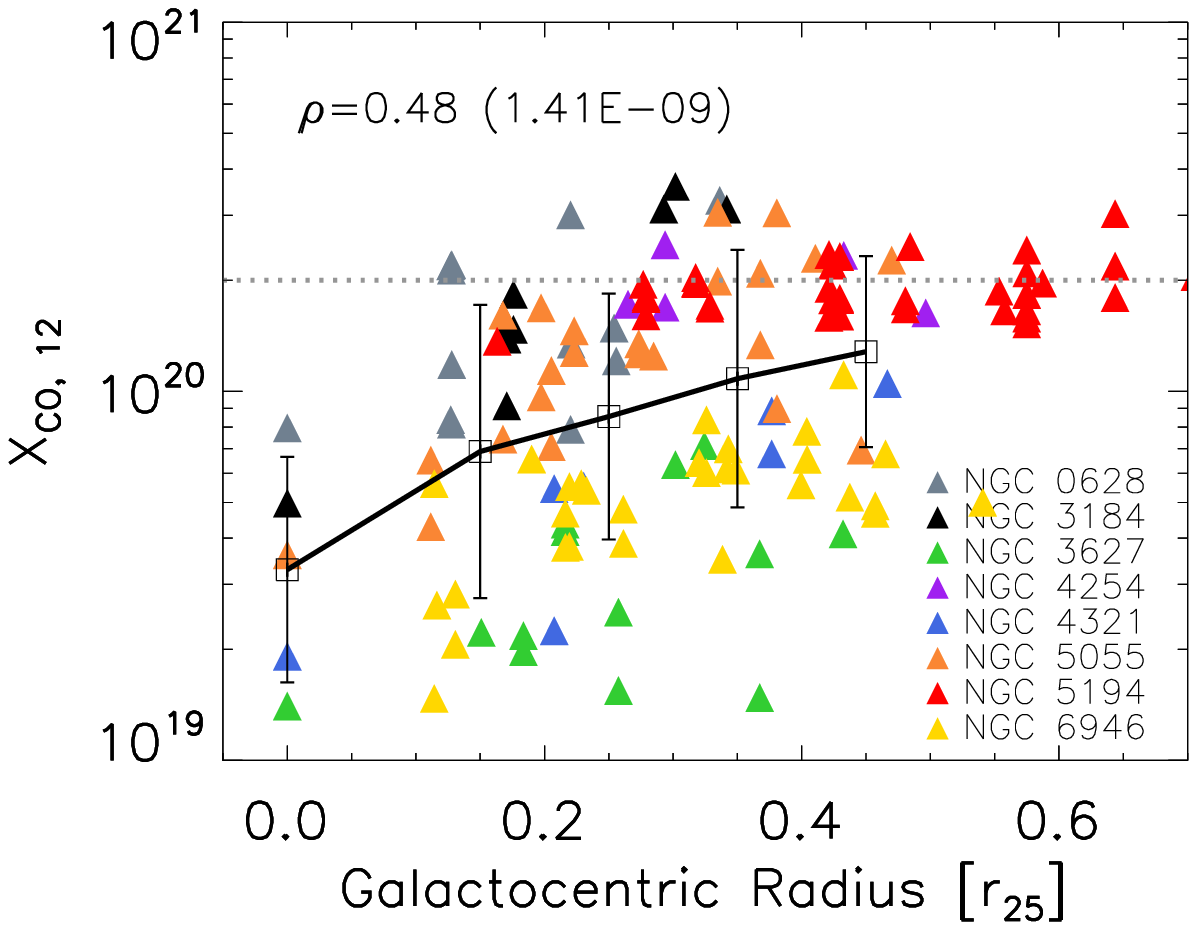}
\vspace{-10pt}
\caption{
\underline{Top panel:}
Empirical \xco factor showing values lower than average
in the galaxy centres and increasing with distance from
the centre. $X_{\rm CO, 13}$ is derived from dust-based
\htwo column densities and \coi emission, in units of cm$^{-2}$\,(K\,\kms)$^{-1}$.
The black open squares and curve show the mean values
and dispersion of all data points within bins of width
$0.1\,r_{25}$.
The horizontal dotted line indicates the typical value of the
$^{12}$CO-to-\htwo conversion factor in the Milky Way.
We also report Spearman's rank correlation coefficients
and their significance in parenthesis.
\underline{Middle panel:}
Same as above, normalized to the mean $X_{\rm CO, 13}$
value in each galaxy.
\underline{Bottom panel:}
Same as the top panel for \com emission.
}
\label{fig:xco13}
\vspace{-3pt}
\end{figure}

\begin{table}
  \caption{The $^{13}$CO-to-\htwo
  conversion factor in different regions of spirals.} 
\begin{center}
\begin{tabular}{lccc}
    \hline\hline
     \vspace{-8pt}\\
    \multicolumn{1}{l}{} & 
    \multicolumn{1}{c}{$X_{\rm CO, 13 - total}$} & 
    \multicolumn{1}{c}{$X_{\rm CO, 13 - centre}$} & 
    \multicolumn{1}{c}{$X_{\rm CO, 13 - disc}$} \\
    \hline
	{average all}			& 10.0 (0.29)	& 3.6 (0.25)	& 10.8 (0.29) \\
	{average nuc.$^{(a)}$}	& 4.9 (0.10)	& 2.2			& 5.3 (0.11) \\
    \hline \hline
\end{tabular}
\end{center}
    \vspace{-8pt}
\begin{minipage}{8.5cm}
Average values, in units of $10^{20}$\,cm$^{-2}$\,(K\,\kms)$^{-1}$.
The standard deviation is given in parenthesis and is in dex.
The averages are measured as the mean of the logarithmic
values of $X_{\rm CO, 13}$.
$(a)$~Galaxies with nuclear starbursts (NGC\,2903,
NGC\,3627, NGC\,4321, and NGC\,6946).
\end{minipage}
\vspace{8pt}
  \label{table:xcoval}
\end{table}

We can derive an empirical \xco factor from \coi emission
by dividing dust-based \htwo column densities (see
section~\ref{sect:xco}) by the \coi intensity:
$X_{\rm CO, 13} = {\rm N(H_2, dust)} / I_{13}$
or $X_{\rm CO, 13} = X_{\rm CO, 12} \times \Re $.
Figure~\ref{fig:xco13} shows values of $X_{\rm CO, 13}$
(absolute and normalized to the average in each galaxy)
as a function of distance to the centre for all galaxies.
Averaged values in different parts of the galaxies are reported
in Table~\ref{table:xcoval}.
For entire galaxies, we find an average value of
$X_{\rm CO, 13} = 1.0\times10^{21}$\,cm$^{-2}$\,(K\,\kms)$^{-1}$
over our sample.
If one were to consider a galactic conversion factor appropriate,
then $X_{\rm CO, 13}$ can just be obtained by scaling \xco
with $\Re$, and the average $X_{\rm CO, 13}$ factor to use
would be $2.2\times10^{21}$\,cm$^{-2}$\,(K\,\kms)$^{-1}$.

As for $X_{\rm CO, 12}$ \citep{sandstrom-2013},
$X_{\rm CO, 13}$ has lower values in the galaxy centres,
by a factor of $\sim3$, and increases with distance. The scatter
is large for absolute $X_{\rm CO, 13}$ values and
reduces to less than a factor of 2 for $X_{\rm CO, 13}$
values normalized to each galaxy average.
Such trends are also observed but less pronounced with
\sigsfr and the stellar surface density decreasing with
increasing $X_{\rm CO, 13}$ because those quantities
have higher values in galaxy centres.
\cite{sandstrom-2013} found that the \com-to-\htwo conversion factor,
$X_{\rm CO, 12}$, varies more within the galaxies dominated
by a central starburst: it is about one order
of magnitude lower than the Galactic value in the centres and three
times lower on average. Such low conversion factors
can be found for extreme starbursts like LIRGs \citep{downes-1998,
kamenetzky-2014,sliwa-2017a}.
The galaxies dominated by a central starburst in \cite{sandstrom-2013}
that are overlapping with our EMPIRE sample are NGC\,3627,
NGC\,4321, and NGC\,6946 (bottom panel of Figure~\ref{fig:xco13}).
Those galaxies have higher $\Re$ values
in their centres and on average. The top panel of Figure~\ref{fig:xco13}
shows that they have lower $X_{\rm CO, 13}$ than the other
galaxies on average, and slightly less variation from centre to outer
disc than with \com because of their declining $\Re$ profiles.
Within galaxies, $X_{\rm CO, 13}$ and $X_{\rm CO, 12}$
show a similar dispersion of $\sim$0.2\,dex, but there are still
large variations from galaxy to galaxy. Taking all galaxies together,
the dispersion per radial bin is about 0.30\,dex for
$X_{\rm CO, 13}$ and 0.35\,dex for $X_{\rm CO, 12}$.

Overall, our results suggest that the observed changes in
$X_{\rm CO, 12}$ cannot be explained purely by $^{12}$CO
optical depth effects. In that case we would expect $X_{\rm CO, 13}$
to stay approximately constant, while $X_{\rm CO, 12}$ changes.
In contrast, our observations favor variations in $X_{\rm CO, 13}$
as well, with lower values in the centres. This suggests that some
combination of varying parameters (abundance, optical depth, etc.)
is also affecting $X_{\rm CO, 13}$. We note that these
conclusions are all tied to \htwo column estimates from the
dust-based method, so they are subject to the same systematic
uncertainties inherent in that technique.

\section{Conclusions}
\label{sect:concl}
We present new observations of the \coi emission from
the EMPIRE survey \citep{bigiel-2016,jimenez-2017c} and
of the \com emission from follow-up programs obtained
with the IRAM 30-m in 9 nearby spiral galaxies (NGC\,0628,
NGC\,2903, NGC\,3184, NGC\,4254, NGC\,4321, NGC\,5055,
NGC\,6946; and NGC\,5194 from PAWS, \citealt{pety-2013}).
\coi is detected at high signal-to-noise across the
entire molecular disc in those galaxies. We summarize
our results as follows:

\begin{itemize}
\item[-] The integrated intensity of \coi is on average 11 times
fainter than that of \com. The \com-to-\coi intensity ratio ($\Re$)
does not vary significantly within or amongst galaxies
in our sample (at a resolution of $\sim$1.5\,kpc), by a factor
of $\sim$2 at most, and the spread in values is largest in the
galaxy centres. On those spatial scales, the \cou-to-\com ratio
varies between 0.3 and 2.
\item[-] We correlate $\Re$ with several physical quantities
on global and resolved scales. We find no strong trend on
global scales. On resolved scales, $\Re$ anti-correlates
weakly to moderately with the \cou/\com ratio, the IR colour,
\sigsfr, and the SFE (\sigsfr/\sightwo).
Anti-correlations are more visible for galaxies with lowest $\Re$
values in their centres (NGC\,4254, NGC\,5055, NGC\,5194),
which we interpret as a local temperature/excitation effect.
However, other conditions such as density/optical depths must
be at play to explain all $\Re$ behaviors.
\item[-] We find that galaxies with starburst-dominated nuclei
(which are also barred galaxies)
have higher $\Re$ values in their centres than the other
galaxies in our sample. In those galaxies, the anti-correlation
of $\Re$ with the $\alpha_{\rm CO}$ conversion factor from
dust is compatible with optical depth effects. Changes in
the isotope abundance ratio from stellar nucleosynthesis
due to the recent central bursts are also possible.
\item[-] Assuming LTE and fixed abundances, we compute
optical depths and column densities for the \coi line.
Beam-averaged optical depths are on the order of $0.1$.
We find that the \htwo column densities derived from \coi
are systematically lower by a factor of $2-3$ than those
derived from \com.
Those discrepancies can be mainly explained by non-LTE
effects (significant diffuse phase in the discs where $^{13}$CO
is sub-thermally excited) or departure of abundances
from nominal galactic values.
\item[-] Assuming \htwo column densities from dust emission,
we calculate an empirical \coi-to-\htwo conversion factor.
The average value found in our sample is
$1.0\times10^{21}$\,cm$^{-2}$\,(K\,\kms)$^{-1}$.
It increases from galaxy centre to outer disc with similar
scatter ($\sim$0.3\,dex) than the \com-to-\htwo conversion
factor (based on \cou data; \citealt{sandstrom-2013}).
Except in the centres, $^{13}$CO does not appear as
a more stable tracer of the molecular gas mass or
star-formation rate than \com in normal star-forming
disc galaxies.
\end{itemize}

Despite being optically thin, the use of \coi emission 
as a tracer of total mass seems to be limited in
normal star-forming disc galaxies, due to the presence
of a significant diffuse phase and variations in physical
conditions of the gas and abundances.
To constrain those, systematic observations of at least
one more $^{13}$CO transitions will be an important next step.

\section*{Acknowledgements}
We thank Sacha Hony, Maud Galametz, Fr{\'e}d{\'e}ric Galliano
for useful discussions, and the referee for a careful reading of
our manuscript.
DC is supported by the European Union's Horizon 2020 research
and innovation programme under the Marie Sk\l{}odowska-Curie
grant agreement No 702622.
DC also acknowledges support from the DAAD/PROCOPE projects
57210883/35265PE.
MJJD and FB acknowledge support from DFG grant BI 1546/1-1.
FB acknowledges funding from the European Union's Horizon 2020
research and innovation programme (grant agreement No 726384 - EMPIRE).
The work of MG and AKL is partially supported by the
National Science Foundation under Grants No. 1615105, 1615109, and 1653300.
ER is supported by a Discovery Grant from NSERC of Canada.
ES acknowledges funding from the European Research Council
(ERC) under the European UnionÕs Horizon 2020 research and
innovation programme (grant agreement No. 694343)
This work is based on observations carried out with the IRAM 
30-m Telescope. IRAM is supported by INSU/CNRS (France), 
MPG (Germany) and IGN (Spain).



\bibliographystyle{mnras}
\bibliography{references-empire-13co}


\appendix
%
\section{EMPIRE \coi observations and radial profiles}
This Appendix presents an atlas of the dust continuum
at 70\,$\mu$m, \coi emission, and $\Re$ in the EMPIRE
galaxies, as well as radial profiles and stacked spectra
of CO lines for each galaxy.

\begin{figure*}
\centering
\includegraphics[width=5.8cm]{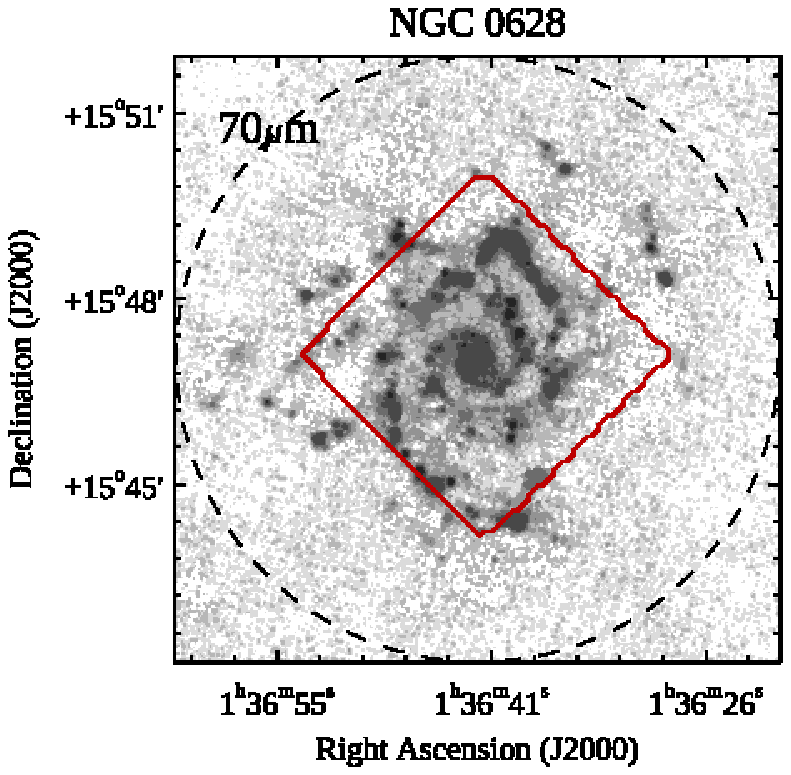}
\includegraphics[width=5.8cm]{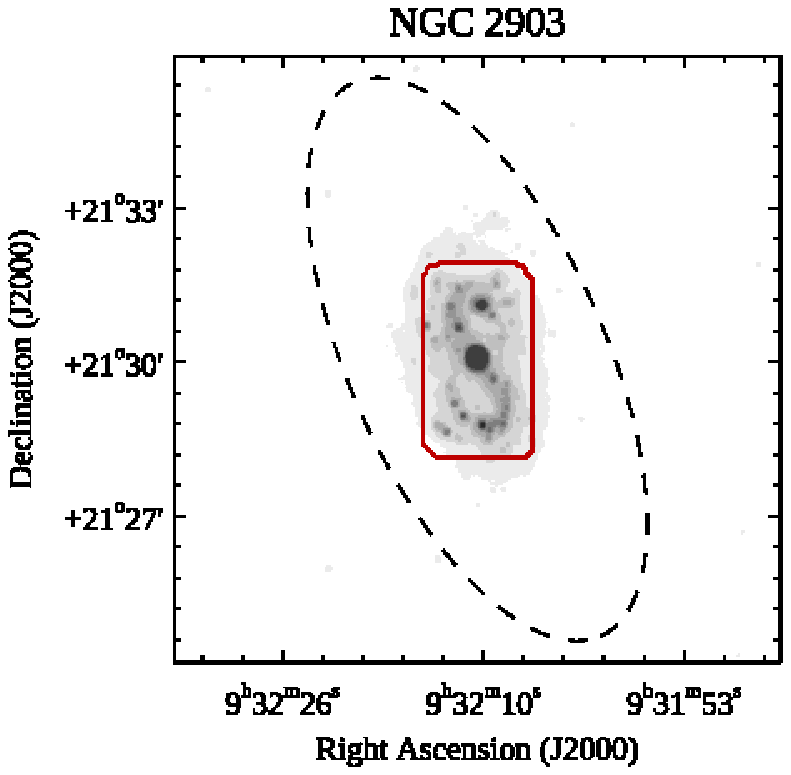}
\includegraphics[width=5.8cm]{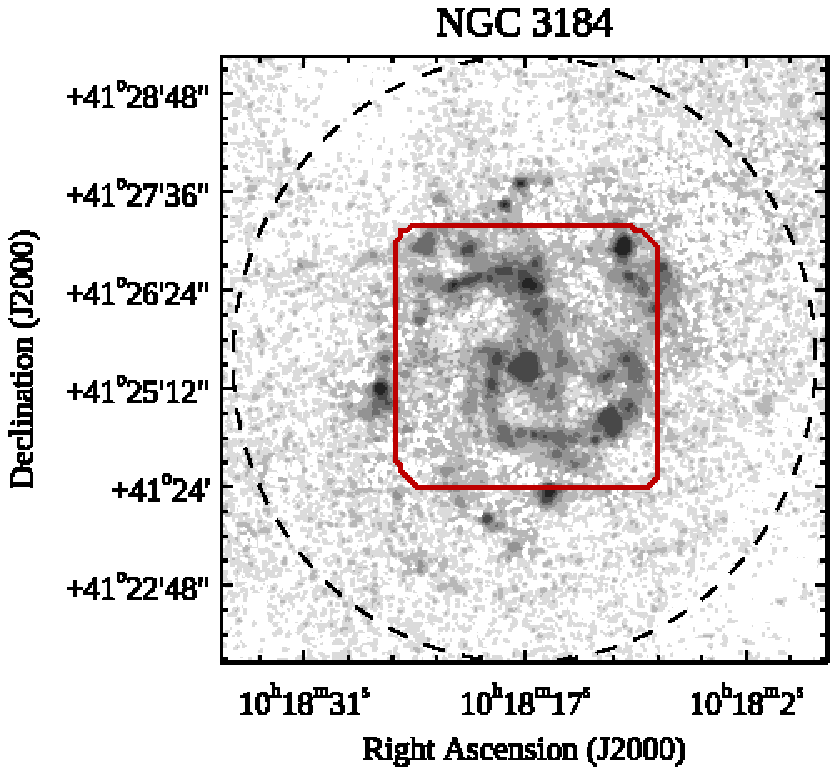} \\
\includegraphics[width=5.8cm]{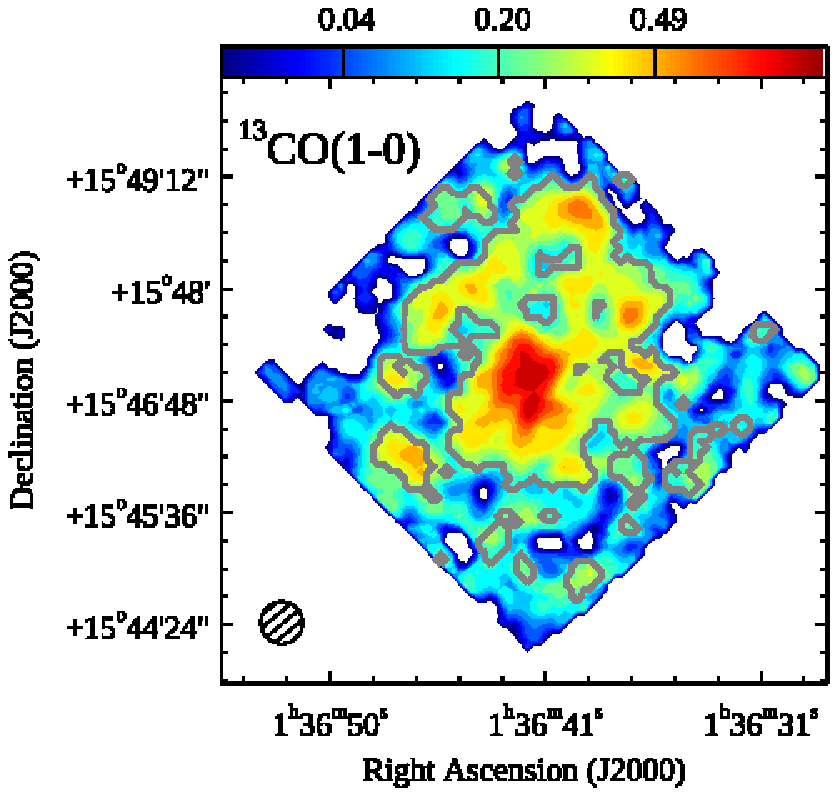}
\includegraphics[width=5.8cm]{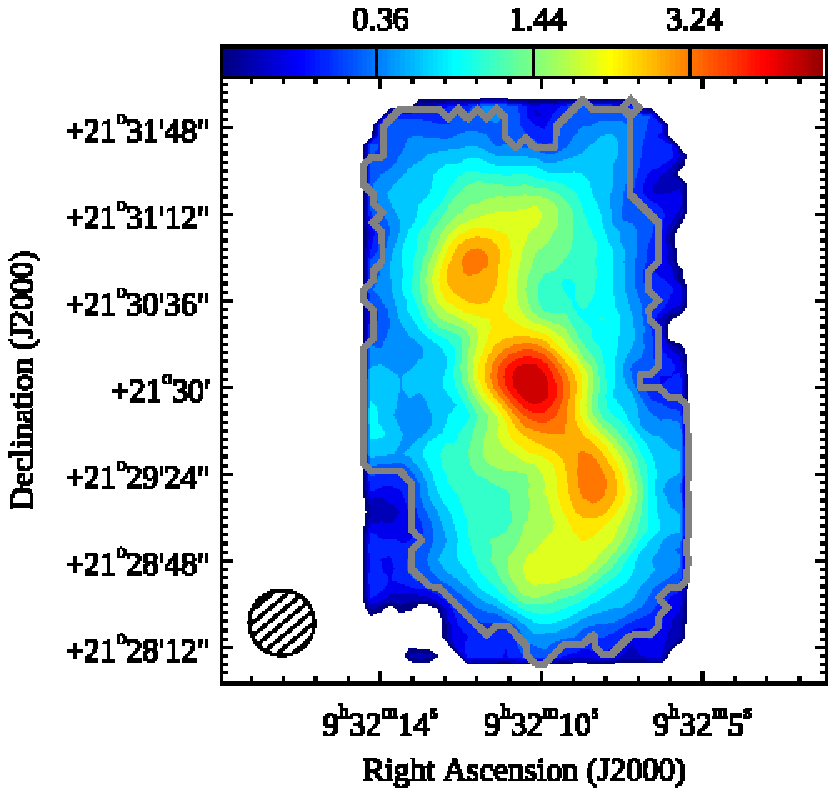}
\includegraphics[width=5.8cm]{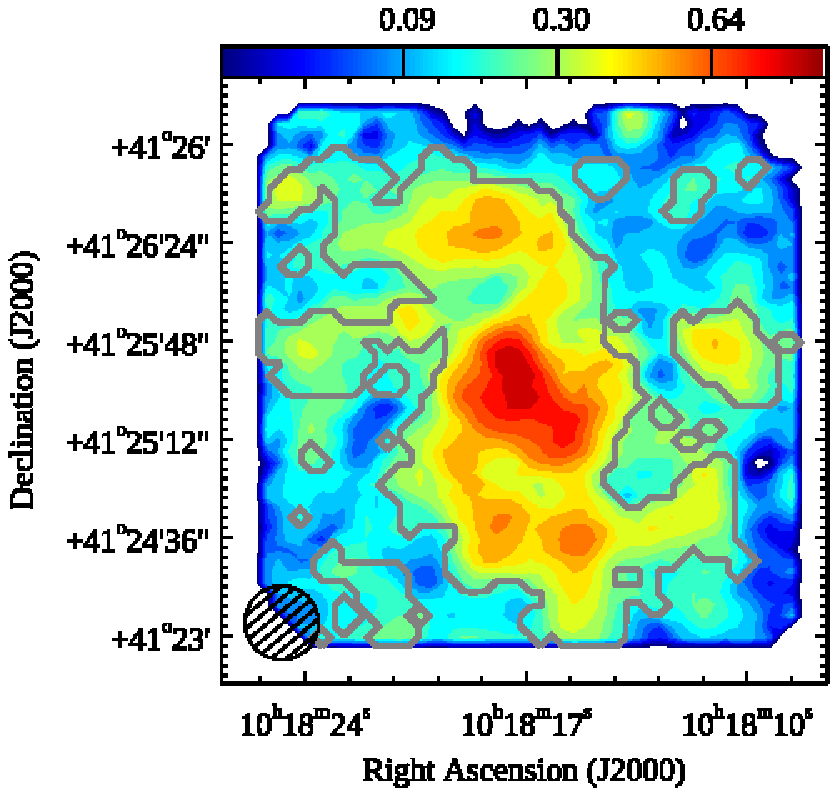} \\
\includegraphics[width=5.8cm]{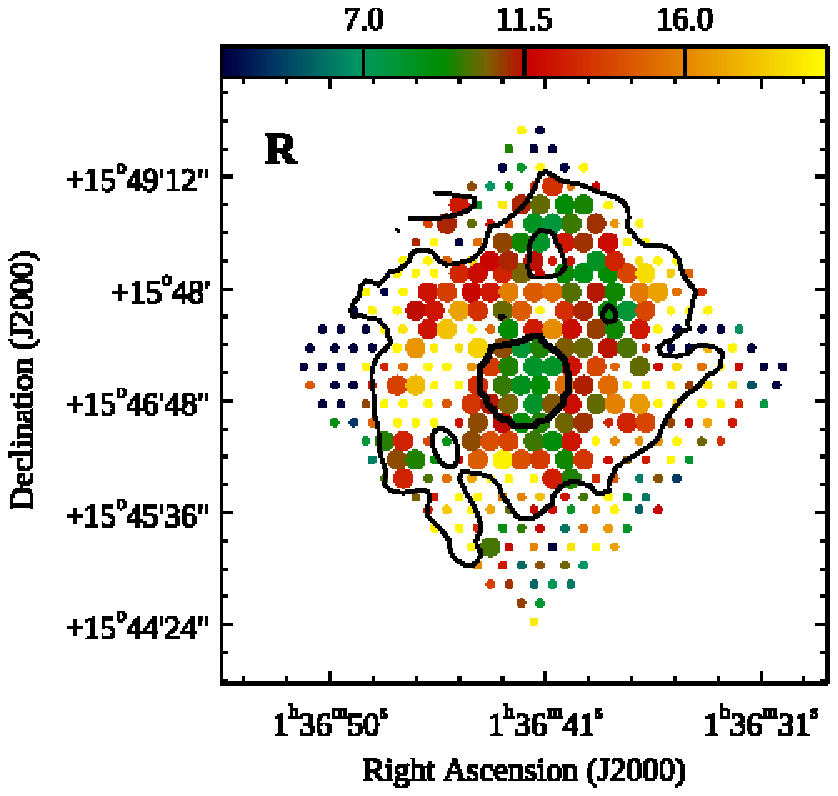}
\includegraphics[width=5.8cm]{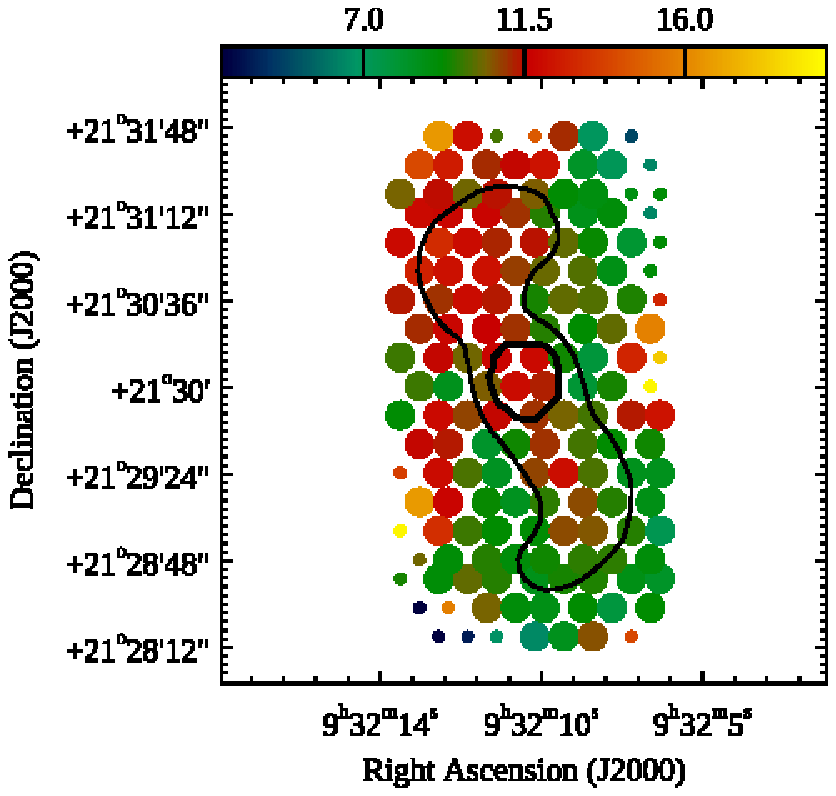}
\includegraphics[width=5.8cm]{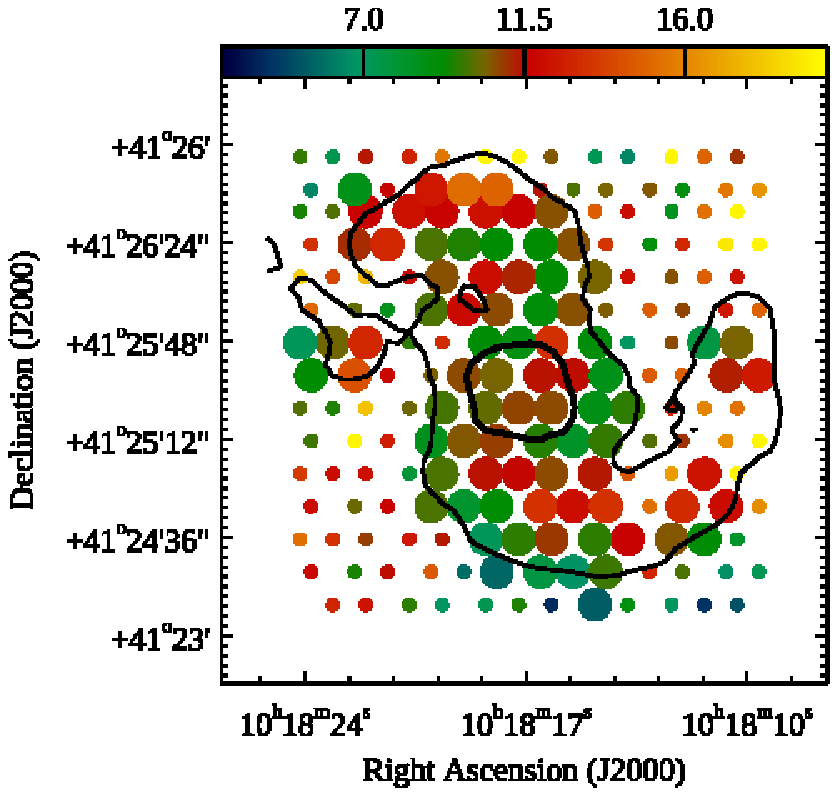} \\
\caption{
Atlas of infrared dust continuum, \coi integrated intensity, and
\com-to\coi ratio ($\Re$) for the EMPIRE galaxies.
\underline{Top panels:}
{\it Herschel} 70\,$\mu$m map ({\it Spitzer} 24\,$\mu$m map
in the case of NGC\,2903), in square root scale.
The dashed ellipse shows the $r_{25}$ radius and
the red contours show the IRAM coverage.
\underline{Middle panels:}
\coi integrated intensity map, in square root scale.
The colourbar is in units of K\,\kms.
The beam size of 27\,arcsec is indicated on the bottom left corner.
Contours at 5$\sigma$ are shown in grey. 
\underline{Bottom panels:}
Map of $\Re$ where we draw a filled circle for each sampling point.
Smaller circles correspond to sampling points with signal-to-noise
ratio of $\Re$ below $5$. 
Black contours delineate the center and arm regions that are
defined by eye on the \com intensity maps and used for stacking
(see section~\ref{sect:envir}).
}
\label{fig:maps}
\vspace{8cm}
\end{figure*}
\begin{figure*}
\centering
\includegraphics[width=5.8cm]{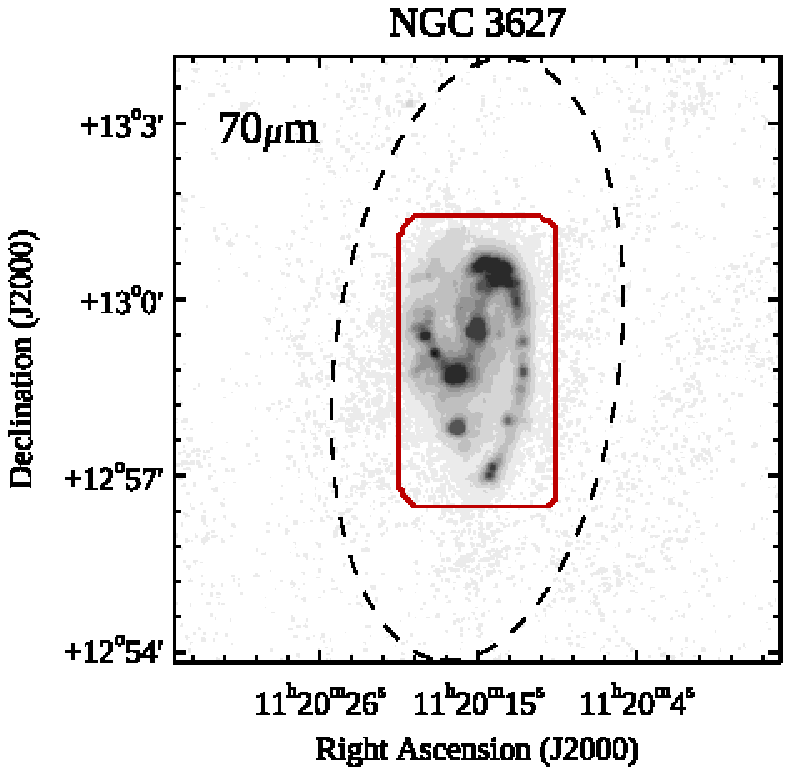}
\includegraphics[width=5.8cm]{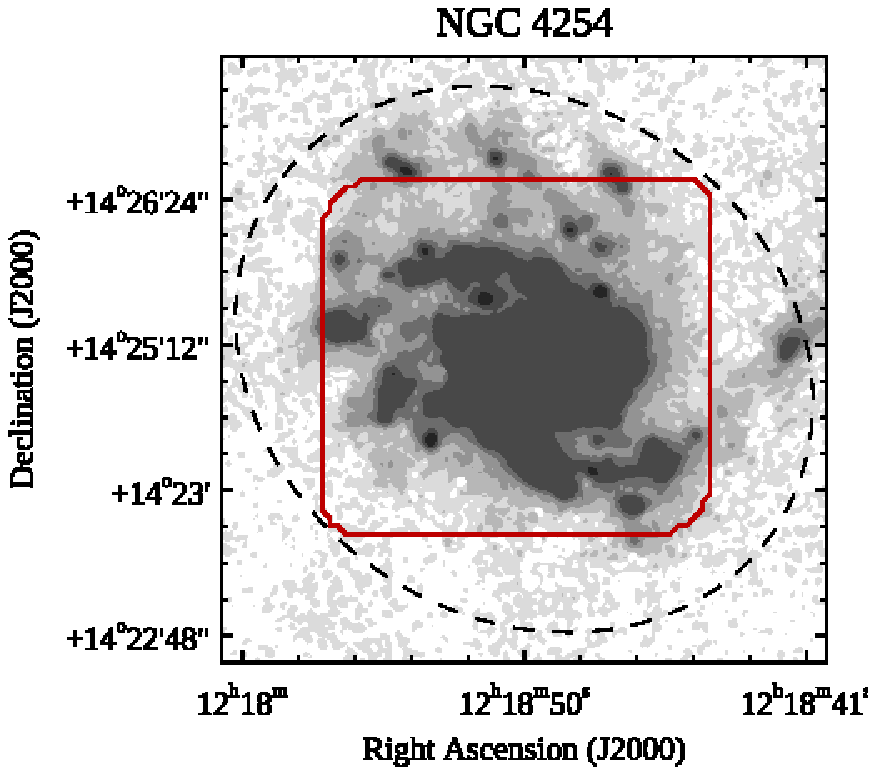}
\includegraphics[width=5.8cm]{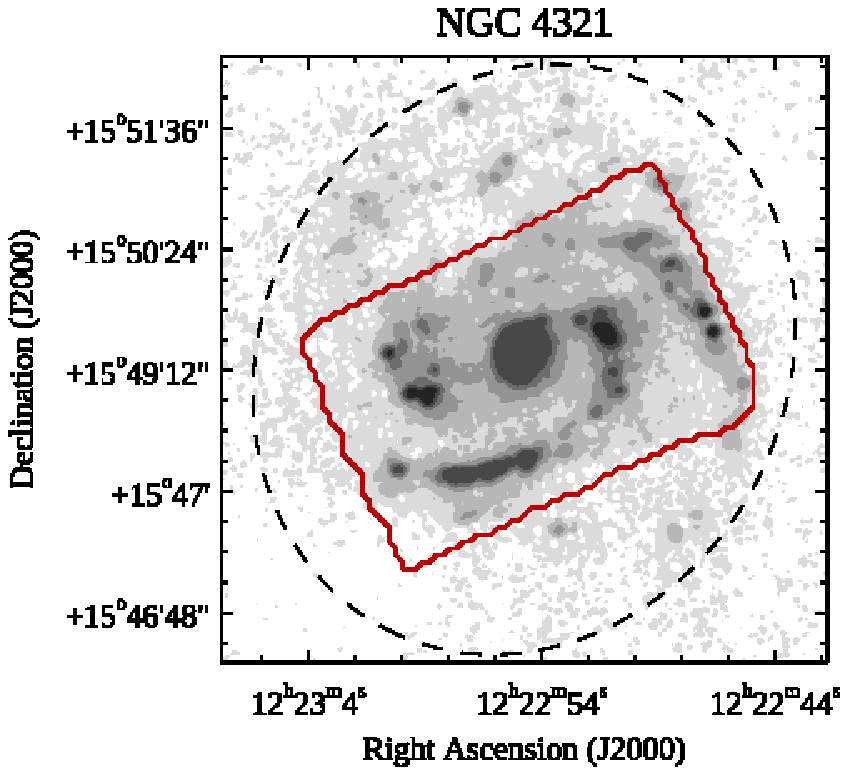} \\
\includegraphics[width=5.8cm]{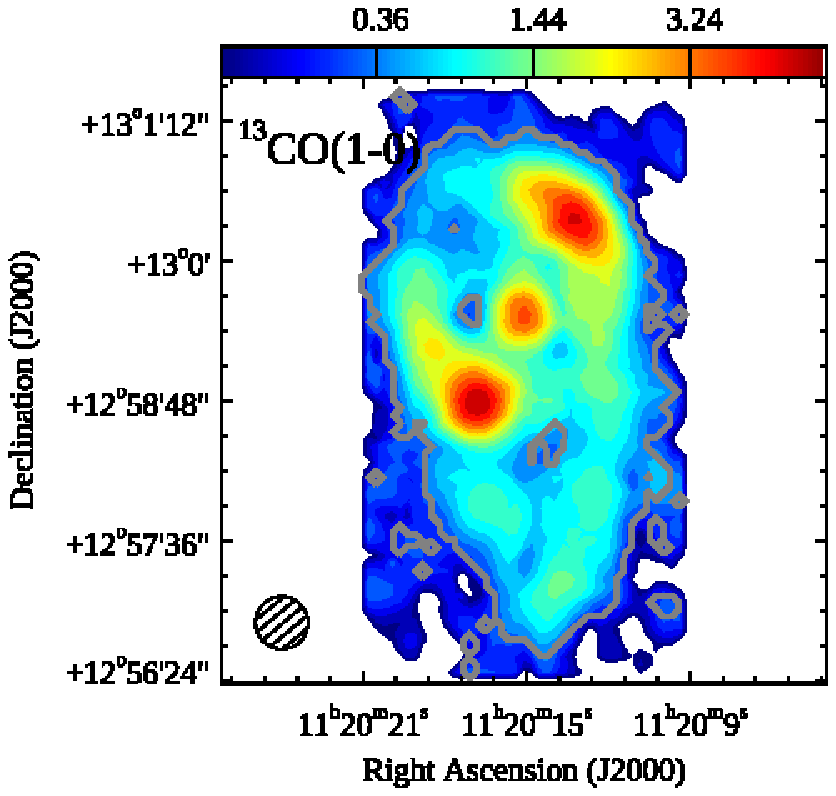}
\includegraphics[width=5.8cm]{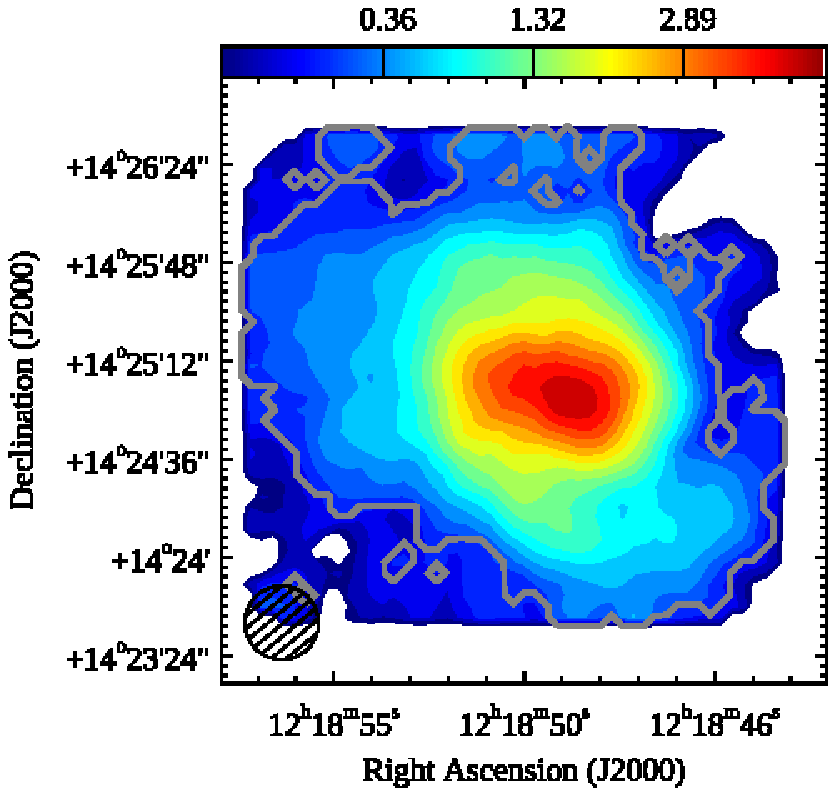}
\includegraphics[width=5.8cm]{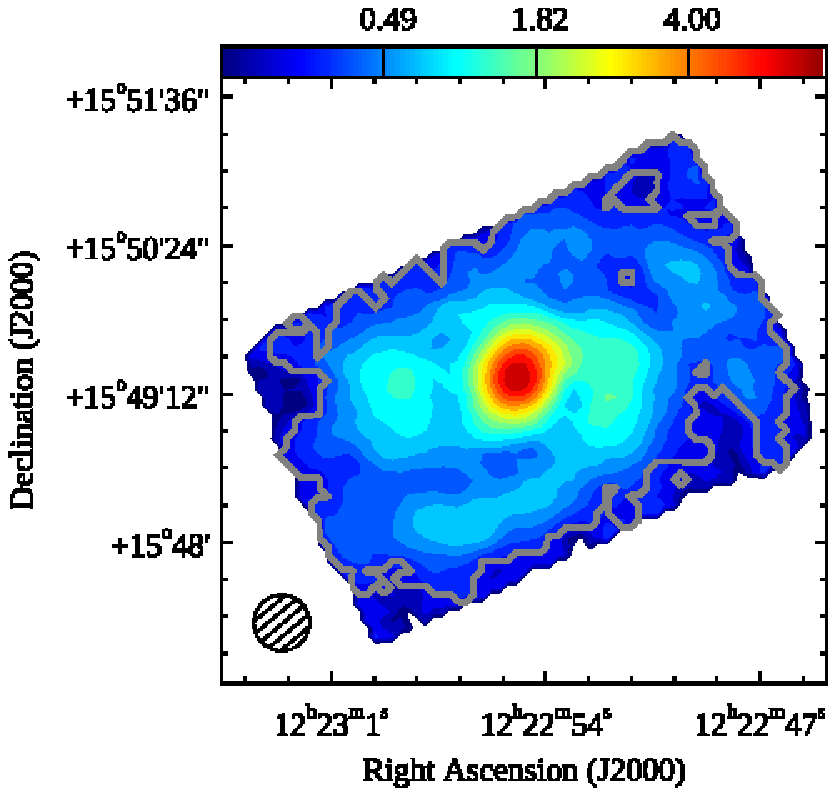} \\
\includegraphics[width=5.8cm]{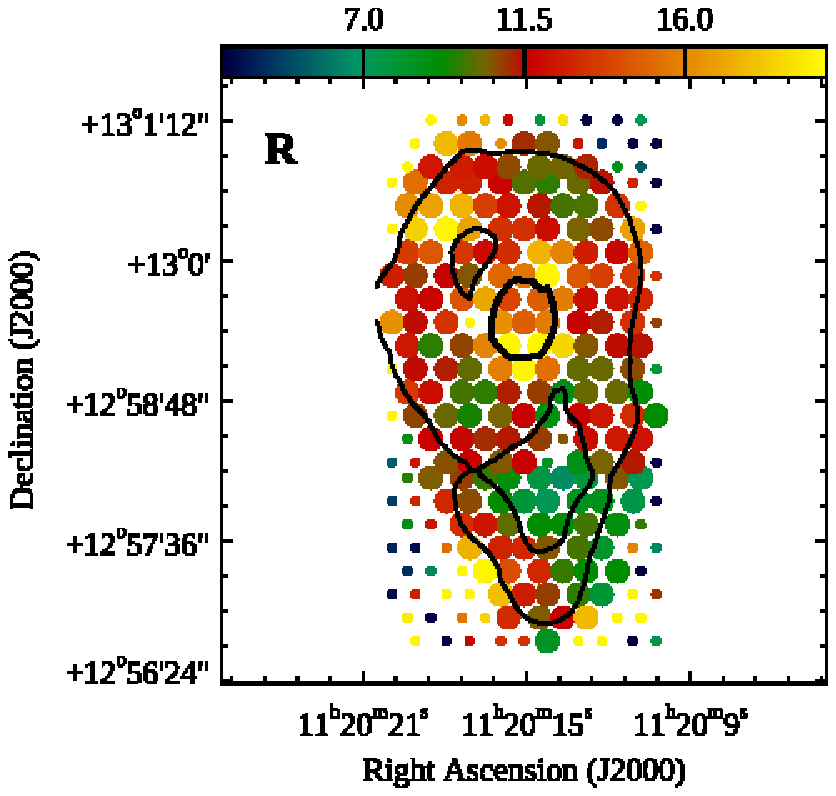}
\includegraphics[width=5.8cm]{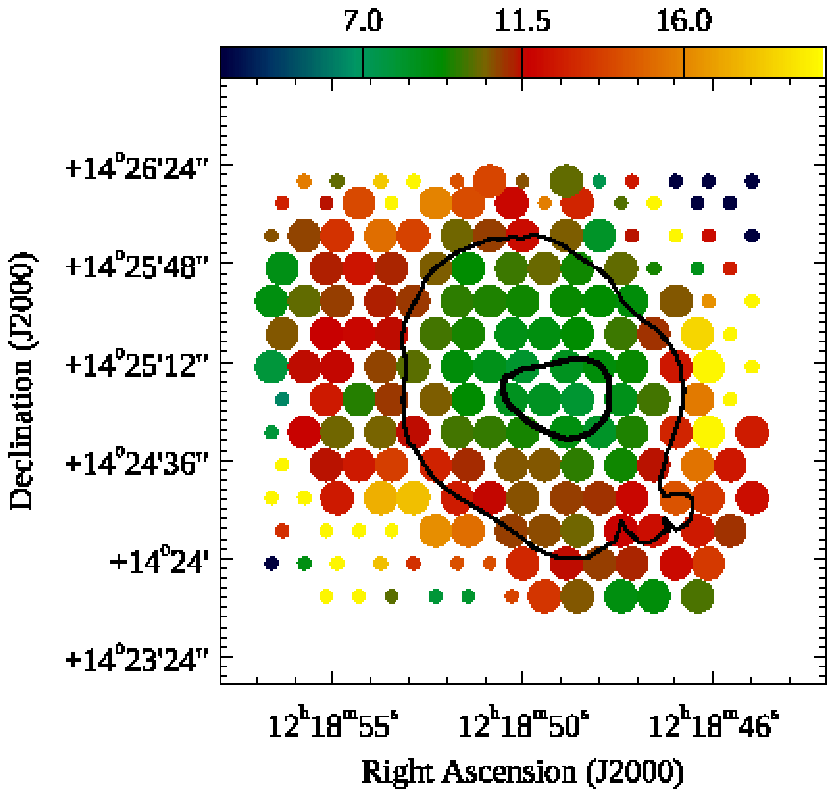}
\includegraphics[width=5.8cm]{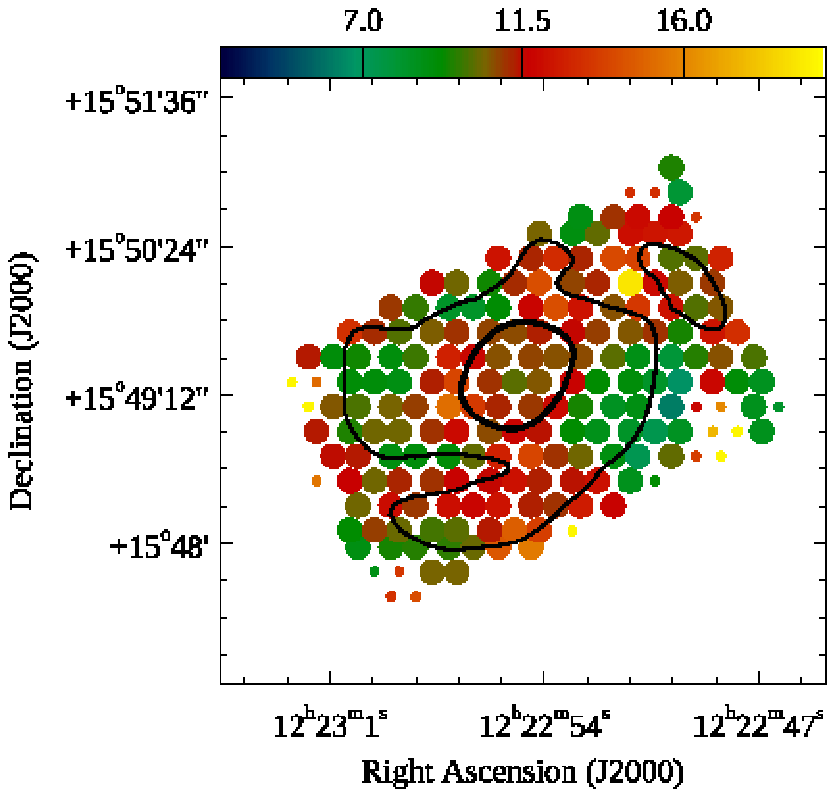} \\
\contcaption{}
\vspace{8cm}
\end{figure*}
\begin{figure*}
\centering
\includegraphics[width=5.8cm]{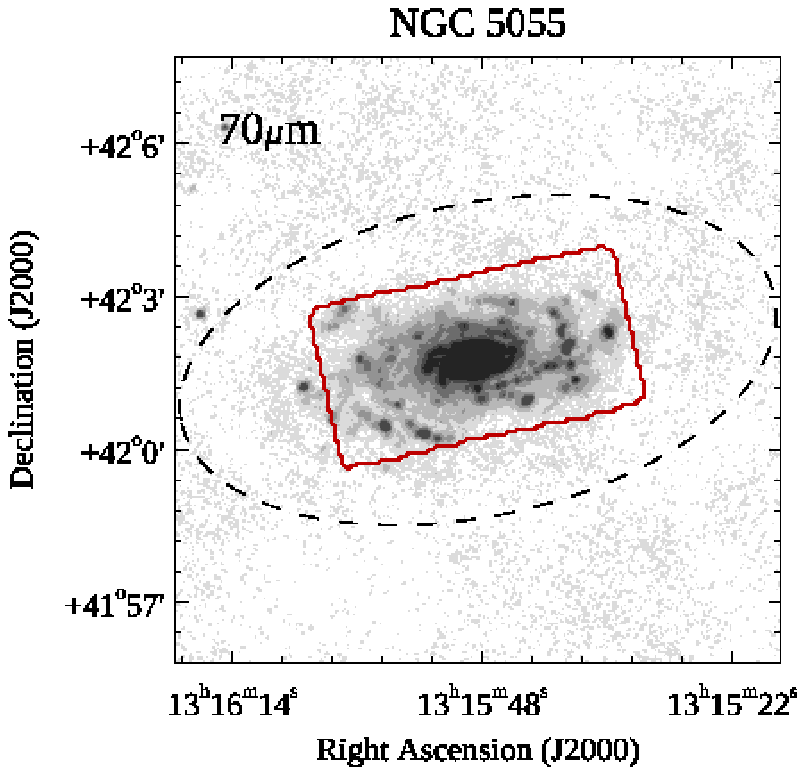} 
\includegraphics[width=5.8cm]{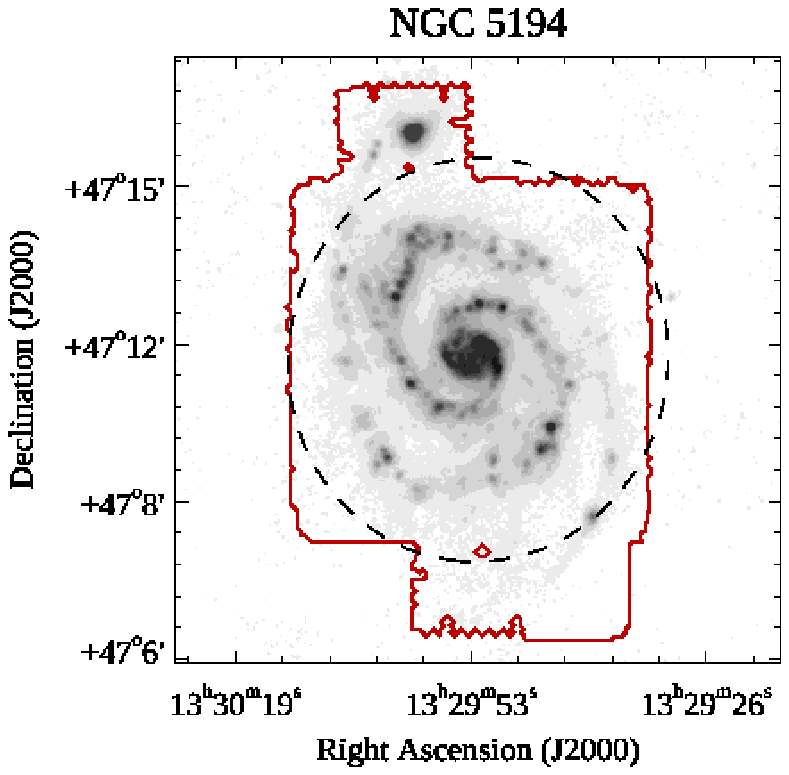}
\includegraphics[width=5.8cm]{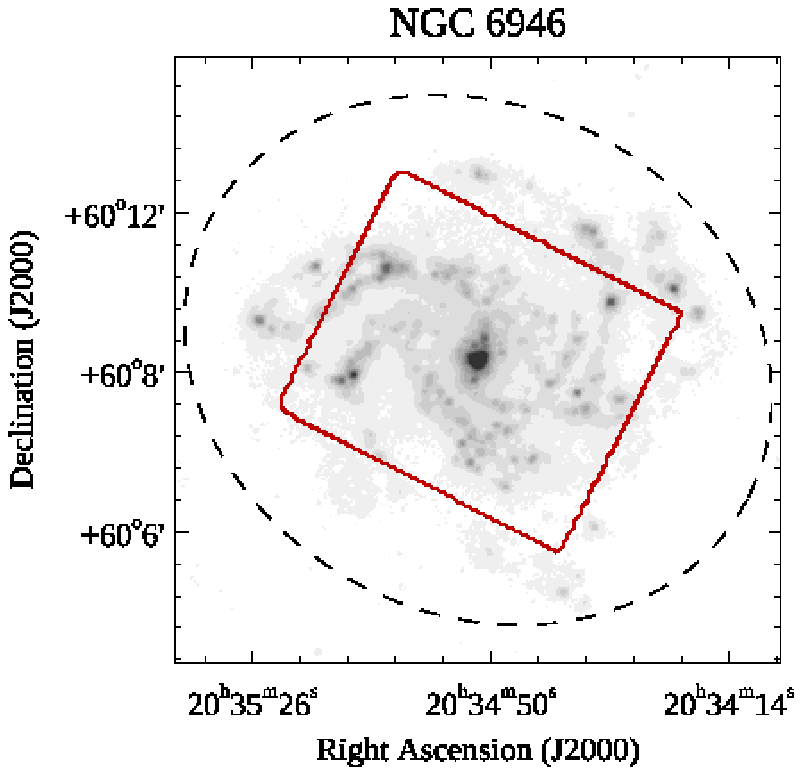} \\
\includegraphics[width=5.8cm]{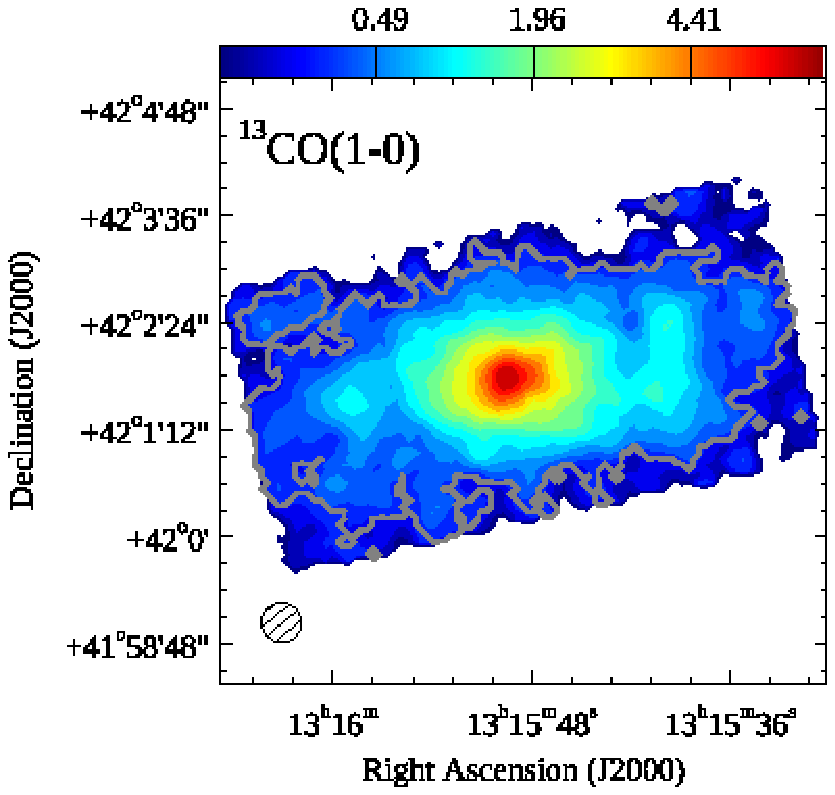} 
\includegraphics[width=5.8cm]{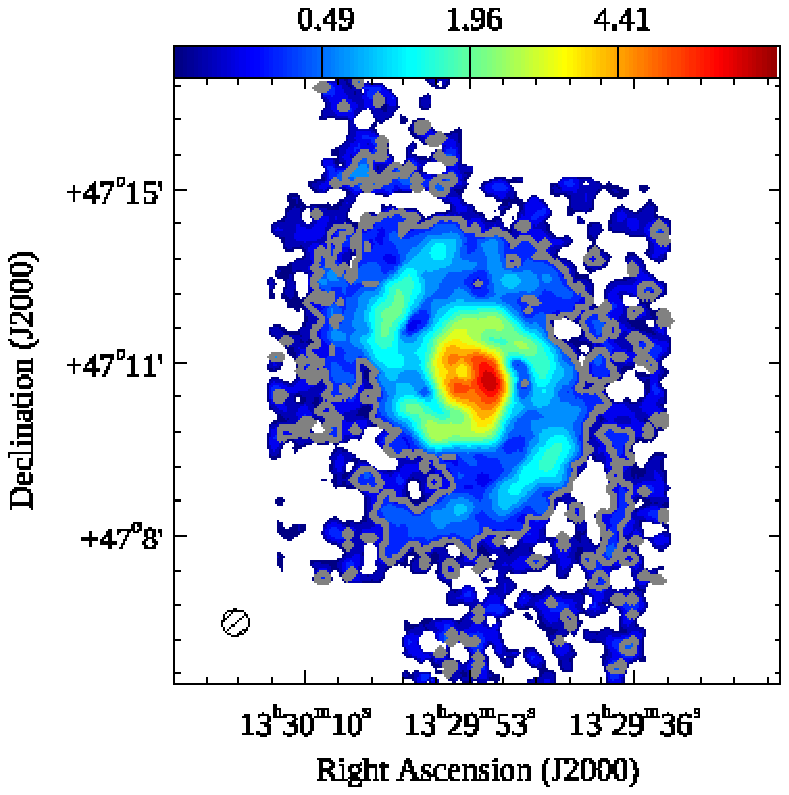}
\includegraphics[width=5.8cm]{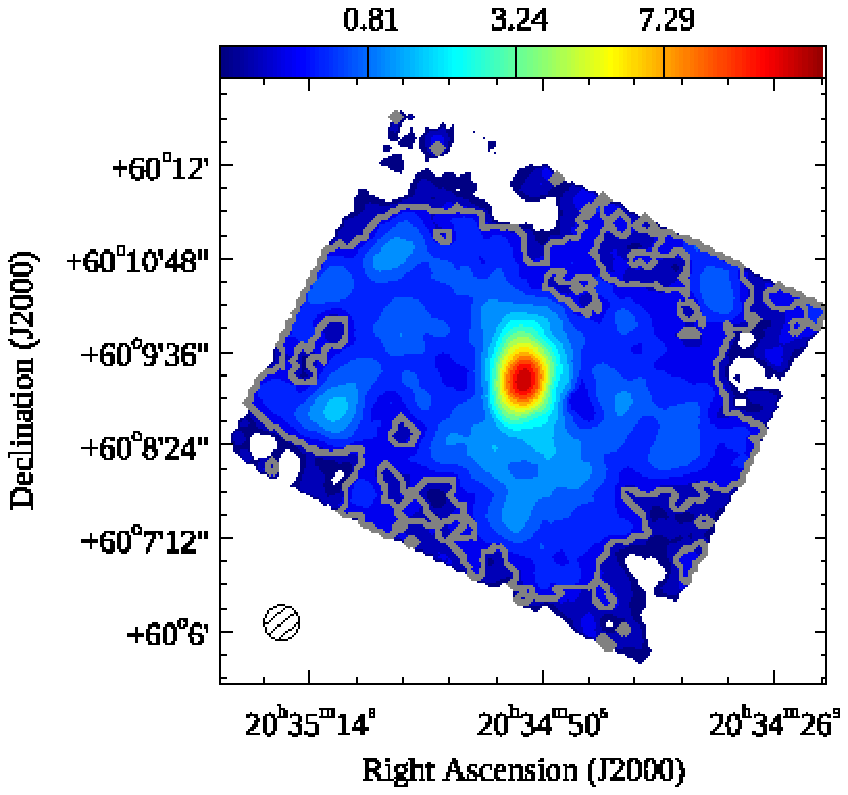} \\
\includegraphics[width=5.8cm]{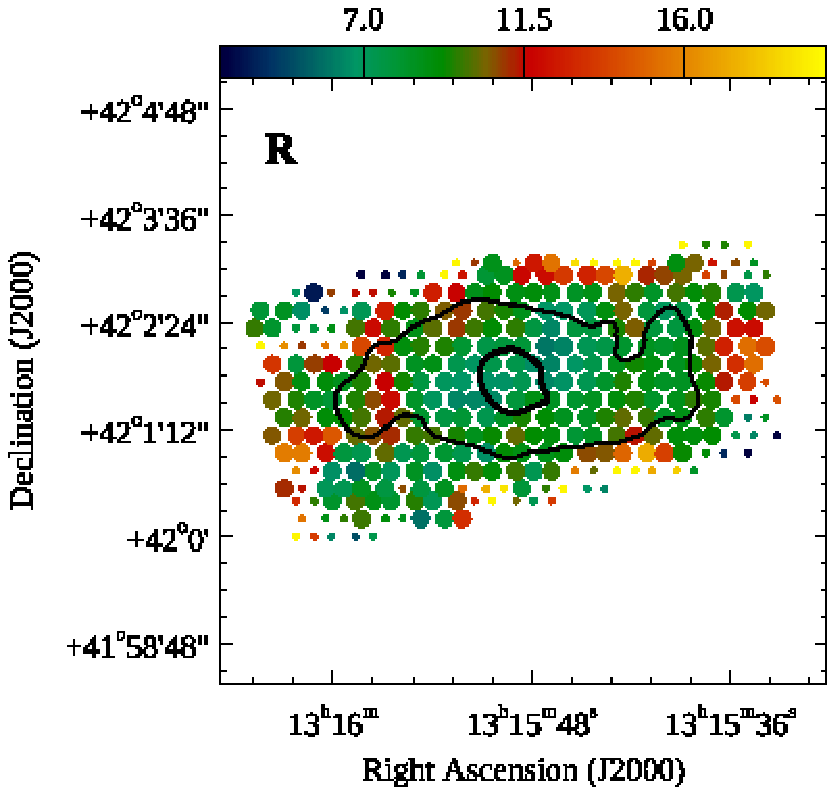} 
\includegraphics[width=5.8cm]{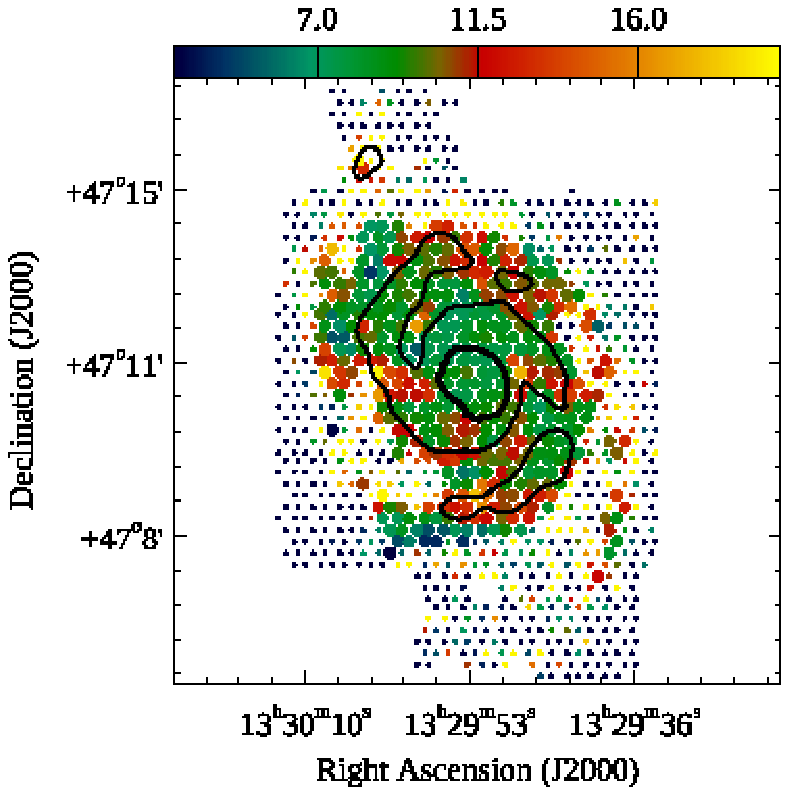}
\includegraphics[width=5.8cm]{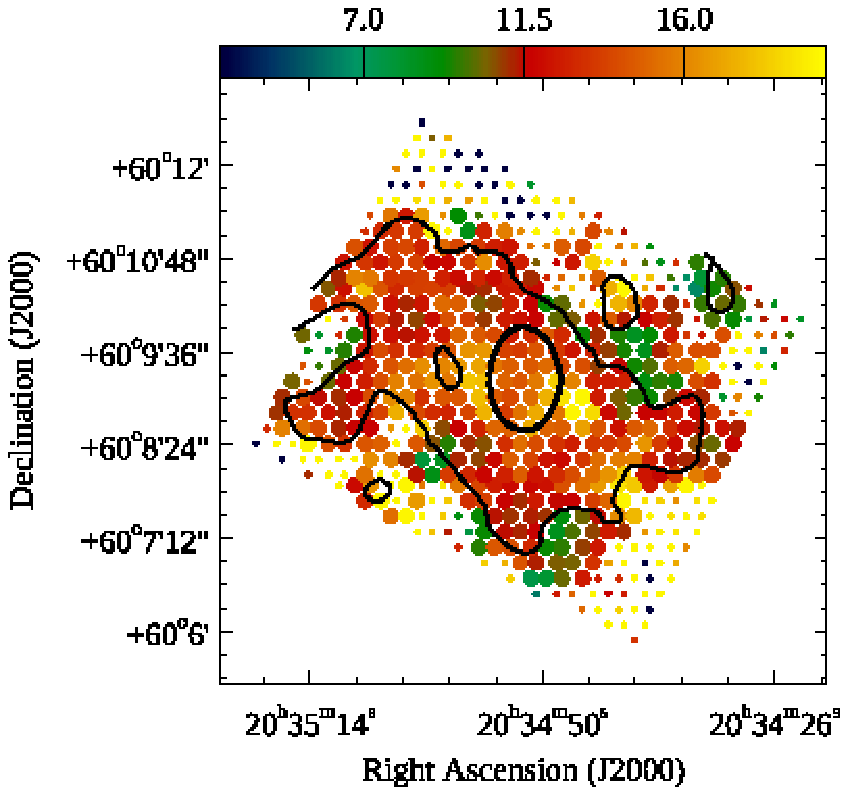} \\
\contcaption{}
\vspace{8cm}
\end{figure*}
%
\begin{figure*}
\centering
\includegraphics[clip,width=5.8cm]{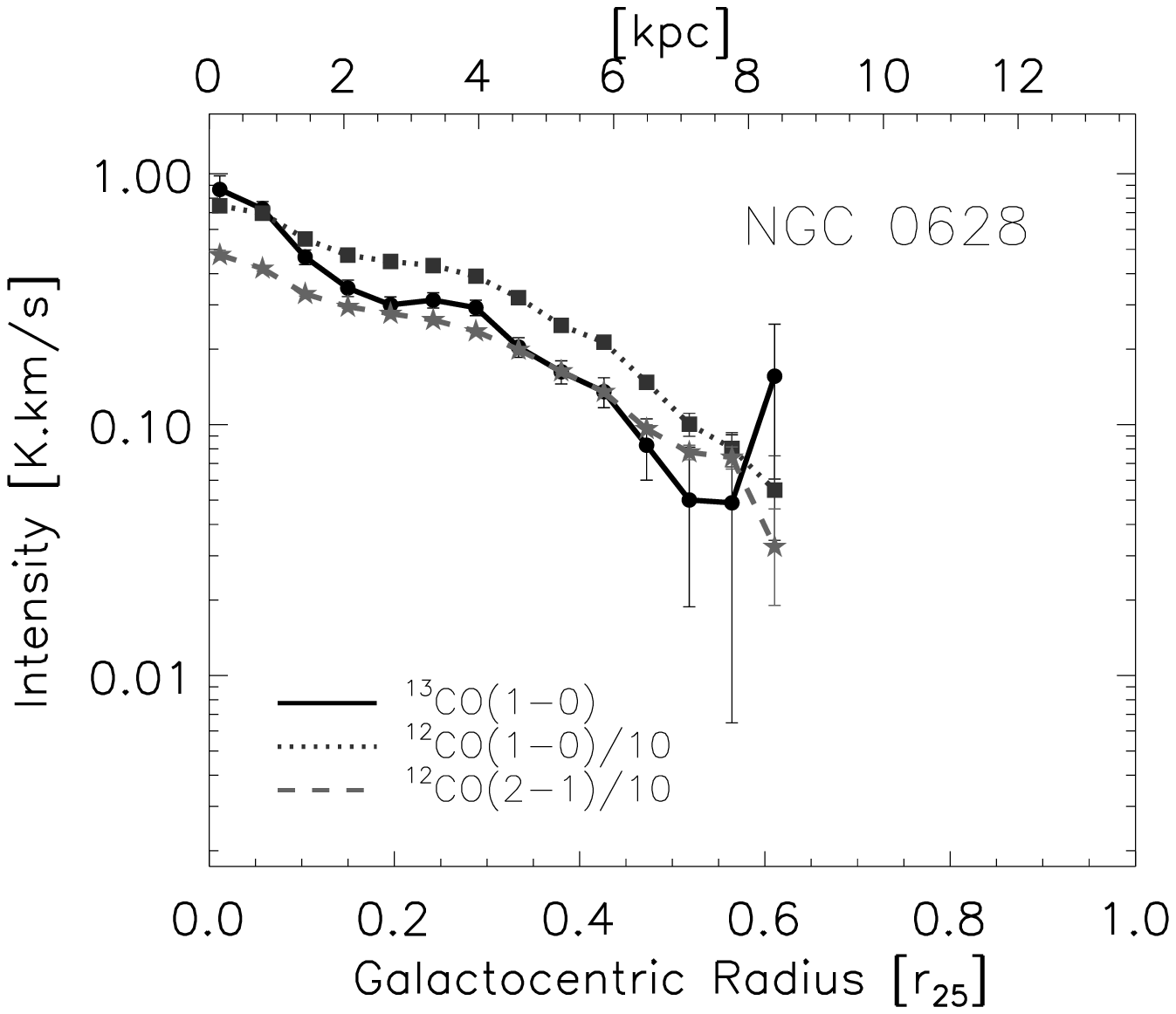}
\includegraphics[clip,width=5.8cm]{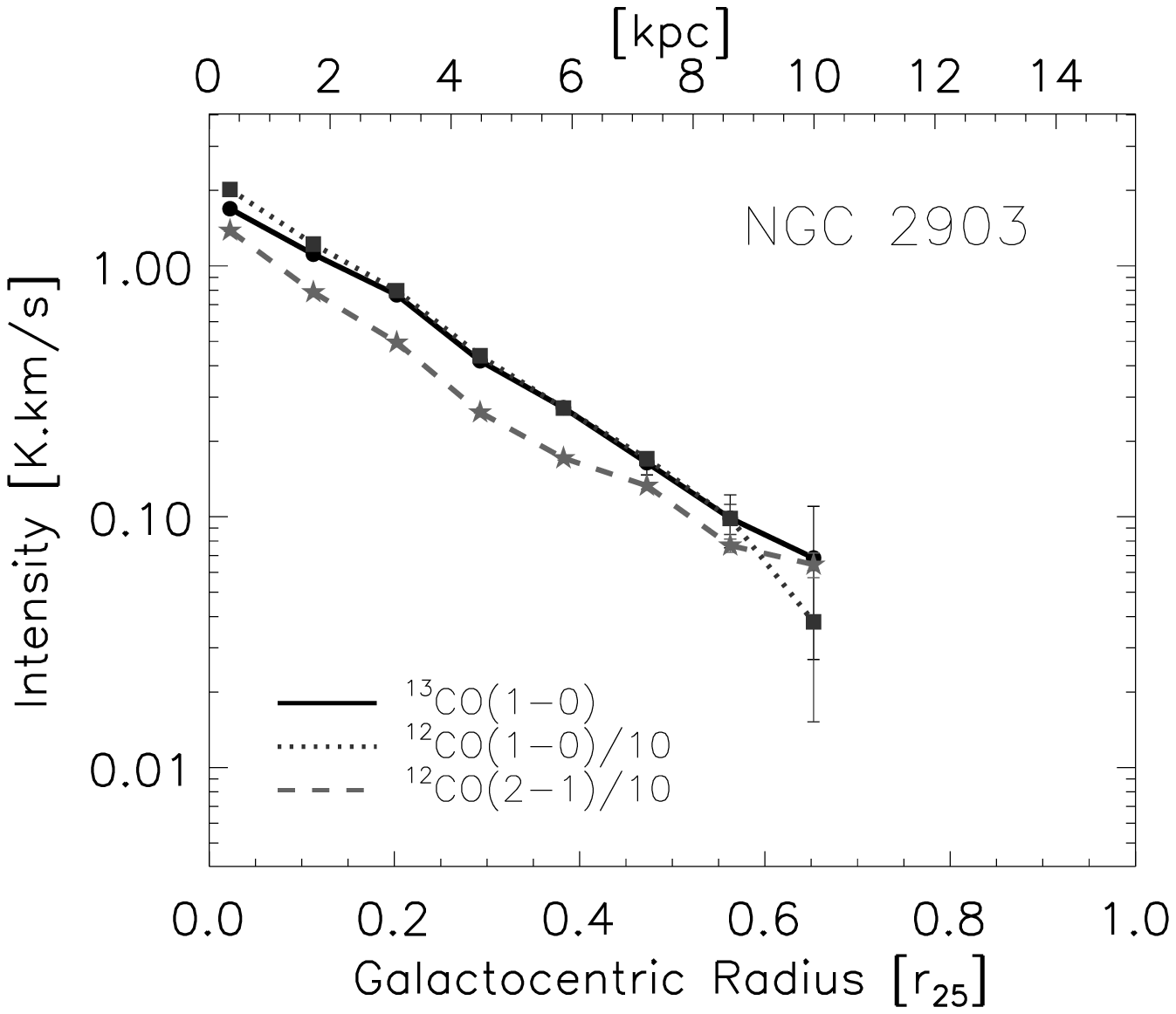}
\includegraphics[clip,width=5.8cm]{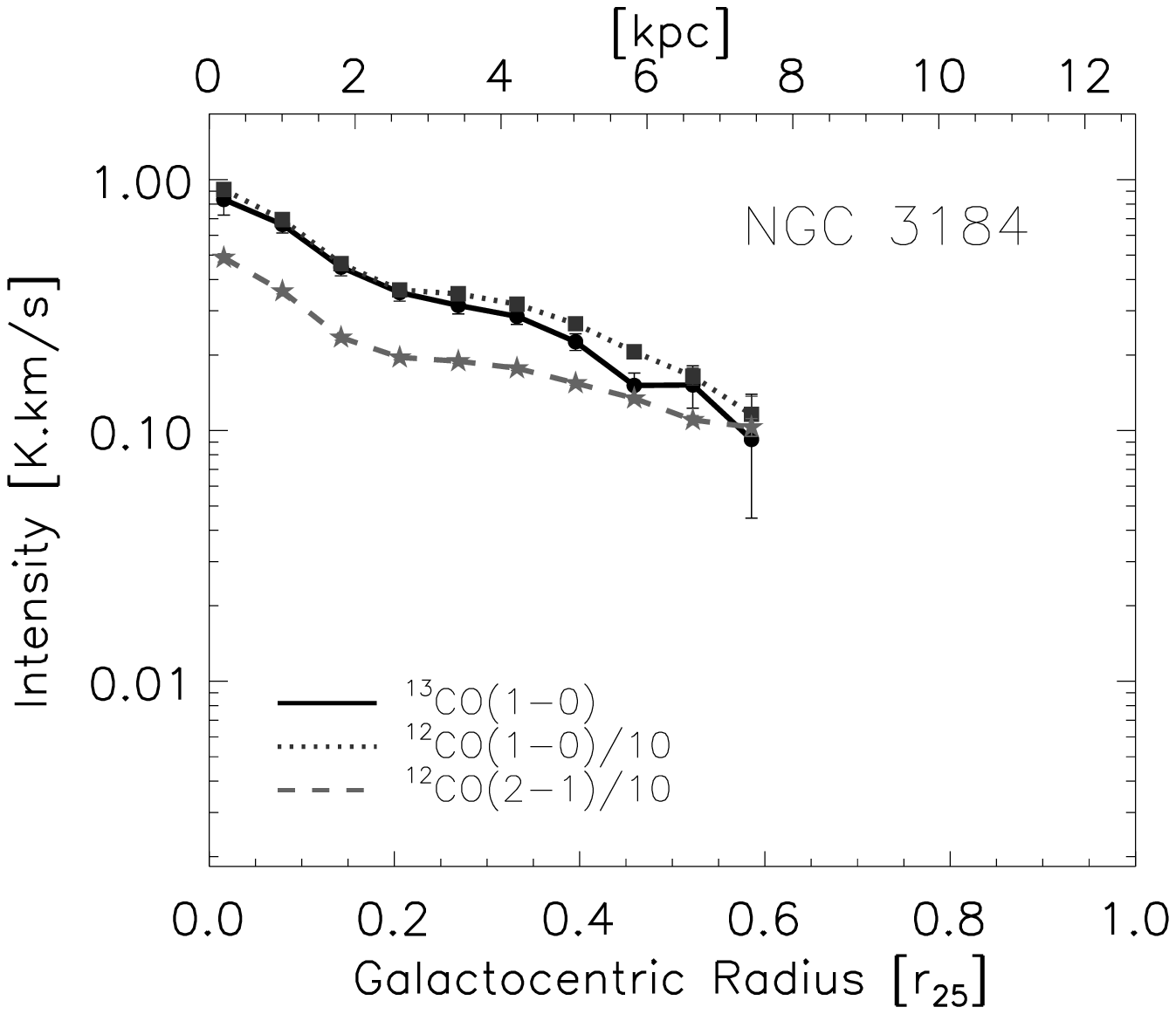} \hfill
\includegraphics[clip,width=5.8cm]{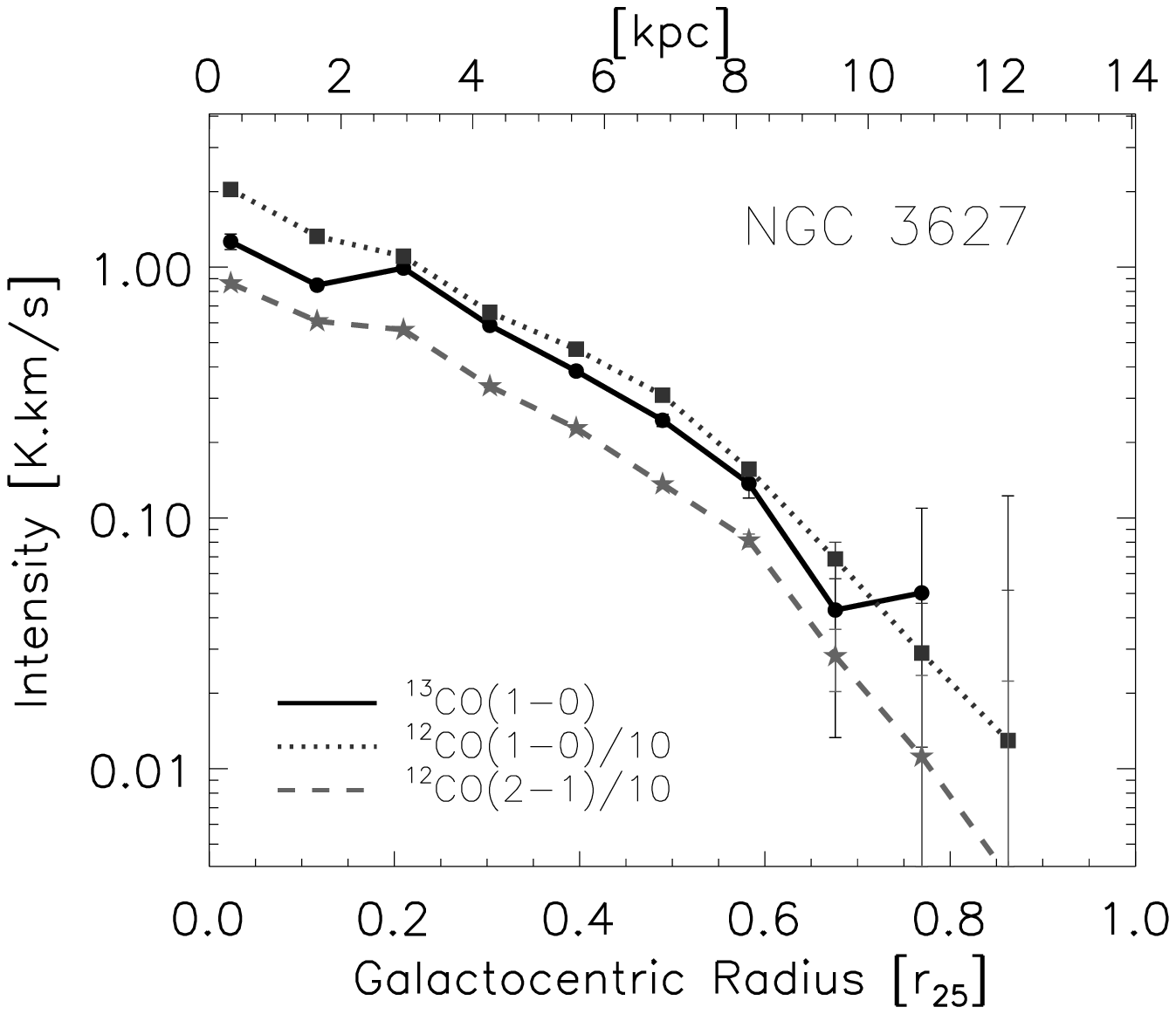}
\includegraphics[clip,width=5.8cm]{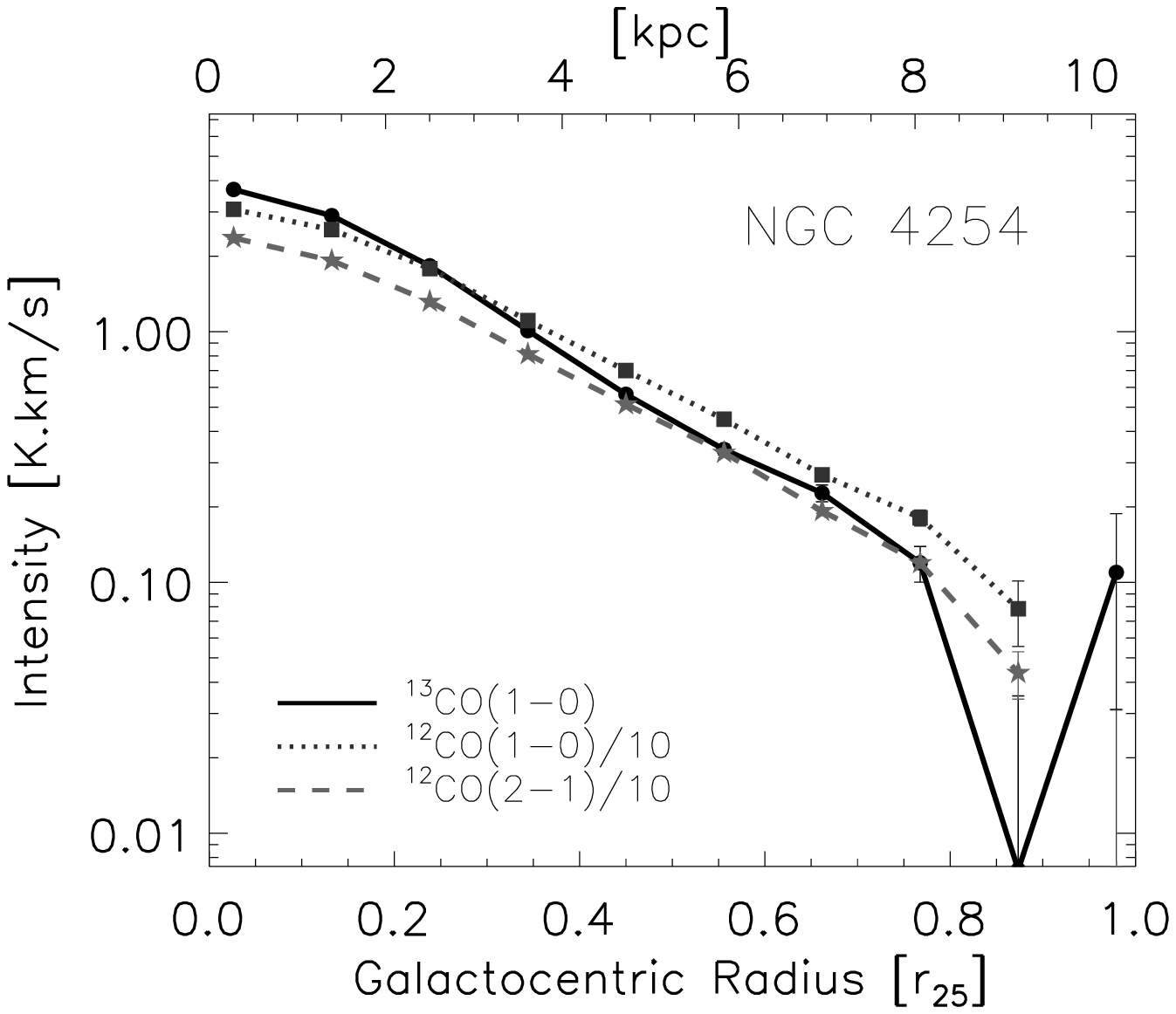}
\includegraphics[clip,width=5.8cm]{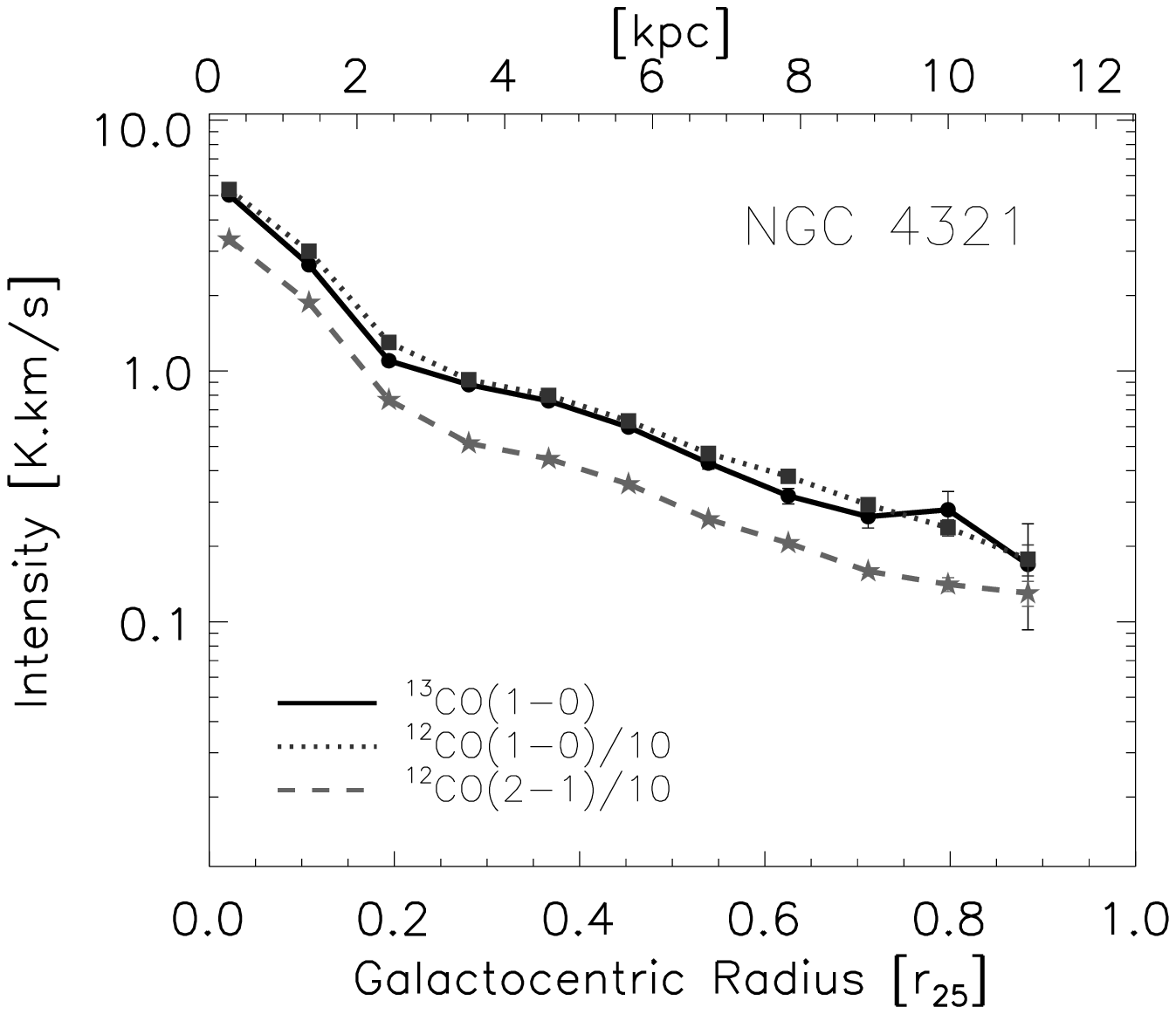} \hfill
\includegraphics[clip,width=5.8cm]{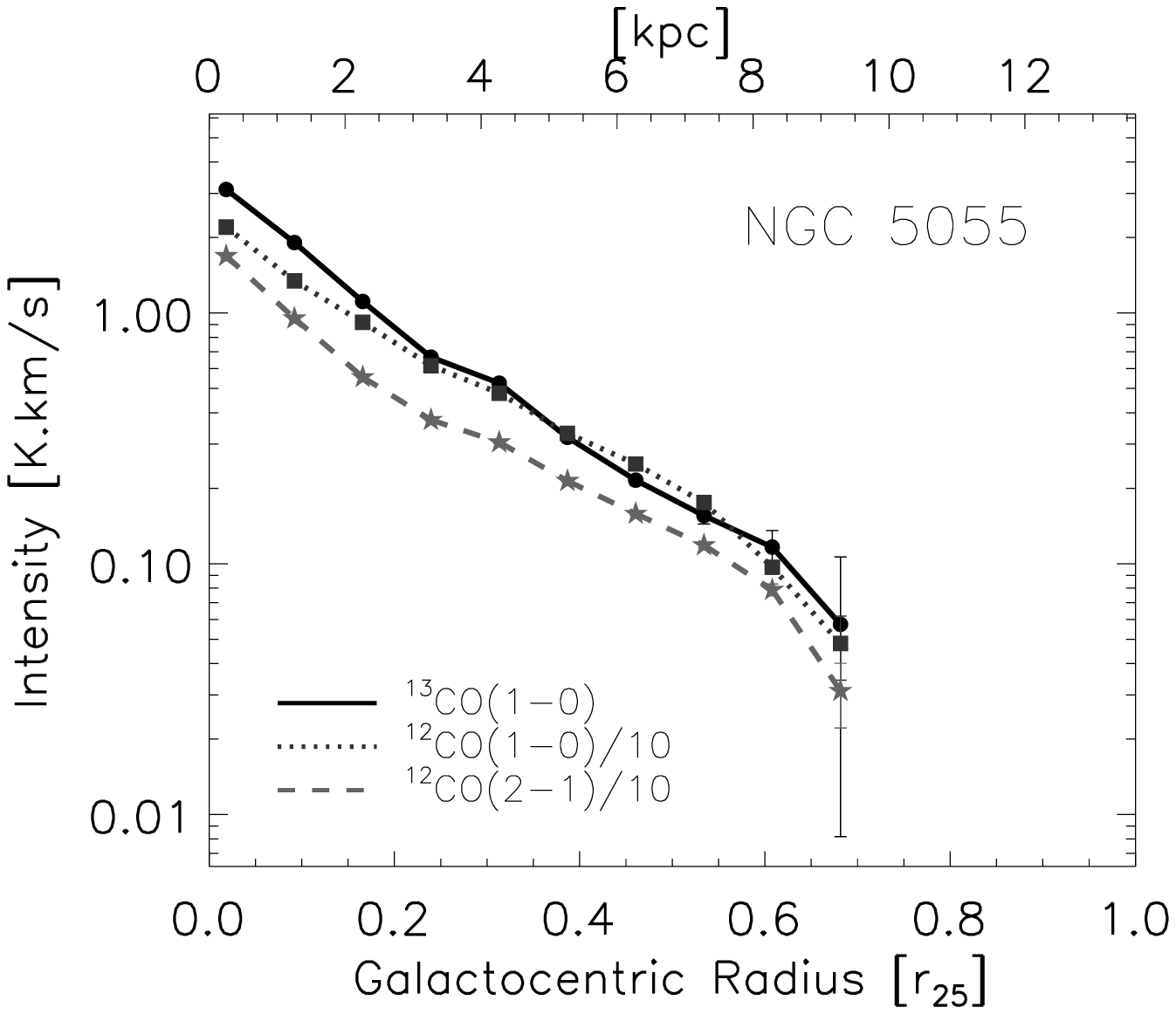}
\includegraphics[clip,width=5.8cm]{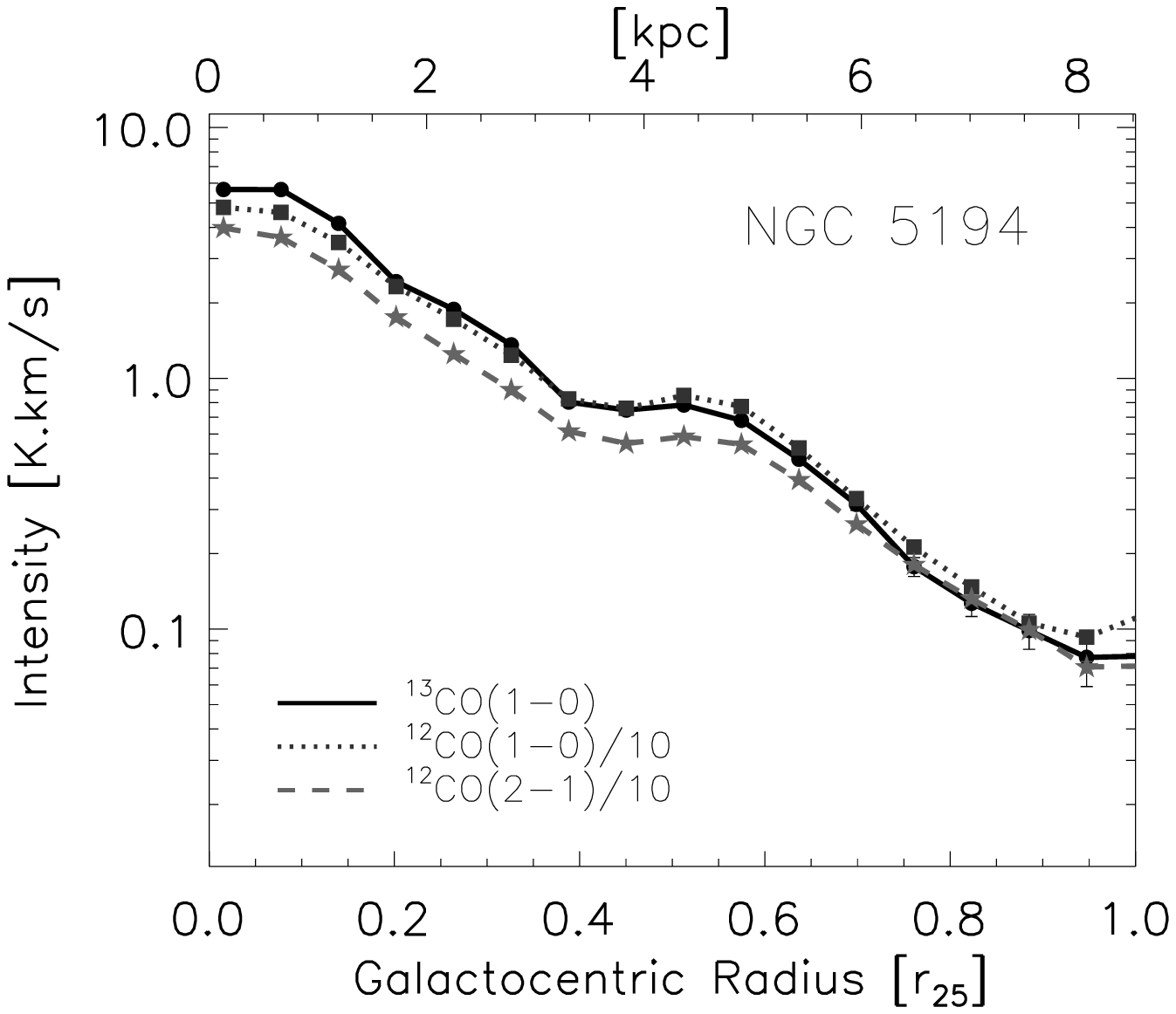}
\includegraphics[clip,width=5.8cm]{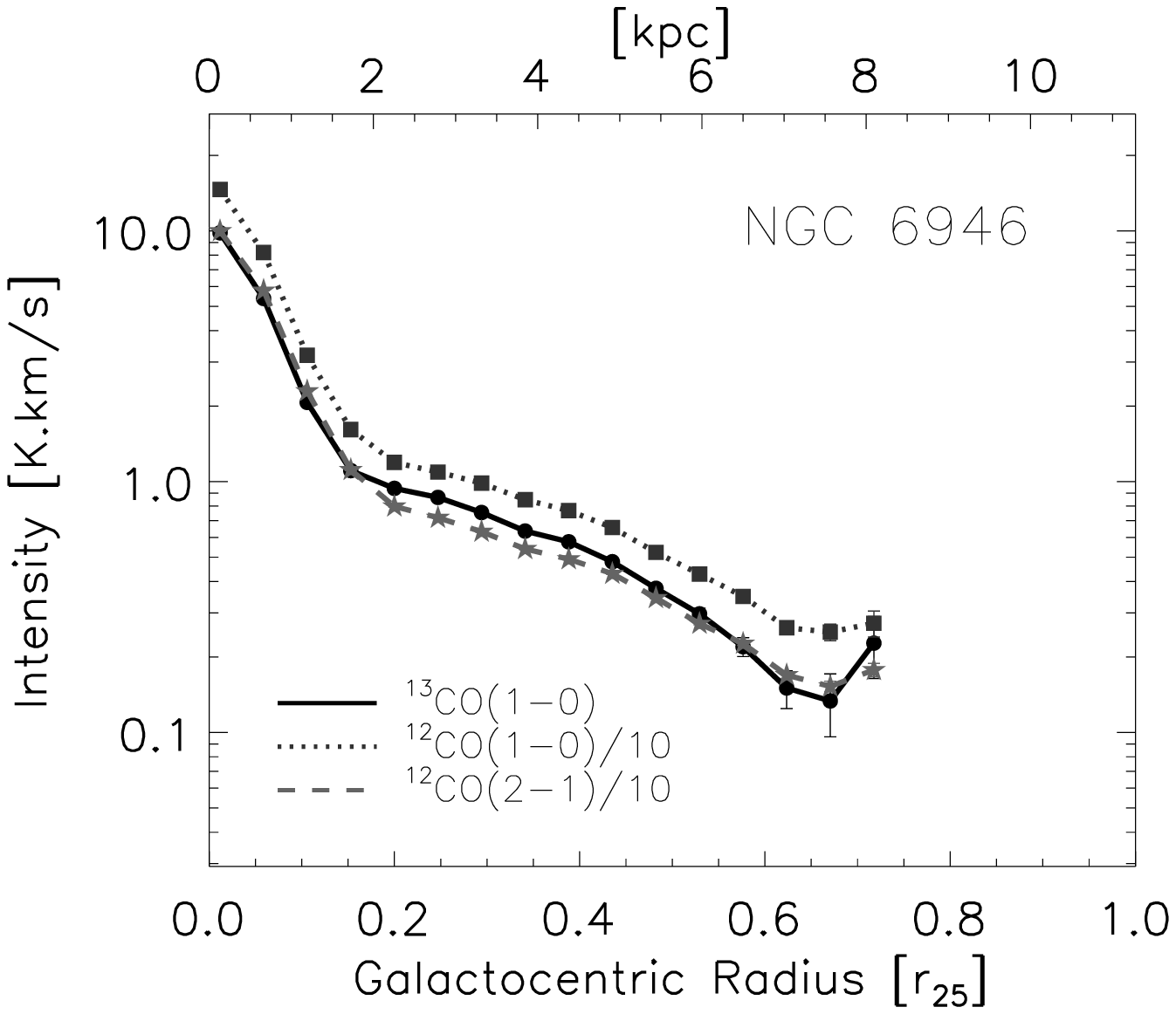}
\caption{
Profiles of the \coi, \com and \cou intensities as a function of galactocentric radius.
}
\label{fig:allprofiles}
\vspace{1cm}
\end{figure*}

\begin{figure*}
\centering
\includegraphics[clip,trim=0 22mm 0 0,width=5.5cm]{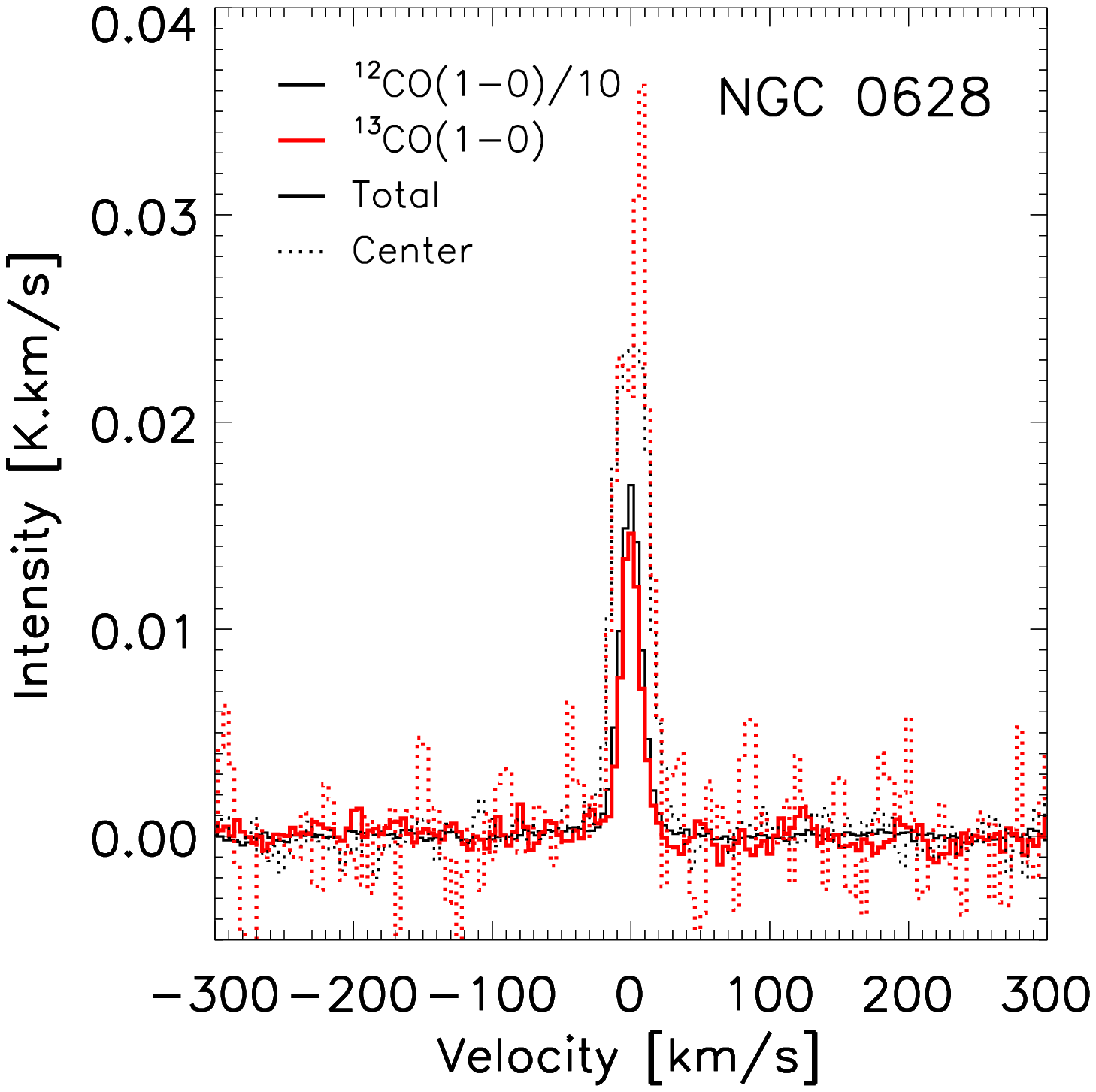}
\includegraphics[clip,trim=0 22mm 0 0,width=5.5cm]{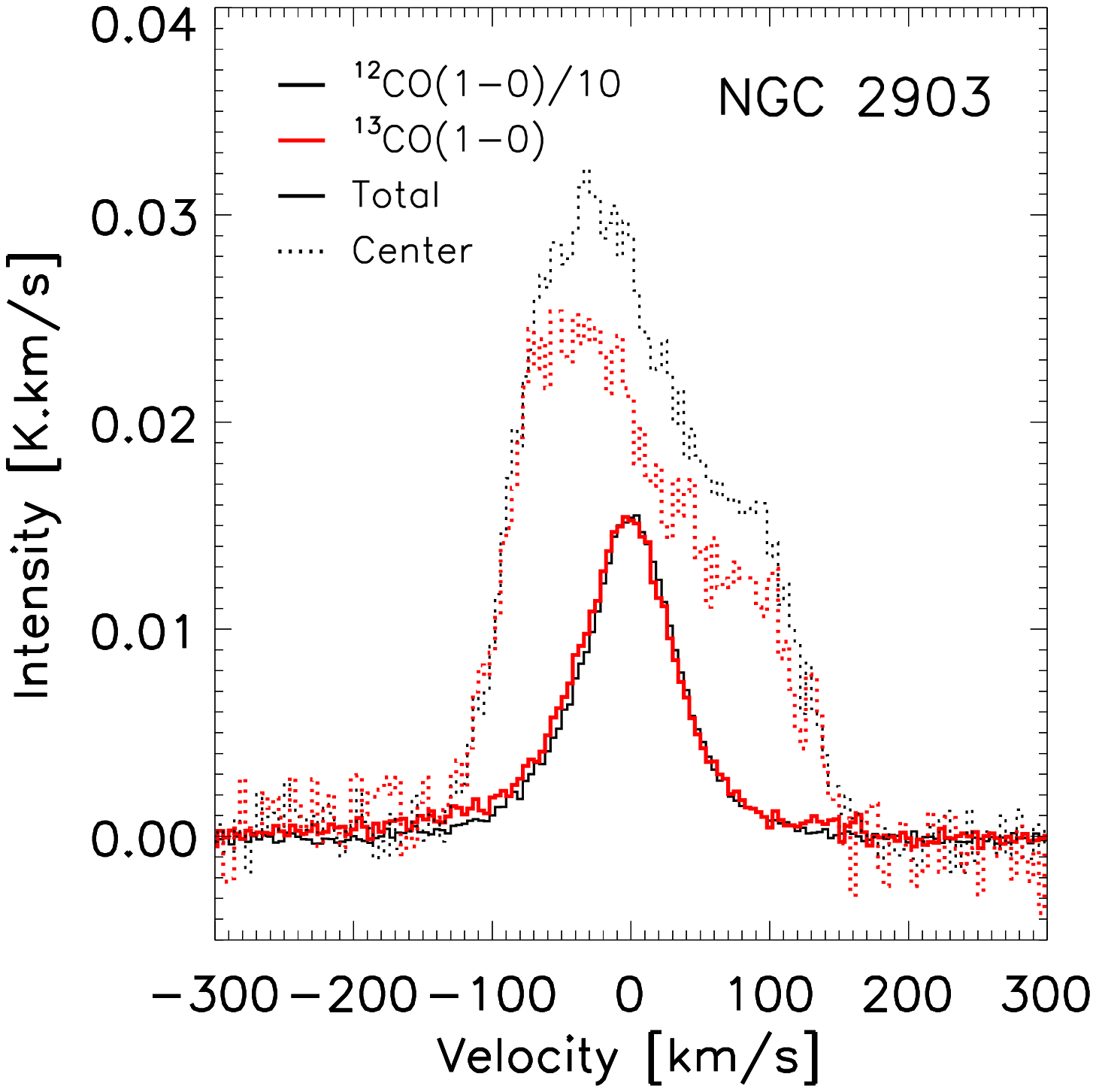}
\includegraphics[clip,trim=0 22mm 0 0,width=5.5cm]{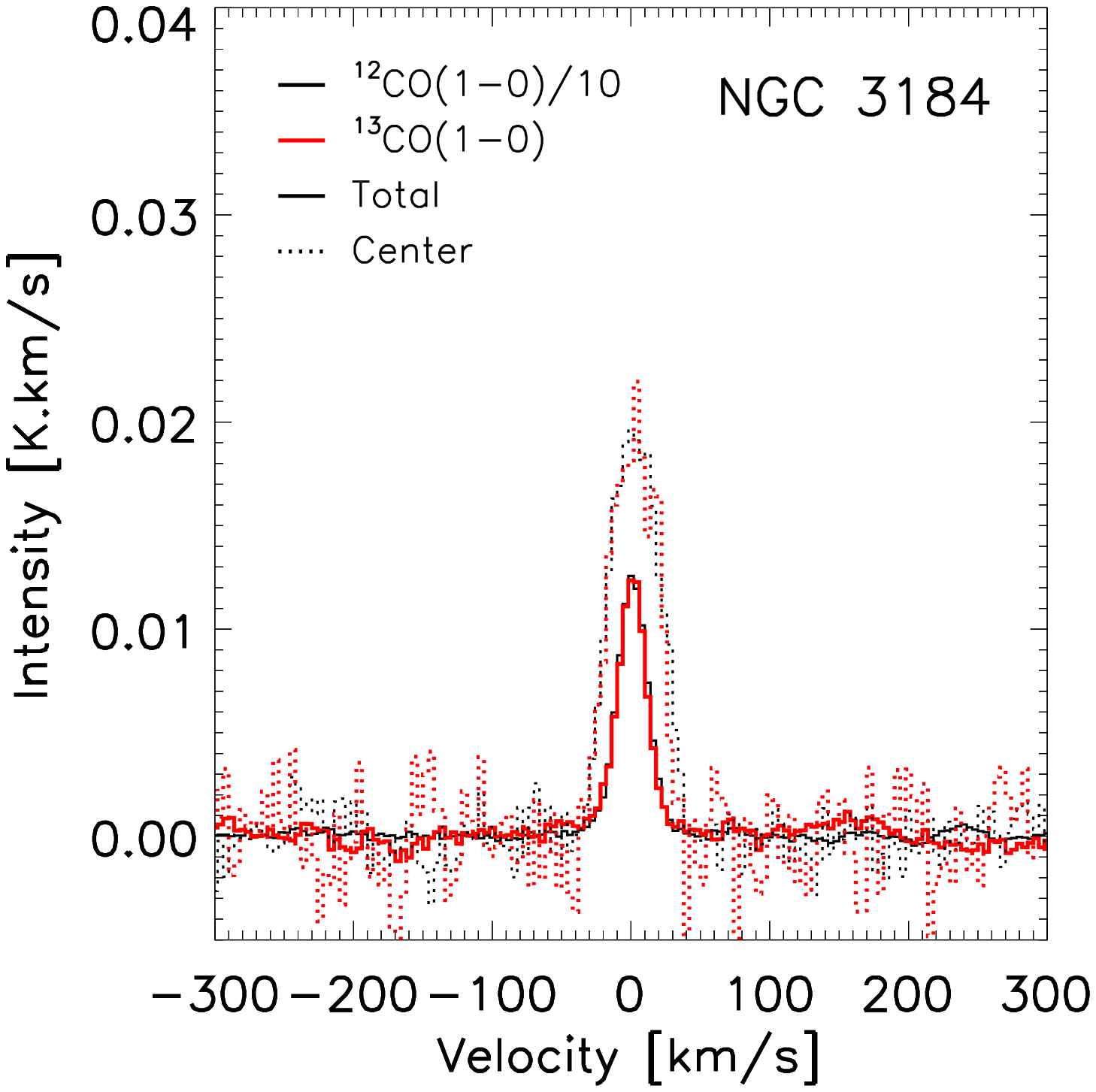}\\
\includegraphics[clip,trim=0 22mm 0 0,width=5.5cm]{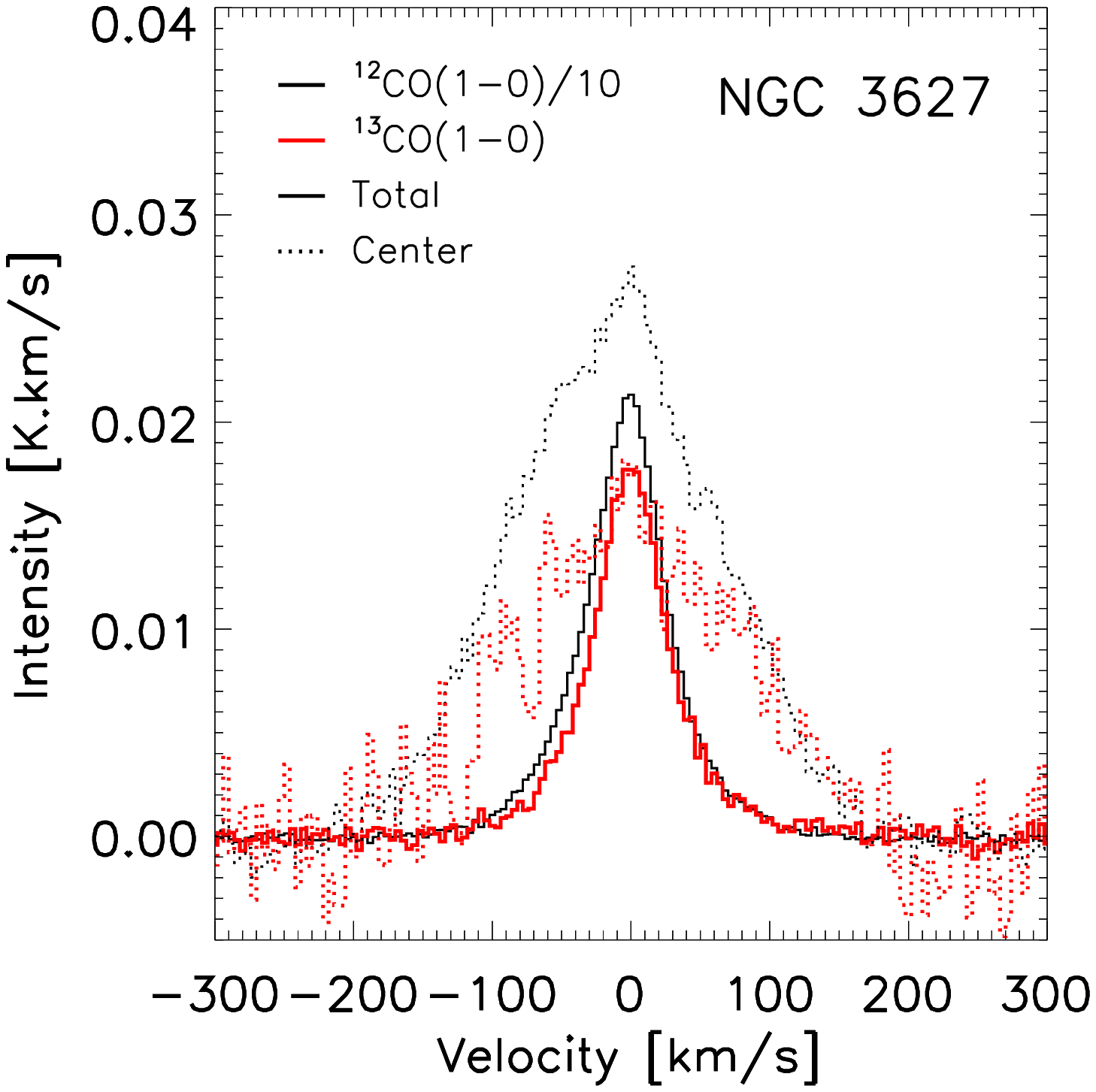}
\includegraphics[clip,trim=0 22mm 0 0,width=5.5cm]{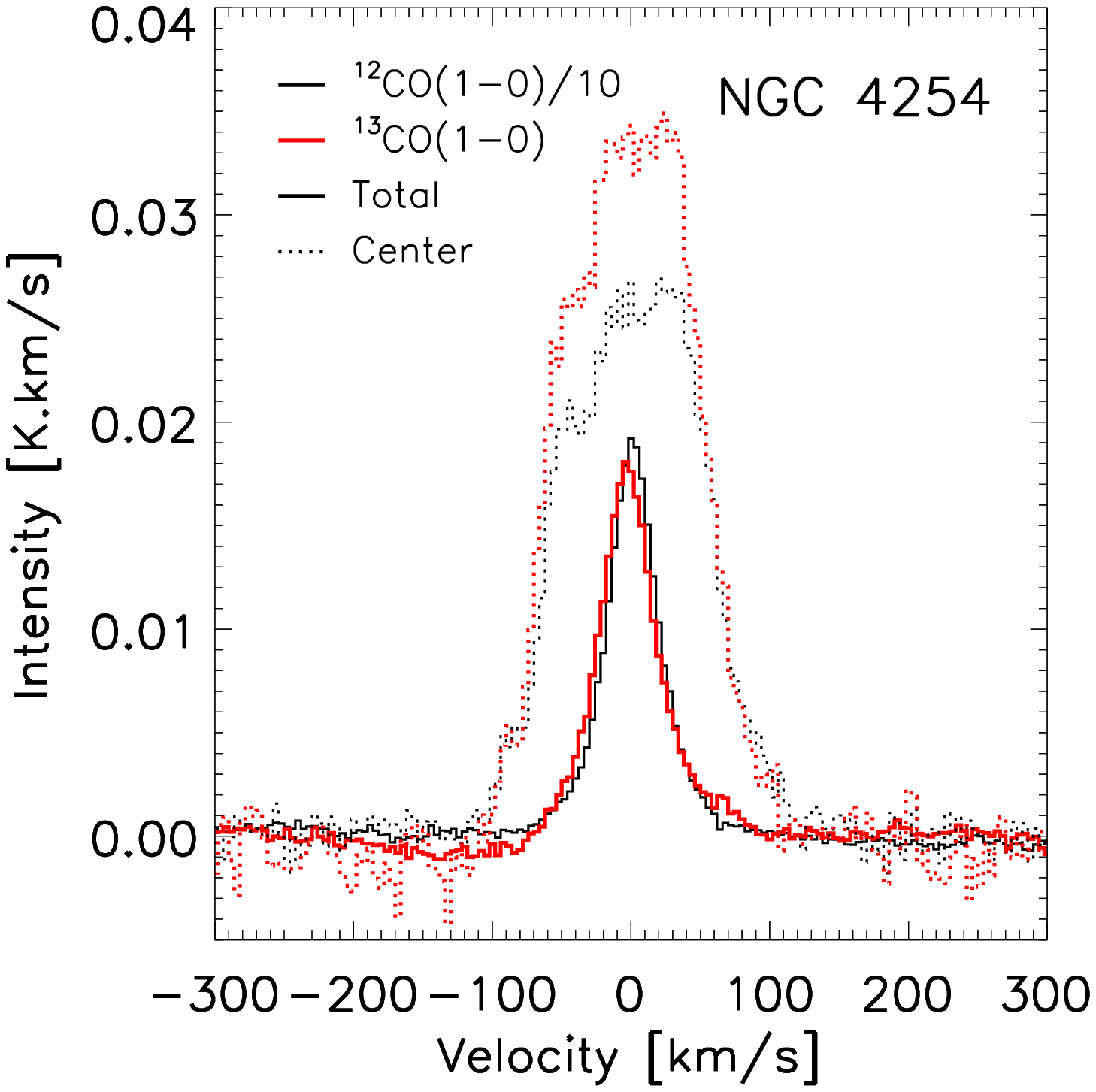}
\includegraphics[clip,trim=0 22mm 0 0,width=5.5cm]{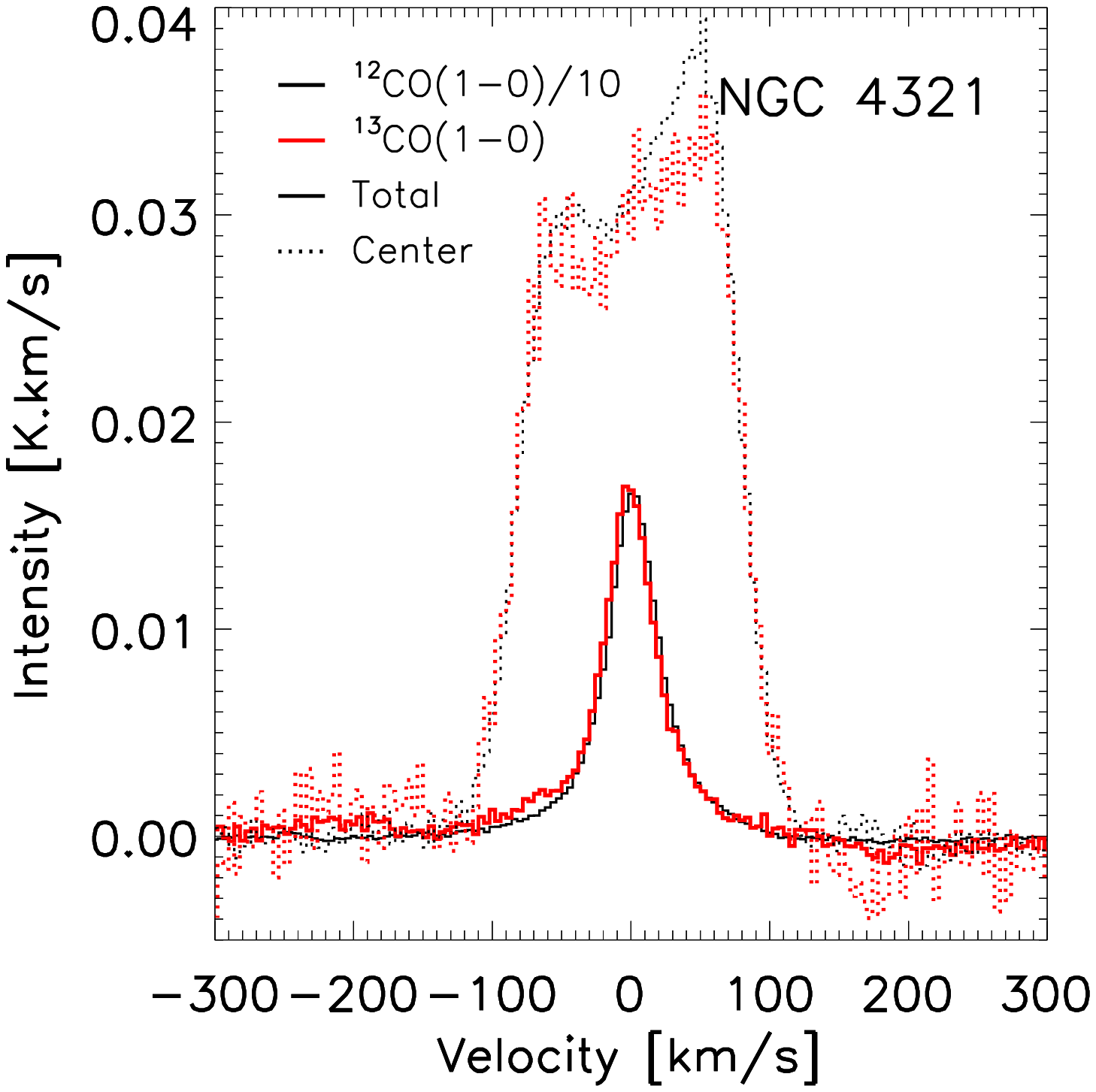}\\
\includegraphics[clip,width=5.5cm]{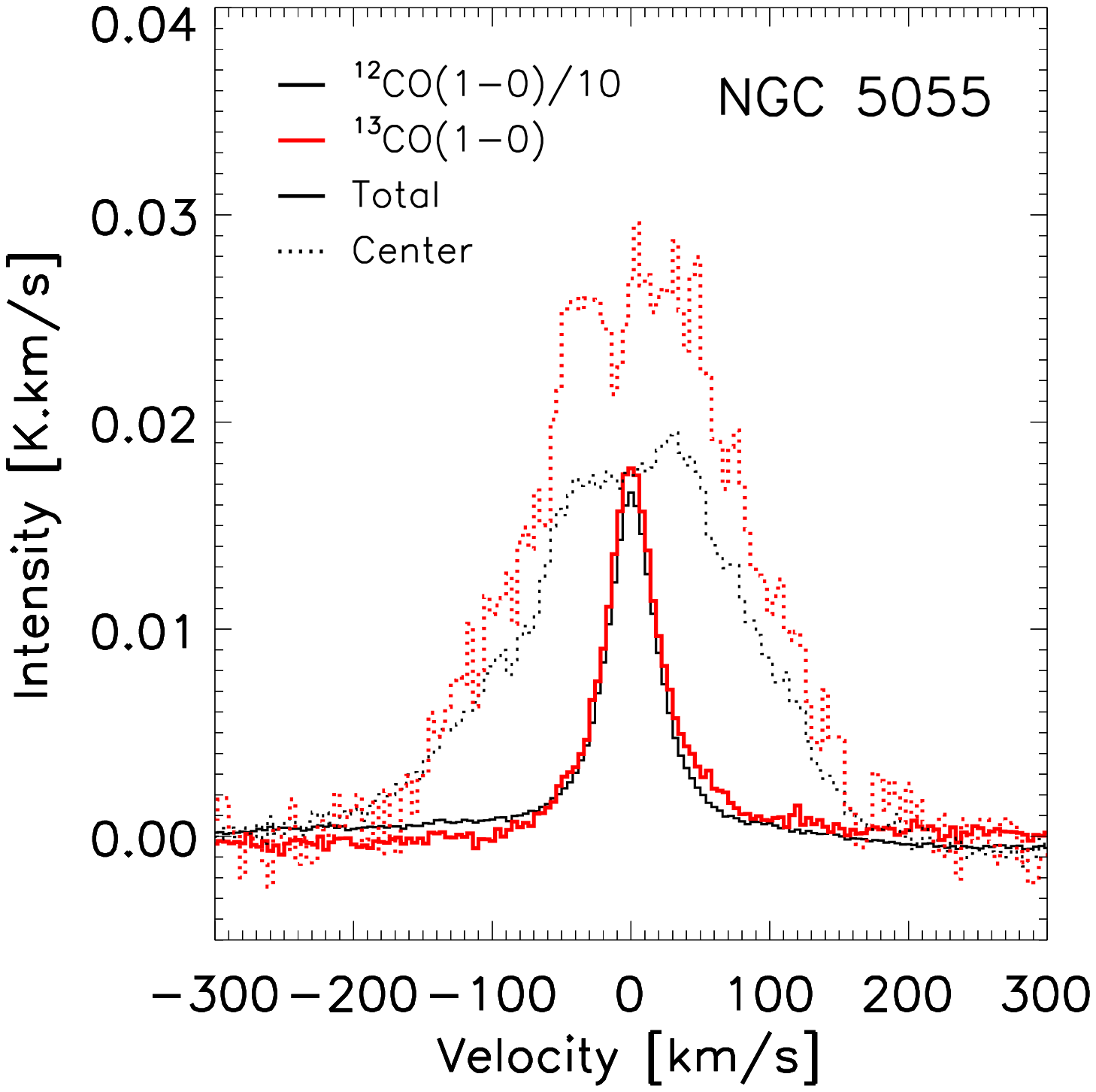}
\includegraphics[clip,width=5.5cm]{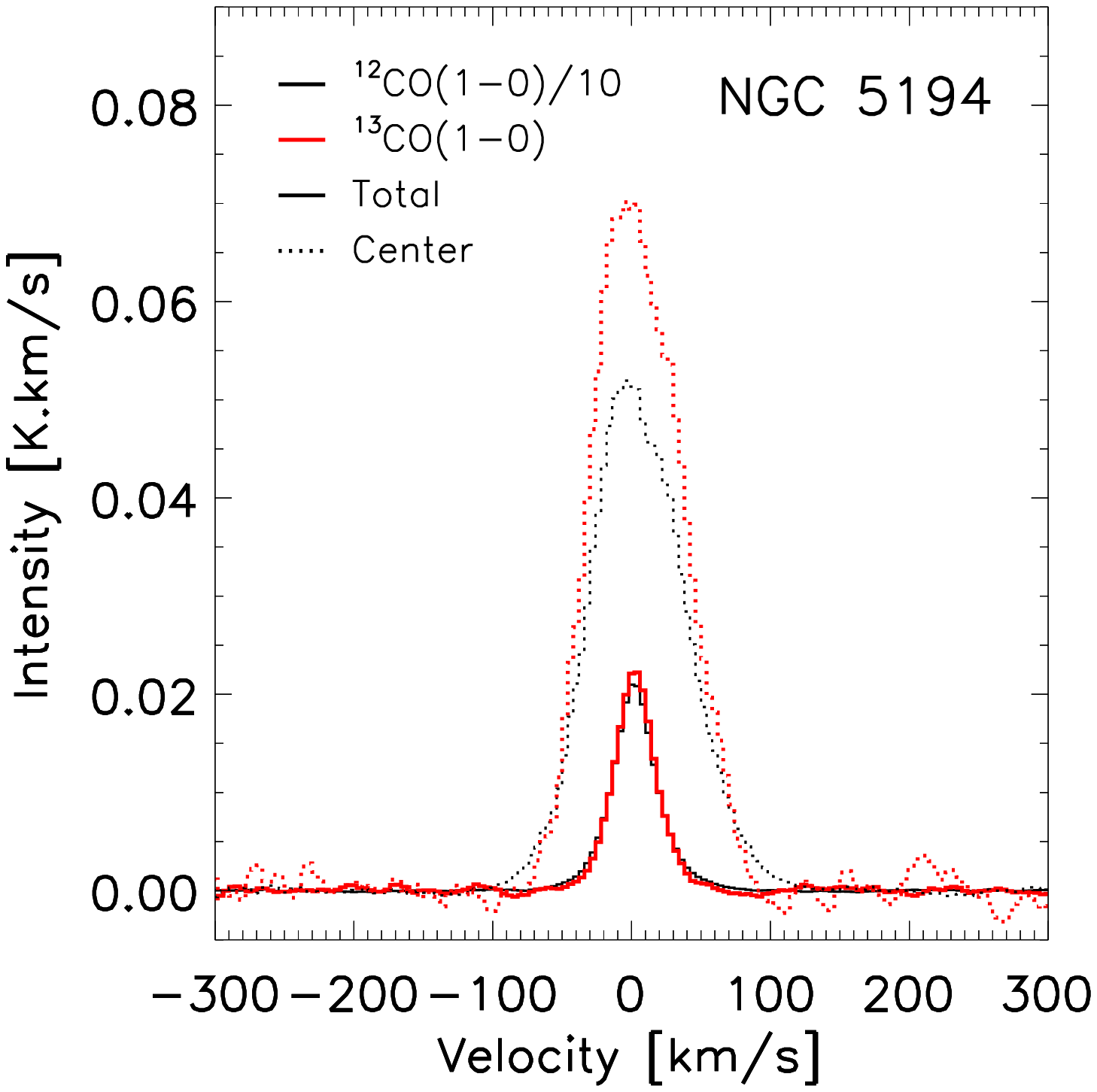}
\includegraphics[clip,width=5.5cm]{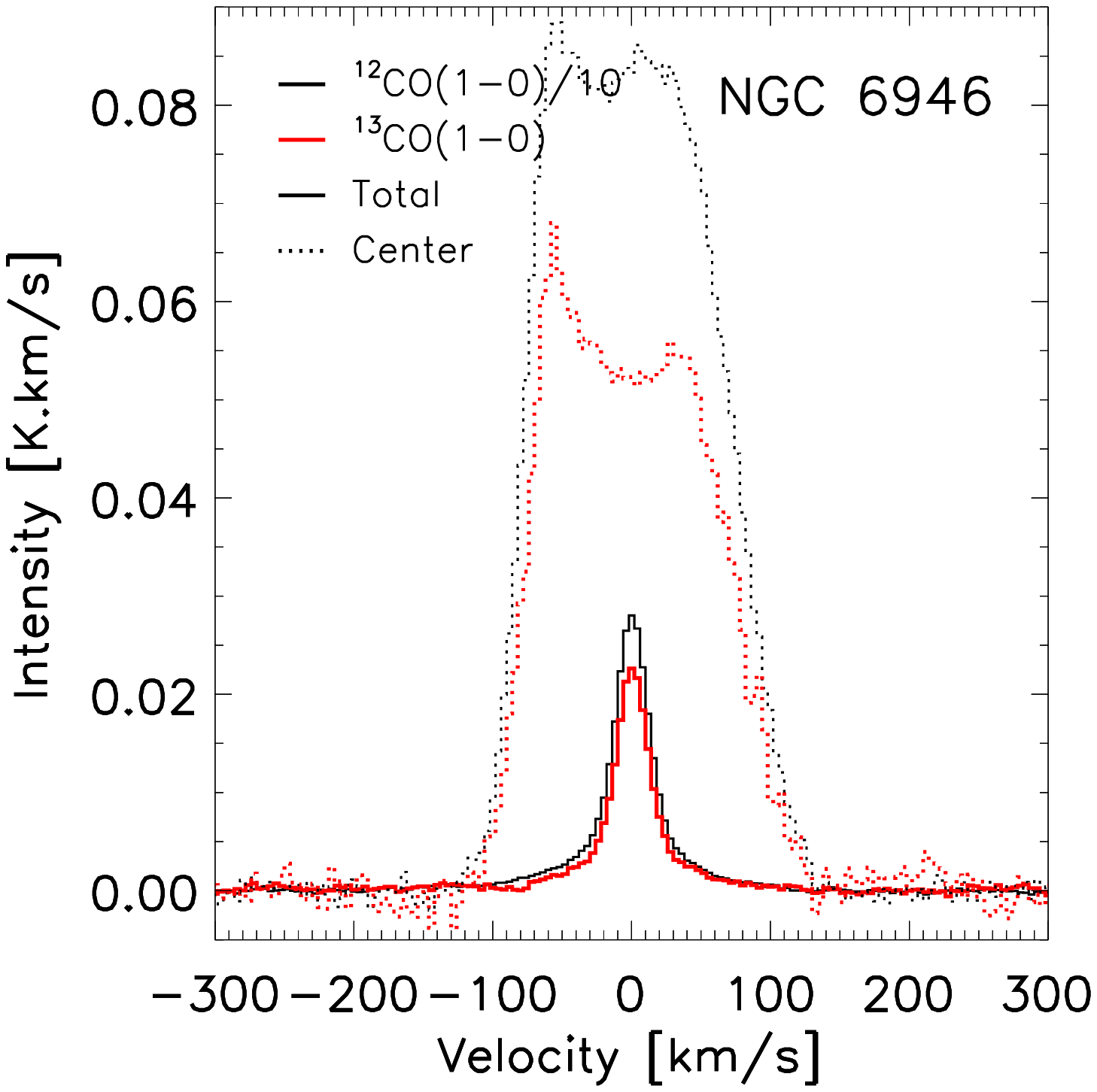}
\caption{
Stacked spectra of \coi and \com over entire galaxies
and in the galaxy centres.
}
\label{fig:specs}
\end{figure*}

\section{Monte-Carlo test of ratio variations}
\label{sect:mctest}
We perform two Monte-Carlo tests regarding the ratios
CO(2-1)/CO(1-0), $\Sigma$(SFR)/CO(1-0), and 1/$\Re$.

\underline{CO(2-1)/CO(1-0) and 1/$\Re$:}
For the first test, we aim to understand whether there
are real physical variations in the line ratios or whether
the scatter in line ratio values can be explained by noise only.
The distribution of lines ratios using noisy data
is non-Gaussian even if the noise on the line measurements
is Gaussian. Here we are not interested in the exact shape
of the distribution of the line ratios but rather whether
the observed range could be driven by the uncertainties
on the line measurements. To this end we compare the
line ratio distribution with Monte-Carlo simulations as follows.
We consider the observed \com intensities as true values and
generate new \coi and \cou intensities considering fixed line
ratios with \com and adding noise based on the observed
uncertainties of each measurement. The value of the fixed line
ratio is set to the median of the observed ratios, though this value
does not really matter for this experiment. We iterate this process
1\,000 times, and, for each iteration, we calculate the standard
deviation of the simulated line ratios. For both $\Re$ and
CO(2-1)/CO(1-0), we find that the peak of the distribution of
those standard deviations is systematically lower by a factor
of $\sim$2 compared to the standard deviation of the observed
line ratios. The peak is closer to the observed value only 
for $\Re$ in NGC\,3184. Hence we conclude that the scatter
in the observed $\Re$ and CO(2-1)/CO(1-0) ratios is
largely physical.

\noindent
For the second test, we aim to verify the robustness of
the correlation coefficients measured in Fig.~\ref{fig:allcorr}a.
For this, we generate new \coi, \com, and \cou intensities
by taking the observed intensities and adding noise
based on the observed uncertainties of each measurement.
We also iterate the process 1\,000 times.
We find that the peak of the distribution of standard deviations
is slightly larger than the standard deviation of the observed ratios
because of the noise added by the simulations.
For each iteration, we measure the correlation coefficient
and slope between $\Re$ and CO(2-1)/CO(1-0).
We find that the distribution of the correlation coefficients
is well peaked around the coefficient measured on the observations.
This indicates that the correlations persist after adding noise.
The distribution of the slopes generally peaks around $0.5$
and it is less well peaked, probably because
the dynamic range in the ratios is not very large.

\underline{$\Sigma$(SFR)/CO(1-0) and 1/$\Re$:}
The same two tests as described above are performed for
$\Sigma$(SFR)/CO(1-0) instead of CO(2-1)/CO(1-0).
Here, we find for all galaxies that the peak of the distribution
of the standard deviations in the simulated $\Sigma$(SFR)/CO(1-0)
values is systematically lower by a factor of $\sim$2 compared
to the observed standard deviation. Hence the scatter
in the observed $\Sigma$(SFR)/CO(1-0) values is physical.

\noindent
The second test indicates that the correlation coefficients
and slope values between $\Re$ and $\Sigma$(SFR)/CO(1-0)
reported in Fig.~\ref{fig:allcorr}d are also robust.
The distribution of the slopes is more peaked than with
CO(2-1)/CO(1-0).


\bsp	
\label{lastpage}
\end{document}